\documentclass[twocolumn]{jpsj3-arXiv}
\usepackage{txfonts}
\usepackage{graphicx}
\usepackage{dcolumn}
\usepackage{bm}
\usepackage{color}
\usepackage{amsmath}
\newcommand{\nn}{\nonumber}
\newcommand{\beq}{\begin{eqnarray}}
\newcommand{\eeq}{\end{eqnarray}}

\begin{document}

\title{Symmetry-Protected Topological Superfluids and Superconductors \\
--- From the Basics to $^3$He ---}

\author{Takeshi Mizushima$^{1}$\thanks{E-mail: mizushima@mp.es.osaka-u.ac.jp}, 
Yasumasa Tsutsumi$^{2,3}$,
Takuto Kawakami$^{4}$, 
Masatoshi Sato$^5$,
Masanori Ichioka$^6$,
and Kazushige Machida$^{7}$}
\inst{
$^1$Department of Materials Engineering Science, Osaka University, Toyonaka, Osaka 560-8531, Japan \\
$^2$Condensed Matter Theory Laboratory, RIKEN, Wako, Saitama 351-0198, Japan \\
$^3$Department of Basic Science, The University of Tokyo, Meguro, Tokyo 153-8902, Japan \\
$^4$International Center for Materials Nanoarchitectonics (WPI-MANA),
National Institute for Materials Science, Tsukuba 305-0044, Japan \\
$^5$Yukawa Institute for Theoretical Physics, Kyoto University, Kyoto 606-8502, Japan \\
$^6$Department of Physics, Okayama University, Okayama 700-8530, Japan \\
$^7$Department of Physics, Ritsumeikan University, Kusatsu 525-8577, Japan }

\date{\today}

\abst{
In this article, we give a comprehensive review of recent progress in research on symmetry-protected topological superfluids and topological crystalline superconductors, and their physical consequences such as helical and chiral Majorana fermions. We start this review article with the minimal model that captures the essence of such topological materials. The central part of this article is devoted to the superfluid $^3$He, which serves as a rich repository of novel topological quantum phenomena originating from the intertwining of symmetries and topologies. In particular, it is emphasized that the quantum fluid confined to nanofabricated geometries possesses multiple superfluid phases composed of the symmetry-protected topological superfluid B-phase, the A-phase as a Weyl superfluid, the nodal planar and polar phases, and the crystalline ordered stripe phase. All these phases generate noteworthy topological phenomena, including topological phase transitions concomitant with spontaneous symmetry breaking, Majorana fermions, Weyl superfluidity, emergent supersymmetry, spontaneous edge mass and spin currents, topological Fermi arcs, and exotic quasiparticles bound to topological defects. In relation to the mass current carried by gapless edge states, we also briefly review a longstanding issue on the intrinsic angular momentum paradox in $^3$He-A. Moreover, we share the current status of our knowledge on the topological aspects of unconventional superconductors, such as the heavy-fermion superconductor UPt$_3$ and superconducting doped topological insulators, in connection with the superfluid $^3$He. 
}





\maketitle

\section{Introduction}
\label{sec:intro}

A new paradigm based on topology has been rapidly and widely expanded to various fields of condensed matter physics. This has offered a new state of matter that cannot be explained by spontaneous symmetry breaking but can be characterized by  topological number/order. The underlying concept was originally initiated by Thouless, Kohmoto, Nightingale, and den Nijs~\cite{tknn,kohmoto} in quantum Hall systems to unveil the interplay between quantized Hall conductivity and the nontrivial topology of underlying fermions. The Hall conductivity does not depend on the details of fermionic spectra but is determined by their global structure, namely, the topological structure. The topological number that characterizes the system is the first Chern number or the Thouless-Kohmoto-Nightingale-den Nijs number, which measures the ``magnetic flux'' penetrating the magnetic Brillouin zone. Hatsugai~\cite{hatsugai1,hatsugai2} uncovered another physical meaning of the topological number: it counts the number of gapless edge states. This is known as the bulk-edge correspondence.

Although the original concept of the topological number/order in quantum Hall systems is independent of symmetry, it is now widely accepted that the marriage of topology with symmetry gives rise to diverse topological quantum phenomena and sheds light on a new facet of quantum matters~\cite{tanakaJPSJ12,qiRMP11,chiu15}. A milestone in the development of the new  paradigm was the topological classifications of insulating and superconducting Hamiltonians in terms of fundamental discrete symmetries, i.e., time-reversal, particle-hole, and chiral symmetries~\cite{schnyderPRB08,schnyderAIP09,kitaev09,ryuNJP10}. The classification clarified that the superfluid $^3$He is a promising candidate topological superconductor~\cite{schnyderPRB08,roy08,qiPRL09,satoPRB09,volovikJETP09v2,volovik}. The noteworthy consequence of topological odd-parity superconductivity is that owing to the particle-hole symmetry, the topologically protected gapless quasiparticles behave as Majorana fermions. The topological classification was further extended to include additional discrete symmetries, such as crystalline symmetry and magnetic $\pi$-rotation symmetry~\cite{fuPRL11,chiuPRB13,morimotoPRB13,shiozakiPRB14,benalcazarPRB14,hsiehPRB14,jadaunPRB13,chiu15}. Furthermore, it has been discussed that additional discrete symmetries enrich the characteristic features of Majorana fermions, including Majorana Ising spins and symmetry-protected non-Abelian anyons~\cite{mizushimaPRL12,uenoPRL13,zhangPRL13,tsutsumiJPSJ13,sato14,shiozakiPRB14,fangPRL14,liuprx14}.

\begin{figure*}[t!]
\begin{center}
\includegraphics[width=170mm]{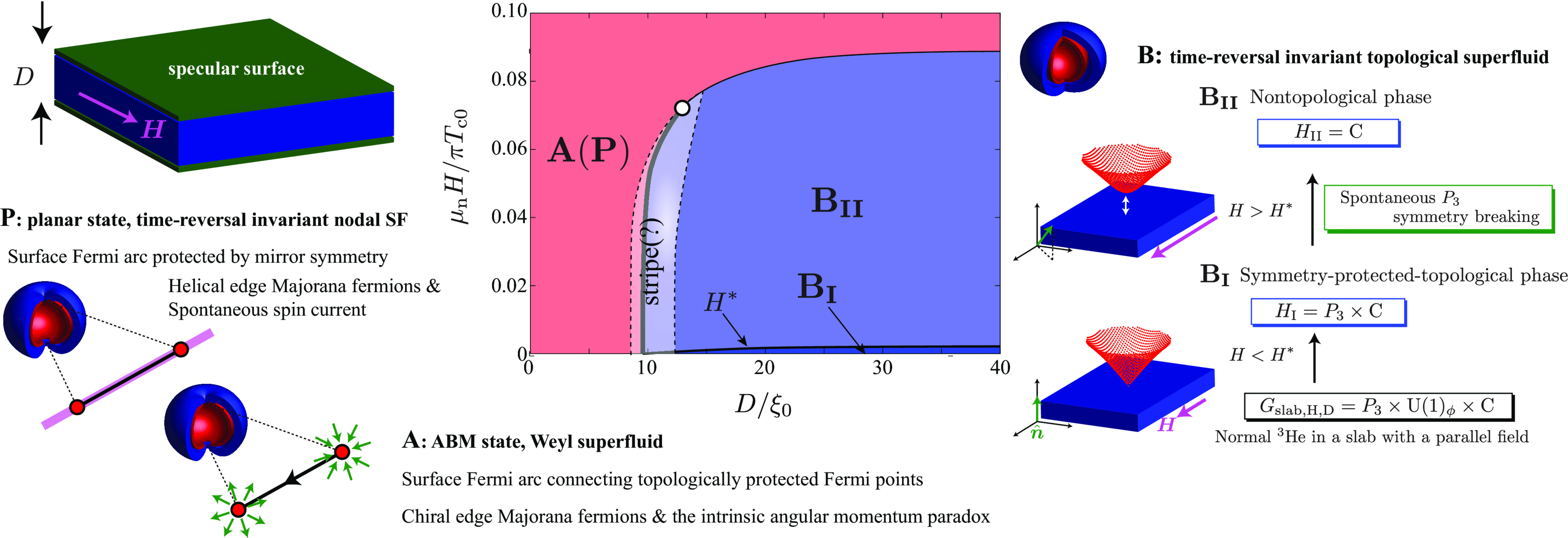} 
\end{center}
\caption{(Color online) Phase diagram of the superfluid $^3$He confined in a slab geometry and remarkable consequences in each phase, where a magnetic field is parallel to the surface (center). The temperature is set to be $T=0.4T_{\rm c0}$, and $D$ and $H$ are the thickness of the slab and the magnitude of the applied field, respectively. The thin (thick) curves in the phase diagram denote the first (second)-order transition line. }
\label{fig:phaseDH}
\end{figure*}

Among the possible candidate topological superconductors, the superfluid phases of $^3$He is the most concrete and ideal platform for studying diverse topological quantum phenomena in the following viewpoints: (i) The system composed of neutral fermions with nuclear spin $1/2$ remains as a quantum liquid down to zero temperature and the normal state independently maintains huge continuous rotation symmetries in spin and coordinate spaces. (ii) The bulk superfluidity of $^3$He has been well established as spin-triplet odd-parity pairing~\cite{leggettRMP,vollhardt}. The A-phase that appears in the high-temperature and high-pressure region is identified as the chiral $p$-wave pairing with spontaneously broken time-reversal symmetry~\cite{abm1,abm2}, and the B-phase is known as a fully gapped pairing with time-reversal symmetry~\cite{bw} (see Figs.~\ref{fig:phaseDH} and ~\ref{fig:phase_bulk}). The superfluid $^3$He, having huge order parameter manifolds, has fascinated many physicists not only as a prototype of unconventional superconductors but also as a treasure box of topology of order parameter manifolds, such as textures, Nambu-Goldstone and Higgs modes, and topological excitations~\cite{volovik,vollhardt,salomaaRMP87,sauls99,erikki}. (iii) Recent developments in nanofabrication techniques enable one to confine the quantum liquid to a variety of geometries, such as a single slab and narrow cylinders with a thickness/radius comparable to the superfluid coherence length~\cite{kawae,miyawakiPRB00,kawasakiPRL04,levitinJLTP10,bennettJLTP10,levitin13,levitinPRL13}. In these geometries, the planar, polar, and crystalline ordered phases become energetically competitive with the A- and B-phases~\cite{fetter,li,haraJLTP88,vorontsovPRL07,aoyamaPRB14}. New phase diagrams have also been proposed for aerogels~\cite{dmitriev,pollanen}. (iv) The surface density of states peculiar to gapless quasiparticle states has already been observed in specific heat measurements and high-precision spectroscopy based on transverse acoustics under well-controlled surface conditions~\cite{choiPRL06,bunkov,bunkov1,davisPRL08,murakawaPRL09,murakawaJPSJ11,okuda}.

Motivated by puzzling issues on the intrinsic angular momentum paradox, investigations on the nontrivial momentum space topology were first initiated in $^3$He by Stone {\it et al.}~\cite{stonePRL85,stoneAP87} and Volovik~\cite{volovik86,balatsky}, independently. In connection with an analogue of a two-dimensional $^3$He-A thin film to the quantum Hall effect and gauge theories, Volovik~\cite{volovik86,volovik87,volovik87v2,volovik88} further uncovered the remarkable fact that the pairwise point nodes on the Fermi surface are protected by the first Chern number as a ``magnetic'' monopole, and low-energy quasiparticles near the Fermi points behave as chiral Weyl fermions. The superfluid $^3$He-A thin film is now widely recognized as a prototype of Weyl superconductors~\cite{volovik,balentsPRB12,sauPRB12,dasPRB13,goswami13,xuPRL14,yangPRL14}, which are accompanied by a zero-energy flatband terminated to pairwise Weyl points~\cite{heikkila,tsutsumi:2010b,tsutsumi:2011b,silaev:2012,silaevJETP14}. 

As mentioned above, recent developments in topological classifications clarified the distinct topological structures between the A- and B-phases: The $^3$He-A thin film is a Weyl superconductor characterized by the first Chern number, while the bulk B-phase possesses topological superfluidity protected by the time-reversal symmetry~\cite{schnyderPRB08,roy08,qiPRL09,satoPRB09,volovikJETP09v2}. Furthermore, it has been proposed that the marriage of the superfluid $^3$He with nanofabrication techniques gives rise to diverse topological phenomena intertwined with symmetry~\cite{mizushimaJPCM15,silaevJETP14,vorontsovPRL07}. As shown in Fig.~\ref{fig:phaseDH}, for instance, confined $^3$He under a magnetic field has a nontrivial phase diagram composed of a variety of topological and nontopological phases: The symmetry-protected topological phase B$_{\rm I}$, the symmetry-broken nontopological phase B$_{\rm II}$, the Weyl superfluid A-phase, the planar phase, and the crystalline ordered ``stripe'' phase. The critical field $H^{\ast}$ in Fig.~\ref{fig:phaseDH} is identified as the topological phase transition concomitant with spontaneous symmetry breaking~\cite{mizushimaPRL12} and is accompanied by noteworthy topological quantum critical phenomena, such as emergent supersymmetry~\cite{grover}. In contrast to the A-phase, the pairwise point nodes in the planar phase are protected by a mirror reflection symmetry, and the zero-energy flatband emergent in the surface exhibits anisotropic magnetic responses~\cite{tsutsumiJPSJ13,mizushimaJPCM15,sasakiPC15}. It is also interesting to note that apart from the topological aspect of $^3$He, there has been a long history of investigations on gapless quasiparticles in the direction of Andreev bound states~\cite{nagaiJPSJ08,kashiwayaRPP}. Nowadays, Majorana fermions are identified as a special type of surface Andreev bound state in the context of topological superconductors~\cite{tanakaJPSJ12}.

In this article, we give a comprehensive review of recent progress in symmetry-protected topological superfluids and topological crystalline superconductors with a special focus on $^3$He. In Sec.~\ref{sec:topology}, we start with the minimal model that captures the essence of the topological aspect of superfluids and superconductors. Using this minimal model, we emphasize that additional discrete symmetries enrich the topological properties of quantum matters. In Sec.~\ref{sec:topology}, we also clarify the remarkable consequences, which are symmetry-protected non-Abelian anyons and Ising spins. In Sec.~\ref{sec:spinless}, we make an overview of the topological aspects and important consequences of a spin-polarized chiral $p$-wave pairing system that gives a prototype of a topological phase transition. 

The central part of this article is devoted to emphasizing that the superfluid $^3$He serves as a rich repository of novel topological quantum phenomena originating from the intertwining of symmetries and topologies. In Sec.~\ref{sec:bulk}, we provide a review of the bulk symmetry and topology of the B-, A-, and planar phases, where we introduce topological invariants relevant to their phases, the three-dimensional winding number, the first Chern number in a sliced momentum plane, and the bulk $\mathbb{Z}_2$ topological number. We also discuss the peculiarities of surface Andreev bound states in $^3$He-B as helical Majorana fermions by explicitly solving the Bogoliubov-de Gennes (BdG) equation. 

\begin{figure}[t!]
\begin{center}
\includegraphics[width=70mm]{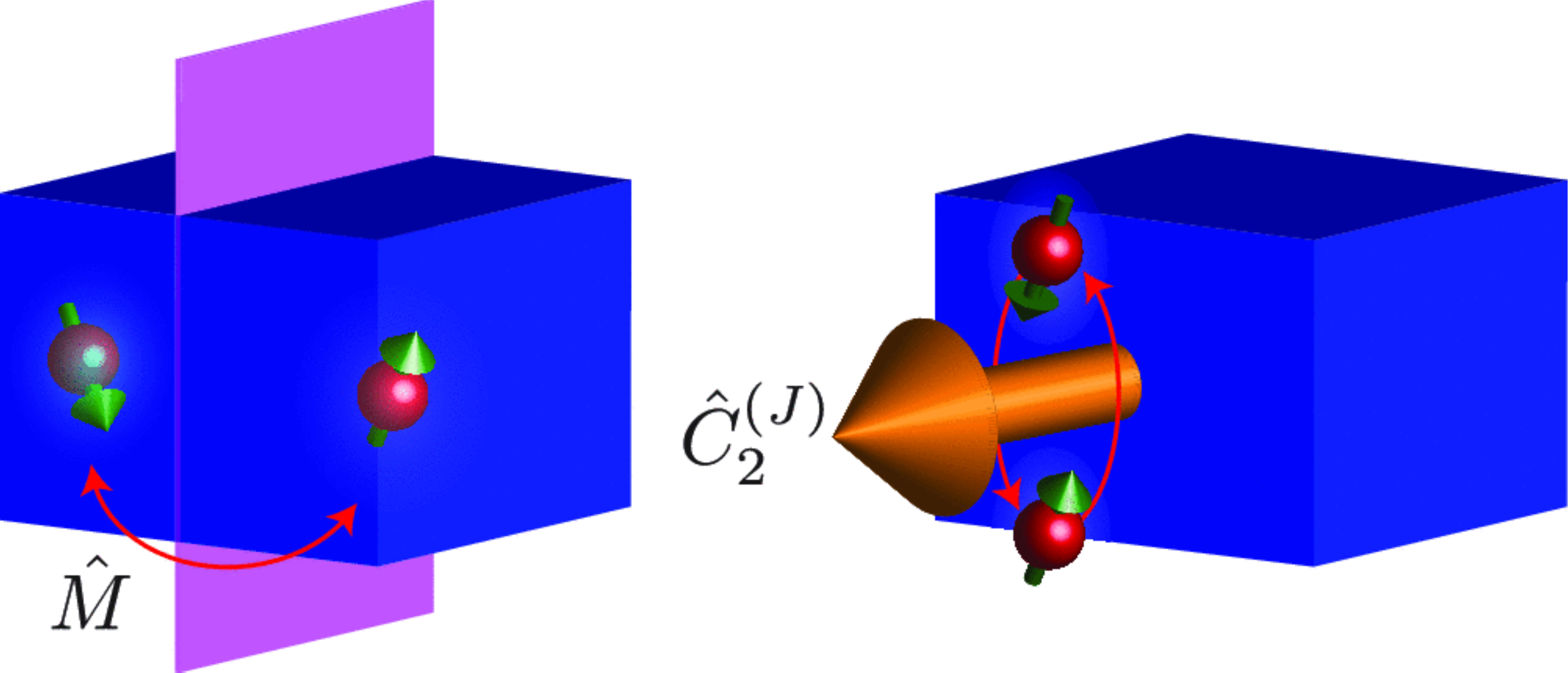} 
\end{center}
\caption{(Color online) Schematic images of the action of discrete rotations: Mirror reflection $\hat{M}$ for $\hat{P}_2\!=\!\hat{T}\hat{M} $ and the magnetic $\pi$-rotation $\hat{C}^{(J)}_2$ for $\hat{P}_3 \!=\! \hat{T}\hat{C}^{(J)}_2$. }
\label{fig:discrete}
\end{figure}

Section~\ref{sec:spt} is devoted to the topology, symmetry, and consequences of the superfluid $^3$He confined in a restricted geometry. We start to summarize numerous efforts toward understanding the superfluid and quasiparticle structures in confined $^3$He on the basis of the Ginzburg-Landau (GL) and quasiclassical theories. Subsequently, we overview the symmetry-protected topological B$_{\rm I}$, nontopological B$_{\rm II}$, A-, and planar phases in a restricted geometry. We emphasize here the essential roles of additional discrete symmetries, $P_2$ and $P_3$, 
\beq
\hat{P}_2 = \hat{T}\hat{M}, \quad 
\hat{P}_3 = \hat{T}\hat{C}^{(J)}_2,
\label{eq:p2p3}
\eeq
whose actions are depicted in Fig.~\ref{fig:discrete}. Here, $\hat{T}$, $\hat{M}$, and $\hat{C}^{(J)}_2$ denote the operators of the time inversion, mirror reflection, and the joint $\pi$ rotation of spin and orbital spaces, respectively. The essential roles of the $P_3$ and $P_2$ symmetries in topological superfluidity and Majorana fermions were first unveiled in Ref.~\citeonline{mizushimaPRL12} for the superfluid $^3$He confined to a restricted geometry and in Ref.~\citeonline{mizushimaNJP13} for quasi-one-dimensional spin-orbit-coupled Fermi gases, respectively. The topological phase transition between B$_{\rm I}$ and B$_{\rm II}$ is accompanied by the anomalous enhancement of the surface spin susceptibility, which is a hallmark of the mass acquisition of surface helical Majorana fermions. In Sec.~\ref{sec:odd}, we discuss the underlying physics of such anomalous magnetic response of the surface states in the light of emergent odd-frequency pairing amplitudes. In Sec.~\ref{sec:exp}, we also provide a brief summary on the recent progress in the observations of surface-bound state in the superfluid $^3$He-B. 

Helical and chiral Majorana fermions bound to the surface of $^3$He-B and A carry macroscopic spin and mass current. In relation to a longstanding issue on the intrinsic angular momentum paradox in $^3$He-A, we briefly review such a macroscopic current in Sec.~\ref{sec:mass}. In Sec.~\ref{sec:vortices}, we summarize the possible types of topological excitations in $^3$He-A and B associated with the order parameter manifold, and overview recent progress in the understanding of symmetry-protected Majorana fermions bound to such topological defects, singular and continuous vortices. Section \ref{sec:materials} is devoted to sharing the current status of our knowledge on possible candidates of topological crystalline superconductors, such as the heavy-fermion superconductor UPt$_3$ and superconducting doped topological insulators, in connection with $^3$He. In Sec.~\ref{sec:summary}, we provide a summary and give some prospects on searching for topological superfluidity and exotic quasiparticles in $^3$He.

Throughout this paper, we set $\hbar \!=\! k_{\rm B} \!=\! 1$ and the repeated Greek (Roman) indices imply the sum over $x, y, z$ (spins $\uparrow$ and $\downarrow$). The Pauli matrices in spin and particle-hole (Nambu) spaces are denoted by  $\sigma _{\mu}$ and $\tau _{\mu}$, respectively.

\section{Roles of Symmetries in Topology and Majorana Fermions}
\label{sec:topology}

We here summarize the topological properties of BdG Hamiltonians relevant to superconductors and superfluids. A general Hamiltonian for spatially uniform superconductors and superfluids is given by 
\beq
\mathcal{H} = \frac{1}{2}\sum _{{\bm k}}
\left( {\bm c}^{\dag}_{{\bm k}}, {\bm c}^{\rm T}_{-{\bm k}}\right)
\mathcal{H}({\bm k}) 
\left( 
\begin{array}{c}
{\bm c}_{{\bm k}} \\ {\bm c}^{\dag}_{-{\bm k}}
\end{array}
\right),
\label{eq:H}
\eeq
where we ignore the constant energy term. In Eq.~(\ref{eq:H}), ${\bm c}_{{\bm k}}$ and ${\bm c}^{\dag}_{-{\bm k}}$ denote a set of $N$ annihilation and creation operators of fermions with momentum ${\bm k}$, respectively. The degrees of freedom for fermions, $N$, include the spin, orbital, sublattice indices, and so on. The BdG Hamiltonian density is given by
\beq
\mathcal{H}({\bm k}) = \left(
\begin{array}{cc}
\epsilon ({\bm k}) & \Delta ({\bm k}) \\
-\Delta^{\ast}(-{\bm k}) & -\epsilon^{\rm T}(-{\bm k})
\end{array}
\right),
\label{eq:Hbdg}
\eeq
where the Hamiltonian matrix $\varepsilon({\bm k})$ describes the dispersion of the normal state, which may contain the diagonal elements of mean-field approximated self-energies and potentials, and the $N \times N$ gap function $\Delta({\bm k})$ satisfies $\Delta({\bm k})=-\Delta(-{\bm k})^{\rm T}$. 


It is convenient to parameterize the $N\times N$ pair potential $\Delta ({\bm k})$ with the spin-singlet component $\psi ({\bm k}) =\psi (-{\bm k})$ and the spin-triplet component $d_{\mu}({\bm k}) = - d_{\mu}(-{\bm k})$ as 
\beq
\Delta ({\bm k}) = i\sigma _y \psi ({\bm k}) + i\sigma  _{\mu}\sigma _y d_{\mu}({\bm k}),
\label{eq:generalD}
\eeq
where $\sigma _{\mu}$ denotes the Pauli matrices in the spin space.
Under spin rotation, $\psi$ is  a scalar, while ${\bm d}({\bm k})$ behaves as a vector in three-dimensional space, 
\beq
U({\bm n},\varphi) i\sigma  _{\mu}\sigma  _y d_{\mu}({\bm k}) U^{\rm T}({\bm n},\varphi) 
= i\sigma _{\mu}\sigma _yR_{\mu \nu}(\hat{\bm n},\varphi)d_{\nu}({\bm k}).
\label{eq:drot}
\eeq
In Eq.~(\ref{eq:drot}), the ${\rm SU}(2)$ matrix, $U(\hat{\bm n},\varphi)=\cos(\varphi/2) - i \sigma _{\mu}\hat{n}_{\mu}\sin(\varphi/2)$, describes the rotation of a spin matrix about the $\hat{\bm n}$-axis by the angle $\varphi$, and $R_{\mu\nu}$ denotes the corresponding ${\rm SO}(3)$ matrix, $R_{\mu\nu}(\hat{\bm n},\varphi) = \cos\varphi\delta _{\mu,\nu} + (1-\cos\varphi)\hat{n}_{\mu}\hat{n}_{\nu} - \epsilon^{\mu\nu\eta}\hat{n}_{\eta}\sin\varphi$~\cite{jj} ($\epsilon^{\mu\nu\eta}$ is the Levi-Civita symbol). These two matrices are related to each other through the identity $U\sigma _{\nu}U^{\dag} = \sigma _{\mu} R_{\mu\nu}$.

The quasiparticle structure in spatially uniform superconductors and superfluids is obtained by diagonalizing the BdG Hamiltonian as
\beq
\mathcal{H}({\bm k}) | u_n ({\bm k}) \rangle = E_n({\bm k}) | u_n ({\bm k}) \rangle,
\label{eq:BdGk}
\eeq
where the eigenfunction $|u_n({\bm k})\rangle$ satisfies the orthonormal condition $\langle u_n ({\bm k}) | u_m ({\bm k})\rangle = \delta _{n,m}$. The suffix ``$n$'' denotes the band index.

The fundamental discrete symmetries that the Hamiltonian may preserve are the particle-hole symmetry (PHS), time-reversal symmetry (TRS), and chiral symmetry,
\beq
\mathcal{C}\mathcal{H}({\bm k})\mathcal{C}^{-1} = -\mathcal{H}(-{\bm k}), \\
\label{eq:PHS}
\mathcal{T}\mathcal{H}({\bm k})\mathcal{T}^{-1} = \mathcal{H}(-{\bm k}), \\
\label{eq:TRS}
{\Gamma}\mathcal{H}({\bm k}){\Gamma}^{-1} = -\mathcal{H}({\bm k}).
\label{eq:chiral}
\eeq
The BdG Hamiltonian in Eq.~\eqref{eq:Hbdg} relevant to superconductors and superfluids naturally holds the PHS $\mathcal{C}=\tau _x K$ with $\mathcal{C}^2\!=\! 1$, where $\tau _{\mu}$ denotes the Pauli matrices in the particle-hole (Nambu) space and $K$ is the complex conjugation operator. The TRS for spin-$1/2$ fermions is given by $\mathcal{T}\!=\! i\sigma  _y K$ with $\mathcal{T}^2\!=\! -1$. The TRS requires the single-particle Hamiltonian density to hold the relation, $\mathcal{T}\varepsilon({\bm k})\mathcal{T}^{-1} = \varepsilon (-{\bm k})$. The general form of the pair potential in Eq.~(\ref{eq:generalD}) holds the TRS, $\mathcal{T}\Delta({\bm k})\mathcal{T}^{-1} = \Delta (-{\bm k})$, when $\psi({\bm k})$ and ${\bm d}({\bm k})$ are real. The chiral symmetry operator $\Gamma$ can be uniquely constructed by a combination of PHS and TRS as 
\beq
\Gamma = e^{i\alpha}\mathcal{T}\mathcal{C},
\eeq
where $\alpha$ is chosen so as to satisfy $\Gamma^2 = +1$. 

When the BdG Hamiltonian in Eq.~\eqref{eq:Hbdg} maintains an additional symmetry, such as spin rotational symmetry, the $2N\times 2N$ BdG Hamiltonian is reduced to $N\times N$ BdG Hamiltonians in each spin sector, and the reduced Hamiltonian may hold different types of PHS and TRS, $\mathcal{C}^2\!=\! -1$ and $\mathcal{T}^2=+1$. In general, band insulators do not have PHS, but they may support TRS and/or chiral symmetry. Hence, Bloch-BdG Hamiltonians for insulators and superconductors can be classified into ten symmetry classes in terms of the presence ($\pm 1$) and absence ($0$) of these fundamental discrete symmetries, which are called the Altland-Zirnbauer symmetry classes.

The discrete symmetries introduced above impose the symmetric relation on the eigenstates of $\mathcal{H}({\bm k})$. First, the PHS in Eq.~(\ref{eq:PHS}) ensures that there is a one-to-one correspondence between the positive and negative energy states, where the quasiparticle state with $E_n({\bm k})>0$ is associated with that with $-E_n(-{\bm k})$ through the relation
\beq
|u_{-E}({\bm k})\rangle = \mathcal{C}| u_E(-{\bm k})\rangle.
\label{eq:PHSu}
\eeq
In addition, the quasiparticle states are twofold degenerate as a consequence of the TRS with $\mathcal{T}^2=-1$ in Eq.~(\ref{eq:TRS}), which form a Kramers pair,
\beq
|u_n({\bm k})\rangle = \mathcal{T}|u_n(-{\bm k})\rangle.
\label{eq:kramers}
\eeq
Here, $\mathcal{T}$ flips the spin of the quasiparticle together with changing the sign of the momentum.

\subsection{Topology and symmetries: A pedagogical model}

We here overview an essential part of the topological classification of quantum systems in terms of discrete symmetries.  
As the minimal model to capture the relation between topology and symmetries, let us consider a $2\times 2$ BdG Hamiltonian on the two-dimensional ${\bm k}$-space, which is parameterized by Pauli matrices in the Nambu space as 
\beq
\mathcal{H}({\bm k}) = m_{\mu}({\bm k})\tau _{\mu}.
\label{eq:toy}
\eeq
The topological properties are intrinsic in the occupied states of eigenstates, $E_-({\bm k})$, and its eigenvector $|u_{-}({\bm k})\rangle$. This requires that two bands, $E_{\pm}({\bm k}) = \pm \sqrt{{\bm m}^2 ({\bm k})}$, are fully gapped in the Fermi level for any ${\bm k}$. The Hamiltonian is often dominated by one of the $m_i({\bm k})$, say, $m_z({\bm k})$, at $|{\bm k}|\rightarrow \infty$. The ${\bm k}$-space is identical to the two-sphere $S^2$ by compactifying the $|{\bm k}| \!\rightarrow\! \infty$ points to one point, since the eigenvectors $|u_{\pm}({\bm k})\rangle$ is identical at $|{\bm k}|\rightarrow \infty$.  

\subsubsection{Chern number and winding number}
\label{sec:chern}

{\it Chern number.}--- Let us consider the topological invariant in the case of $d\!=\! 2$, where the ${\bm k}$-space is compactified on $S^2$. It is well known that the quantum state acquires a phase accumulation when it adiabatically moves along a trajectory in the ${\bm k}$-space. The nontrivial phase accumulation of the occupied state along the closed loop in the ${\bm k}$-space is called the Berry phase, 
\beq
\gamma (C) = i\oint _{C} \mathcal{A}_{\mu}({\bm k})dk_{\mu} .
\label{eq:berry}
\eeq
The integrand $\mathcal{A}_{\mu}({\bm k})$ is the vector potential associated with the ${\rm U}(1)$ phase,
\beq
\mathcal{A} = \langle u_-| d u_- \rangle = \mathcal{A}_{\mu}({\bm k})dk_{\mu},
\label{eq:aaa}
\eeq
which is called the Berry connection. The Berry connection is not invariant under the gauge transformation: The gauge transformation, $|u_{-}({\bm k})\rangle\!\rightarrow\! e^{i\varphi({\bm k})}|u_{-}({\bm k})\rangle$, transforms the Berry connection as $\mathcal{A}_{\mu}({\bm k})\!\rightarrow\! \mathcal{A}_{\mu}({\bm k}) + i \partial _{k_{\mu}}\varphi ({\bm k})$, where $e^{i\varphi ({\bm k})}$ is an arbitrary smooth and single-valued function. Thus, the Berry phase has an ambiguity of $2\pi \times $ integer as 
$\gamma (C) \rightarrow \gamma (C)  + 2\pi N$, where $N=\oint _C \partial _{k_{\mu}}\varphi(k) dk_{\mu} \in Z$. Thus, only $e^{i\gamma(C)}$ is gauge-invariant.

Another gauge-invariant quantity is obtained by integrating the Berry curvature $\mathcal{F}$ over $S^2$,
\beq
{\rm Ch}_1 = \frac{i}{2\pi}\int _{S^2} \mathcal{F},
\label{eq:1stchern}
\eeq
which is called the first Chern number or the Thouless-Kohmoto-den Njis-Nightingale (TKNN) number~\cite{tknn,kohmoto}. The Berry curvature is defined as 
\beq
\mathcal{F} = d\mathcal{A} 
\label{eq:curvature}
\eeq 
with the connection 1-form $\mathcal{A}$ in Eq.~\eqref{eq:aaa}. The Chern number measures the ``magnetic flux'' of $\mathcal{A}_{\mu}({\bm k})$ penetrating the ${\bm k}$-space, or equivalently, the charge of the Dirac monopole inside $S^2$. If the Berry connection $\mathcal{A}_{\mu}({\bm k})$ is nonsingular in the entire $S^2$, the Stokes theorem shows that ${\rm Ch}_1$ vanishes. This implies that ${\rm Ch}_1$ is nonzero only when the connection ${\mathcal A}_{\mu}({\bm k})$ with a fixed gauge has a singularity in $S^2$ and the Stokes theorem is not applicable to the global ${\bm k}$-space. 

Following the arguments by Kohmoto~\cite{kohmoto}, let us show that the Chern number \eqref{eq:1stchern} is an integer. We start by dividing the momentum space $S^2$ into two patches to avoid a singularity, and the gauge in each patch is chosen so that ${\mathcal A}_{\mu}({\bm k})$ is a single-valued smooth function. $S^2$ is covered by two neighborhoods $S_{\rm I}$ and $S_{\rm II}$, which are defined as the northern and southern hemispheres, respectively: $S_{\rm I} \!\equiv\! \{ {\bm k}\!\in\! S^2 | 0 \!\le\! \theta _{\bm k} \!\le\! \pi/2 + \delta\}$ and $S_{\rm II} \!\equiv\! \{{\bm k}\!\in\! S^2 | \pi/2 - \delta \!\le\! \theta _{\bm k} \!\le\! \pi \}$ with an arbitrary small constant $\delta$. 
Let $S_{\rm I}$ be the region where the wave function $|u^{({\rm I})}_-({\bm k})\rangle$ is well-defined. However, $|u^{({\rm I})}_-({\bm k})\rangle$ may have a singularity in the region $S_{\rm II}$.
Then, we introduce $|u^{({\rm II})}_-({\bm k})\rangle$ as the ${\rm U}(1)$ gauge transformation of $|u_-({\bm k})\rangle$, which is a smooth function in the region $S_{\rm II}$. In the overlap region $S_{\rm I} \cap S_{\rm II}$, since the two eigenvectors must be smoothly connected to each other, we define the function $g$ that glues two eigenvectors as 
\beq
g({\bm k})=\langle u_-^{({\rm II})}({\bm k})|u_-^{({\rm I})}({\bm k})\rangle=e^{i\chi({\bm k})}.
\eeq
In other words, two eigenvectors that are well-defined in $S_{\rm I}$ and $S_{\rm II}$ are linked to each other in the overlap region by the ${\rm U}(1)$ gauge transformation, $|u^{({\rm II})}_-({\bm k})\rangle \!=\! e^{i\chi ({\bm k})}|u^{({\rm I})}_-({\bm k})\rangle$. At the boundary, the gluing function $g$ transforms the connection $\mathcal{A}^{({\rm I})}_{\mu}$ in $S_{\rm I}$ to $\mathcal{A}^{({\rm II})}_{\mu}$ in $S_{\rm II}$ as
\beq
\mathcal{A}^{({\rm II})}_{\mu} = \mathcal{A}^{({\rm I})}_{\mu} + i \partial _{k_{\mu}}\chi ({\bm k}).
\label{eq:glue}
\eeq
To evaluate the Chern number, we separate the integration of Eq.~\eqref{eq:1stchern} to each hemisphere $S_{\rm I}$ and $S_{\rm II}$ and utilize the Stokes theorem. Then, the integral of Eq.~\eqref{eq:1stchern} is recast into
$\frac{i}{2\pi}\left[\oint _{C} \mathcal{A}^{({\rm I})}_{\mu} dk_{\mu}
- \oint _{C} \mathcal{A}^{({\rm II})}_{\mu} dk_{\mu}\right]$, where the loop $C$ denotes the boundary between $S_{\rm I}$ and $S_{\rm II}$, {\it i.e.}, the equator on $S^2$. The gauge transformation in Eq.~\eqref{eq:glue} and the single-valuedness of the eigenvector on $C$ require the Chern number to be an integer value,
\beq
{\rm Ch}_1 = \frac{1}{2\pi}\oint _{C} {\bm \nabla}\chi({\bm k})\cdot d{\bm k} \in \mathbb{Z}.
\label{eq:ch1}
\eeq
Equation \eqref{eq:ch1} indicates that the Chern number can be regarded as the winding number of the ${\rm U}(1)$ gauge transformation on the boundary of patches in which the wavefunctions are smoothly defined. 

{\it Bulk-edge correspondence.}--- It is now worth mentioning that the Chern number gives two important consequences: Quantization of transport quantities and bulk-edge correspondence. Both consequences of the nontrivial Chern number were first unveiled in quantum Hall systems. Using the Nakano-Kubo formula, TKNN~\cite{tknn,kohmoto} clarified that the Hall conductivity $\sigma _{\rm H}$ of two-dimensional electrons under a magnetic field can be written as the first Chern number ${\rm Ch}_1$ of the ${\rm U}(1)$ bundle over the magnetic Brillouin zone as $\sigma _{\rm H} \!=\! (e^2/h){\rm Ch}_1$ ($h$ is Planck's constant). 
On the other hand, it was elucidated by Halperin~\cite{halperin} that, if the system has a boundary, gapless edge states appear whose wave functions are localized at the boundary. The gapless edge states are essential for the nontrivial transport of the quantized Hall current. He found that there are $n_{\rm edge}$ branches of gapless edge states propagating along the circumference of the bulk insulator when the $(n_{\rm edge}-1)$th Landau level is occupied. This gives an alternative explanation of the quantized Hall conductance in terms of the number of gapless edge states as $\sigma _{\rm H} \!=\! (e^2/h)n_{\rm edge}$. 

A direct connection between these two different interpretations of the Hall conductance was given by Hatsugai~\cite{hatsugai1,hatsugai2}. Using a tight-binding model, Hatsugai showed that the number of gapless edge states should be the same as the nontrivial Chern number, $n_{\rm edge} \!=\! {\rm Ch}_1$. This leads to the important consequence that, if the bulk has a nontrivial Chern number, corresponding gapless states appear on the boundary.

Such a relation between the bulk topological number and the boundary gapless state is known as the bulk-edge correspondence. Recently, this correspondence has been extended to more general systems. Qi {\it et al.}~\cite{qiPRB06} proved the correspondence between the bulk topological number and the edge states in two-dimensional insulators with the twisted boundary condition~\cite{niu84}. Fukui {\it et al.}~\cite{fukui} revisited the bulk-edge correspondence in the light of the index theorem for time-reversal-broken topological insulators. Teo and Kane~\cite{teoPRB10} and Sato {\it et al.}~\cite{satoPRB11} also provided a proof of the bulk-boundary correspondence for a class of insulators and superconductors. The topological invariant and the bulk-edge correspondence were also formulated for more generic systems by Volovik~\cite{volovik} and Essin and Gurarie~\cite{essin} on the basis of the Green's function formalism by Matsuyama, Ishikawa, and Volovik~\cite{ishikawa86,ishikawaNPB87,matsuyamaPTP87,volovik}.

{\it Winding number.}--- In general, the topological classification of band insulators and superconductors is provided by a homotopy of mappings from a base space defined by ${\bm k}\!\in\! S^{d}$ (for $d$ dimensions) to a target space of $\mathcal{H}({\bm k})$, $\mathcal{M}$~\cite{avronPRL83}. If the maps are smoothly connected to each other without closing the energy gap, they are in the same topological phase, but if not, they are in topologically different phases.

For superconductors, the target space is spanned by the set of eigenvectors $| u_n({\bm k})\rangle $ in Eq.~\eqref{eq:BdGk}. Since we suppose that occupied bands with $E_n\!<\! 0$ are separated by a bulk excitation gap from positive energy states, the energy levels can be flattened by continuously transforming $\mathcal{H}({\bm k})$ to the $Q$-matrix~\cite{schnyderPRB08},
\beq
Q({\bm k}) \!=\! \sum _{E_n>0} | u_n({\bm k})\rangle \langle u_n({\bm k}) | 
- \sum _{E_n<0} | u_n({\bm k})\rangle \langle u_n({\bm k}) |.
\label{eq:Qmat}
\eeq
The $Q$-matrix satisfies the conditions $Q^2({\bm k})=+1$ and $Q^{\dag}({\bm k})=Q({\bm k})$. All the positive (negative) eigenvalues of $\mathcal{H}({\bm k})$ are replaced by $+1$ ($-1$). The state $| u_n({\bm k})\rangle$ is a simultaneous eigenvector of $\mathcal{H}({\bm k})$ and $Q({\bm k})$, and the $Q$-matrix retains the symmetries held by the BdG Hamiltonian. Therefore, the $Q$-matrix can be regarded as the mapping of $S^d \rightarrow \mathcal{M}$, and the set of mappings is represented by the homotopy group $\pi _2(\mathcal{M})$.

For the $2\times 2$ BdG Hamiltonian in two dimensions ($N=d=2$), therefore, the $Q$-matrix can be regarded as the mapping of $S^2 \rightarrow \mathcal{M}$, and the set of mappings is represented by the homotopy group $\pi _2(\mathcal{M})$. The flattened Hamiltonian $Q$ is diagonalized by using the unitary matrix $U(k)\in {\rm U}(2)$ as $Q(k) \!=\! U(k)\Lambda U^{\dag}(k)$, where $\Lambda\!\equiv\! {\rm diag}(+1,-1)$. Since the occupied and empty states are invariant under the ${\rm U}(1)$ symmetry, the matrix $U(k)$ is uniquely determined up to an element of ${\rm U}(1)\times {\rm U}(1)$. Therefore, the set of eigenvectors, or equivalently, the flattened Hamiltonian, constitutes the space 
\beq
\mathcal{M} = {\rm U}(2)/[{\rm U}(1)\times {\rm U}(1)] = S^2.
\eeq
The homotopy group of the topological space provides the equivalence classes that can be continuously deformed into one another without closing the energy gap. For the case of two dimensions, the homotopy group is nontrivial, $\pi _2 (\mathcal{M}) \!=\! \pi _2 (S^2) \!=\! \mathbb{Z}$.
The nontrivial homotopy group, $\pi _2(S^2) \!=\! \mathbb{Z}$, enables one to introduce a topological invariant, i.e., the winding number $w_{\rm 2d}$ that characterizes the mapping from $S^2$ to the target space $S^2$. For the Hamiltonian \eqref{eq:toy}, the flattened Hamiltonian $Q({\bm k})$ is expressed by the two-dimensional spinor $\hat{m}_{\mu}({\bm k})\!=\!m_{\mu}({\bm k})/|{\bm m}({\bm k})|$ as
\beq
{Q}({\bm k}) = \hat{m}_{\mu}({\bm k})\tau _{\mu},
\label{eq:Qform}
\eeq
which spans the topological space $S^2$. The target space is thus parameterized with two variables $\theta \!\in\! [0,\pi]$ and $\phi \!\in\! [0,2\pi]$, where $\hat{m}_{\mu}({\bm k})\!=\! [\cos\phi ({\bm k})\sin\theta ({\bm k}),\sin\phi({\bm k})\sin\theta({\bm k}),\cos\theta({\bm k})]$. These variables define the images of the ${\bm k}$-space onto the target space $\mathcal{M} \!=\! S^2$, where the surface element of $S^2$ is given by $d\omega = \sin\theta d\theta \wedge d\phi$. The winding number that characterizes the map of $S^2\mapsto S^2$ is given by $w_{\rm 2d} = \frac{1}{4\pi}\int _{S^2} d\omega$, which is recast into
\begin{align}
w_{\rm 2d} 
 = - \frac{1}{8\pi}\int d^2{\bm k} \epsilon ^{ijl} \hat{m}_i({\bm k})
\partial _{k_x}\hat{m}_j({\bm k}) \partial _{k_y} \hat{m}_l({\bm k}).
\end{align}
This determines how much the target space warps the two-dimensional momentum space. The nontrivial winding number $w_{\rm 2d}$ implies that a two-dimensional skyrmion texture of $\hat{\bm m}$ emerges in the momentum space~\cite{volovik,read}.

It is remarkable that the two-dimensional winding number is equivalent to the first Chern number. First, we notice that the unit vector $\hat{\bm m}({\bm k})$ is associated with the Berry connection $\mathcal{A}_{\mu}({\bm k})$ as $\epsilon ^{\mu \nu}\partial _{k_{\mu}}\mathcal{A}_{\nu}({\bm k}) = -\frac{1}{4}\epsilon ^{\mu\nu}\epsilon ^{ijl}\hat{m}_i({\bm k})
\partial _{k_{\mu}}\hat{m}_j({\bm k})\partial _{k_{\nu}}\hat{m}_l({\bm k})$~\cite{bussei}. By using this relation, the first Chern number turns out to be equivalent to the winding number,
\beq
{\rm Ch}_1 = w_{\rm 2d}.
\eeq
In Sec.~\ref{sec:spinless}, we will share the topological aspect of a spin-polarized chiral $p$-wave superconducting state as a specific model having nontrivial $w_{\rm 2d}\!=\! {\rm Ch}_1$. 



\subsubsection{Topology subject to discrete symmetries}
\label{sec:topology2}

Naively, any one-dimensional closed loop $S^1$ cannot cover the target space $S^2$. Thus, a generic $2\times 2$ Hamiltonian in one dimension cannot provide a stable topological structure. However, discrete symmetries of H in Eq.~\eqref{eq:Hbdg} impose strong constraints on the spinor $\hat{\bm m}$, and thus nontrivial topological numbers can be introduced even in one dimension, as illustrated below.

\begin{figure}
\includegraphics[width=80mm]{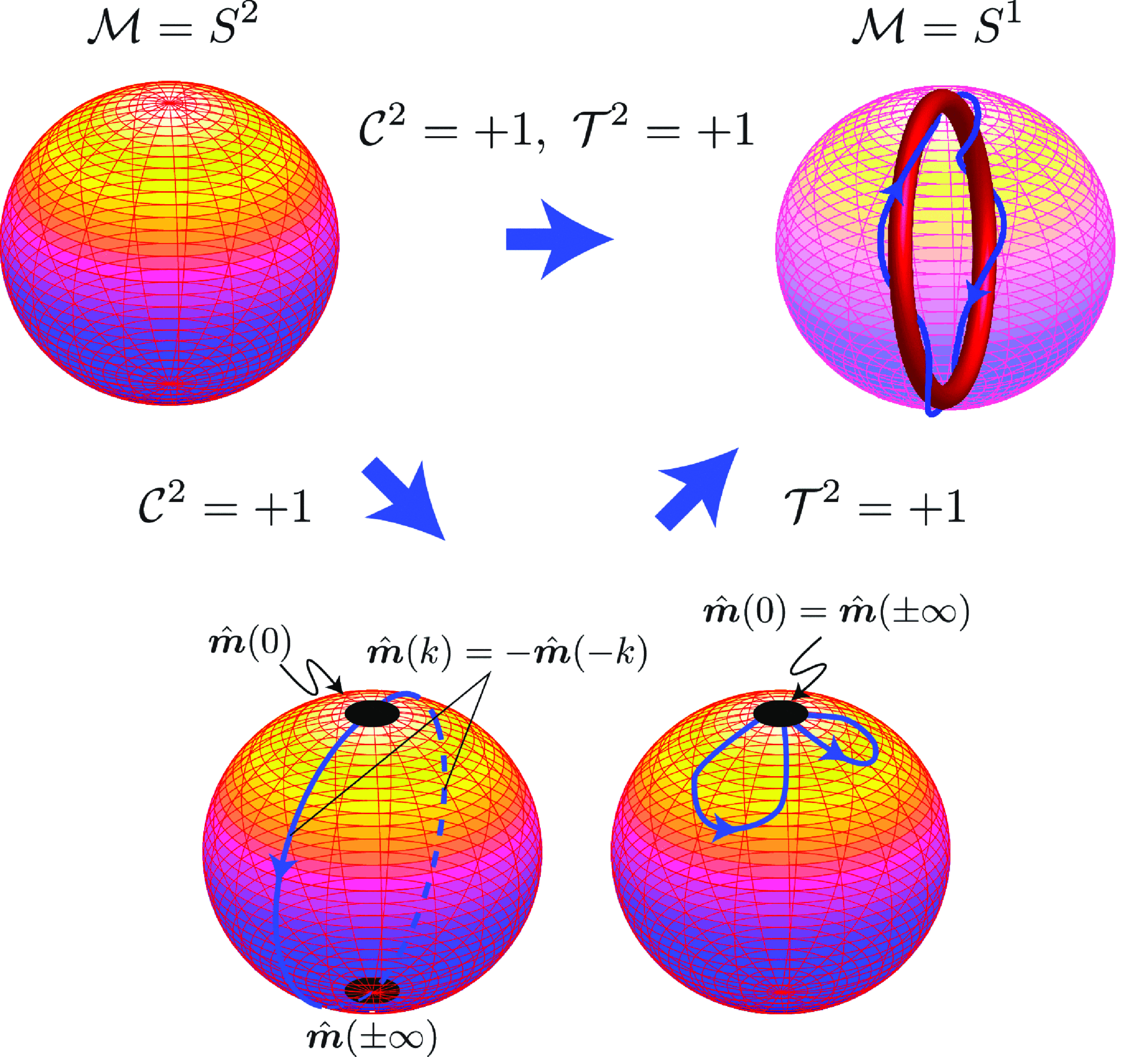}
\caption{(Color online) Target spaces $\mathcal{M}$ subject to discrete symmetries $\mathcal{T}$ and $\mathcal{C}$. Possible trajectories of $\hat{\bm m}$ on $\mathcal{M}$ with $\mathcal{C}^2\!=\! +1$ are also depicted. } 
\label{fig:target}
\end{figure}

{\it Particle-hole symmetry $\mathcal{C}^2\!=\! +1$ (class D).}--- 
Let us first suppose that the minimal Hamiltonian \eqref{eq:toy} holds the PHS $\mathcal{C} \!=\! \tau _x K$ ($\mathcal{C}^2\!=\! +1$). The operation of PHS changes the spinor $\hat{\bm m}({\bm k})$ to $[-\hat{m}_x(-{\bm k}),-\hat{m}_y(-{\bm k}),\hat{m}_z(-{\bm k})]$. For the momentum space characterized by $S^2$, there are two particle-hole invariant momenta, ${\bm k}\!=\!{\bm 0}$ and $|{\bm k}|=\infty$, where the infinite points are identical to a single point.
At the particle-hole invariant momenta, the spinor $\hat{\bm m}$ must point to the north or south pole on $S^2$. Therefore, we have two different situations: One is that the spinors $\hat{\bm m}$ at ${\bm k}={\bm 0}$ and $|{\bm k}|={\bm \infty}$ point in the same direction, 
\beq
\hat{m}_z({\bm 0})= \hat{m}_z({\bm \infty}),
\eeq
and the other is that they point in opposite directions,
\beq
\hat{m}_z({\bm 0})=-\hat{m}_z({\bm \infty}).
\eeq
The corresponding two trajectories of $\hat{\bm m}$ are not deformable to each other, as illustrated in Fig.~\ref{fig:target}. These two situations define trivial and nontrivial $\mathbb{Z}_2$ topological phases, respectively.


It is worth mentioning that for $\mathcal{C}^2\!=\! -1$, which is relevant to spin-singlet superconductors, the PHS imposes the constraint $m_{\mu}({\bm k}) \!=\! m_{\mu}(-{\bm k})$. Hence, the corresponding trajectory of $\hat{\bm m}$ is always $\mathbb{Z}_2$ trivial.

We now present an explicit form of the $\mathbb{Z}_2$ number in the above. We first notice that the Berry phase $\gamma$ in Eq.~\eqref{eq:berry} is not invariant under gauge transformation, and the gauge transformation $|u_n({\bm k})\rangle \!\mapsto\! e^{i\varphi ({\bm r})} |u_n({\bm k})\rangle$ shifts the value of $\gamma$ as~\cite{tanakaJPSJ12}
\beq
\gamma \mapsto \gamma - 2\pi \times ({\rm integer}).
\label{eq:gammagauge}
\eeq
Therefore, the Berry phase can be set in the range of $0\!\le\! \gamma \!< \! 2\pi$.

Qi {\it et al.}~\cite{qiPRB08} showed that the PHS discretizes the Berry phase $\gamma/\pi$ to an integer: The PHS links the Berry curvature constructed from the occupied states, $\mathcal{A}^{(-)}_{i} \!\equiv\! \langle u_-({\bm k})| \partial _{k_{i}} u_-({\bm k}) \rangle$, to that constructed from the empty states, $\mathcal{A}^{(+)}_{i} \!\equiv\! \langle u_+({\bm k})| \partial _{k_{i}} u_+({\bm k}) \rangle$, as $\mathcal{A}^{(-)}_{i}(-{\bm k}) \!=\! -\mathcal{A}^{(+)\ast}_{i}(-{\bm k})$. Therefore, the Berry phase is recast into $\gamma \!=\! \frac{i}{2}\sum _{\pm}\int _{S_1} dk _{i}\mathcal{A}^{(\pm)}_{i}({\bm k})$,
\beq
\nu \equiv \frac{\gamma}{\pi} = \frac{i}{2\pi}\int _{S^1} {\rm tr}\mathcal{A},
\label{eq:z21d}
\eeq
where ${\rm tr}\mathcal{A}=\sum _{\pm}\langle u_{\pm}({\bm k})|\partial _{k_{i}} u_{\pm}({\bm k})\rangle dk_i={\rm tr}[U^{\dag}(k)\partial _k U (k)]dk$.
We also notice that the Berry curvature is expressed in terms of the $2\times 2$ unitary matrix $U(k)$, where $U^{\rm T}(k) \!=\! (|u_-({\bm k})\rangle,|u_+({\bm k})\rangle)$. The quantity $\nu$ is then reduced to 
\begin{align}
\nu = \frac{i}{2\pi}\int dk_i {\rm tr} \left[ U^{\dag}({\bm k})\partial _{k_i} U ({\bm k}) \right] 
= \frac{i}{2\pi}\int dk_i \partial _{k_i} \ln \det U({\bm k}).
\label{eq:nuz2}
\end{align}
Since the determinant of the unitary matrix $U(k)$ is written as ${\rm det}U({\bm k})=e^{i\chi({\bm k})}$, the integral \eqref{eq:nuz2} can be evaluated as $\nu=\int dk \partial _k \chi(k)/2\pi$. Because ${\rm det}U({\bm k})$ is a single-valued function on $S_1$, we have 
\beq
\nu  \in \mathbb{Z} \quad {\rm mod}~2,
\eeq
which provides an analytic expression for the $\mathbb{Z}_2$ number. Here, the mod 2 ambiguity comes from the gauge ambiguity of the Berry phase \eqref{eq:gammagauge}. 



The $\mathbb{Z}_2$ number is related to the first Chern number ${\rm Ch}_1$ by the dimensional reduction. Consider ${\rm Ch}_1$ defined for the BdG Hamiltonian $\mathcal{H}(k_1, k_2)$ on $S^2$. Then the dimensional reduced Hamiltonian $\mathcal{H}(k_1, 0)$ with PHS defines a $\mathbb{Z}_2$ number $\nu$ (here, note that the $\mathbb{Z}_2$ number for $\mathcal{H}(k_1, \infty)$ is trivial). The $\mathbb{Z}_2$ number is found to have the same parity as the Chern number as
\beq
(-1)^{\nu}=(-1)^{{\rm Ch}_1}.
\eeq
The $\mathbb{Z}_2$ number is applied to full-gapped spin-triplet and odd-parity superconductors without or with time-reversal invariance. In Refs.~\citeonline{satoPRB09} and \citeonline{satoPRB10}, Sato revealed the intrinsic connection between their topological phases and the Fermi surface structures in the normal states. The efficient calculation combined with the first Chern number and winding number succeeded in uncovering the intrinsic relations between the Fermi surface topology and the gapless surface states in the superconducting states.

{\it Winding number for chiral symmetric classes.}--- 
When the Hamiltonian holds the chiral symmetry \eqref{eq:chiral}, another one-dimensional topological invariant can be defined as a winding number of the $Q$-matrix. To be specific, let us consider the $2\times 2 $ Hamiltonian that holds TRS ($\mathcal{T} \!=\! K$), PHS ($\mathcal{C} \!=\! \tau _x K$), and the chiral symmetry $\Gamma \!=\! \mathcal{CT} \!=\! \tau _x$, which belongs to class BDI in Table~\ref{table1}. This represents the Hamiltonian for a time-reversal-invariant superconductor with conserved $S_z$. 

It turns out that $\mathcal{T}^2=+1$ and $\mathcal{C}^2=+1$ change the spinor $\hat{\bm m}(k)$ to $[-\hat{m}_x(k),\hat{m}_y(k),\hat{m}_z(k)]$, and thus they impose a constraint on $\mathcal{M}$ that $\hat{m}_x$ must vanish for $\forall k$. As shown in Fig.~\ref{fig:target}, the target space $\mathcal{M}$ reduces to $S^1$ and is parameterized with $\phi\!\in\! [0,2\pi]$ as $\hat{\bm m}(k) \!\equiv\! [\hat{m}_1(k),\hat{m}_2(k)] \!=\![\cos\phi(k),\sin\phi (k)]$, where we set $m_1 \!=\! m_y$ and $m_2 \!=\! m_z$. The mapping of the one-dimensional ${\bm k}$-space $S^1$ to $\mathcal{M} \!=\! S^1$ is characterized by the fundamental group $\pi _{1}(S^1)\!=\! \mathbb{Z}$ and the relevant topological invariant is the one-dimensional winding number
\beq
w_{\rm 1d} = \frac{1}{2\pi}\int _{S^1}d\phi
= - \frac{1}{2\pi}\int^{\infty}_{-\infty}\epsilon ^{ij}\hat{m}_{i}(k)
\partial _k \hat{m}_j(k)dk, 
\label{eq:windingnumber}
\eeq
where we use $d\phi = \epsilon ^{ij}\hat{m}_i\partial _k \hat{m}_jdk$ ($i,j \!=\! 1,2$). The integral is simplified to the sum at ${ k}_0$ that satisfies $\varepsilon(k_0) \!=\! 0$~\cite{satoPRB09,satoPRB11,tanakaJPSJ12},
\beq
w_{\rm 1d} = -\frac{1}{2}\sum _{k_0} {\rm sgn}[\Delta (k)] {\rm sgn}[\partial _k \varepsilon (k)] \in \mathbb{Z}. 
\label{eq:w1d}
\eeq
As mentioned in Sec.~\ref{sec:az}, the winding number is also rewritten in terms of the $Q$-matrix. The index theorem for chiral symmetric classes with $w_{\rm 1d}$ is discussed in Sec.~\ref{sec:index}.


The $\mathbb{Z}_2$ topological number in Eq.~(\ref{eq:z21d}) remains well-defined in chiral symmetric classes as long as $\mathcal{C}^2=+1$ is maintained. We note that the parity of the winding number coincides with the $\mathbb{Z}_2$ number 
\beq
(-1)^{\nu} = (-1)^{w_{\rm 1d}}.
\label{eq:z2w1d}
\eeq
This implies that $w_{\rm 1d}$ can be nonzero even when $\nu$ is trivial but the opposite is not true in the symmetry class with $\mathcal{C}^2\!=\!+1$ and $\mathcal{T}^2\!=\!+1$. Therefore, the actual number of zero-energy states is determined by $w_{\rm 1d}$ unless the TRS is broken. Once the TRS is broken, however, $\nu$ in Eq.~\eqref{eq:z21d} determines the topological stability of the gapless states. 

This winding number $w_{\rm 1d}$ was first introduced in a different context by Wen and Zee~\cite{wen} for the topological stability of bulk gapless excitations of an electron-hopping Hamiltonian in a magnetic field, $\mathcal{H}=\sum _{ij} t_{ij}c^{\dag}_i c_j u_{ij}$. It can be proven that $w_{\rm 1d}$ is a topological invariant, implying that $w_{\rm 1d}$ is invariant under any continuous deformation of $\mathcal{H}$ without breaking the chiral symmetry \eqref{eq:chiral}. Using the winding number, Sato and Fujimoto uncovered the nontrivial topological properties of noncentrosymmetric superconductors under a spatially uniform magnetic field~\cite{satoPRB09-2}. The winding number is now widely applied to reveal the topological aspects of various systems (see for example, Refs.~\citeonline{tanakaJPSJ12} and \citeonline{mizushimaJPCM15}).

\subsubsection{Index theorem for chiral symmetric systems}
\label{sec:index}

Following  Ref.~\citeonline{satoPRB11}, we briefly mention the index theorem and bulk-edge correspondence for chiral symmetric systems with $w_{\rm 1d}$, which connect the winding number to the number of zero-energy states bound at the edge of the system.
First of all, let $|v^{\pm}_n({\bm k})\rangle$ be an eigenstate of $\mathcal{H}^2({\bm k})$,
\beq
\mathcal{H}^2({\bm k})\left| v_n({\bm k})\right\rangle = E^2_n \left| v_n({\bm k})\right\rangle.
\eeq
Then, there exists a one-to-one correspondence between the eigenstates $|u_n({\bm k})\rangle$ and $|v_n({\bm k})\rangle$~\cite{satoPRB11}. For a finite-energy state $E^2_n\neq 0$, the eigenvector $|v_n({\bm k})\rangle$ is associated with the eigenstate of $\mathcal{H}({\bm k})$ as 
$| u_n({\bm k})\rangle = c(\mathcal{H}({\bm k})+E_n) | v_n({\bm k})\rangle$ for $(\mathcal{H}({\bm k})+E_n)|v_n({\bm k})\rangle\neq 0$
and as $| u_n({\bm k})\rangle = \Gamma |v_n({\bm k})\rangle$ for $(\mathcal{H}({\bm k})+E_n)|v_n({\bm k})\rangle=0$, where $c$ is the normalization constant. For the zero-energy state, one finds $|u_0({\bm k})\rangle = |v_0({\bm k})\rangle$. 

The chiral symmetry ensures that the chiral operator $\Gamma$ is commutable with $\mathcal{H}^2({\bm k})$, 
\beq
[\Gamma,\mathcal{H}^2({\bm k})]=0.
\eeq 
This indicates that $|v_n({\bm k})\rangle$ is the simultaneous eigenstate of $\Gamma$ and $\mathcal{H}^2({\bm k})$, 
\beq
\Gamma |v^{\pm}_n({\bm k})\rangle = \pm |v^{\pm}_n({\bm k})\rangle .
\eeq
It turns out that the eigenvector $|v^+_n({\bm k})\rangle$ is constructed from the counterpart $|v^-_n({\bm k})\rangle$ as 
\beq
|v^+_n({\bm k})\rangle = c^{\prime}\mathcal{H}({\bm k})|v^-_n({\bm k})\rangle, 
\label{eq:vn}
\eeq 
for $E^2_n \neq 0$, where $c^{\prime}$ is the normalization constant. Hence, the eigenstate with $E^2_n \neq 0$ forms a chiral pair, where the quasiparticle with the chirality $\Gamma = +1$ is always paired with the quasiparticle with the opposite chirality $\Gamma = -1$. 

For zero-energy states, however, the solution does not form a pair in general since the right-hand side of Eq.~(\ref{eq:vn}) vanishes. Although they are eigenstates of $\Gamma$, the number of zero-energy states with $\Gamma=+1$, $n_+$, is not the same as the number of  $\Gamma=-1$ zero-energy states, $n_-$.  As proved in Ref.~\citeonline{satoPRB11}, the nonzero value of $w_{1d}$ in Eq.~(\ref{eq:w1d}) is identical to the difference in the number of zero-energy states in each chiral subsector,
\beq 
|w_{\rm 1d}| = |n_- - n_+|.
\eeq 
Hence, at least $|w_{\rm 1d}|$ zero-energy states exist as long as the chiral symmetry is preserved.

\subsection{The basic concept of topological classification}
\label{sec:az}

The basic idea in the pedagogical model can be generalized to any number of dimensions. The topological classification subject to the Altland Zirnbauer symmetries was first introduced by Schnyder {\it et al.}~\cite{schnyderPRB08}. The random matrices of Bloch and BdG Hamiltonians are categorized to the tenfold way and their topological properties are characterized by 0, $\mathbb{Z}$, $\mathbb{Z}_2$, depending on the symmetry classes and the momentum dimension $d$. Subsequently, Kitaev~\cite{kitaev09}, based on an elegant mathematical framework, the Clifford algebra and $K$-theory, generalized the topological classification with Bott periodicity. The {\it periodic} topological table was obtained for any momentum dimension, which is found to be mod 2 or mod $8$~\cite{kitaev09,ryuNJP10}. 

The periodic topological table was further extended by Teo and Kane~\cite{teoPRB10} to Bloch and BdG Hamiltonians with topological defects, including edge, surface, vortices, dislocations, and so forth. Such defects induce spatial gradient in the phase and amplitude of pair potentials. We present in Table~\ref{table1} the periodic topological table of Bloch and BdG Hamiltonians with the Altland Zirnbauer symmetries, where $\delta=d-D$ is the difference between the momentum dimension $d$ and the dimension $D$ of the surface enclosing the topological defect.

To accomplish the topological classification of Hamiltonians with topological defects, Teo and Kane~\cite{teoPRB10} started with the semiclassical approximation where the Hamiltonian varies slowly in the real-space coordinate. Then, the spatial modulation due to a topological defect can be considered as adiabatic changes in the Hamiltonian as a function of the real-space coordinate surrounding the defect, ${\bm R} \!=\! (R_1,R_2,\cdots,R_{D})$. The Hamiltonian is obtained in the base space, $({\bm k},{\bm R})$, as~\cite{volovik,teoPRB10,shiozakiPRB14}
\beq
\mathcal{H}({\bm k},{\bm R}).
\label{eq:hamiltonian}
\eeq
The momentum ${\bm k}$ is defined in the $d$-dimensional Brillouin zone and the real-space coordinate ${\bm R}$ is characterized by the $D$-dimensional sphere $S^D$ surrounding the defect. Hence, the base space is $T^d\times S^D$, where $T^d$ is a $d$-torus. Without losing generality, however, we here consider a simple space $S^{d+D}$ rather than the exact base space. The choice of $S^{d+D}$ misses ``weak'' topological indices. As pointed out in Refs.~\citeonline{teoPRB10} and \citeonline{shiozakiPRB14}, however, ``weak'' topological invariants can be obtained as ``strong'' topological invariants in lower dimensions.

The semiclassical approximation with classical variables (${\bm k}$, ${\bm R}$) is valid when the characteristic length scale of the spatial inhomogeneity is sufficiently longer than that of the quantum coherence. While a realistic Hamiltonian does not always satisfy this semiclassical condition, the full quantum Hamiltonian can be smoothly deformed into the semiclassical Hamiltonian $\mathcal{H}({\bm k},{\bm R})$ while maintaining the bulk gap and symmetries of the system. This implies that zero energy states in the semiclassical $\mathcal{H}({\bm k},{\bm R})$ survive in the full quantum Hamiltonian.

\begin{table*}[t!]
\begin{center}
\begin{tabular}{ccccc|cccccccc}
\hline\hline
$s$ & AZ class & TRS & PHS & CS & $\delta =0$ & $\delta =1$ & $\delta =2$ & $\delta =3$ & $\delta =4$ 
& $\delta =5$ & $\delta =6$ & $\delta =7$ \\
\hline
0 & A & 0 & 0 & 0 & $\mathbb{Z}$ & ${0}$ & $\mathbb{Z}$ & ${0}$ & $\mathbb{Z}$ & ${0}$ & $\mathbb{Z}$ & ${0}$ \\
1 & AIII & 0 & 0 & 1 & ${0}$ & $\mathbb{Z}$ & $0$ & $\mathbb{Z}$ & ${0}$ & $\mathbb{Z}$ & ${0}$ & $\mathbb{Z}$ \\
\hline
0 & AI   & $1$  & $0$  & $0$ & $\mathbb{Z}$ & 0 & 0 & 0 & $2\mathbb{Z}$ & 0 & $\mathbb{Z}_2$ & $\mathbb{Z}_2$ \\
1 & BDI  & $1$  & $1$  & $1$ & $\mathbb{Z}_2$ & $\mathbb{Z}$ & 0 & 0 & 0 & $2\mathbb{Z}$ & 0 & $\mathbb{Z}_2$ \\
2 & D    & $0$  & $1$  & $0$ & $\mathbb{Z}_2$ & $\mathbb{Z}_2$ & $\mathbb{Z}$ & 0 & 0 & 0 & $2\mathbb{Z}$ & 0 \\
3 & DIII & $-1$ & $1$  & $1$ & $0$ & $\mathbb{Z}_2$ & $\mathbb{Z}_2$ & $\mathbb{Z}$ & 0 & 0 & 0 & $2\mathbb{Z}$ \\
4 & AII  & $-1$ & $0$  & $0$ & $2\mathbb{Z}$ & $0$ & $\mathbb{Z}_2$ & $\mathbb{Z}_2$ & $\mathbb{Z}$ & 0 & 0 & 0 \\
5 & CII  & $-1$ & $-1$ & $1$ & $0$ & $2\mathbb{Z}$ & $0$ & $\mathbb{Z}_2$ & $\mathbb{Z}_2$ & $\mathbb{Z}$ & 0 & 0 \\
6 & C    & $0$  & $-1$ & $0$ & 0 & $0$ & $2\mathbb{Z}$ & $0$ & $\mathbb{Z}_2$ & $\mathbb{Z}_2$ & $\mathbb{Z}$ & 0 \\
7 & CI   & $1$  & $-1$ & $1$ & 0 & 0 & $0$ & $2\mathbb{Z}$ & $0$ & $\mathbb{Z}_2$ & $\mathbb{Z}_2$ & $\mathbb{Z}$ \\
\hline\hline
\end{tabular}
\caption{Periodic table for the classification of BdG and Bloch Hamiltonians with a topological defect, $\mathcal{H}({\bm k},{\bm R})$, subject to the time-reversal symmetry (TRS), the particle-hole symmetry (PHS), and the chiral symmetry (CS)~\cite{teoPRB10}. The second column gives the names of the Altland-Zirnbauer (AZ) classes and third and fourth columns indicate the absence ($0$) or presence ($\mathcal{C}^2\!=\! \pm 1$, $\mathcal{T}^2\!=\! \pm 1$, and $\Gamma^2\!=\! 1$) of PHS, TRS, and CS, respectively. The other columns list the topological class of quantum ground states in relative dimensions $\delta = d-D$. The topologically nontrivial class is characterized by $\mathbb{Z}$ (integer) or $\mathbb{Z}_2$ ($\{0,1\}$). When additional discrete symmetries are absent, the classes D and DIII host chiral and helical Majorana fermions.
}
\label{table1}
\end{center}
\end{table*}

The classifying space of $\mathcal{H}({\bm k},{\bm R})$ is spanned by the set of eigenvectors $| u_n({\bm k},{\bm R})\rangle $, which are given by solving the BdG equation
\beq
\mathcal{H}({\bm k},{\bm R}) | u_n({\bm k},{\bm R})\rangle  = E_n({\bm k},{\bm R}) | u_n({\bm k},{\bm R})\rangle.
\eeq
For fully gapped superconductors and superfluids, occupied bands with $E_n\!<\! 0$ are separated by a bulk excitation gap from positive energy states, and thus ${\mathcal H}({\bm k}, {\bm R})$ can be flattened into the $Q$-matrix, $Q({\bm k},{\bm R})$, in the same manner as Eq.~\eqref{eq:Qmat}. The state $| u_n({\bm k},{\bm R})\rangle$ is a simultaneous eigenvector of $\mathcal{H}({\bm k},{\bm R})$ and $Q({\bm k},{\bm R})$, and the $Q$-matrix retains the symmetries held by the BdG Hamiltonian. The $Q$-matrix is the projector that maps the base space $({\bm k},{\bm R})$ onto the target space $\mathcal{M}$ spanned by $| u_n({\bm k},{\bm R})\rangle$. 


The basic idea of the topological classification is provided by the homotopy of mappings from a base space defined by $({\bm k},{\bm R})\in S^{d+D}$ to the classifying space of ${Q}({\bm k},{\bm R})$, $\mathcal{M}$, subject to a set of discrete symmetries. If the maps are smoothly connected to each other, they belong to the same topological phase, but if not, they are in topologically different phases. For generic Hamiltonians having $N$ occupied and $M$ empty bands, the target space is the complex Grassmannian $\mathcal{M} \!=\! G_{M,N+M} (\mathbb{C}) \!=\! U(N+M)/[U(N)\times U(M)]$, and the relevant homotopy group is $\pi_{d+D}({\mathcal M})$, which is $\mathbb{Z}$ for $d+D=2n$ and $0$ for $d+D=2n+1$. The topological invariant associated with $\pi _{2n}(G_{M,N+M} (\mathbb{C}))$ in $2n$ dimensions is the $n$th Chern number~\cite{nakahara,teoPRB10,shiozakiPRB14}
\beq
{\rm Ch}_n = \frac{1}{n!} \int _{S^{d}\times S^{D}} {\rm tr}
\left( 
\frac{i\mathcal{F}}{2\pi}
\right)^n \in \mathbb{Z},
\label{eq:nchern}
\eeq
where the Berry curvature $\mathcal{F}$ is generally defined as $\mathcal{F}=d\mathcal{A}+\mathcal{A}\wedge\mathcal{A}$ with the Berry connection $\mathcal{A}$ constructed from occupied bands. The $n$-th Chern number relevant to $\pi _{2n}(G_{M,N+M} (\mathbb{C}))$ is the generalization of the first Chern number in Sec.~\ref{sec:chern}. The $\mathbb{Z}$ topological numbers in $2n$ dimensions in Table~\ref{table1} are given by the $n$-th Chern number, although for real Hamiltonians (AI, AII, D, and C classes), the Chern number is subject to discrete symmetries such as TRS and PHS.

When the Hamiltonian holds the chiral symmetry \eqref{eq:chiral}, the Chern number must vanish, ${\rm Ch}_n\!=\! 0$. For chiral symmetric Hamiltonians, however, the target space $\mathcal{M}$ is reduced to ${\rm U}(n)$, and another $\mathbb{Z}$ topological invariant can be defined in the same manner as in Sec.~\ref{sec:topology2}. 
The mapping relevant to $\pi _{2n+1}({\rm U}(n)) = \mathbb{Z}$ is characterized by the winding number $w_{2n+1}$, 
\beq
w_{2n+1} = \frac{(-1)^nn!}{2(2\pi i)^{n+1}(2n+1)!}\int _{S^d\times S^{D}}{\rm tr}[\Gamma QdQ^{\dag}]^{2n+1},
\eeq
where $2n+1\!\equiv\! d+D$. By using the Hamiltonian, the winding number is recast into 
\beq
w_{2n+1} = \frac{n!}{2(2\pi i)^{n+1}(2n+1)!}\int _{S^d\times S^{D}}{\rm tr}\Gamma 
(\mathcal{H}^{-1}d\mathcal{H})^{2n+1}.
\label{eq:windingn}
\eeq
The winding number for $n\!=\! 0$ is reduced to Eq.~\eqref{eq:windingnumber}. An alternative expression for the winding number in terms of the Green's function is given by Volovik~\cite{volovik}. 

The winding number characterizes the nontrivial topological properties of class AIII in odd dimensions, while it is subject to discrete symmetries in real chiral symmetric classes (BDI, DIII, CI, and CII). One of the remarkable consequences shown in Table~\ref{table1} is that the bulk superfluid $^3$He-B is a promising candidate for three-dimensional time-reversal-invariant topological superfluids, which is categorized into class DIII. The bulk topological superfluid is accompanied by helical Majorana fermions bound to the surface. The detailed topological properties will be discussed in Sec.~\ref{sec:bulk}. 

In Table~\ref{table1}, the $\mathbb{Z}$ topological invariants for $\delta \!=\! 2n$ and $2n+1$ are given by ${\rm Ch}_n$ in Eq.~\eqref{eq:nchern} and $w_{2n+1}$ in Eq.~\eqref{eq:windingn}, respectively. For real nonchiral classes with $(s,\delta) \!=\! (2n+2,2n+1)$ ($n\ge 0$), the $\mathbb{Z}_2$ topological invariant is given by the integral of the Chern-Simons form over the base space as~\cite{shiozakiPRB14}
\beq
\nu _{2n+1} = \frac{2}{(n+1)!}\left( \frac{i}{2\pi}\right)^{n+1}
\int _{S^{d+D}} {\rm CS}_{2n+1} \in \mathbb{Z} \quad \mbox{mod 2},
\label{eq:z2cs}
\eeq 
when the dimension of the base space is $d+D\!=\!2n+1$. Here, the Chern-Simons 1- and 3-forms are given by 
\begin{gather}
{\rm CS}_1 = {\rm tr}\mathcal{A}, \quad
{\rm CS}_3 = {\rm tr}\left( \mathcal{A} d\mathcal{A} + \frac{2}{3}\mathcal{A}^3\right). 
\label{eq:cs}
\end{gather}
The $\mathbb{Z}_2$ Chern-Simons number can also be interpreted as the dimensional reduction of ${\rm Ch}_{n+1}$ in $(2n+2)$ dimensions.~\cite{qiPRB08} As discussed below, the $\mathbb{Z}_2$ topological number associated with the Chern-Simons form describes the nontrivial topological properties of class D topological superconductors with a vortex. This $\mathbb{Z}_2$ number is a generalization of Eq.~\eqref{eq:z21d}.

The other $\mathbb{Z}_2$ numbers can be defined under a constraint of the TRS. This was first introduced for band insulators by Fu and Kane~\cite{fuRPB06} in the context of time-reversal polarization analogous to Berry's phase formulation of the charge polarization.



When the original Hamiltonian maintains discrete symmetries such as crystalline symmetries and $P_2$ and $P_3$ symmetries [see Eq.~\eqref{eq:p2p3} and Fig.~\ref{fig:discrete}], the base and classifying spaces are subject to additional symmetries. Let us assume that the normal state holds the mirror symmetry $M \varepsilon ({\bm k},{\bm R}) M^{\dag} = \varepsilon (\underline{\bm k}_{\rm M},\underline{\bm R}_{\rm M})$. Here, we define the operator $M = i({\bm \sigma}\cdot\hat{\bm o})$ that denotes the mirror reflection in the plane perpendicular to the $\hat{\bm o}$-axis. The mirror operator changes the spin ${\bm \sigma}\rightarrow -{\bm \sigma}+2\hat{\bm o}({\bm \sigma}\cdot\hat{\bm o})$, the momentum ${\bm k}\rightarrow\underline{\bm k}_{\rm M} = {\bm k}-2\hat{\bm o}({\bm k}\cdot\hat{\bm o})$, and the coordinate ${\bm R}\rightarrow\underline{\bm R}_{\rm M} = {\bm R}-2\hat{\bm o}({\bm R}\cdot\hat{\bm o})$. Then, the mirror operator in the Nambu space, $\mathcal{M}\equiv {\rm diag}(M,M^{\rm \ast})$, acts on the BdG Hamiltonian as
\beq
\mathcal{M}\mathcal{H}({\bm k},{\bm R})\mathcal{M}^{\dag} = \mathcal{H}(\underline{\bm k}_{\rm M},\underline{\bm R}_{\rm M})
\label{eq:mirrorop}
\eeq
when $M\Delta ({\bm k},{\bm R})M^{\rm T}=\Delta (\underline{\bm k}_{\rm M},\underline{\bm R}_{\rm M})$ is satisfied. Next, we suppose that the single-particle energy and the pair potential are invariant under the joint spin-orbit rotation about the $\hat{\bm a}$-axis by the angle $\varphi=\pi$, $U_S(\hat{\bm a},\pi)\varepsilon ({\bm k},{\bm R})U^{\dag}_S(\hat{\bm a},\pi)=\varepsilon (R^{(L)}{\bm k},R^{(L)}{\bm R})$ and $U_S(\hat{\bm a},\pi)\Delta ({\bm k},{\bm R})U^{\rm T}_S(\hat{\bm a},\pi)=\Delta (R^{(L)}{\bm k},R^{(L)}{\bm R})$, where $U_S$ denotes the ${\rm SU}(2)$ rotation matrix associated with the ${\rm SO}(3)$ rotation matrix $R^{(L)}$. The BdG Hamiltonian then has the rotation symmetry for $\mathcal{U}_S={\rm diag}(U_S,U^{\rm \ast}_S)$,
\beq
\mathcal{U}_S(\hat{\bm a},\pi)\mathcal{H}({\bm k},{\bm R})\mathcal{U}^{\dag}_S(\hat{\bm a},\pi) = \mathcal{H}(\underline{\bm k}_{\rm M},\underline{\bm R}_{\rm M}).
\eeq
The $P_2$ and $P_3$ symmetries introduced in Eq.~\eqref{eq:p2p3} (see also Fig.~\ref{fig:discrete}) are defined as a combination of these symmetry operators with the time-reversal operator as
\beq
\mathcal{P}_2 = \mathcal{T}\mathcal{M}, \quad \mathcal{P}_3 = \mathcal{T}\mathcal{U}_S.
\label{eq:p2p3v2}
\eeq
For $\hat{\bm a}=\hat{\bm o}$, the combination of $P_2$ and $P_3$ symmetries defines the $P_1$ symmetry, which is a combination of the inversion and discrete ${\rm U}(1)$ phase rotation operators. It is remarkable to notice that, even if each symmetry is broken, the combined symmetries, $P_2$ and $P_3$, may be maintained and responsible for topological invariants. Since all these additional symmetries flippes the coordinate and momentum spaces, they impose constraints on both base and target spaces. 

Fu~\cite{fuPRL11} introduced a concept of topological crystalline insulators, where the nontrivial topology of band insulators can be protected by some crystalline symmetries. Using this concept, Hsieh {\it et al.}~\cite{hsieh} predicted that SnTe possesses gapless surface states protected by the mirror Chern number defined on a mirror-invariant plane in the Brillouin zone that satisfies ${\bm k}=\underline{\bm k}_{\rm M}$. In the superfluid $^3$He under a magnetic field, Mizushima {\it et al.}~\cite{mizushimaPRL12} demonstrated the existence of a new topological phase that is protected by the $P_3$ symmetry, a combination of the time-reversal operation $\mathcal{T}$ and the magnetic $\pi$-rotation. The topology of the bulk $^3$He-B is characterized by the homotopy group $\pi _3(\mathcal{M}=S^3)=\mathbb{Z}$, while the B phase under a magnetic field possesses the $P_3$ symmetry which reduces both the base and target spaces to $S^3\rightarrow S^1$.  Other examples of symmetry-protected topological superfluids and superconductors are an array of one-dimensional spin-orbit-coupled Fermi gases and a superconducting nanowire~\cite{mizushimaNJP13,tewariPRL12,tewariPRB12}. The topological phase becomes nontrivial when the $P_2$ symmetry, a combination of $\mathcal{T}$ and mirror reflection symmetry, is maintained. 

The Altland-Zirnbauer classification of topological band insulators and superconductors was extended by Chiu {\it et al.}~\cite{chiuPRB13} to include additional reflection symmetry. Morimoto and Furusaki~\cite{morimotoPRB13} developed a systematic method based on the representation theory of Clifford algebras and $K$-theory. Using this mathematical framework, Shiozaki and Sato~\cite{shiozakiPRB14} proposed a complete classification of topological phases and defects for noninteracting fermions in the presence of additional order-two symmetry. In addition to the Bott periodicity, they found periodicity in the number of flipped coordinates under the additional symmetry. Further topological classifications were accomplished by several authors~\cite{benalcazarPRB14,hsiehPRB14,jadaunPRB13,chiu15}. In Sec.~\ref{sec:MF}, \ref{sec:spt}, and \ref{sec:vortices}, we clarify the roles of additional order-two symmetry on the topological superfluidity and Majorana fermions in $^3$He. This includes $^3$He-B protected by the $P_3$ symmetry, the $P_2$-symmetry-protected planar phase, and the $^3$He-A thin film protected by mirror reflection symmetry. Furthermore, it has been discussed that additional order-two symmetries play a crucial role in the characteristic features of Majorana fermions~\cite{mizushimaPRL12,uenoPRL13,zhangPRL13,tsutsumiJPSJ13,sato14,shiozakiPRB14,fangPRL14,liuprx14}. The details are described in Sec.~\ref{sec:mirror}, \ref{sec:chiral}, and \ref{sec:IQV}.

\subsection{Majorana fermion and non-Abelian statistics}
\label{sec:MF}

In relativistic field theory, self-conjugate Dirac fermions are called Majorana fermions. They are represented by the quantized field ${\bm \Psi}$ that satisfies the constraint of the self-charge-conjugation,~\cite{semenoff,wilczek,wilczek14,jackiw14,chamon}.  
\beq
{\bm \Psi}({\bm r}) = \mathcal{C}{\bm \Psi}({\bm r}).
\label{eq:selfcharge}
\eeq
The self-charge conjugation relation \eqref{eq:selfcharge} for Majorana fermions is satisfied only when the antiunitary particle-hole (or charge conjugation) operator $\mathcal{C}$ is 
\beq
\mathcal{C}^2 = +1.
\label{eq:self}
\eeq
As mentioned above, the condition \eqref{eq:self} can be fulfilled by odd-parity superconductors and superfluids. In spin-singlet even-parity superconductors without spin-orbit coupling, the pair potential $\Delta ({\bm k})$ is always invariant under the spin rotation, and the particle-hole operator is given by $\mathcal{C}^2\!=\! -1 $, which cannot satisfy Eq.~\eqref{eq:selfcharge} and cannot be Majorana fermions. The presence of a strong spin-orbit interaction, however, may mix the spin-singlet and triplet pairings, which enable even spin singlet superconductors to host Majorana fermions~\cite{satoPRB09-2,satoPRL09,satoPRB10-2}.

To capture the consequences of Majorana fermions, we expand the quantized field ${\bm \Psi}$ in terms of the energy eigenstates. The energy eigenstates are generally obtained by diagonalizing the mean-field approximated Hamiltonian in the coordinate space, 
$\mathcal{H} = \int d{\bm r}_1\int d{\bm r}_2 {\bm \Psi}^{\dag}({\bm r}_1)\mathcal{H}({\bm r}_1,{\bm r}_2){\bm \Psi}({\bm r}_2)$. 
The energy eigenstates are determined using the BdG equation,
\beq
\int d{\bm r}_2 \mathcal{H}({\bm r}_1,{\bm r}_2){\bm \varphi}_{E}({\bm r}_2) = E{\bm \varphi}_E({\bm r}_1).
\label{eq:Hbdgr}
\eeq
As shown in Eq.~(\ref{eq:PHS}), the BdG Hamiltonian density holds PHS, 
$\mathcal{C}\mathcal
{H}({\bm r}_1,{\bm r}_2)\mathcal{C}^{-1}=-\mathcal{H}({\bm r}_1,{\bm
r}_2)$, and thus the eigenstates may satisfy 
\beq
{\bm \varphi}_E({\bm r}) = \mathcal{C} {\bm \varphi}_{-E}({\bm r})
\label{eq:PHS2}. 
\eeq
When the BdG Hamiltonian has no zero-energy eigenstates, ${\bm \Psi}$ is in general expanded in terms of the positive and negative energy states as 
${\bm \Psi}({\bm r}) = \sum_{E>0}\left[ {\bm \varphi}_E({\bm r})\eta_E + {\bm \varphi}_{-E}({\bm r})\eta_{-E}\right]$. The PHS in Eq.~\eqref{eq:PHS} ensures the one-to-one correspondence between negative and positive energy eigenstates, i.e., ${\bm \varphi}_{-E}({\bm r}) \!=\! \mathcal{C}{\bm \varphi}_{E}({\bm r})$. From the orthogonality and completeness relations, 
$\int d{\bm r} {\bm \varphi}^{\dag}_E({\bm r}) {\bm \varphi}_{E^{\prime}}({\bm r}) \!=\! \delta _{E,E^{\prime}}$ and
$\sum _E {\bm \varphi}_E({\bm r}_1){\bm \varphi}^{\dag}_E({\bm r}_2) \!=\! \delta ({\bm r}_1-{\bm r}_2)$, the quasiparticle operator $\eta_{E>0}$ satisfies the anticommutation relations, $\{ \eta _E, \eta^{\dag}_{E^{\prime}}\} \!=\! \delta _{E,E^{\prime}}$ and $\{\eta _E, \eta _{E^{\prime}} \}= \{\eta^{\dag}_E, \eta ^{\dag}_{E^{\prime}} \} \!=\! 0$. The self-charge conjugation relation \eqref{eq:selfcharge} then implies that the quasiparticle annihilation operator with a positive energy is equivalent to the creation with a negative energy as
\beq
\eta _{E} = \eta ^{\dag}_{-E}. 
\eeq
Therefore, ${\bm \Psi}$ in the context of superconductors and superfluids can be expanded only in terms of positive energy states as 
\beq
{\bm \Psi}({\bm r}) = \sum_{E>0}\left[ {\bm \varphi}_E({\bm r})\eta_E + \mathcal{C}{\bm \varphi}_{E}({\bm r})\eta^{\dag}_{E}\right].
\eeq



Now, let us suppose that $n$ zero-energy states ${\bm \varphi}^{(a)}_{E=0}({\bm r})$ ($a=1,\cdots,n$) exist. For topological superconductors, such zero energy states exist at boundaries or defects such as vortices. Then, we can rewrite the quantized field ${\bm \Psi}$ as 
\beq
{\bm \Psi}({\bm r}) = \sum^n_{a=1} {\bm \varphi}^{(a)}_{0}({\bm r}) \gamma^{(a)}
+ \sum _{E>0} \left[ 
{\bm \varphi}_E({\bm r}) \eta _E + \mathcal{C}{\bm \varphi}_E({\bm r}) \eta^{\dag}_E
\right],
\label{eq:PsiM}
\eeq
where we have used $\gamma^{(a)}$, instead of $\eta_{E=0}$, to distinguish these zero modes.
Owing to the PHS in Eq.~(\ref{eq:PHS}), the zero energy states are composed of equal contributions from the particle-like and hole-like components of quasiparticles. The self-conjugate constraint in Eq.~(\ref{eq:selfcharge}) imposes the following relation: 
\beq
\gamma^{(a)} = \gamma^{(a)\dag}.
\label{eq:majo}
\eeq
Since the zero modes satisfy the self-conjugate constraint, they are called the Majorana zero modes.
The operator of the Majorana zero modes has the following relation:
\beq
\gamma^{(a)2} = 1,\hspace{3mm} \{ \gamma^{(a)}, \eta _{E} \} = \{ \gamma^{(a)}, \eta^{\dag}_E \} = 0.
\label{eq:anti}
\eeq 
As we discuss below, this unusual relation gives rise to a significant feature inherent to the Majorana field.

\subsubsection{Majorana fermions: Their basic properties}
\label{sec:nonabelian}

{\it Density operator.}--- 
Let us consider the situation that $n$ Majorana zero modes exist in a surface or defect of topological superconductors. The local density operator in the Nambu space is defined in terms of the field operator \eqref{eq:PsiM} as 
$\rho ({\bm r}) \!=\! {\bm \Psi}^{\dag}({\bm r}) \tau _z {\bm \Psi}({\bm r})/2$. For $n$ Majorana zero modes, it is given by~\cite{shiozakiPRB14} 
\beq
\rho ({\bm r}) = \frac{1}{2}\sum^{n}_{a,b = 1} \left[ 
\gamma^{(a)}, \gamma^{(b)}
\right] {\bm \varphi}^{(a)\dag}_0({\bm r}) {\bm \varphi}^{(b)}_0({\bm r}), 
\eeq
where the contributions from quasiparticles with a finite $E$ are omitted. First, the orthogonality condition of the zero modes implies that the total density operator of the $n$ Majorana zero modes vanishes, 
\beq
\int d{\bm r}\rho ({\bm r}) = 0.
\label{eq:MFcharge}
\eeq
For $n=1$, it is also obvious that the local density operator is identically zero, 
\beq
\rho ({\bm r}) = 0 . 
\label{eq:density}
\eeq
This indicates that the Majorana zero modes cannot be coupled to the local density fluctuation and thus are very robust against nonmagnetic impurities. 

The local density operator of $n$ Majorana zero modes is not necessarily zero when $n\ge 2$. Nevertheless, an additional antiunitary symmetry can protect the characteristic feature in Eq.~\eqref{eq:density} even for $n \ge 2$. For instance, if the system holds TRS, the zero-energy modes form Kramers pairs. If a single Kramers pair exits, 
the local density operator is identically zero.~\cite{qiPRL09,shiozakiPRB14}

{\it Fermion parity.}--- In the case that two Majorana zero modes exist ($n=2$), the minimal representation of the algebra in Eq.~(\ref{eq:anti}) is two-dimensional. The two-dimensional representation is built up by defining the new fermion operators $c$ and $c^{\dag}$ as
\beq
c = \frac{1}{\sqrt{2}} (\gamma^{(1)}+i\gamma^{(2)}), \hspace{3mm}
c^{\dag} = \frac{1}{\sqrt{2}} (\gamma^{(1)}-i\gamma^{(2)}).
\eeq 
It is obvious that this {\it complex fermion} operator obeys the standard anticommutation relations, $\{c,c^{\dag}\}=1$ and $\{c,c\}=\{c^{\dag},c^{\dag}\}=0$. The two degenerate vacuums $|\pm \rangle$ are defined as the eigenstates of the fermion parity. We here assume $| + \rangle$ ($| - \rangle$) to be the empty (occupied) state of the complex fermion, 
\beq
c|-\rangle = c^{\dag}|+\rangle = 0.
\eeq
The state $| + \rangle$ ($| - \rangle$) has even (odd) fermion parity. Two Hilbert spaces are now spanned by using the vacuum state $|\pm \rangle$ and excited states that are constructed as $\eta^{\dag}_{E}\eta ^{\dag}_{E^{\prime}}\eta ^{\dag}_{E^{\prime\prime}} \cdots |\pm \rangle $. The complex zero mode operators $c$ and $c^{\dag}$ connect two Hilbert spaces with different parities as
\beq
c | + \rangle = | - \rangle, \hspace{3mm}
c^{\dag} | - \rangle = | + \rangle .
\eeq

We notice that the complex fermion is indispensable for the preservation of the fermion parity~\cite{semenoff,chamon}. Indeed, $\gamma^{(a)}$ itself can be diagonalized as $\gamma^{(a)} |a \pm \rangle = \pm  | a \pm \rangle$, where
\begin{gather}
\left|1 \pm \right\rangle = \frac{1}{\sqrt{2}}(\left| + \right\rangle \pm \left| - \right\rangle), 
\label{eq:majovector0} \\
\left|2 \pm \right\rangle = \frac{1}{\sqrt{2}}(e^{i\frac{\pi}{4}}\left| + \right\rangle 
\pm e^{-i\frac{\pi}{4}}\left| - \right\rangle).
\label{eq:majovector}
\end{gather}
It turns out that $| a \pm \rangle$ are superpositions of two ground states with opposite fermion parities, and thus they cannot be physical states preserving the fermion parity. 

{\it Non-Abelian statistics.}--- 
Equations~(\ref{eq:majovector0}) and \eqref{eq:majovector} represent two novel quantum phenomena: (i) nonlocal correlation~\cite{semenoff,semenoff06} and (ii) non-Abelian statistics~\cite{ivanovPRL01}. These two phenomena require Majorana zero modes to be spatially separated and well isolated from other quasiparticle states with higher energies. 

Let us start to consider a quantum vortex, where the pair potential is given by the Fourier transformation $\Delta ({\bm r},{\bm r}_{12}) = \int \frac{d{\bm k}}{(2\pi)^3}\Delta ({\bm k},{\bm r})e^{i{\bm k}\cdot{\bm r}_{12}}$ as
\beq
\Delta ({\bm r},{\bm r}_{12}) = e^{i\kappa \phi}\Delta ({\bm r})\Phi ({\bm r}_{12}),
\label{eq:vortex}
\eeq
where ${\bm r}\!=\!({\bm r}_1+{\bm r}_2)/2$ is the center-of-mass coordinate of the Cooper pair and $\Phi ({\bm r}_{12})$ is the gap function on the relative coordinate, ${\bm r}_{12} \!=\! {\bm r}_1-{\bm r}_{2}$. For simplicity, we take the vorticity $\kappa$ as $\kappa \!=\! 1$. As clarified in Secs.~\ref{sec:spinless} and \ref{sec:vortices}, a quantum vortex with a singular core in topological superconductors can host topologically protected Majorana zero modes. The vortices with Majorana zero modes behave as non-Abelian anyons when each vortex has a single Majorana zero mode.~\cite{ivanovPRL01} The representations of the braiding operation of vortices are obtained as a discrete set of the unitary group that manipulates the occupation of complex fermions~\cite{ivanovPRL01,nayakRMP08}. Since the zero modes are topologically protected against quantum decoherence, they offer a promising platform to realize topological quantum computation~\cite{ivanovPRL01,ohmi,nayakRMP08,kitaev,freedman,sarmaPRL05,tewariPRL07-1,zhangPRL07}.

\begin{figure}
\includegraphics[width=80mm]{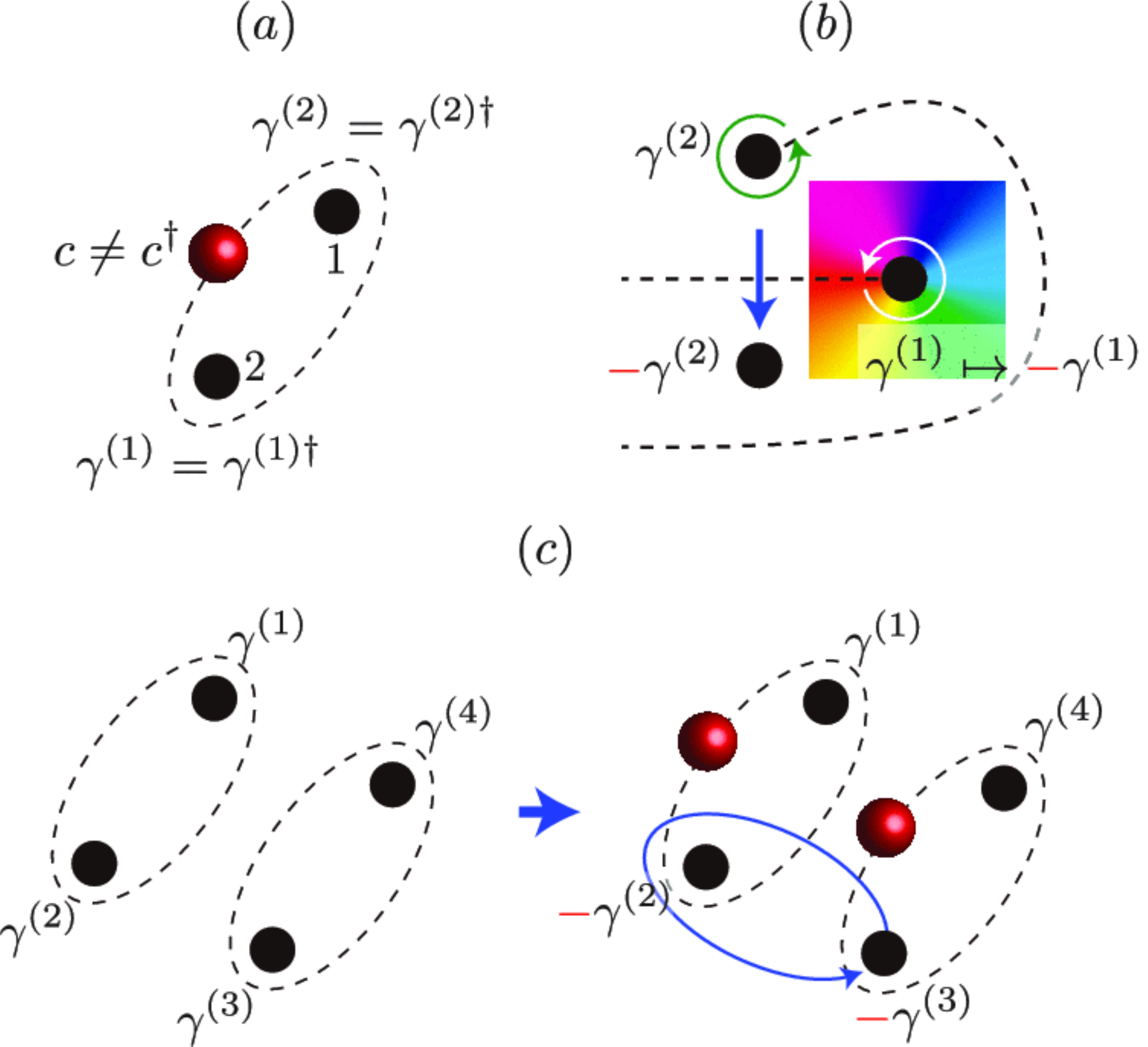}
\caption{(Color online) (a) Schematic picture of Majorana fermions $\gamma$ and complex fermion $c$ bound to the vortices labeled by ``1'' and ``2''. Schematic pictures of (b) the consequence of moving a vortex across the branch cut and (c) braiding vortices in the case of four vortices.} 
\label{fig:braiding}
\end{figure}

We now summarize the braiding rule of two vortices labeled by ``1'' and ``2'', which host Majorana zero modes $\gamma^{(1)}$ and $\gamma^{(2)}$. As illustrated in Fig.~\ref{fig:braiding}(b), each vortex is accompanied by a ``branch cut'' in which the pair potential has a $2\pi$-phase change attributable to the vortex phase winding. When a quasiparticle crosses the branch cut, it generally acquires the $\pi$-phase shift. The Bogoliubov quasiparticle operator is expressed in terms of ${\bm \Psi}\!=\!(\psi,\psi^{\dag})$ as
$\eta _{E} = \int \left[ u^{\ast}_E({\bm r})\psi ({\bm r}) + v^{\ast}_E({\bm r})\psi^{\dag}({\bm r}) \right] d{\bm r}$,
where $u_{E}$ and $v_{E}$ are the particle and hole components of quasiparticles with energy $E$, i.e., ${\bm \varphi}_E({\bm r})=[u_E({\bm r}),v_E({\bm r})]^{\rm T}$. As depicted in Fig.~\ref{fig:braiding}(b), the global phase rotation of the Cooper pair potential by $\varphi$ involves the phase shift $\varphi /2$ ($-\varphi/2$) in the particle (hole) component. As a result, the quasiparticle operator crossing the branch cut changes $\eta _E\mapsto -\eta _E$, regardless of $E$, and the zero modes behave as
\beq
\gamma^{(1)} \mapsto -\gamma^{(1)}, \quad \gamma^{(2)} \mapsto -\gamma^{(2)}
\label{eq:braiding}
\eeq
when the vortex ``2'' encircles the vortex ``1''. After this operation, the complex fermion changes the phase as $c\mapsto -c$.

Now, consider the four vortices illustrated in Fig.~\ref{fig:braiding}(c), where Majorana zero modes bound to two vortices form complex fermions as $c_{\rm L} \!\equiv\! (\gamma^{(1)}+i\gamma^{(2)})/\sqrt{2}$ and $c_{\rm R} \!\equiv\! (\gamma^{(3)}+i\gamma^{(4)})/\sqrt{2}$. When the vortex ``3'' adiabatically encircles the vortex ``2'', both Majorana zero mode operators acquire the $\pi$ phase shift, $\gamma^{(2)}\mapsto -\gamma^{(2)}$ and $\gamma^{(3)}\mapsto -\gamma^{(3)}$ in the same manner as Eq.~\eqref{eq:braiding}. Therefore, the braiding operation of the vortices changes the complex fermion operators $c_{\rm L}$ and $c_{\rm R}$ as
\beq
c_{\rm L} \mapsto c^{\dag}_{\rm L}, \quad c_{\rm R} \mapsto c^{\dag}_{\rm R},
\eeq
with $c_{\rm L}|00\rangle=c_{\rm R}|00\rangle=0$. 
This implies that, if the vacuum of complex fermions, $|00\rangle$, is initially prepared, the above braiding generates a pair of complex fermions and transforms the initial state to the completely different state $|11 \rangle $. The final state is orthogonal to the initial state, $\langle 11 | 00 \rangle = 0$. 

The braiding rule can be generalized to $2N$ vortices with $2N$ Majorana zero modes. The $2N$ Majorana zero modes are combined into $N$ complex fermions, which give rise to the $2^{N-1}$-fold degeneracy of ground states while preserving fermion parity. As discussed above, when the $i$-th vortex is exchanged with the $(i+1)$-th vortex, the zero modes behave as 
\beq
\gamma _i \mapsto \gamma _{i+1}, \quad \gamma _{i+1} \mapsto -\gamma _i.
\label{eq:braiding2}
\eeq
Here, note that in Eq.~\eqref{eq:braiding}, we have considered the process that a vortex encircles another one, which corresponds to the two successive vortex exchange processes of Eq.~\eqref{eq:braiding2}. 
The explicit formulas for the representation of the braiding operation that satisfies Eq.~\eqref{eq:braiding2} were given by Ivanov~\cite{ivanovPRL01} in terms of the zero mode operators as
\beq
\tau _i = \exp\left( \frac{\pi}{4}\gamma _{i+1}\gamma _i\right) = \frac{1}{\sqrt{2}}(1+\gamma _{i+1}\gamma _i),
\eeq
where a phase factor is omitted. For $N=1$, there is only a single ground state in each sector with definite fermion parity, and the exchange of two vortices results in the global phase of the ground state by $e^{i\pi /4}$.

For four vortices ($N=2$), twofold degenerate ground states exist, $|00 \rangle \!\equiv\! |{\rm vac}\rangle$ and $|11\rangle \!=\! c^{\dag}_{\rm L}c^{\dag}_{\rm R}|{\rm vac}\rangle$ in the sector of even fermion parity, and $|10\rangle \!=\! c^{\dag}_{\rm L}|{\rm vac}\rangle$ and $|01\rangle \!=\! c^{\dag}_{\rm R}|{\rm vac}\rangle$ exist in the sector of odd fermion parity, respectively. For the even-parity sector, the representation matrix for the intravortex exchange [$1\leftrightarrow 2$ and $3\leftrightarrow 4$ in Fig.~\ref{fig:braiding}(c)] is given by 
\beq
\tau _1 = \tau _3 = e^{-i\frac{\pi}{4}} |00\rangle \langle 00 |
+ e^{i\frac{\pi}{4}} |11\rangle \langle 11 |.
\eeq
This merely rotates the global phase of the ground state similarly to the $N=1$ case. In contrast, the representation matrix for the intervortex exchange [$2\leftrightarrow 3$ in Fig.~\ref{fig:braiding}(c)] has the mixing terms of the two degenerate ground states $|00\rangle$ and $|11\rangle$,
\beq
\tau _2 = \frac{1}{\sqrt{2}} \left[ 
| 00 \rangle \langle 00| - i | 00 \rangle \langle 11 | 
+ | 11 \rangle \langle 11| - i | 11 \rangle \langle 00 | 
\right].
\eeq
This implies that, if the state $|00\rangle$ is initially prepared, the braiding operation transforms the ground states under $\tau _2$ as
\beq
| 00 \rangle \stackrel{\tau_2}{\rightarrow} | 00 \rangle -i |11\rangle \stackrel{\tau_2}{\rightarrow}
|11\rangle
\eeq
up to a global phase factor. Hence, the braiding operation creates a pair of complex fermions, implying the non-Abelian statistics of vortices with Majorana zero modes. 

Ivanov~\cite{ivanovPRL01} clarified the non-Abelian anyonic behaviors of Majorana fermions from the viewpoint of unitary transformations acting on the Hilbert space. The unitary transformations associated with vortex interchange are found to be identical to those derived by Nayak and Wilczek~\cite{nayakNPB96} with conformal field theory. In addition to these arguments, Stern {\it et al.}~\cite{sternPRB04} gave a physical picture of the effect of braiding vortices on the manifold of topologically degenerate ground states. They revealed that the non-Abelian statistics originates from the quantum entanglement and the accumulation of a geometric phase by braiding vortices. For spinless chiral $p$-wave pairing with $2N$ vortices, a $2^{N}$-dimensional Fock space can be constructed from a set of quasiparticle states bound to the vortex cores. They found ground states to be entangled superpositions of all possible occupations of the core-bound states. The entangled ground state acquires a geometric phase under the adiabatic motion of vortices, depending on the occupancy. This clarifies that braiding vortices give rise to the change in the relative phase difference between different components of a superposed ground state.

An effective realization of non-Abelian statistics requires Majorana fermions without internal degrees of freedom. Possible realizations include axion strings~\cite{satoPL03}, fermionic cold atoms with a $p$-wave Feshbach resonance~\cite{read,gurarieAP07,mizushimaPRL08,tewariPRL07-2}, proximity-induced superconductivity on the surface of a topological insulator~\cite{fuPRL08}, an $s$-wave superconductor with the Rashba spin-orbit interaction and the Zeeman field~\cite{satoPRL09,satoPRB10-2,sauPRL10}, one-dimensional nanowire systems,~\cite{lutchynPRL10,oregPRL10,aliceaNP11} and ferromagnetic atomic chains on a bulk superconductor~\cite{nadjperge,nadjperge14}. In addition, the low-energy physics of half-quantized vortices in spinful chiral $p$-wave superconductors is describable as a spinless chiral superconductor~\cite{ivanovPRL01}. In Sec.~\ref{sec:vortices}, we will examine the thermodynamic stability of half-quantum vortices in a rotating $^3$He-A thin film.



\subsubsection{Mirror-symmetry-protected Majorana fermions}
\label{sec:mirror}

Although spin-polarized Majorana zero modes offer a good platform for realizing topologically protected quantum computation, it is difficult to realize such zero modes in the real materials listed above. As we will discuss in Sec.~\ref{sec:HQV}, for instance, the configuration of the half-quantum vortex is rather thermodynamically unstable~\cite{chungPRL07,kawakamiPRB09,vakaryukPRL09,kawakamiJPSJ10,kawakamiJPSJ11,kondoJPSJ12,nakaharaPRB14}, and its realization remains as an experimentally challenging task in both superfluid $^3$He-A thin films and the spin-triplet superconductor Sr$_2$RuO$_4$~\cite{yamashitaPRL08,budakian}. For a superconducting nanowire, although a zero-bias conductance peak was observed by several experimental groups~\cite{mourik12,deng12,das12,finck13,churchill13}, the issue of the Majorana nature has not been settled yet~\cite{stanescuJPCM13}. In addition, the spin-polarized $p$-wave superfluidity has not been realized yet in ultracold experiments, since the $p$-wave bound pairs of neutral atoms have a much shorter lifetime than the typical timescale in which a superfluid phase transition is accomplished~\cite{inadaPRL08,chinRMP10}. 

Symmetry-protected non-Abelian anyons in spinful superconductor and superfluids were discussed in Refs.~\citeonline{ueno13}, \citeonline{sato14}, and \citeonline{fangPRL14} for class D and Ref.~\citeonline{liuprx14} for one-dimensional DIII superconductors~\cite{sunPRB14}. In the former (latter) case, the mirror reflection symmetry (TRS) plays an essential role on the topological protection of non-Abelian nature. Contrary to the standard wisdom, the mirror reflection symmetry protects a pair of topologically stable Majorana zero modes in integer quantum vortices of spinful superconductors and superfluids~\cite{ueno13,sato14}. From the Majorana zero modes protected by the mirror symmetry, the integer quantum vortex obeys the non-Abelian anyon statistics~\cite{sato14}. The non-Abelian statistics of vortices having multiple Majorana fermions has also been discussed in Refs.~\citeonline{yasuiPRB11} and \citeonline{hironoPRB12} in the context of high-energy physics.

{\it Mirror symmetry and topological invariant.}--- 
To demonstrate symmetry-protected non-Abelian anyons in spinful systems, we begin with a generic form of the BdG Hamiltonian for superconductors with the spontaneous breaking of TRS. Without losing generality, we suppose that both $\varepsilon ({\bm k})$ and $\Delta ({\bm k})$ are $2N \times 2N$ matrices composed of the $1/2$ spin ($\uparrow$ and $\downarrow$) and $N$ orbitals. Let us now assume that the normal state holds the mirror symmetry with respect to the $xy$-plane, which can be associated with crystalline symmetry, $M_{xy} \varepsilon ({\bm k},{\bm R}) M^{\dag}_{xy} = \varepsilon (\underline{\bm k}_{\rm M},\underline{\bm R}_{\rm M})$. The mirror operator is defined in Eq.~\eqref{eq:mirrorop}.

In the Nambu space, the mirror operator is naturally extended as
${\mathcal{M}} = 
{\rm diag}( e^{i\chi}M , e^{-i\chi} M^{\ast} )$
by taking into account the ${\rm U}(1)$ gauge symmetry $e^{i\chi}$ and the ambiguity of the overall phase $e^{i\varphi}$. It turns out that only when $\Delta ({\bm k})$ has a definite parity under $M$ as~\cite{ueno13,sato14}
\beq
M\Delta ({\bm k},{\bm R})M^{\rm T} = \eta \Delta (\underline{\bm k}_{\rm M},\underline{\bm R}_{\rm M}), \quad
\eta = \pm
\label{eq:dmirror}
\eeq 
the BdG Hamiltonian is invariant under the mirror reflection, 
\beq
{\mathcal{M}}^{\eta}\mathcal{H}({\bm k},{\bm R}){\mathcal{M}}^{\eta \dag} 
= {\mathcal{H}}(\underline{\bm k}_{\rm M},\underline{\bm R}_{\rm M}).
\label{eq:Hmirror}
\eeq
Here, the mirror operator in the Nambu space is given by
\beq
{\mathcal{M}}^{\eta} = 
\left( 
\begin{array}{cc}
M & 0 \\ 0 & \eta M^{\ast}
\end{array}
\right), \quad ({\mathcal{M}}^{\eta})^2 = -1.
\label{eq:mirror}
\eeq

We first consider the case of $D\!=\! 0$, where the topological defects in $d\!=\! 2$ and $d\!=\! 3$ correspond to the edge and surface, respectively. For ${\bm k}_{\Lambda}$ with ${\bm k}_{\Lambda}=\underline{\bm k}_{\Lambda,{\rm M}}$, 
the BdG Hamiltonian is commutable with the mirror reflection operator as
\beq
\left[ {\mathcal{M}}^{\eta}, \mathcal{H}({\bm k}_{\Lambda}) \right] = 0.
\label{eq:commut}
\eeq
Therefore, in the mirror-invariant plane, ${\bm k}\in {\bm k}_{\Lambda}$, the BdG Hamiltonian $\mathcal{H}({\bm k})$ is block-diagonal, $\mathcal{H}({\bm k}) \!=\! {\rm diag}[\mathcal{H}^{(+i)}({\bm k}),\mathcal{H}^{(-i)}({\bm k})]$, in the diagonal basis of $\mathcal{M}^{\eta}$ with eigenvalues of $\mathcal{M}^{\eta}=\pm i$.
The block-diagonal Hamiltonians in each mirror subsector $\lambda \!=\! \pm i$, $\mathcal{H}^{(\lambda)}({\bm k})$, have eigenvectors and eigenvalues $|u^{(\lambda)}_n({\bm k})\rangle$ and $E^{(\lambda)}({\bm k})$, respectively.

On a two-dimensional mirror-invariant plane ${\bm k}_{\Lambda} \!\in\! S^2$, the nontrivial topological properties are characterized by the first Chern number. Using the eigenvectors of occupied states in each mirror subsector, one can construct the gauge field in the momentum space as $\mathcal{A}^{(\lambda)}_{\mu} ({\bm k}_{\Lambda}) \!=\! \sum _{E^{(\lambda)}_n<0}\langle u^{({\lambda})}_n({\bm k}_{\Lambda})| \partial _{k_{\mu}}u^{({\lambda})}_n({\bm k}_{\Lambda})\rangle$. Then, for $d=2$ in class D, the first Chern number is well-defined in each mirror subsector, 
\beq
{\rm Ch}^{(\lambda)}_1 = \frac{i}{2\pi}\int _{S^2} \mathcal{F}^{(\lambda)} = \mathbb{Z}.
\label{eq:mirrorCh}
\eeq
The integral is taken over the mirror-invariant plane ${\bm k}_{\Lambda} \!\in\! S^2$. The Berry curvature is defined in each mirror subsector as
$\mathcal{F}^{(\lambda)} = d\mathcal{A}^{(\lambda)}$. The nonzero value of ${\rm Ch}^{(\lambda)}_1$ ensures the existence of the zero energy edge state in the $\lambda$ subsector.

For a vortex state, the pair potential is generally given by the Fourier transformation of Eq.~\eqref{eq:vortex} as
\beq
\Delta ({\bm k},{\bm r}) = e^{i\kappa \phi}\Delta ({\bm r})\Phi ({\bm k}),
\eeq
where $\phi$ is the azimuthal angle around the vortex core. When the vortex line is perpendicular to the mirror reflection plane, $\phi$ is mirror-invariant. Hence, the BdG Hamiltonian encircling the vortex line, which is given by $\mathcal{H}({\bm k},{\bm R})=\mathcal{H}({\bm k}_{\Lambda}, \phi)$ on the mirror invariant momentum, is also block-diagonal in the mirror eigen basis.  
Each mirror sector corresponds to $d=2$ and $D=1$, and thus it is categorized into $\delta=1$ in class D in Table~\ref{table1}.  The relevant topological invariant is the mirror $\mathbb{Z}_2$ number given as the integral of the Chern-Simons 3-form over the base space $S^2\times S^1$ as
\beq
\nu^{(\lambda)} = \left(\frac{i}{2\pi}\right)^2 \int _{S^2\times S^1} {\rm CS}^{(\lambda)}_3 
= \mathbb{Z} \quad {\rm mod}~2,
\eeq
where ${\rm CS}^{(\lambda)}_3$ is the Chern-Simons 3-form defined in each mirror subsector. The quasiparticles in the case of  $\nu^{(\lambda)}\!=\! 0$ are topologically trivial, while $\nu^{(\lambda)}\!=\! 1$ indicates the existence of a single zero energy state bound to the vortex core in the $\lambda$ subsector.

{\it Mirror Majorana fermions.}--- From the bulk-edge correspondence, the mirror Chern number is equal to the number of gapless edge states unless the mirror symmetry is broken macroscopically. Using the mirror Chern number, Hsieh {\it et al.}~\cite{hsieh} succeeded in revealing the nontrivial topology of the semiconductor SnTe in the rocksalt structure. Although this material has an even number of band inversions, and thus the $\mathbb{Z}_2$ number as an ordinary topological insulator is trivial, an even number of Dirac cones are topologically stable owing to the mirror Chern number. This is the first discovered {\it topological crystalline insulator}~\cite{fuPRL11} where the topological property is protected by crystalline symmetry. 

The superconducting counterparts are Sr$_2$RuO$_4$ and the $^3$He-A thin film~\cite{ueno13,sato14}. This superconductor and superfluid, however, display essentially different aspects at the same time. The PHS in Eq.~\eqref{eq:PHS} imposes the self-charge conjugation property on the field operator, as in Eq.~\eqref{eq:selfcharge}. The self-charge-conjugation property is a key to understanding the non-Abelian Majorana nature of topologically protected zero modes. We here overview the generic argument of the relation between the mirror symmetry and the Majorana fermions. In Secs.~\ref{sec:abmslab} and \ref{sec:IQV}, it is clarified that the mirror Chern number manifests the connection between the ${\bm d}$-vector orientation and the non-Abelian Majorana fermions.


As mentioned above, the $^3$He-A thin film and Sr$_2$RuO$_4$ may be accompanied by multiple Majorana zero modes owing to the spin degrees of freedom. To realize the non-Abelian statistics in multiple Majorana zero modes, Refs.~\citeonline{ueno13} and \citeonline{sato14} emphasized the role of the PHS in each mirror subsector. A subsector of $\mathcal{H}({\bm k})$ with a definite eigenvalue of $\mathcal{M}^{\eta}=\pm i$ does not always have the PHS within the subsector because the symmetry can exchange a pair of subsectors. Only when the subsectors hold the PHS, Majorana fermions exist as non-Abelian anyons in spinful superconductors and superfluids. Below, we refer to such Majorana fermions as mirror Majorana fermions.

Let us now elucidate the condition for the mirror subsector to host its own PHS and mirror Majorana fermions. When the mirror operator is commutable with $\mathcal{H}$, the eigenvector of $\mathcal{H}$ is simultaneously the eigenvector of the mirror reflection operator $\mathcal{M}^{\eta}$ at ${\bm k}_{\Lambda}$,
\beq
\mathcal{H} | u^{(\lambda)}_n ({\bm k}_{\Lambda})\rangle 
= E^{(\lambda)}_n ({\bm k}_{\Lambda}) | u^{(\lambda)}_n ({\bm k}_{\Lambda})\rangle 
\eeq
and 
\beq
\mathcal{M}^{\eta} | u^{(\lambda)}_n ({\bm k}_{\Lambda})\rangle  = \lambda | u^{(\lambda)}_n ({\bm k}_{\Lambda})\rangle .
\eeq
Hence, all quasiparticles are categorized into two mirror subsectors with $\lambda = \pm i$. In general, the PHS maps $| u^{(\lambda)}_n ({\bm k}_{\Lambda})\rangle$ in mirror subsector $\lambda$ to $\mathcal{C}|u^{(\lambda^{\prime})}_n ({\bm k}_{\Lambda})\rangle$. The self-charge conjugation condition \eqref{eq:selfcharge} is satisfied when the mapped state has the same eigenvalue of $\mathcal{M}^{\eta}$ as the original state, $
\mathcal{M}^{\eta} [
\mathcal{C}| u^{(\lambda)}_n ({\bm k}_{\Lambda})\rangle]  
= \lambda [
\mathcal{C}| u^{(\lambda)}_n ({\bm k}_{\Lambda})\rangle]$.
This leads to the condition 
\beq
\left\{ \mathcal{C}, \mathcal{M}^{\eta}\right\} = 0.
\label{eq:mirrorMF}
\eeq
In accordance with the notation in Ref.~\citeonline{shiozakiPRB14}, $\mathcal{M}^{\eta}$ is labeled as $U^{-}_-$, and the topological class is categorized into class D subject to the additional discrete symmetry $U^-_-$. 




\begin{figure}
\includegraphics[width=85mm]{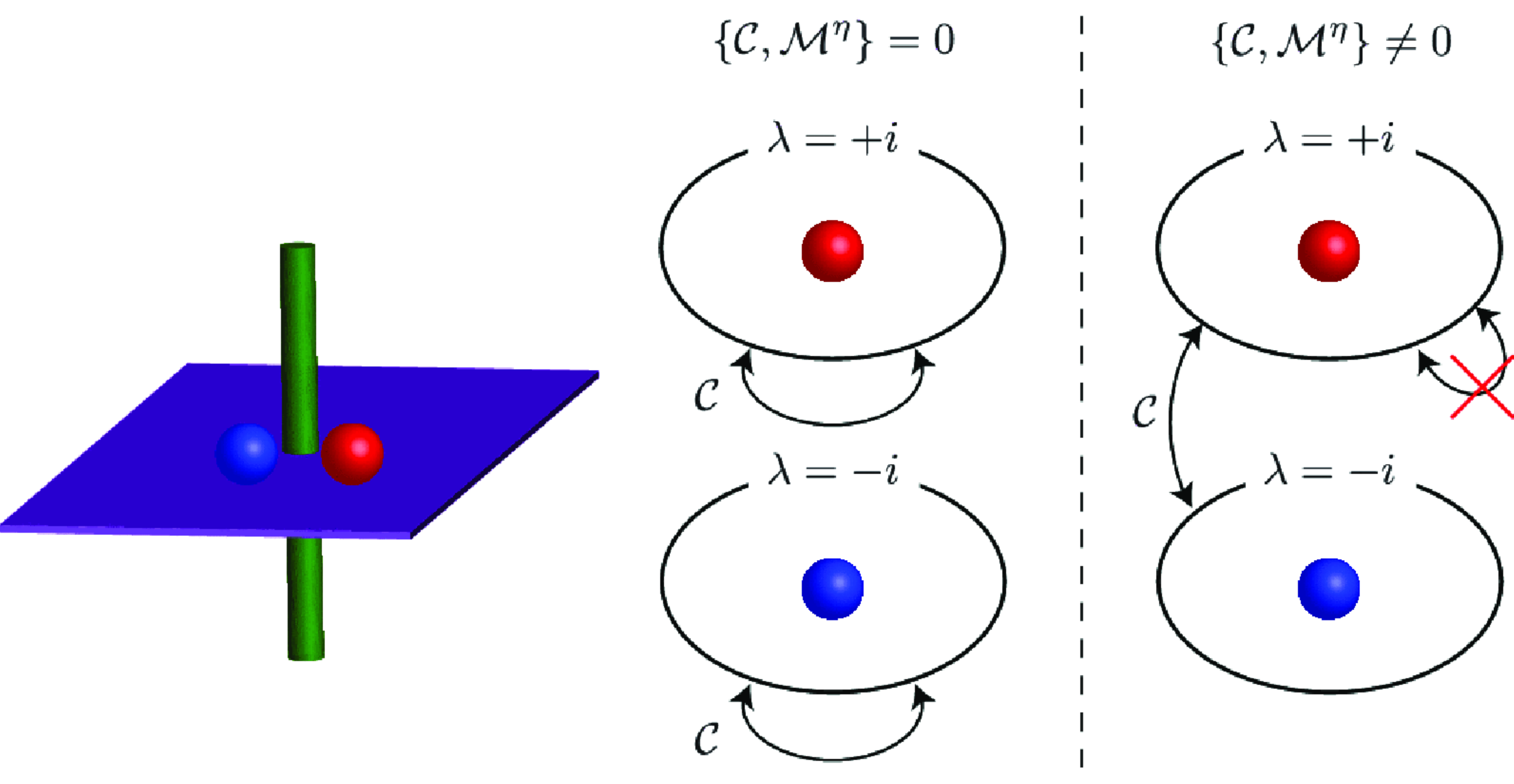}
\caption{(Color online) Schematic picture of mirror Majorana fermions. The PHS is present in each mirror subsector (class D) when $\mathcal{C}$ is anticommutable with $\mathcal{M}^{\eta}$, otherwise, the PHS is absent (class A).} 
\label{fig:mirror}
\end{figure}

Equation \eqref{eq:mirrorMF} implies that, if the condition is satisfied, the PHS exists in each mirror subsector (see Fig.~\ref{fig:mirror}), and the existence of the Majorana fermion is ensured by the mirror reflection symmetry. In this situation, each mirror subsector is regarded as class D, and the topological property is characterized by the mirror Chern number ${\rm Ch}^{(\lambda)}_1$ in Eq.~\eqref{eq:mirrorCh}. It is worth mentioning that the condition \eqref{eq:mirrorMF} is associated with the orientation of the ${\bm d}$-vector in spin-triplet superconductors and superfluids. We clarify the connection of the mirror symmetry with the order parameter configuration in Sec.~\ref{sec:abmslab} and examine the non-Abelian statistics of mirror Majorana fermions hosted by integer quantum vortices in Sec.~\ref{sec:IQV}. Even though integer quantum vortices are accompanied by spinful Majorana fermions, the mirror symmetry protects multiple Majorana fermions as non-Abelian anyons. 

For the mirror operator that does not satisfy Eq.~\eqref{eq:mirrorMF}, as shown in Fig.~\ref{fig:mirror}, the PHS maps a state $|u^{(\lambda)}_n({\bm k})\rangle$ to an eigenstate in a different subsector. This implies that PHS is absent in each mirror subsector and the system belongs to class A similarly to a quantum Hall state. In this case, only Dirac fermions can be realized even if the mirror Chern number and mirror $\mathbb{Z}_2$ number are nontrivial.

\subsubsection{Chiral-symmetry-protected Majorana fermions}
\label{sec:chiral}

Apart from the non-Abelian statistics, Majorana fermions possess another remarkable facet, that is, an intrinsic anisotropy of the local spin operator, when the system holds the chiral symmetry \eqref{eq:chiral}. The local spin operator for spin $1/2$ fermions in the Nambu space is defined as 
\beq
S_{\mu}({\bm r}) \equiv \frac{1}{4}{\bm \Psi}^{\dag}({\bm r})
\left( 
\begin{array}{cc}
\sigma _{\mu} & 0 \\ 0 & - \sigma^{\rm T} _{\mu}
\end{array}
\right){\bm \Psi}({\bm r}).
\eeq
It is obvious that, if only a single Majorana zero mode exists, i.e., $N=1$, the local spin operator is identically zero as well as the local density operator, $\rho ({\bm r}) \!=\! S_{\mu}({\bm r}) \!=\! 0$. Hence, the single Majorana zero mode cannot yield the coupling to both the local density fluctuation and the magnetic response. 

For spinful systems, however, multiple Majorana zero modes may appear. As discussed above, if the system holds the mirror reflection symmetry, the Majorana zero modes can possess non-Abelian properties. Here, we show that the chiral symmetry is responsible for the Ising-like magnetic response of Majorana fermions as
\beq
\rho({\bm r}) = 0, \quad {\bm S} = S({\bm r})\hat{\bm a}.
\label{eq:MIS1}
\eeq
The direction of the Ising magnetic response is fixed to be parallel to the spin rotation axis $\hat{\bm a}$ associated with the magnetic point group symmetry. 

The so-called Majorana Ising spin in Eq.~\eqref{eq:MIS1} was originally revealed by analytically solving the BdG equation within the Andreev approximation for topological phases of noncentrosymmetric superconductors~\cite{satoPRB09-2} and $^3$He-B~\cite{chungPRL09,volovikJETP09,nagatoJPSJ09}. Shindou {\it et al.}~\cite{shindouPRB10} examined the coupling of a spin-$1/2$ magnetic impurity to Majorana fermions bound at the edge of two-dimensional topological superconductors. Owing to quantum dissipation from the Majorana Ising spin, the quantum impurity spin yields a strongly anisotropic and singular magnetic response, where the electron spin resonance may serve as a local probe for Majorana Ising spins.


It has recently been recognized that the Majorana Ising spin \eqref{eq:MIS1} is a consequence of the symmetry-protected topological phase associated with the chiral symmetry~\cite{mizushimaPRL12,shiozakiPRB14}. Following Refs.~\citeonline{mizushimaPRL12} and \citeonline{shiozakiPRB14}, let us now prove the generic result for the Majorana Ising spin \eqref{eq:MIS1}. Time-reversal-invariant superconductors and superfluids may hold the magnetic point group symmetry that is obtained by combining the time-reversal $\mathcal{T}$ and either twofold spin rotation, mirror reflection, or twofold rotation, even if each discrete symmetry is independently broken. The latter operators can be described with the ${\rm SU}(2)$ spin matrix $U(\hat{\bm a},\pi)\!=\! -i{\bm \sigma}\cdot\hat{\bm a}$, where $\hat{\bm a}$ is the spin rotation axis. The antiunitary operator relevant to the magnetic point group (or $P_3$) symmetry is then obtained from Eq.~\eqref{eq:p2p3v2} as
\beq
A _{\rm spin} = \mathcal{T}U(\hat{\bm a},\pi) .
\label{eq:aspin}
\eeq
Since $\mathcal{T} \!=\! i\sigma _yK$ for spin $1/2$ fermions, one finds that $A^2 \!=\! 1$. The antiunitary operator is extended to the Nambu space, which typically forms $A \!=\! {\rm diag}(A_{\rm spin}, \pm A^{\ast}_{\rm spin})$, which is recast into ${A} =  {\rm A}_{\rm spin}\tau _0$ or $ {A} = {\rm A}_{\rm spin}\tau _z $, depending on the gap function. Following the notation in Ref.~\citeonline{shiozakiPRB14}, the antiunitary operator acting on the Nambu space is labeled as $A^{+}_{\epsilon _{\mathcal{C}}}$, where $A^{+}_{\epsilon _{\mathcal{C}}} \mathcal{C} \!=\! \epsilon _{\mathcal{C}}\mathcal{C}A^{+}_{\epsilon _{\mathcal{C}}}$ with $\epsilon _{\mathcal{C}}\!=\! \pm 1$.

To prove the Majorana Ising property, we note two relations that the zero energy states must satisfy. First, PHS \eqref{eq:PHS} imposes the following relation on the zero energy state:
\beq
\mathcal{C} {\bm \varphi}^{(a)}_0 ({\bm r}) =  {\bm \varphi}^{(a)}_0 ({\bm r}).
\label{eq:PHS0}
\eeq
We also have the chiral symmetry obtained by combining PHS and the magnetic point group symmetry as 
\beq
\Gamma = e^{i\alpha}\mathcal{C}A^+_{\epsilon _{\mathcal{C}}},
\label{eq:CS0}
\eeq
where $\alpha$ is chosen so as to obey $\Gamma^2\!=\! +1$. Since the chiral symmetry $\Gamma$ is commutable with $\mathcal{C}$, the zero energy state is the simultaneous eigenstate of $\mathcal{C}$ and $\Gamma$, 
\beq
\Gamma  {\bm \varphi}^{(a)}_0 ({\bm r}) = \lambda _{\Gamma}  {\bm \varphi}^{(a)}_0 ({\bm r}),
\eeq
where $\lambda _{\Gamma} \!=\! \pm 1$. As clarified in Sec.~\ref{sec:index}, the winding number $w_{\rm 1d}$ gives the number of  zero energy states in the $\lambda _{\Gamma} \!=\! + 1$ or $-1$ sector.

The PHS in Eq.~\eqref{eq:PHS0} allows one to parameterize ${\bm \varphi}^{(a)}_0({\bm r})$ as $
{\bm \varphi}^{(a)}_0({\bm r}) = [
\chi^{(a)}({\bm r}), \chi^{(a)\ast}({\bm r})]^{\rm T}$ with two-dimensional spinors $\chi^{(a)}({\bm r})$.
In addition, the chiral symmetry \eqref{eq:CS0} imposes the following constraint on the spinor
$
\chi^{(a)}({\bm r}) = \epsilon _{\mathcal{C}}\lambda _{\Gamma} e^{i\alpha} U_{\hat{\bm a}}(\pi) \mathcal{T}
\chi^{(a)}({\bm r}) $.
These two constraints due to PHS and chiral symmetry imply the intrinsic relation between Nambu and spin spaces in the Majorana zero modes, i.e., the equivalence between $\psi _{\sigma} ({\bm r})$ and $\psi^{\dag}_{\sigma^{\prime}}({\bm r})$ up to a ${\rm U}(1)$ phase factor. As a result, Majorana fermions that hold the magnetic point group symmetry possess the relation
\beq
\left(
\begin{array}{c}
\psi _{\uparrow}({\bm r}) \\ \psi _{\downarrow}({\bm r}) 
\end{array}
\right) = \epsilon _{\mathcal{C}}\lambda _{\Gamma} e^{i\alpha} U_{\hat{\bm a}}(\pi) 
\left( 
\begin{array}{c}
\psi^{\dag} _{\downarrow} ({\bm r}) \\ - \psi^{\dag}_{\uparrow}({\bm r})
\end{array}
\right).
\label{eq:miscond}
\eeq
Using this relation, one finds that the local spin operator of Majorana fermions is recast into
\beq
S_{\mu} ({\bm r}) = \frac{1}{4}\epsilon _{\mathcal{C}}\lambda _{\Gamma} e^{i\alpha}
{\bm \psi}^{\rm T}({\bm r})
\sigma _y\left\{ {\bm \sigma}\cdot\hat{\bm a}, \sigma _{\mu} \right\} {\bm \psi}({\bm r}),
\eeq
where ${\bm \psi} \!= \! (\psi _{\uparrow},\psi _{\downarrow})$. This is recast into the Majorana Ising spin in Eq.~\eqref{eq:MIS1}.

The {\it symmetry-protected} Majorana Ising spin was first demonstrated in $^3$He-B confined in a restricted geometry under a parallel magnetic field~\cite{mizushimaPRL12,mizushimaPRB12}. The topological superfluidity in this system is protected by the magnetic point group symmetry, where $A_{\rm spin}$ in Eq.~\eqref{eq:aspin} is associated with the joint rotation of spin and orbital about a particular axis, ${\rm SO}_{{\bm L}+{\bm S}}(2)$. This is called the hidden ${\bm Z}_2$ symmetry, which protects the Majorana Ising spin \eqref{eq:MIS1} in $^3$He-B even in the presence of a magnetic field.

The theory of Majorana Ising spins protected by the chiral symmetry was further extended to time-reversal-invariant superconductors, where the magnetic point group symmetry arises from other discrete operations, such as mirror reflection.~\cite{mizushimaJPCM15} The role of mirror reflection was first clarified in quasi-one-dimensional fermionic gases with a synthetic gauge field~\cite{mizushimaNJP13} and then applied to the heavy-fermion superconductor UPt$_3$~\cite{tsutsumiJPSJ13,mizushimaPRB14} and the superconducting doped topological insulator Cu$_x$Bi$_2$Se$_3$~\cite{sasakiPC15}. The role of the chiral symmetry in superconducting nanowires was revealed by Tewari and colleagues~\cite{tewariPRL12,tewariPRB12,Dumitrescu}. Shiozaki and Sato~\cite{shiozakiPRB14} clarified the more generic condition for the Majorana Ising spin.

\section{Topology and BCS-BEC Evolution in Spin-Polarized Chiral Superconductors}
\label{sec:spinless}

In this section, we consider a two-dimensional time-reversal-breaking superconductor that preserves $S_z$ as a pedagogical example to capture the topological aspect. After briefly overviewing the topology of the BdG Hamiltonian within the Andreev approximation, we examine the properties of zero-energy states bound to a topological defect in the vicinity of the BCS-BEC topological phase transition.

\subsection{Jackiw-Rebbi Index theorem for chiral superconductors}
\label{sec:index2}

Let us provide an overview of the generic properties of the Andreev approximation, which offers another facet to the distribution of zero-energy solutions and their properties~\cite{kashiwayaRPP}. The BdG equation within the Andreev approximation is usually mapped onto the one-dimensional Dirac equation,
\beq
\left[
-i v_{\rm F}\partial _y {\tau}_x  + m (y){\tau}_z
\right]\tilde{\bm \varphi}(y) = E \tilde{\bm \varphi}(y),
\label{eq:dirac}
\eeq
where a spatially inhomogeneous mass $m(y)$ is associated with the pair potential that incoming and outgoing quasiparticles are subjected to, and the boundary condition $m(y\rightarrow \pm \infty) ={\rm const.}$ is imposed. 
Jackiw and Rebbi \cite{jackiw} clarified the topologically nontrivial structure of the one-dimensional Dirac equation (\ref{eq:dirac}). The eigenfunction of the zero-energy state is obtained by integrating Eq.~(\ref{eq:dirac}) with $E\!=\! 0$ as
\beq
\tilde{\bm \varphi}(y) = N \exp\left( 
- \frac{1}{v_{\rm F}}\int^{y}_0 m (y^{\prime})dy^{\prime}
\right)\left( 
\begin{array}{c}
1 \\ i
\end{array}
\right),
\label{eq:diracwf}
\eeq
where $N$ is a normalization constant. By assuming that the mass term approaches a uniform value in the limit of $y\rightarrow \pm \infty$, it is obvious that one of the zero-energy solutions is normalizable only when the mass term changes its sign at $y\rightarrow \pm \infty$ as
\beq
\arg m(+\infty) - 
\arg m(-\infty) 
= (2n+1) \pi,
\label{eq:piphase}
\eeq
where $n\!\in\! \mathbb{Z}$. The consequence is that at least one zero-energy solution exists when the mass term $m (y)$ changes its sign at $y\rightarrow\pm\infty$ and the stability of the zero energy state is independent of the detailed structure of the interface.

Equation~(\ref{eq:dirac}) serves as an effective theory for describing the low-lying electronic states of various systems. This includes the one-dimensional Peierls system~\cite{brazovskii,mertsching,horovitz,takayama,yamamoto,nakaharaPRB1981}, spin density waves~\cite{machidaPRB1984-2}, the spin-Peierls system~\cite{fujitaJPSJ1984}, the stripes in high-$T_{\rm c}$ cuprates~\cite{machida1989}, superconducting junction systems~\cite{kashiwayaRPP}, and Fulde-Ferrell-Larkin-Ovchinnikov states~\cite{machidaPRB1984,yoshii}. The simplest form that satisfies the condition \eqref{eq:piphase} is a single-kink solution, e.g., $m(y) \!\propto \! \tanh(y/\xi)$. Using the hypergeometric function, Nakahara~\cite{nakahara:1986} derived the dispersion and wave functions of quasiparticles that are bound to the chiral domain wall of the superfluid $^3$He-A film, where each domain has a different chirality. The self-consistent pair potential is well described with a single-kink shape. 

The counterpart of spatially inhomogeneous fermionic condensates in quantum field theory corresponds to the Gross-Neveu model~\cite{gross} and the Nambu-Jona-Lasinio model,~\cite{nambuPR1961} which give an effective theory for dynamical chiral symmetry breaking~\cite{thies}. Recently, the general solutions to Eq.~(\ref{eq:dirac}) have been proposed by using a technique of the Ablowitz-Kaup-Newell-Segur hierarchy, which is well known in integrable systems~\cite{akns}. These include a single-kink state~\cite{hu,bar-sagi,shei}, multiple kinks (kink-anti-kink and kink-polaron states)~\cite{campbell,okuno,feinberg2003,feinberg2004}, and complex kinks and their crystalline states~\cite{basar1,basar2,basar3,takahashi} as well as a single-kink solution.

Let us now apply the index theorem to the surface Andreev bound state in a chiral $\ell$-wave pairing state \eqref{eq:chirall}. For simplicity, we also do not take into account the spin degrees of freedom. The pair potential of a chiral $\ell$-wave pairing state is then given by
\beq
\Delta (\hat{\bm k},{\bm r}) = \Delta _0 (\hat{k}_x + i\hat{k}_y)^{\ell},
\label{eq:deltaell}
\eeq
where $\ell \!=\! 1$, $2$, and $3$ correspond to chiral $p$-, $d-$, and $f$-wave states, respectively. Since chiral pairing spontaneously breaks the TRS, the topological properties are characterized by the first Chern number. When the surface is perfectly specular and parallel to the $\hat{\bm z}$-axis, the Chern number is well-defined in a two-dimensional plane $(k_x,k_y)$ sliced at each $k_z$ and one finds that
\beq
{\rm Ch}_1(k_z) = \ell, \quad \mbox{for $|k_z|< k_{\rm F}$}.
\label{eq:ch1chiral}
\eeq
This ensures the existence of the flat-band zero-energy states, forming the topologically protected Fermi arc extending in the $k_z$-direction.

The Andreev equation can be derived from the BdG equation \eqref{eq:bdg3} by decomposing the quasiparticle wavefunction ${\bm \varphi}_{i}({\bm r})$ to the slowly varying part $\tilde{\varphi}$ and the rapid oscillation part with the Fermi wavelength $k^{-1}_{\rm F}$,~\cite{andreev} ${\bm \varphi}_{i}({\bm r}) \!=\! \sum _{\alpha = \pm }C_{\alpha} \tilde{\bm \varphi}_{\alpha}({\bm r})e^{i{k}_{\alpha}\cdot{\bm r}}$. ${\bm k}_{\alpha} \!=\! k_{\rm F}(\cos\phi _{\bm k}\sin\theta _{\bm k},\alpha\sin\phi _{\bm k}\sin\theta _{\bm k},\cos \theta _{\bm k})$ denotes the momentum of incoming ($\alpha \!=\! +$) and outgoing ($\alpha \!=\! -$) quasiparticles. The rigid boundary condition at ${\bm r} = {\bm R}$, ${\bm \varphi}({\bm R})=0$, leads to $C_+=-C_-$ and the continuity condition $\tilde{\varphi}_{+}({\bm R}) = \tilde{\varphi}_-({\bm R})$, where we set ${\bm R}=(R_x,0,R_z)$. Substituting the wavefunction into the BdG equation, one obtains the Andreev equation for $\tilde{\varphi}_{\alpha}({\bm r})$ as
\beq
\left[
-i {\bm v}_{{\rm F}}(\hat{\bm k}_{\alpha}) \cdot {\bm \nabla} {\tau}_z  + \underline{v} + \underline{\Delta} (\hat{\bm k}_{\alpha},{\bm r})
\right]\tilde{\bm \varphi}_{\alpha}({\bm r}) = E \tilde{\bm \varphi}_{\alpha}({\bm r}), 
\label{eq:a}
\eeq
with $\hat{\bm k}_{\alpha} \!\equiv\! {\bm k}_{\alpha}/|{\bm k}_{\alpha}| \!\approx\! {\bm k}_{\alpha}/k_{\rm F}$. The normalization condition is imposed on $\tilde{\bm \varphi}_{\alpha}({\bm r})$ as $\sum _{\alpha}\int d{\bm r} \tilde{\bm \varphi}^{\dag}_{\alpha}({\bm r}) \tilde{\bm \varphi}_{\alpha}({\bm r}) \!=\! 1$. This Andreev approximation holds within the weak-coupling regime, $k_{\rm F}\xi = 2E_{\rm F}/\Delta _0 \gg 1$. In Eq.~\eqref{eq:a}, we have introduced the $4\times 4$ matrix form of the diagonal self-energy and the pair potential, $\underline{v}$ and $\underline{\Delta} (\hat{\bm k},{\bm r})$ (for the details, see Eq.~\eqref{eq:underdelta}).
We introduce the coordinate $\tilde{y}_{\pm}$, corresponding to the distance along the classical trajectory:
$\tilde{y}_{\alpha } \equiv y/v_{\rm F}\sin (\alpha \phi _{\bm k})\sin\theta _{\bm k}$~\cite{stone,mizushimaPRB12}. 
By introducing the new axis 
$\tilde{y}_+ \mapsto \rho > 0$ and 
$-\tilde{y}_- \mapsto \rho < 0$, Eq.~\eqref{eq:a} is reduced to the one-dimensional Dirac equation \eqref{eq:dirac} with  
$m(\rho)=\tilde{D}(\theta _{\bm k},\rho) e^{-i{\sigma}_z\vartheta (\rho)}$.
The phase $\vartheta (\rho)$ is given as
$\vartheta (\rho) =
\phi _{\rm R} \equiv \ell \phi _{\bm k}$ for $\rho \!>\! 0$ and 
$\vartheta (\rho) =\phi _{\rm L} \equiv -\ell \phi _{\bm k}$ for $\rho \!<\! 0$, where $\phi _{\rm L}-\phi _{\rm R} \!\in\! [0,2\pi]$ is required. According to the Jackiw-Rebbi index theorem in Eq.~\eqref{eq:piphase}, the chiral $\ell$-wave pairing state has $|\ell|$ gapless points at~\cite{mizushimaJPCM15}
\beq
\hat{k}_x = \sin\!\theta _{\bm k}\cos\left[ \left( n - \frac{1}{2}\right) \frac{\pi}{ |\ell| }\right] ,
\quad n = 1, 2, \cdots, |\ell| .
\label{eq:kx}
\eeq
This is a generic consequence of the Andreev equation for chiral $\ell$-wave superconductors. 

The Fermi arc appears in the momentum space $(k_x,k_z)$ parallel to the surface, which is terminated at the projection of two point nodes at $k_x = 0$ and $k_z = \pm k_{\rm F}$. In Sec.~\ref{sec:spt}, we will mention that the Fermi arc in time-reversal-invariant nodal superfluids is protected by the combined $P_2$ symmetry, while time-reversal-breaking superfluids can be accompanied by the Fermi arc protected by the topological properties of the point nodes.


Owing to the spontaneous breaking of the TRS, the dispersion of the surface Andreev bound states can be asymmetric in $k_x$, i.e., $E_{\rm surf}(k_x,k_z) = -E_{\rm surf}(-k_x,k_z)$, which is called the chiral edge state. Since the negative energy part of the branch is occupied in the ground state, the surface Andreev bound states carry the spontaneous mass current in equilibrium~\cite{stone,sauls:2011,tsutsumiJPSJ12,mizushimaPRL08,furusaki,mizushimaPRA10}. In Sec.~\ref{sec:mass}, we will discuss more details of the spontaneous mass and spin flows in both time-reversal-invariant and -breaking superfluids.


The Jackiw-Rebbi index theorem can be generalized to zero-energy states bound to a vortex core of chiral superconductors. The generalization was made by Tewari {\it et al.}~\cite{tewariPRL07}, who demonstrated that a single Majorana zero mode always appears in the vortex core of a chiral $p$-wave superconductor when the vorticity is odd. They started with the mean-field Hamiltonian with the vortex order parameter with vorticity $\kappa \!\in\! \mathbb{Z}$ 
\beq
\Delta ({\bm r}_1,{\bm r}_2) = e^{i\kappa\phi}\Delta ({\bm r})\Phi ({\bm r}_{12}),
\label{eq:Dpvoretx}
\eeq
where ${\bm r}$ and ${\bm r}_{12}$ are the center-of-mass and relative coordinate, respectively, and $\phi$ is the azimuthal angle of ${\bm r}$. The function $\Phi ({\bm r}_{12})$ is the Fourier transform of the chiral $p$-wave pair $(k_x + ik_y)$ and $\Phi({\bm r}_{12}) \!=\! -\Phi (-{\bm r}_{12})$. 

Using Eq.~\eqref{eq:Dpvoretx}, Tewari {\it et al.}~\cite{tewariPRL07} demonstrated that the Hamiltonian in the sector of the zero energy state is reduced to the Majorana Hamiltonian 
\beq
\mathcal{H}_{\rm M} = \int dx \left[ -iv_{\rm F}\chi^{\dag}\sigma _z\partial _x \chi
+ m(x) \chi^{\dag}\sigma _x \chi\right].
\label{eq:Hmajo}
\eeq
The effective mass term exhibits the $\pi$-phase shift at $x=0$, 
\beq
m(x) = -m(-x),
\label{eq:mass}
\eeq 
only when the vorticity $\kappa$ is odd, 
\beq
\frac{\kappa + 1}{2} \in \mathbb{Z}.
\label{eq:condition}
\eeq
The two-component spinor $\chi$ is given by $\chi (x) = [\psi (x), \psi^{\dag}(-x)]^{\rm T}$ with a spin-polarized fermion field $\psi(x)$. Expressing the Bogoliubov quasiparticle operator as
$\gamma^{\dag} = \int dx [ 
\varphi _1 (x) \psi^{\dag}(x) + \varphi _2 (x) \psi (-x)
]$,
we obtain the BdG (or Andreev) equation for the vortex-bound state from $[\mathcal{H},\gamma^{\dag}]=E\gamma^{\dag}$ as
$[ -iv_{\rm F}\sigma _z \partial _x + m(x)\sigma _x ] {\bm \varphi} (x) = E{\bm \varphi}(x)$,
where ${\bm \varphi} (x) \!=\! [\varphi _1 (x), \varphi _2 (x)]^{\rm T}$. The BdG equation is equivalent to the one-dimensional Dirac (or Majorana) equation with a mass domain wall. In accordance with the Jackiw-Rebbi index theorem, the Dirac equation with the mass term in Eq.~\eqref{eq:mass} always has a single zero-energy state, and the zero-energy solution is given by 
Eq.~\eqref{eq:diracwf} with an appropriate global phase factor, $e^{i\pi/4}$. If the vorticity is odd, therefore, there exists a Majorana zero mode,
\beq
\gamma = \gamma^{\dag}.
\label{eq:majo2}
\eeq

Hence, the index theorem indicates that, if the condition \eqref{eq:condition} is satisfied, there exists a single Majorana zero mode bound to an odd-vorticity vortex. This is in contrast to the index theorem for zero energy eigenstates of the relativistic Dirac Hamiltonian\cite{weinberg,rossi}. The relativistic Dirac Hamiltonian with the vorticity $\kappa$ has the $\kappa$ zero energy states. 

The index theorem was extended by Sato and Fujimoto~\cite{satoPRB09-2} to the vortex state of a noncentrosymmetric superconductor where the pairing interaction mediated by a strong spin-orbit interaction intrinsically mixes the spin-singlet and -triplet components. We would like to emphasize again that the index theorem established by Tewari {\it et al.}~\cite{tewariPRL07} assumes the linear dispersion of the normal state and thus is valid only for the weak coupling limit. Gurarie and Radzihovsky~\cite{gurariePRB07}  generally demonstrated that, if the total vorticity of the order parameter is odd, there is only one Majorana fermion mode. In the vicinity of the topological phase transition, the shape of the zero-energy wavefunction markedly changes and spreads over the entire sample. This can be interpreted as a manifestation of the topological phase transition at which the bulk excitation becomes gapless. The change in the zero-energy wavefunction was numerically confirmed in Ref.~\citeonline{mizushimaPRA10}, where the self-consistent calculation was employed. Further details will be discussed below.

\subsection{A minimal model for the topological phase transition}

The spin-polarized chiral $p$-wave pairing with a single quantum vortex offers a minimal model that describes a topological phase transition in the presence of a topological defect. We here begin with the Hamiltonian for spin-polarized fermions interacting through an interaction potential $\mathcal{V}({\bm k},{\bm k}^{\prime})$,
\begin{align}
\mathcal{H} = \sum _{\bm k}\varepsilon ({\bm k})c^{\dag}_{\bm k}c_{\bm k} 
+ \frac{1}{2}\sum _{{\bm k},{\bm k}^{\prime},{\bm q}}\mathcal{V}({\bm k},{\bm k}^{\prime})
b^{\dag}_{{\bm k},{\bm q}}
b_{{\bm k}^{\prime},{\bm q}},
\label{eq:hamiltonianp}
\end{align}
where we omit the spin index and we introduce the paired operator $b_{{\bm k},{\bm q}}\!\equiv\!c_{-{\bm k}+{\bm q}/2}c_{{\bm k}+{\bm q}/2}$. The interaction potential $\mathcal{V}({\bm k},{\bm k}^{\prime})$ is defined as the Fourier transformation of an isotropic interaction potential $\mathcal{V}(r_{12})$ in the relative coordinate, $\mathcal{V}({\bm k},{\bm k}^{\prime}) \!=\! \int d{\bm r}_{12}e^{-i({\bm k}-{\bm k}^{\prime})\cdot{\bm r}_{12}}\mathcal{V}(r_{12})$. 

For an isotropic interaction potential in two-dimensional systems, $\mathcal{V}({\bm k},{\bm k}^{\prime})$ is generally expanded in terms of the angular momentum $\ell \!=\! 0, \pm 1, \pm 2, \cdots$ as~\cite{botelho}
$\mathcal{V}({\bm k},{\bm k}^{\prime}) = \sum _{\ell}
\mathcal{V}_{\ell}(k,k^{\prime})\Phi _{\ell}(\hat{\bm k})\Phi^{\ast}_{\ell}(\hat{\bm k}^{\prime}) $.
Using the expansion form of the plane wave in terms of the $\ell$-th Bessel function, $J_{\ell}(r)$, $e^{i{\bm k}\cdot{\bm r}} \!=\! \sum _{\ell}i^{\ell}J_{\ell}(kr)e^{i\ell\tilde{\phi}}$, one finds that the function $\Phi _{\ell}(\hat{\bm k})$ is obtained as the eigenstate of the angular momentum, $\Phi _{\ell}(\hat{\bm k})\!=\! e^{i\ell \tilde{\phi}}$, where $\tilde{\phi}\!=\! \cos^{-1}({\bm k}\cdot{\bm r})$. The details of the potential $\mathcal{V}_{\ell}(k,k^{\prime})$ are determined by the potential in real space, $\mathcal{V}(r)$, through the relation $\mathcal{V}_{\ell}(k,k^{\prime}) \!=\! 2\pi \int dr rJ_{\ell}(kr)J_{\ell}(k^{\prime}r)$. From the asymptotic expression of the Bessel function, one finds that the potential $\mathcal{V}_{\ell}(k,k^{\prime})$ reduces to $\sim\!k^{\ell}k^{\prime\ell}$ for small $k$ and the decaying envelope $\sim \! k^{-1/2}k^{\prime -1/2}$ for large $k$. Hence, it is natural to assume the separable form of the potential as $\mathcal{V}_{\ell}(k,k^{\prime}) \!=\! \Gamma _{\ell}(k)\Gamma _{\ell}(k^{\prime})$, where the function $\Gamma _{\ell}(k) \!=\! (k/k_1)^{\ell}/(1+k/k_0)^{1/2}$ interpolates the asymptotic behaviors in the long- and short-wavelength limits. The constants $k_0$ and $k_1$ denote the detailed structure of $\mathcal{V}(r)$.

Let us employ the standard procedure of the mean-field approximation in the Hamiltonian \eqref{eq:hamiltonianp} with the separable form of the interaction, 
\beq
\mathcal{V}({\bm k},{\bm k}^{\prime}) = \sum _{\ell} g_{\ell}\Gamma _{\ell}(k)\Gamma _{\ell}(k^{\prime})\Phi _{\ell}(\hat{\bm k})\Phi^{\ast}_{\ell}(\hat{\bm k}^{\prime}), 
\eeq
where $ g_{\ell} \!<\! 0$ is the interaction strength of the $\ell$-wave channel. We here assume that only a particular angular momentum channel has the dominant contribution to the scattering process of two fermions, which allows us to retain only one of the $\ell$ terms in $\mathcal{V}({\bm k},{\bm k}^{\prime})$. This implies the chiral $\ell$-wave ($\ell\!\le\!0$) state in the pair potential,
\beq
\Delta ({\bm k},{\bm r}) = \Delta ({\bm r}) \left( k_x + ik_y \right)^{\ell},
\label{eq:chirall}
\eeq
where $\Delta ({\bm r})$ denotes the spatial profile of the pair potential in the center-of-mass coordinate. The BdG Hamiltonian in  real space is then obtained by a $2\times 2$ Hermitian matrix as
\beq
\mathcal{H}({\bm r}) = \left( 
\begin{array}{cc}
\varepsilon ({\bm r}) & \frac{1}{2}\{ \Delta ({\bm r}), \Phi _{\ell}(-i{\bm \partial})\} \\ 
-\frac{1}{2}\{ \Delta^{\ast}({\bm r}), \Phi^{\ast}_{\ell}(-i{\bm \partial})\} & - \varepsilon ({\bm r})
\end{array}
\right),
\label{eq:Hchiral}
\eeq
where ${\bm k} =(k_x,k_y)$. 
The pair potential in real space is then obtained as~\cite{mizushimaPRA10}
\beq
\Delta ({\bm r}) = g_{\ell}
\Phi^{\ast}_{\ell}(-i{\bm \partial}_{12})\mathcal{F}({\bm r}_1,{\bm r}_2)\bigg|_{{\bm r}_{12}\rightarrow 0},
\label{eq:chiralgap}
\eeq
where $\mathcal{F}({\bm r}_1,{\bm r}_2) \!=\! \lim _{\eta \rightarrow 0}T\sum _n \mathcal{F}({\bm r}_1,{\bm r}_2;\omega _n)e^{i\omega _n \eta}$ is the Cooper pair amplitude and $\Phi _{\ell}(-i{\bm \partial})$ is obtained by replacing $(\hat{k}_x,\hat{k}_y)$ with $(-i\partial _x,-i\partial _y)/k_0$. Taking the limit ${\bm r}_{12} \!\rightarrow\! 0$ in Eq.~\eqref{eq:chiralgap} corresponds to the zero-range approximation of the Cooper pair size. 

The gap equation \eqref{eq:chiralgap} involves two divergence terms proportional to $E_{\rm c}$ and $\ln E_{\rm c}$ \cite{randeria,gurarieAP07,botelho}, where $E_{\rm c}$ is the cutoff energy. The ultraviolet divergence can be removed by replacing the bare coupling constant with the renormalized one 
\begin{eqnarray}
\frac{1}{g_{\ell}} = - \frac{1}{S} \sum _{\bm k} 
\frac{\left|\Phi _{\ell} ({\bm k})\right|^2}{2 \tilde{\varepsilon} ({\bm k}) - E_{\rm b}},
\label{eq:lambda}
\end{eqnarray}
where $S$ denotes the volume of the system and $\tilde{\varepsilon} ({k}) \!\equiv\! k^2/2M$ is the single-particle energy in vacuum. Here, $E_{\rm b}$ is an eigenvalue of the Schr\"{o}dinger equation for two fermions interacting via the pairing potential $V$ \cite{randeria,botelho}. $E_{\rm b}$ is real and regarded as the two-body bound state energy in vacuum when $E_{\rm b}$ is negative, while it has an imaginary part for positive $E_{\rm b}$. However, it is known that this imaginary part is negligible in the vicinity of a $p$-wave resonance \cite{gurarieAP07}. Hence, the coupling constant $g_{\ell}$ parameterized by $E_{\rm b}$ remains real and negative for all values of $E_{\rm b}$, which can remove the leading term of the ultraviolet divergence in the gap equation (\ref{eq:chiralgap}). 

The minimal model for describing the topological phase transition is accomplished by adding a constraint on the total particle number conservation. The total number is defined as the spatial average of the particle density $\rho ({\bm r})$,
\beq
\rho ({\bm r}) = \lim _{\eta \rightarrow 0}T\sum _n \mathcal{G}({\bm r},{\bm r};\omega _n)e^{i\omega _n \eta}. 
\label{eq:rhochiral}
\eeq
The chemical potential in $\varepsilon ({\bm r})$ is determined so as to fix the total number $N$. 

Let us now turn into the bulk topology of Eq.~\eqref{eq:Hchiral}, where $\Delta ({\bm r})$ is assumed to be spatially uniform. For odd $\ell$, the bulk Hamiltonian preserves only the PHS with $\mathcal{C}^2\!=\! +1$ and breaks the TRS. As shown in Eq.~\eqref{eq:ch1chiral}, the corresponding topological number is characterized by the first Chern number ${\rm Ch}_1$, which can be nontrivial when the chemical potential $\mu$ is positive,
\beq
{\rm Ch}_1 = \left\{
\begin{array}{ll}
\ell & \mbox{ for $\mu > 0$} \\
\\
0 & \mbox{ for $\mu < 0$}
\end{array} 
\right. .
\eeq
As mentioned in Sec.~\ref{sec:chern}, the Chern number is equivalent to the two-dimensional winding number $w_{\rm 2d}$, which characterizes the nontrivial mapping from the $k$-space $S^2$ to the target space $\mathcal{M} \!=\! S^2$.  In the current situation, the $\hat{\bm m}$-vector is obtained by the elements of the BdG Hamiltonian as
\beq
\hat{\bm m}({\bm k}) = \left( \Delta \cos \ell \phi _{\bm k}, - \Delta \sin \ell \phi _{\bm k}, k^2/2m - \mu \right),
\eeq
where we assume $\varepsilon ({\bm k})\!=\! k^2/2m - \mu$. For $\mu \!>\! 0$, the $\hat{\bm m}$-vector at ${\bm k}\!=\! {\bm 0}$ always points to the south pole of the target space $S^2$, while $\hat{\bm m}(\infty)$ points to the north pole. Since the nonzero chirality $\ell$ induces the texture to $m_x$ and $m_y$ components, the $\hat{\bm m}$-vector can cover the entire target space $S^2$, implying the nontrivial value of ${\rm Ch}_1\!=\! w_{\rm 2d}\!=\! \ell$. This is called the topological BCS phase, and the quasiparticle excitation gap is determined by the dissociation energy of the Cooper pairs, $2|\Delta|$. When the Fermi surface is absent, i.e., $\mu \!<\! 0$, however, $\hat{\bm m}$ covers only the northern hemisphere. This leads to the topologically trivial BEC phase ${\rm Ch}_1\!=\! w_{\rm 2d}\!=\! 0$. The energy gap in this phase is given by $2|\mu|$, which reflects the fact that all fermions form bosonic ``molecules'' in the ground state. Hence, in contrast to the BCS-BEC crossover in the $s$-wave case, these two regimes are not smoothly connected but give rise to the topological phase transition at which the topological invariant relevant to this system changes~\cite{read,volovik,gurarieAP07}. The topological phase transition at  $\mu \!=\! 0$ is always accompanied by closing the bulk excitation gap $\min E({\bm k}) \!=\! 0$ at ${\bm k}\!=\! {\bm 0}$~\cite{read,gurarieAP07}. Multiple signatures of topological transitions have been predicted in Ref.~\citeonline{chanPRB15} by analyzing the Kitaev chain.

The self-consistent theory described above is not a toy model but an effective model for neutral Fermi gases with a $p$-wave Feshbach resonance and for low-energy quasiparticles in various topological systems listed in Sec.~\ref{sec:nonabelian}. 
$p$-wave Feshbach resonances have recently been observed by sweeping magnetic fields in experiments with $^6$Li and $^{40}$K atoms~\cite{zhangPRA04,schunckPRA05,jinPRL07,valePRA08,inadaPRL08,mukaiyamaPRA13}. The Feshbach resonance of colliding atoms into an $\ell$-wave bound state allows one to manipulate an $\ell$-wave interatomic interaction~\cite{chinRMP10}. In the weakly interacting regime, the fermionic atoms form Cooper pairs, where low-energy quasiparticles are characterized by $\ell$-wave Cooper pair potential. In contrast, $\ell$-wave diatomic molecules in the BEC phase have an isotropic gap uniquely determined by the binding energy of molecules. In the BEC limit, the set of the self-consistent theory based on the BdG equation can be mapped onto the Gross-Pitaevskii equation for bosonic molecules~\cite{iskinPRA06}. The experimental observation of $p$-wave Feshbach resonances is the first step toward the further realization of topological phase transitions~\cite{chengPRL05,ohashiPRL05,hoPRL05,gurariePRL05,iskinPRL06,gurarieAP07}.

\subsection{Topology of vortex-bound states in BCS-BEC evolution}
\label{sec:vortex1}

We now discuss spin-polarized chiral $p$-wave superfluids with a definite vorticity $\kappa \!\in\! \mathbb{Z}$ as shown in Eq.~\eqref{eq:Dpvoretx}. In the weak coupling BCS regime ($k_{\rm F}\xi \!\gg\! 1$), there exist low-energy quasiparticles embedded inside the threshold energy of the bulk excitation. As discussed in Sec.~\ref{sec:index2}, the phase of the order parameter rotates by $2\pi\kappa$ around a quantum vortex, where the single-valuedness of $\Delta$ requires $\kappa \!\in\! \mathbb{Z}$. Hence, the quasiparticles traveling across the vortex core experience an abrupt shift of the phase, leading to the so-called Caroli-de Gennes-Matricon (CdGM) state~\cite{CdGM}, a special types of Andreev bound state~\cite{kashiwayaRPP}. The dispersion is obtained by analytically solving the BdG equation within $|m| \!\ll\! k_F\xi$~\cite{stone,mizushimaPRA10} as
\begin{eqnarray}
E^{\rm vortex}_{m} = - \left( m - \frac{\kappa +1}{2}\right) \omega _0 .
\label{eq:cdgm}
\end{eqnarray}
The level spacing is $\Delta^2_0/E_{\rm F}$, where $\Delta _0$ is the pair potential far from the vortex center, $\Delta _0\!\equiv\! \Delta (|{\bm \rho}|\!\rightarrow\! \infty)$.

In the quantum regime ($\xi \!\approx k^{-1}_{\rm F}$) of $s$-wave superfluids, it was first clarified by Hayashi {\it et al.}~\cite{hayashiPRL98,hayashiJPSJ98} that the energy levels of the CdGM state become discrete, leading to the strong depletion of the particle density around the core. In the $s$-wave case, Eq.~\eqref{eq:cdgm} is modified to $E^{\rm vortex}_{m} \!=\! - ( m - 1/2) \omega _0$, where the zero energy state is always absent. The discretized core level was clearly observed in the anisotropic spin-singlet superconductor YNi$_2$B$_2$C by using scanning tunneling spectroscopy with an unprecedented $0.1$~nm spatial resolution~\cite{kaneko}. The analysis of the CdGM state and quantum depletion has been theoretically extended to the BCS-BEC crossover regime, \cite{bulgac, feder, mmachida1, mmachida2, sensarma, levin} and depletions in the particle density were experimentally observed in rotating Fermi gases with an $s$-wave resonance \cite{mit1,mit2} as a hallmark of superfluidity. Furthermore, the structures of giant vortices with $|\kappa| \!>\! 1$ have been extensively studied in $s$-wave superfluids~\cite{virtanen, ktanaka, mizushimaPRL05, takahashi, hu1, hu2, suzuki}.

It is now natural to expect that the extremely strong coupling regime of $p$-wave superfluids is accompanied by an intriguing vortex structure. As mentioned in Sec.~\ref{sec:index2}, the vortex-bound quasiparticle in the BCS limit $k_{\rm F}\xi \gg 1$ is well describable with the one-dimensional Dirac equation at the domain wall, and the index theorem ensures the existence of a Majorana zero mode. However, the chiral $p$-wave pairing involves the topological phase transition at $\mu \!=\! 0$. In this regime, the coherence length of the order parameter $\xi$ becomes comparable to the mean interparticle distance $k^{-1}_{\rm F}$, and thus the quasiparticle structure around the vortex might be markedly modified by the quantum effect as well as the change in topology.

For the topological properties of vortex-bound states in two-dimensional chiral $\ell$-wave superfluids, the Hamiltonian relevant to the systems is obtained as $\mathcal{H}(k_x,k_y,\varphi)$ in the base space ${S}^d\times{S}^D = S^1\times S^2$, where $\varphi$ is the azimuthal angle of a path enclosing a vortex (or vortices). The topological invariant for $\delta = 1$ of class D in Table~\ref{table1} is the $\mathbb{Z}_2$ number. As discussed in Eq.~\eqref{eq:z2cs}, the $\mathbb{Z}_2$ number is obtained by the dimensional reduction of the second Chern number in the suspension of the base space, $\Sigma (S^2\times S^1)$, which is given by the integral of the Berry curvature $\mathcal{F}\wedge \mathcal{F}$ over the suspension. The integrand, $\mathcal{F}\wedge \mathcal{F}$, is related to the derivative of the Chern-Simons form as ${\rm tr}[\mathcal{F}\wedge\mathcal{F}] \!=\! dQ_3$. 
For odd $\ell$, the $\mathbb{Z}_2$ number is sensitive to the sign of the chemical potential $\mu$~\cite{teoPRB10},
\begin{align}
{\rm Ch}_2 = - \frac{1}{4\pi^2} \int _{S^2\times S^1} {\rm CS}_3
= 
\left\{ 
\begin{array}{ll}
\kappa~ {\rm mod} ~ 2 & \mbox{for $\mu > 0$} \\
\\ 
0 & \mbox{for $\mu < 0$} 
\end{array}
\right. .
\label{eq:cs3}
\end{align}
This implies that an odd-vorticity vortex for $\mu > 0$ is always accompanied by a single zero energy state, which is consistent with the Jackiw-Rebbi index theorem. In contrast, the BEC regime is topologically trivial, and thus the zero energy state is absent. 

\begin{figure*}[t!]
\includegraphics[width=160mm]{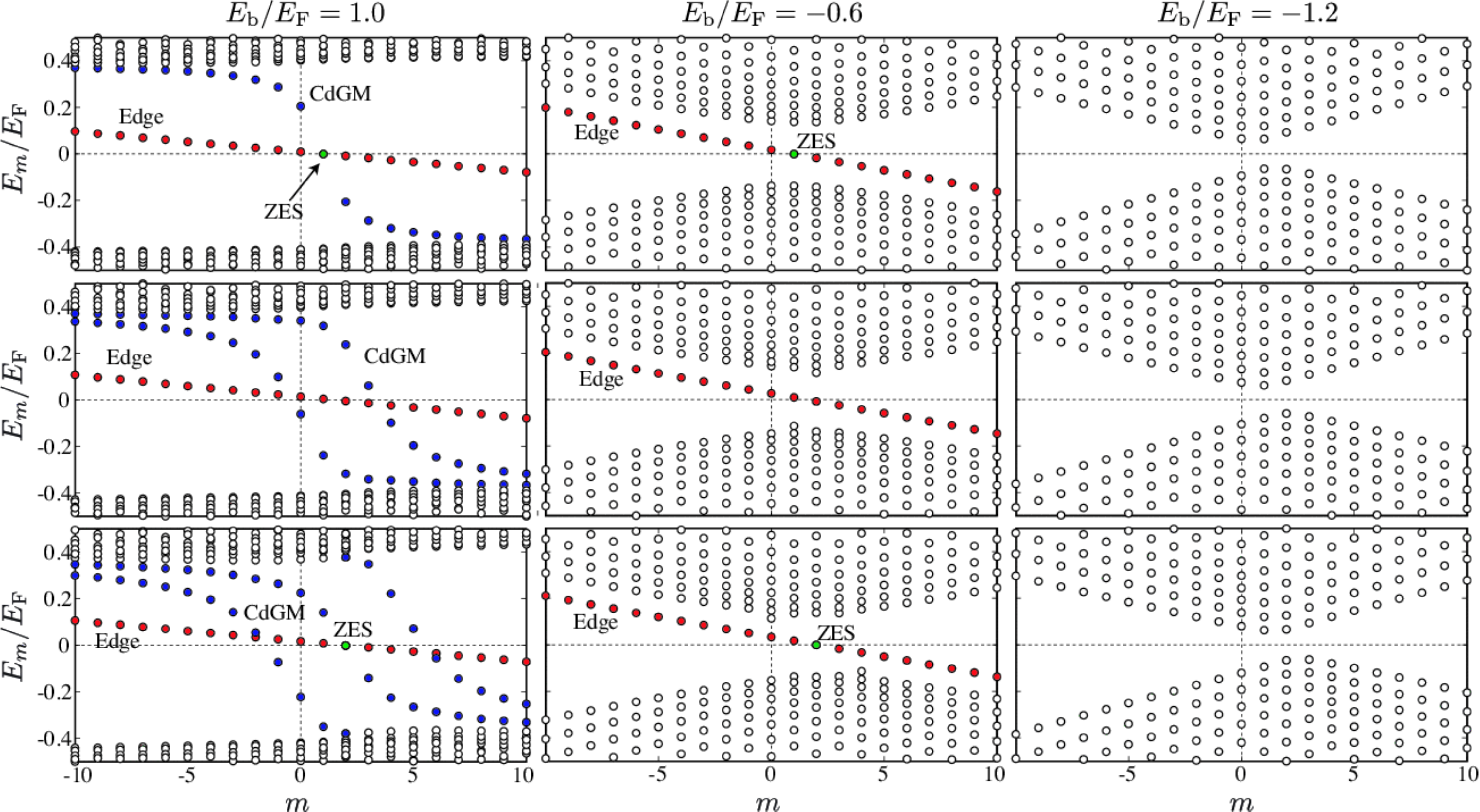}
\caption{(Color online) Quasiparticle excitation spectra as a function of the azimuthal quantum number $m$ at $E_b/E_F \!=\! 1.0$ (left column), $-0.6$ (center column), and $-1.2$ (right column) in the vortex states of chiral $k_x+ik_y$ pairing with $\kappa  \!=\! 1$ (top row), $\kappa  \!=\! 2$ (middle row), and $\kappa \!=\! 3$ (bottom row). The chemical potential is estimated as $\mu/E_{\rm F}=-1.0$, $0.0$, and $1.0$ for $E_b/E_F \!=\! 1.0$, $-0.6$, and $-1.2$, respectively. ``CdGM'', ``Edge'', and ``ZES'' denote the branches of the CdGM, edge, and zero-energy states, respectively. Figures adapted from Ref.~\citeonline{mizushimaPRA10}.
}
\label{fig:spectra}
\end{figure*}

In Fig.~\ref{fig:spectra}, we present quasiparticle spectra in vortex states with vorticity $\kappa  \!=\! +1, +2, +3$ in three different regimes: The weak coupling BCS regime ($E_{\rm b} /E_{\rm F} \!=\! 1.0$ and $\mu \sim E_{\rm F}$), the vicinity of the BCS-to-BEC transition point ($E_{\rm b} /E_{\rm F} \!=\!-0.6$ and $\mu \sim 0$), and the BEC regime ($E_{\rm b} /E_{\rm F} \!=\!-1.2$ and $\mu < 0$). The spectra are obtained by self-consistently solving the BdG equation~\eqref{eq:Hchiral} and the gap equation~\eqref{eq:chiralgap} with a fixed total particle number $N$ estimated with Eq.~\eqref{eq:rhochiral}~\cite{mizushimaPRA10}. Here, the chemical potential $\mu$ shifts from positive to negative as the two-body bound-state energy in vacuum, $E_{\rm b}$, decreases. 
In the weak coupling BCS regime ($k_{\rm F}\xi \!\gg\! 1$), two branches are embedded inside the threshold energy of continuum excitations, $E \!=\! \pm \Delta _0 \!\approx\! 0.4E_{\rm F}$: Edge- and vortex-bound states. The branch labeled as ``CdGM'' in Fig.~\ref{fig:spectra} is given by Eq.~\eqref{eq:cdgm}, while the ``Edge'' branch is obtained by the same dispersion as ~\eqref{eq:cdgm} with $\omega _0 = \Delta _0/k_{\rm F}R$. The number of CdGM branches uniquely depends on $\kappa$ in the weak coupling BCS regime, {\it e.g.}, three CdGM branches appear in the $\kappa \!=\! +3$ vortex state.

In the vicinity of the topological phase transition $\mu \!\sim \! 0$, the bulk energy gap is characterized as $\min|E_{\rm bulk}| \!=\! |\mu|$. In this regime with $\Delta _0 \!\approx\! E_{\rm F}$, the energy spacing of the CdGM states becomes comparable to the Fermi energy, {\it i.e.}, $\omega _0 \!\approx\! E_{\rm F}$ in Eq.~(\ref{eq:cdgm}), and thus the zero energy state bound to the vortex is isolated from the higher CdGM states as shown for $E_{\rm b}/E_{\rm F} \!=\!-0.6$ of Fig.~\ref{fig:spectra}. Beyond the topological phase transition $E_{\rm b}/E_{\rm F} \!=\! -1.0$, the excitation gap is uniquely characterized by $\min|E_n| \!=\! |\mu|$, and any bound states disappear even in the presence of a vortex. The numerical results in Fig.~\ref{fig:spectra} coincide with the index theorem in Eq.~\eqref{eq:condition} and the topological invariant in Eq.~\eqref{eq:cs3}.

The spatial profile of the zero-energy wavefunction reflects the topological phase transition at $\mu \!=\! 0$. To derive the zero energy solution of Eq.~\eqref{eq:Hchiral}, we employ the low-energy approximation in the function $\Phi _{\ell}({\bm k})$ as $\Phi _{\ell}({\bm k})\!\approx\!(k/k_0)^{|\ell|}e^{i\ell\phi _{\bm k}}$. The zero energy solution for the chiral $p$-wave pairing state is given as~\cite{gurariePRB07,mizushimaPRA10}
\begin{align}
{\bm \varphi}_{m}(\rho)
\propto e^{i\frac{\kappa+1}{2}\phi\sigma _z}
\left[
\begin{array}{c}
f_{(\kappa+1)/2}(\rho) \\
f_{-(\kappa+1)/2}(\rho)
\end{array}
\right] \exp\left[-\frac{M}{k_{\rm F}}\int^{\rho}_0\Delta (\rho^{\prime})d\rho^{\prime}\right]
\label{eq:cdgmwf}
\end{align}
when $\kappa$ is odd. The wavefunction consists of the exponential decay factor and $f_{m}$ which describes the quantum oscillation on the scale of $k^{-1}_{\mu} \!\equiv\!(2M\mu)^{-1/2}$,
\beq
f_{m}(\rho)  = J_{m}\left(\rho k_{\mu}\sqrt{1-\frac{1}{(k_{\mu}\xi_0)^2}} \right),
\label{eq:bessel}
\eeq 
for $k_{\mu}\xi_0 \!>\! 1$. This reduces to the Bessel function $J_{m}(k_{\rm F} \rho )$ in the BCS limit $k_{\mu}\xi _0\!\approx\! k_{\rm F}\xi _0 \!\gg \! 1$. The exponential decay factor has the characteristic length $\xi _0 \!=\! k_{\rm F}/M\Delta _0 $, which is the superfluid coherence length approximately corresponding to the diameter of the vortex core. Hence, the envelope function in the BCS regime is $\cos(k_{\rm F}\rho)e^{-\rho /\xi _0}$.

In Ref.~\citeonline{gurariePRB07}, Gurarie and Radzihovsky clarified that in the regime of $\mu \!>\! 0$ and $k_{\mu}\xi_0 \!<\! 1$, the rapid oscillation associated with $J_{m}(k_{\rm F} \rho )$ disappears. In this regime corresponding to the BCS regime close to the topological phase transition, the function $f_m(\rho)$ of the zero energy state is modified to 
\beq
f_{m}(\rho) = I_{m}\left(\rho k_{\mu}\sqrt{\frac{1}{(k_{\mu}\xi_0)^2}-1} \right),
\label{eq:modified}
\eeq 
where $I_{m}(\rho)$ denotes the $m$-th-order modified Bessel function. Equation \eqref{eq:modified} indicates that the exponential decay factor $\sim\!e^{-\rho/\xi _0}$ is canceled out by the modified Bessel function for a large $\rho$ as $f_{m}(\rho) \!\approx\! \frac{1}{\sqrt{\rho}}e^{-(1-\lambda)\rho/\xi _0}$, where $\lambda \!=\! \sqrt{1-(k_{\mu}\xi _0)^2}$. The wavefunction is extended over $\xi _0$, which manifests the topological phase transition at $k_{\mu}\xi _0 \!\rightarrow\! 0$.

The CdGM wavefunction with $\ell$ satisfies $u_{m}(r) \!\propto\! J_{m}(k_{\mu}r)\!\approx\! r^{|m|}$ and $v_{m}(r) \!\approx\! r^{|m - \kappa -1|}$ at $r\!\rightarrow\! 0$. Hence, the asymptotic wavefunctions of the zero-energy states that satisfy the condition \eqref{eq:condition} exhibits $u_{m} (r) \!=\! v_{m}(r) \!\approx\! r^{|m +1|/2}$. The asymptotic behaviors of the CdGM states indicate that they play an important role in determining the particle density around the core. Indeed, the quantum depletion in the particle density around the core is sensitive to the vorticity $\kappa$ and markedly changes in the vicinity of the topological phase transition~\cite{mizushimaPRL08}.

\begin{figure}[t!]
\includegraphics[width=75mm]{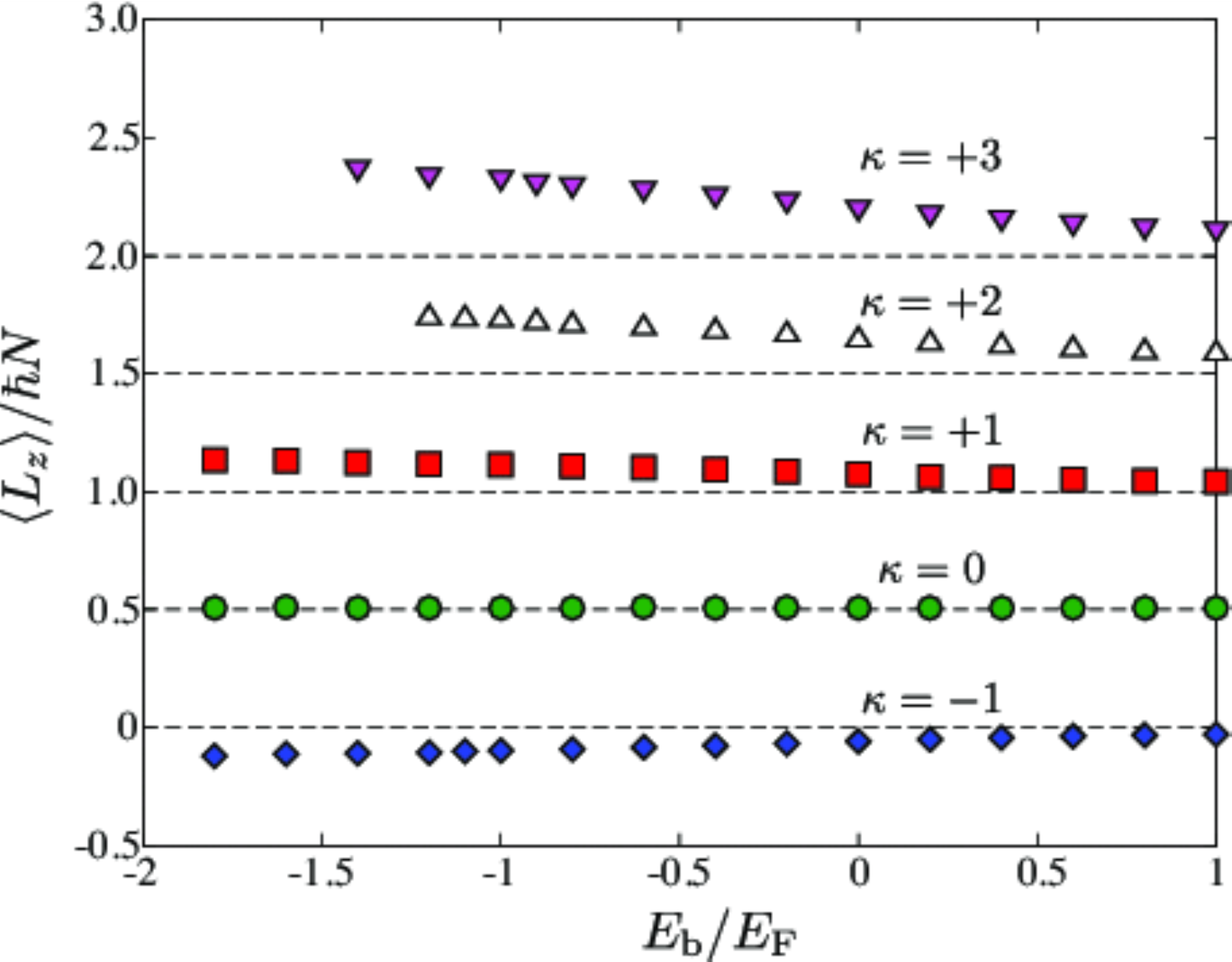}
\caption{(Color online) Total angular momentum $\langle L_z \rangle$ as a function of $E_{\rm b}$ in various vortex states with $\kappa  \!=\! -1$, $0$, $+1$, $+2$, and $+3$. The dashed lines denote Eq.~(\ref{eq:lz}). Figures adapted from Ref.~\citeonline{mizushimaPRA10}.
}
\label{fig:lz}
\end{figure}

Even in the BEC phase, as shown in Fig.~\ref{fig:spectra}, the spectrum is asymmetric with respect to $m$. This reflects the fact that the BdG equation (\ref{eq:Hchiral}) maintains the PHS, regardless of the sign of $\mu$. In the cylindrical systems that we consider here, the PHS indicates that there is a one-to-one correspondence between the quasiparticle states ${\bm \varphi}_m({\bm r})$ with $E_{m}$ and $\mathcal{C}{\bm \varphi}_{-m+\kappa+1}({\bm r})$ with $-E_{-m+\kappa +1}$. This implies that the quasiparticle spectrum intrinsically has asymmetry with respect to $m$ even in the BEC phase. Although no low-lying bound states exist, this asymmetry with respect to $m$ gives rise to the nontrivial net angular momentum that is carried by the continuum states. In Ref.~\citeonline{mizushimaPRA10}, the total angular momentum per particle was estimated by fully self-consistent calculations as 
\beq
\frac{\langle L_z \rangle}{\hbar N} \!\approx\! \frac{\kappa}{2} + \frac{1}{2},
\label{eq:lz0}
\eeq
in the chiral $p$-wave pairing state, as shown in Fig.~\ref{fig:lz}. The first term in Eq.~\eqref{eq:lz0} is the contribution from the vorticity $\hbar\kappa/2$, while the second term reflects the angular momentum carried by the edge current. The total angular momentum represented in Eq.~\eqref{eq:lz0} is not topologically protected but sensitive to the details of the system and pairing state~\cite{sauls:2011,tada:arXiv,volovikJETP15,kallinPRB14}. Further details are discussed in Sec.~\ref{sec:mass} in connection with the paradox of intrinsic angular momentum in the superfluid $^3$He-A.


\subsection{Intervortex tunneling of Majorana zero modes}

Let us now discuss the Majorana zero modes on vortex lattices. The topological properties of spin-polarized chiral $p$-wave superfluids with $N_{\rm v}$ vortices are still characterized by the $\mathbb{Z}_2$ number. Since systems having $N_{\rm v}$ vortices can be continuously connected to a giant vortex with vorticity $\kappa \!=\! N_{\rm v}$, the relevant topological number is given by replacing the vorticity $\kappa$ to $N_{\rm v}$ in Eq.~\eqref{eq:cs3} as
\beq
{\rm Ch}_2 = N_{\rm v} \quad {\rm mod}~2
\eeq 
for $\mu \!>\! 0$ and ${\rm Ch}_2 \!=\! 0$ for $\mu \!<\! 0$. This implies that only a single Majorana zero mode can be topologically protected when the number of vortices is odd. The other cases can be gapped out by any disturbances, e.g., intervortex tunneling.

We notice that a vortex lattice realized in the interface between a topological insulator and an $s$-wave superconductor can be characterized by the $\mathbb{Z}$ topological number rather than the $\mathbb{Z}_2$ number when the chemical potential is fine-tuned to the Dirac point of the surface state emergent in the topological insulator~\cite{chiuPRB15}. At this point, the effective Hamiltonian for the proximity-induced superconducting Dirac fermions with $N_{\rm v}$ vortices is reduced to  the Jackiw-Rossi model~\cite{chiuPRB15,rossi}, which brings about topologically protected $N_{\rm v}$ Majorana zero modes. Hence, the system with a fine-tuned chemical potential maintains ``exact'' Majorana zero modes that are unaffected by the splitting due to the intervortex tunneling~\cite{chengPRB10}. 

Let us now focus our attention on the splitting of Majorana zero modes characterized by the $\mathbb{Z}_2$ number. As two vortices approach each other, it is expected that the wavefunctions of Majorana zero modes overlap with each other and the quantum interference lifts the degeneracy from zero energy to $E_+$ and $E_-$. Understanding the splitting of the zero-energy states is indispensable for the implementation of the non-Abelian braiding of vortices, since the characteristic time scale of braiding operation must satisfy 
\beq
\omega^{-1}_0 \ll T \ll \delta E^{-1},
\eeq
where $\omega _0 \!\approx\! \Delta^2_0/E_{\rm F}$ denotes the level spacing of CdGM states and $\delta E \!=\! E_+-E_-$ defines the width of the splitting or the width of the ``Majorana band'' attributable to the intervortex tunneling. The first inequality is indispensable for neglecting the contributions of the higher CdGM states. The second one indicates the time scale for which the two splitting levels can be superposed during vortex operation. A practical scheme for realizing quantum computation operated by the braiding of Majorana zero modes has been discussed for ultracold fermionic gases~\cite{tewariPRL07-2}. The issue on the splitting of Majorana zero modes was addressed by a large number of authors~\cite{chengPRL09,mizushimaPRA10-2,chengPRB10,silaevPRB13,hungPRB13,zhouEPL13}. The nonadiabatic effect associated with tunneling splitting and transitions to higher CdGM states was also studied by Cheng {\it et al.}~\cite{chengPRB11}.


We now consider the low-lying quasiparticles in the presence of an array of $N_{\rm v}$ vortices.
The order parameter in the $(\hat{k}_x+i\hat{k}_y)$ pairing state is given by the ansatz 
$\Delta ({\bm r}) = \Delta _0 \prod^{N_{\rm v}}_{j=1} e^{i\kappa _j\bar{\theta}_j}\tanh(|\bar{\bm \rho}_j|/\xi _0)$, 
where $\bar{\bm \rho}_j \!\equiv\! {\bm \rho}-{\bm R}_j \!=\! \bar{\rho}_j (\cos\bar{\theta}_j,\sin\bar{\theta}_j)$ denotes the coordinate centered at the $j$-th vortex core ${\bm R}_j$ and $\bar{\theta}_j \!\equiv\! \tan^{-1}(\bar{y}_j / \bar{x}_j)$ is the polar angle. Without loss of generality, we take the vorticity of the $j$-th vortex as $\kappa_{j}\!=\! -1$. The distance between neighboring vortices is characterized by $D_{\rm v} \!\equiv\! |{\bm R}_{j} - {\bm R}_{j+1}|$. The BCS-BEC evolution is parameterized by varying the set of $\Delta_0$ and $\mu$. When two vortices are well separated, i.e., $D_{\rm v} \!\gg\! \xi _0$, the variational wavefunctions ${\bm \varphi}_{\pm}({\bm \rho})$ are defined as 
${\bm \varphi}_{\pm} \!\equiv\! 
[{\bm \varphi}_{m=0,j\!=\!1}\mp i{\bm \varphi}_{m \!=\! 0,j\!=\!2}]/\sqrt{2}$.
We also suppose that the Majorana zero modes are well isolated from the higher energy CdGM states, i.e., $\Delta _0\!\sim\! E_{\rm F}$. The function ${\bm \varphi}_{m \!=\! 0,j}$ describes the wavefunctions of the zero energy states bound at the vortex position ${\bm R}_{j}$, which is obtained from Eq.~(\ref{eq:cdgmwf}) as 
\beq
{\bm \varphi}_{m,j}({\bm \rho}) \!=\! e^{im\bar{\theta}_j}e^{i\frac{\Omega _j}{2}\hat{\tau}_z} {\bm \varphi}_{m}({\bm \rho}).
\label{eq:varphi}
\eeq 
The Majorana zero mode bound to the $j$-th vortex is subjected to the ${\rm U}(1)$ phase of the order parameter  $\Omega _j$, where $\Omega _j \!=\! \sum_{k\!\neq\! j}\bar{\theta}_k ({\bm R}_j)$ is the sum of all the contributions of the ${\rm U}(1)$ phase shift from the other vortices. For $N_{\rm v}\!=\! 2$, the ${\rm U}(1)$ phase factor does not play an important role.

The eigenenergy $E_+$ of the hybridized wavefunction ${\bm \varphi}_{\pm}$ obeys the BdG equation $\mathcal{H}({\bm \rho}){\bm \varphi}_{\pm}({\bm \rho}) \!=\! E_{\pm} {\bm \varphi}_{\pm}({\bm \rho})$, where $\mathcal{H}({\bm \rho})$ is given in Eq.~(\ref{eq:Hchiral}) and the PHS satisfies $E_+\!=\! -E_-$. For $k_{\mu}\xi _0\!>\! 1$, the splitting energy due to the hybridization ${\bm \varphi}_{+}$ is obtained in the weak coupling limit with $k_{\mu}\xi _0 \!\approx \! k_{\rm F}\xi _0\!\gg\! 1$ as~\cite{chengPRL09,mizushimaPRA10-2}
\begin{align}
E_{+} \approx - \frac{\Delta _0}{\sqrt{2\pi k_{\rm F}D_{\rm v}}}
\cos\left(k_{\rm F}D_{\rm v} + \frac{\pi}{4}\right)e^{-D_{\rm v}/\xi _0}.
\label{eq:E+1}
\end{align}
The rapid oscillation on the scale of the Fermi wavelength $k^{-1}_{F}$ reflects the quantum oscillation associated with the Bessel function $J_{\nu}(k_{\rm F}\rho)$ in Eq.~(\ref{eq:cdgmwf}). For $k_{\mu}\xi _0 \!<\! 1$, the splitting energy is evaluated as~\cite{mizushimaPRA10-2,chengPRB10}
\beq
E_{+} \approx  \frac{\Delta _0}{\sqrt{\pi}}\frac{\lambda^{3/2}\sqrt{1-\lambda}}{k_{\rm F}\xi _0}
e^{-(1-\lambda)D_{\rm v}/\xi _0},
\label{eq:E+2}
\eeq
where we introduce $\lambda \!\equiv\! \sqrt{1-(k_{\mu}\xi _0)^2}$. This reflects the characteristic form of the zero-energy wavefunction \eqref{eq:modified} that the quantum oscillation term is absent and the exponential decay factor is canceled out by the modified Bessel function. The latter effect indicates that the wavefunction is extended beyond the vortex core region $\xi _0$ and thus, for $k_{\mu}\xi _0 \!\rightarrow\! 0$, the Majorana zero modes can tunnel between neighboring vortices over a distance much larger than the coherence length $\xi _0$. This is a manifestation of the topological phase transition at $\mu \!=\! 0$.

\begin{figure}[t!]
\includegraphics[width=85mm]{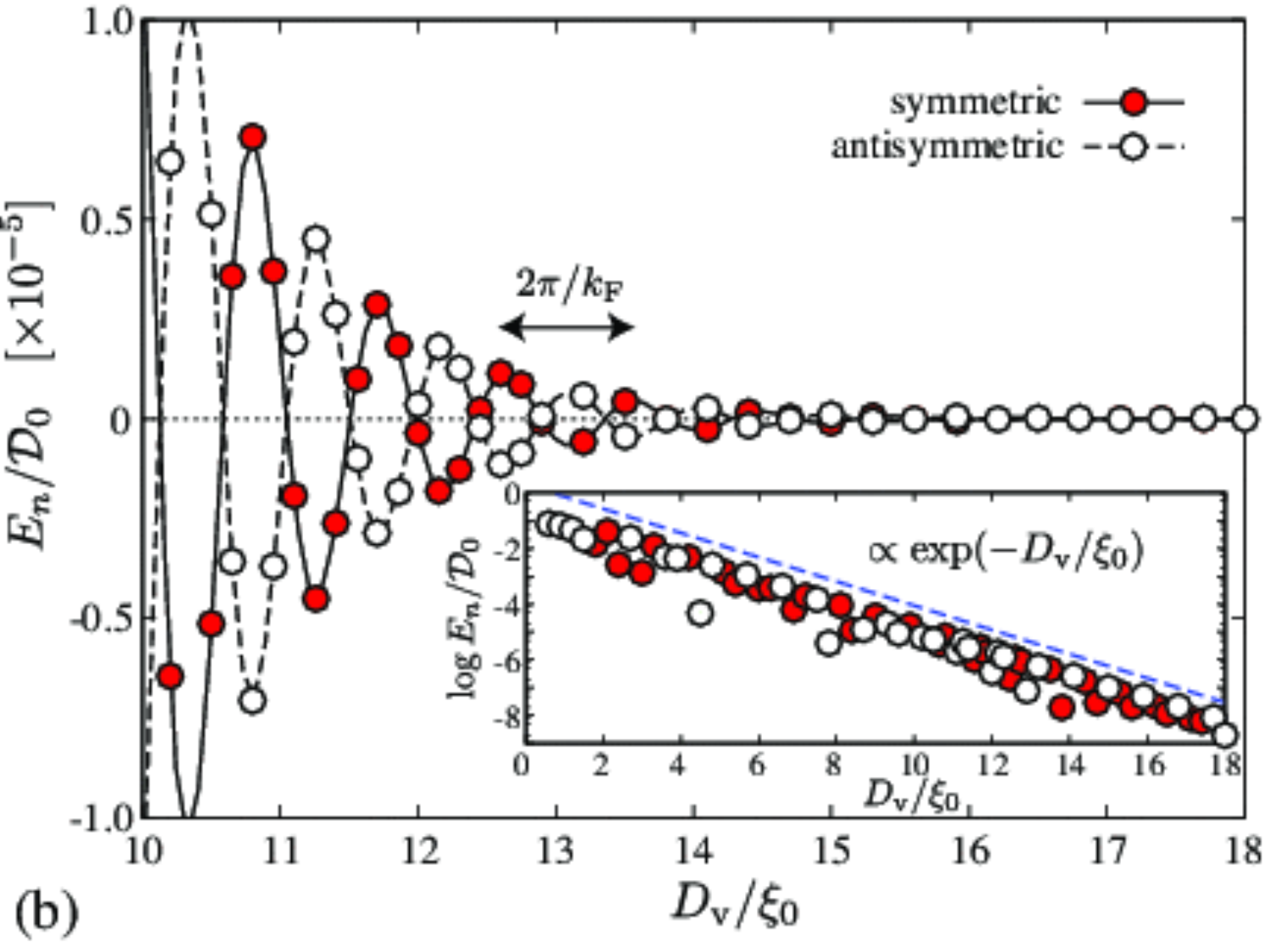}
\caption{(Color online) The lowest eigenenergies are plotted as a function of $D_{\rm v}/\xi _0 \!\in\! [10, 18]$, where the filled (open) circles denote the energies of the symmetric (antisymmetric) state. The inset shows the positive eigenenergies in the range of $D_{\rm v}/\xi _0 \!\in\! [0,18]$ with the logarithmic scale. The dashed line in the inset depicts $\exp(-D_{\rm v}/\xi _0)$. Figure adapted from Ref.~\citeonline{mizushimaPRA10-2}.
}
\label{fig:d0.3}
\end{figure}

Understanding Majorana fermions on a vortex lattice is difficult and the resultant structure is not trivial. A key to solving such a problem was offered by Biswas~\cite{biswasPRL13}, who emphasized the role of the relative difference between phases at two neighboring vortices. To construct an effective model for this issue, let us begin with a tight-binding Hamiltonian for vortex-bound states, $\mathcal{H}_{\rm eff} \!=\! \int d{\bm r}{\bm \Psi}^{\dag}({\bm r})\mathcal{H}({\bm r}){\bm \Psi}({\bm r})$. We here focus on only the subspace of the Majorana zero modes, where the fermionic field operator is expanded in terms of Majorana zero modes bound to vortices as ${\bm \Psi}({\bm r}) \!=\! \sum _j {\bm \varphi}_{m\!=\!0,j}({\bm r})\gamma^{(j)}$ with Eq.~\eqref{eq:varphi}. Then, the tunneling energy between neighboring vortices is given as $t _{ij} \!=\! \int d{\bm r}{\bm \varphi}^{\dag}_{m\!=\!0,i}({\bm r})\mathcal{H}({\bm r}){\bm \varphi}_{m\!=\!0,j}({\bm r})$ and depends on the background ${\rm U}(1)$ phases $\Omega _i$ and $\Omega _j$ at the two vortices. The tunneling amplitude has the phase factor $i \sin[(\Omega _i-\Omega _j)/2]$~\cite{biswasPRL13}, which leads to the antisymmetric relation $t_{ij} \!=\! -t_{ji}$. As a result, the minimal model for vortex lattices is given by the tight-binding Hamiltonian of Majorana fermions,
\beq
\mathcal{H}_{\rm eff} = it\sum _{\langle i,j\rangle} s_{ij} \gamma^{(i)} \gamma^{(j)},
\label{eq:Hmf}
\eeq
where $\langle i,j\rangle$ are nearest neighbors and $\gamma ^{(i)}$ denotes the Majorana operator on the $i$-th vortex. The function $s_{ij}$ is antisymmetric, $s_{ij}\!=\! -s_{ji}$, and reflects the effect of the relative phase difference $\sin[(\Omega _i-\Omega _j)/2]$. The tunneling amplitude $t$ is associated with the splitting energy of Majorana zero modes, $E_+$ in Eqs.~\eqref{eq:E+1} and \eqref{eq:E+2}. The Hamiltonian is also equivalent to that mapped from the Kitaev model, which describes spin $1/2$ fermions on the vertices of a honeycomb lattice. 

It is important to emphasize that the effective Hamiltonian \eqref{eq:Hmf} has the local $\mathbb{Z}_2$ gauge invariance. The self-hermitian condition on $\gamma^{(i)}$ does not prohibit the local $\mathbb{Z}_2$ gauge transformation, which changes the sign of $\gamma^{(i)}$ to $-\gamma^{(i)}$. Since this transformation makes $s_{ij}$ gauge-dependent, Majorana fermions moving along a closed path on vortex lattices accumulate a nontrivial phase factor analogous to the Peierls phase factor. The Majorana fermions residing in a background of  $\mathbb{Z}_2$ magnetic flux bring about intriguing transport properties~\cite{grosfeldPRB06} and rich competing topological phases~\cite{ludwigNJP11,lahtinenPRB12,lahtinenPRB14}. In particular, as shown in Eq.~\eqref{eq:E+1} and Fig.~\ref{fig:d0.3}, the tunneling amplitude $t$ is rapidly oscillating as a function of the vortex separation $D_{\rm v}$ and sensitive to the background ${\rm U}(1)$ phase of the order parameter. In that sense, this situation may be mapped onto Majorana fermions on a distorted vortex lattice or the Kitaev model with random hopping energy. It has been found that the disorder induces unconventional subgap modes~\cite{krausNJP11}. 

Furthermore, the interaction between Majorana fermions can be mediated by a long-range interparticle interaction, such as the Coulomb interaction. Notice that spin-polarized Majorana fermions do not have definite charge density, i.e., $\rho({\bm r}) \!=\! 0$, as shown in Eq.~\eqref{eq:MFcharge}. Nevertheless, the long-range interaction acting on local densities, $H_{\rm int} \!=\! \frac{1}{2}\int d{\bm r}_1\int d{\bm r}_2 \rho ({\bm r}_1)U({\bm r}_{12})\rho({\bm r}_2)$, is mapped to the effective interaction Hamiltonian in the subspace composed of Majorana fermions as
\beq
H_{\rm int}  = \sum _{ijkl} g_{ijkl} \gamma ^{(i)}\gamma ^{(j)}\gamma ^{(k)}\gamma ^{(l)}.
\eeq 
This describes the interaction between two complex fermions, each of which is formed by two paired Majorana fermions on vortex cores. Using interacting Majorana fermion models on specific lattices, Chiu {\it et al.}~\cite{chiuPRB15} revealed that the interaction induces a rich topological phase diagram.

\section{Superfluid $^3$He: Bulk Symmetries and Topologies}
\label{sec:bulk}

\begin{figure}[tb!]
\begin{center}
\includegraphics[width=75mm]{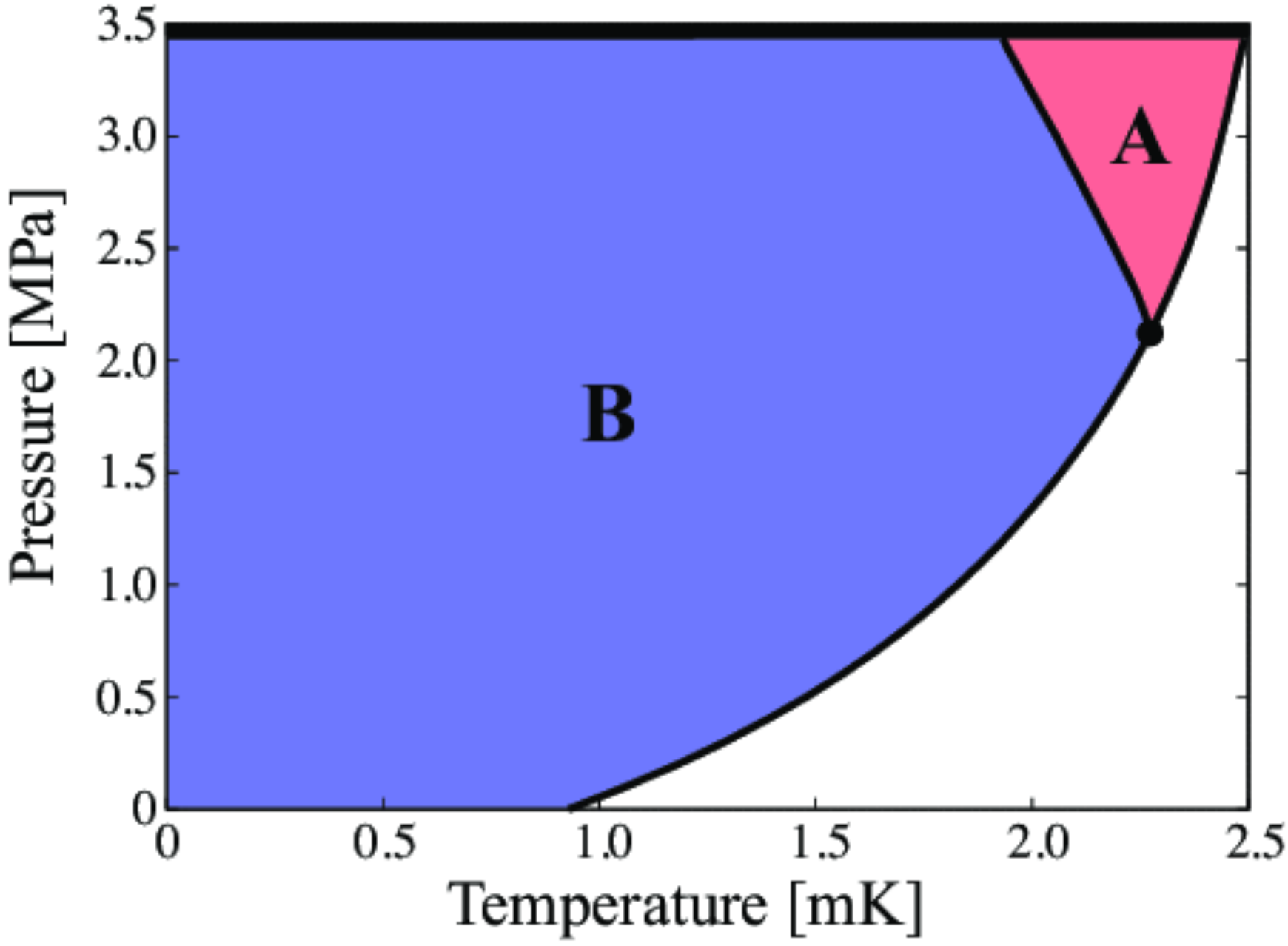} 
\end{center}
\caption{(Color online) Superfluid phase diagram of the bulk $^3$He in the plane of the temperature $T$ and pressure $P$.}
\label{fig:phase_bulk}
\end{figure}

The $^3$He atom is a neutral atom having nuclear spin $1/2$ and zero electron spin. The system remains in the liquid phase down to zero temperature, and possesses the typical properties of a strongly correlated Fermi liquid with the effective mass $m^{\ast}=(1+F^{\rm s}_1/3)m_3$, where $m_3$ is the mass of a $^3$He atom and the spin-independent Landau parameter varies from $F^{\rm s}_1=5.39$ at $P=0$MPa and to $F^{\rm s}_1=14.56$ at $P=3.4$MPa~\cite{vollhardt}. Since the quantum liquid preserves the continuous rotational symmetry in spin and coordinate spaces independently, the liquid $^3$He in normal states is characterized using the huge symmetry group $G$,
\beq
G = {\rm SO}(3)_{\bm L} \times {\rm SO}(3)_{\bm S} \times {\rm U}(1)_{\phi} \times {\rm T} \times {\rm P}.
\label{eq:full}
\eeq
The symmetry group contains the continuous-symmetry groups, which are the group of three-dimensional rotations of coordinate ${\rm SO}(3)_{\bm L}$, the rotation group of spin spaces ${\rm SO}(3)_{\bm S}$, and the global phase transformation group ${\rm U}(1)_{\phi}$. The total group $G$ is also composed of discrete-symmetry groups: The time-reversal ${\rm T}$ and the space parity ${\rm P}$. As shown in Fig.~\ref{fig:phase_bulk}, two distinctive superfluid phases, called the A- and B-phases, are energetically competitive in the bulk. The B-phase is identified as the Balian-Werthamer state~\cite{bw}, which is the spin-triplet $p$-wave pairing with TRS, while the A-phase is established as the Anderson-Brinkman-Morel state, which spontaneously breaks TRS. The symmetry and topology of the bulk BW and ABM states will be summarized in Secs.~\ref{sec:bw} and \ref{sec:abm}.

In a realistic situation of $^3$He, however, there is a magnetic dipole-dipole interaction, which originates from the magnetic moment of $^3$He nuclei. This induces the spin-orbit interaction and reduces the rotational symmetries to the joint rotation ${\rm SO}(3)_{{\bm L}+{\bm S}}$ in $G$. Since the magnetic dipole-dipole interaction is six orders weaker than the dominant inter-atomic interaction having the ${\rm SO}(3)_{\bm L} \times {\rm SO}(3)_{\bm S}$ symmetry, the dipole interaction is typically negligible in the classification of the ground states. We, however, emphasize the role of the nuclear-dipole interaction in the topological superfluidity of $^3$He in a restricted geometry in Sec.~\ref{sec:spt}.


All the thermodynamic and low-lying quasiparticle properties of the bulk $^3$He can be derived from the mean-field Hamiltonian introduced in \eqref{eq:H} with the $4\times 4$ BdG Hamiltonian density $\mathcal{H}({\bm k})$ in Eq.~\eqref{eq:Hbdg}. The relevant pair potential for spin-$1/2$ neutral fermions interacting through the potential $\mathcal{V}^{c,d}_{a,b}({\bm k},{\bm k}^{\prime})$ is obtained in the $2\!\times \! 2$ matrix form as
\beq
\Delta _{ab} ({\bm k}) = \mathcal{V}^{c,d}_{a,b}({\bm k},{\bm k}^{\prime})\langle c_{{\bm k}^{\prime},c} c_{{\bm k}^{\prime},d}\rangle.
\label{eq:gap_bulk}
\eeq 
From now on, we take only a spin-triplet $p$-wave (odd-parity) channel of the interaction potential, which obeys $\mathcal{V}^{c,d}_{a,b} \!=\! \mathcal{V}^{d,c}_{a,b} \!=\! \mathcal{V}^{c,d}_{b,a}$ and $\mathcal{V}^{c,d}_{a,b}({\bm k},{\bm k}^{\prime}) \!=\! -\mathcal{V}^{c,d}_{a,b}(-{\bm k},{\bm k}^{\prime}) \!=\! -\mathcal{V}^{c,d}_{a,b}({\bm k},-{\bm k}^{\prime})$. 
Owing to the emergence of the symmetry breaking term due to the order parameter $\Delta ({\bm k})$, the $4\times 4$ BdG Hamiltonian density $\mathcal{H}({\bm k})$ in Eq.~\eqref{eq:Hbdg} has the remaining symmetry $H \subset G$ and the order parameter manifold is characterized by the coset $R=G/H$.

Following Refs.~\citeonline{salomaaRMP87} and \citeonline{volovik90}, let us now summarize the action of the elements of $G$ on the creation and annihilation operators of $^3$He atoms, $c_{{\bm k}a}$ and $c^{\dag}_{{\bm k}a}$, and $\Delta ({\bm k})$ in Eq.~(\ref{eq:Hbdg}). The continuous rotational groups in the coordinate space and spin space, ${\rm SO}(3)_{\bm L}$ and ${\rm SO}(3)_{\bm S}$, have the generators, the orbital angular momentum operator $\hat{\bm L}\!\equiv\! \hat{\bm r}\times \hat{\bm p}$, and the spin angular momentum operator $\hat{\bm S}$, which act on the field operators of $^3$He atoms as
\beq
\hat{\bm L} c_{{\bm k}a} = -i{\bm k}\times \partial _{\bm k} c_{{\bm k}a}, \hspace{3mm}
\hat{\bm S} c_{{\bm k}a} = \frac{1}{2}{\bm \sigma}_{ab}c_{{\bm k}b}.
\label{eq:LS}
\eeq
The element of ${\rm U}(1)_{\phi}$ rotates the phase of the creation by $\phi$ and annihilation operators as
$\hat{U}_{\phi} c_{{\bm k}a} = e^{-i\phi \hat{N}} c_{{\bm k}a} e^{i\phi \hat{N}} 
= e^{i\phi} c_{{\bm k}a}$,
where $\hat{N} = \sum _{\bm k,a} c^{\dag}_{{\bm k}a}c_{{\bm k}a}$ denotes the number operator of particles. The time-inversion operator $\hat{T}$ transforms a $^3$He atom with the momentum ${\bm k}$ and spin $\sigma$ to an atom with momentum $-{\bm k}$ and spin rotated by $\pi$ by a unitary transformation, 
$\hat{T} {c}_{{\bm k}a} = \Theta _{ab} {c}_{-{\bm k}b}$ with 
$\Theta = i\sigma _y$.
This transformation exchanges the direction of the spin and momentum, $c_{{\bm k}\uparrow} \mapsto c_{-{\bm k}\downarrow}$ and $c_{{\bm k}\downarrow} \mapsto -c_{-{\bm k}\uparrow}$. The space inversion operator $\hat{P}$ rotates the momentum ${\bm k}$ by $\pi$ as 
$\hat{P} c_{{\bm k}a} = c_{-{\bm k}a}$. 
The superfluid and superconducting states also maintain the PHS, whose element $\hat{C}$ exchanges the particle and hole as
\beq
\hat{C}c_{{\bm k}a} = c^{\dag}_{-{\bm k}a}.
\label{eq:ph}
\eeq
Note that the definition of the particle-hole conversion $\hat{C}$ is different from that in Ref.~\citeonline{fishmanPRB85}, which rotates the spin by $\pi$ in addition to the momentum.

The pair potential for bulk superfluids and superconductors is defined as Eq.~(\ref{eq:gap_bulk}) in the momentum space, where the interaction potential $V^{c,d}_{a,b}({\bm k},{\bm k}^{\prime})$ is assumed to be invariant under the full symmetry group $G$. Each element of $G$ acts on $\Delta ({\bm k})$ as 
\begin{gather}
\hat{L}_{\mu}\Delta ({\bm k}) 
= -i \epsilon ^{\mu\nu\eta} k_{\nu} \partial _{k_{\eta}} \Delta ({\bm k}), 
\label{eq:L} \\
\hat{S}_{\mu}\Delta ({\bm k}) = \frac{1}{2}(\sigma _{\mu})\Delta ({\bm k}) 
+ \frac{1}{2}\Delta ({\bm k})\sigma^{\rm T}_{\mu},
\end{gather}
$\hat{U}_{\phi}\Delta ({\bm k}) = e^{2i\phi}\Delta ({\bm k})$,
$\hat{T} \Delta _{ab}({\bm k}) = \Theta \Delta^{\ast}(-{\bm k}) \Theta^{\rm T}$, 
$\hat{P} \Delta ({\bm k}) = \Delta (-{\bm k})$, 
and $\hat{C} \Delta ({\bm k}) = \Delta^{\ast}(-{\bm k})$,
where we utilize the basic properties of $\mathcal{V}^{c,d}_{a,b}({\bm k},{\bm k}^{\prime})$.

\subsection{Symmetry and topology of the bulk $^3$He-B}
\label{sec:bw}

{\it Order parameter.}---
The irreducible representations of $G$ are characterized by the values of orbital and spin moments $L$ and $S$, the space parity $P$, and the quantum number $N$ associated with the number of paired particles. We here focus our attention to the superfluid $^3$He-B, that is, spin-triplet $p$-wave pairing states having $L=S=1$, $P=-1$, $N=2$. 

The B-phase is identified as the Balian-Werthamer (BW) state~\cite{bw}. The order parameter having $S=L=1$ is classified with the quantum numbers $J = 0$, $1$, and $2$, where $J$ denotes the total angular momentum operator and the generator of the joint rotational symmetry group ${\rm SO}(3)_{{\bm L}+{\bm S}}$. The simplest form of ${\bm d}({\bm k})$ is categorized into the $J=0$ sector, 
\begin{align}
\Delta_0 ({\bm k}) &= \Delta _{\rm B} \left[
\hat{k}_{-} \left| \uparrow\uparrow \right\rangle
+ \hat{k}_z \left| \uparrow\downarrow + \downarrow \uparrow\right\rangle
+ \hat{k}_+ \left| \downarrow\downarrow \right\rangle
\right],
\label{eq:diag}
\end{align}
where $\hat{k}_{\pm} \!=\! \pm (\hat{k}_x\pm i\hat{k}_y)$ are the eigenstates of $L\!=\! 1$ and $L_z \!\pm \! 1$. This form of $\Delta _0({\bm k})$ implies that the ${\bm d}$-vector is parallel to the ${\bm k}$-vector, $d_{\mu}({\bm k}) = \Delta _0\hat{k}_{\mu}$. The BW state holds the maximal symmetry group of the subset of $G$, 
\beq
{H}_{\rm B} = {\rm SO}(3)_{{\bm L}+{\bm S}} \times {\rm T} \times {\rm P}{\rm U}_{\pi/2}, 
\label{eq:symmetrybw}
\eeq
where ${\rm P}{\rm U}_{\pi/2}$ denotes the combined discrete symmetry of the inversion ${\rm P}$ and the $\pi/2$-phase rotation ${\rm U}_{\pi/2}$. The broken symmetry $R$ of the BW state is then given by 
\beq
\mathcal{R}=G/H_{\rm B} = {\rm SO}(3)_{{\bm L}-{\bm S}} \times {\rm U}(1)_{\phi},
\label{eq:R}
\eeq
where ${\rm SO}(3)_{{\bm L}-{\bm S}} $ is the relative rotation of the spin and orbital spaces. Hence, the BW state is regarded as the spontaneous breaking phase of the spin-orbit symmetry~\cite{leggettRMP}, and the spin-orbit interaction emerges as a result of the spontaneous breaking.


As shown in Eq.~(\ref{eq:R}), the degeneracy space is characterized in Eq.~(\ref{eq:R}) by the continuous rotations ${\rm SO}(3)_{{\bm L}-{\bm S}}$ and ${\rm U}(1)_{\phi}$. Therefore, the generic form of the BW state is obtained by rotating the ${\bm d}$-vector from the ${\bm k}$-vector as
\beq
d_{\mu}({\bm k}) = \Delta _{\rm B} e^{i\phi} R_{\mu\nu}(\hat{\bm n},\varphi) \hat{k}_{\nu},
\label{eq:dvecB}
\eeq
where the rotation matrix $R_{\mu\nu}$ is associated with the spin-orbit symmetry breaking ${\rm SO}(3)_{{\bm L}-{\bm S}}$ and $\phi$ arises from the breaking of the ${\rm U}(1)_{\phi}$ symmetry. In the bulk BW state without the magnetic field, the angle $\varphi$ is fixed to the so-called Leggett angle, 
\beq
\varphi _{\rm L} = \cos^{-1}\left( -\frac{1}{4}\right),
\label{eq:leggett}
\eeq
while the orientation $\hat{\bm n}$ is not uniquely determined. The $\hat{\bm n}$-vector is, however, affected by confinement and a magnetic field, as shown in Sec.~\ref{sec:spt}. The equilibrium order parameter of the BW state is the eigenfunction of the total twisted angular momentum~\cite{makiJLTP74,fishmanPRB86}
\beq
J_{\mu} =L_{\mu} + S_{\nu}R_{\nu\mu}(\hat{\bm n},\varphi).
\eeq
This implies that the generic form of the BW state is expressed as a combination of all the total angular momentum sectors, $J = 0$, $1$, and $2$. The BdG Hamiltonian with the generic form of ${\bm d}({\bm k})$ in Eq.~\eqref{eq:dvecB} is associated with $\mathcal{H}_0({\bm k})$ with the diagonal representation \eqref{eq:diag} as 
\beq
\mathcal{U}(\hat{\bm n},\varphi)\mathcal{H}_0({\bm k})\mathcal{U}^{\dag}(\hat{\bm n},\varphi) = \mathcal{H}({\bm k}), 
\label{eq:HB}
\eeq
where $\mathcal{U}(\hat{\bm n},\varphi) \!=\! {\rm diag}[U(\hat{\bm n},\varphi),U^{\ast}(\hat{\bm n},\varphi)]$ is the ${\rm SU}(2)$ matrix extended to the Nambu space.

The BW order parameter is composed of three components of the ${\bm d}$-vectors. This implies that one of the ${\bm d}$-vectors is always parallel to an applied magnetic field, while the others are not responsible for the suppression of the spin susceptibility. Hence, the magnetic response of the BW state is isotropic and the spin susceptibility at zero temperature is given by $2/3$ times the spin susceptibility of the normal $^3$He when the Fermi liquid correction is ignored. However, since the liquid $^3$He is a strongly interacting Fermi liquid, the Fermi liquid correction gives rise to the screening (or enhancement) effect of the applied magnetic field. The effective magnetic field that the liquid $^3$He is subjected to is deviated from the applied field ${\bm H}$ by the ``exchange'' interaction to a polarized medium. In the bulk BW state, therefore, the renormalized spin susceptibility is given by 
\beq
\chi _{\mu \nu} = \delta _{\mu\nu}\frac{(1+F^{\rm a}_0)[2+Y(T)]}{3+F^{\rm a}_0[2+Y(T)]}\chi _{\rm N},
\label{eq:chibw}
\eeq
where $Y(T)$ is the Yosida function at the temperature $T$~\cite{vollhardt}. The renormalized spin susceptibility in the normal $^3$He is given by
$\chi _{\rm N} = \frac{1}{2}\frac{\gamma^2\mathcal{N}_{\rm F}}{1+F^{\rm a}_0}$,
where $\mathcal{N}_{\rm F}$ is the density of states at the Fermi level. 
The Fermi liquid parameter $F^{\rm a}_0$ describes the ``exchange'' interaction to a polarized medium, and for $^3$He at low temperatures, the Fermi liquid parameter is $F^{\rm a}_0 \!\sim\! -0.7$, slightly depending on the pressure~\cite{vollhardt}. In the normal state, therefore, the Fermi liquid correction has the ferromagnetic exchange interaction, which enhances the polarization of the medium. Notice that the Stoner instability occurs at $F^{\rm a}_0 \!=\! -1$, which cannot be accomplished by the liquid $^3$He. In the bulk BW state, as shown in Eq.~\eqref{eq:chibw}, the ``ferromagnetic exchange interaction'' with $F^{\rm a}_0< 0$ enhances the suppression of $\chi _{\mu\nu}$. 


{\it Symmetries.}---
The remaining symmetry in the BW state, ${\rm SO}(3)_{{\bm L}+{\bm S}}$, simultaneously rotates the momentum and the ${\bm d}$-vector as $\hat{k}_{\mu} \mapsto \hat{k}^{\prime}_{\mu}=R^{(L)}_{\mu\nu}\hat{k}_{\nu}$ and $d_{\mu} \mapsto d^{\prime}_{\mu}=R^{(S)}_{\mu\nu}d_{\nu}$, where $R^{(S)}_{\mu\nu} =(RR^{(L)}R^{-1})_{\mu\nu}$ denotes the rotational matrix in the spin space. 
Then, the ${\rm SU}(2)$ representation of the ${\rm SO}(3)_{{\bm L}+{\bm S}}$ symmetry in the bulk BW state is given as 
\beq
\mathcal{U}_{ S}\mathcal{H}({\bm k})\mathcal{U}^{\dag}_{ S} = \mathcal{H}(R^{(L)}{\bm k}),
\label{eq:SU2}
\eeq
where we introduce $\mathcal{U}_{ S} = \mathcal{U}(\hat{\bm n},\varphi)\mathcal{U}_{ L} \mathcal{U}^{\dag}(\hat{\bm n},\varphi)$, and $\mathcal{U}_{ L}\equiv {\rm diag}(U_{ L},U^{\ast}_{ L})$ is the ${\rm SU}(2)$ representation of the rotation matrix associated with $R^{(L)}$. 

The BW state holds the discrete symmetries that play crucial roles in determining the topological properties: (i) PHS and (ii) TRS. The quasiparticle states are twofold degenerate as a consequence of the TRS with $\mathcal{T}^2=-1$, which form a Kramers pair as shown in Eq.~\eqref{eq:kramers}. In addition, since the inversion operator $\hat{P}$ acts on the B-phase pair potential as $\hat{P}\Delta ({\bm k})=-\Delta (-{\bm k})$, the inversion symmetry is realized up to the $U(1)_{\phi=\pi/2}$ gauge symmetry as
\beq
\mathcal{P}\mathcal{H}({\bm k})\mathcal{P}^{\dag} = \mathcal{H}(-{\bm k}),
\hspace{3mm} \mathcal{P} = \tau _z.
\label{eq:inversion}
\eeq
In contrast to the TRS in Eq.~(\ref{eq:TRS}), this operation changes the sign of the momentum without changing the spin state. The inversion symmetry in Eq.~(\ref{eq:inversion}) results in the relation between quasiparticle states with ${\bm k}$ and $-{\bm k}$,
$|u_n({\bm k})\rangle = \mathcal{TP}|u_n({\bm k})\rangle $.
The twofold degenerate energy spectrum in the BW state is obtained by diagonalizing $\mathcal{H}({\bm k})$ as
\beq
E({\bm k}) = \pm \sqrt{[\varepsilon ({\bm k})]^2 + \Delta^2_{\rm B}},
\label{eq:EkB}
\eeq
which yields a fully gapped excitation at the Fermi level.



{\it Bulk topology.}--- 
As mentioned in Sec.~\ref{sec:topology}, the topological properties of superfluids and superconductors are determined by the global structure of the Hilbert space spanned by the eigenvectors of the occupied band, $|u_n({\bm k})\rangle$. Since the BdG Hamiltonian of the BW state holds the chiral symmetry Eq.~\eqref{eq:chiral} as a combination of TRS and PHS, the BW state is categorized into class DIII in the AZ table~\cite{schnyderPRB08}. 


The topological property of class DIII is obtained from the $Q$-matrix introduced in Eq.~\eqref{eq:Qmat}. The $Q$-matrix maps the three-dimensional ${\bm k}$-space ($S^3$) onto the target space $\mathcal{M}$ spanned by the eigenvectors $|u_n({\bm k})\rangle$. As clarified in Sec.~\ref{sec:topology}, on the basis that the chiral operator is diagonal $\Gamma \!=\! {\rm diag}(+1,-1)$, the $Q$-matrix becomes off-diagonal and is reduced to an element of the unitary group ${\rm U}(2)$, $q({\bm k})$. The relevant homotopy group for the projector
$Q({\bm k})$ in three dimensions is given by $\pi _3 [{\rm U}(2)] = \mathbb{Z}$. 
The topological invariant that characterizes the
classes of topologically distinct $q$-configurations in bulk
superconductors and superfluids is defined as the three-dimensional
winding number
\begin{align}
w_{\rm 3d} 
&= \int\frac{d{\bm k}}{24\pi^3}\epsilon ^{\mu\nu\eta}{\rm tr}
\left[
(q^{\dagger}\partial _\mu q)
(q^{\dagger}\partial _\nu q)
(q^{\dagger}\partial _\eta q)
\right]. 
\label{eq:w3d} \\
&=-\int\frac{d{\bm k}}{48\pi^3}\epsilon^{\mu\nu\eta}
{\rm tr}\left[\Gamma (Q\partial_\mu Q)(Q\partial_\nu Q)(Q\partial_\eta Q)\right].
\label{eq:windingQ} 
\end{align}
For the DIII class with $\mathcal{T}^2=-1$ and $\mathcal{C}^2=+1$, the
winding number can be an arbitrary integer value. 

A generic form of a $4\times 4$ hermitian matrix can be expanded in terms of the five $\gamma$-matrices and their ten commutators in addition to the unit matrix as
$\mathcal{H}({\bm k}) = m_0({\bm k}) 
+ \sum _j m_j({\bm k})\gamma _j
+ \sum _{ij} m_{ij}({\bm k})\gamma _{ij}$. 
We here introduce the $\gamma$-matrices that satisfy the relations 
$\{ \gamma _i, \gamma _j \} = 2\delta _{ij}$, where $i,j = 1, 2, \cdots, 5$. 
We also introduce their commutator $\gamma _{ij}$, which is defined as 
$\gamma _{ij} = \frac{1}{2i}[\gamma _i, \gamma _j] $. Since the BW state holds the inversion symmetry and TRS, we choose the $\gamma$ matrix to be even under $\mathcal{P}\mathcal{T}$, $\mathcal{PT} \gamma _j \mathcal{T}^{-1}\mathcal{P}^{-1}=\gamma _j$. Then, the five $\gamma$ matrices are given as 
$(\gamma _1, \gamma _2, \gamma _3, \gamma _4) \!=\!  
(- \sigma_z \tau _x, - \tau _y,  \sigma_x \tau _x, \tau _z)$.
Using this expression, one finds that the inversion symmetry and TRS require all $d_{ij}({\bm k})$ to be zero. In addition, $m_0({\bm k})$ and $m_5({\bm k})$ must vanish because of PHS \eqref{eq:PHS} and the ${\rm SO}(3)_{{\bm L}+{\bm S}}$ symmetry, respectively. As a result, the Hamiltonian with PHS, TRS, inversion symmetry, and ${\rm SO}(3)$ symmetry is generally parameterized with four Dirac $\gamma$ matrices and the four-dimensional vector $[m_1({\bm k}),m_2({\bm k}),m_3({\bm k}),m_4({\bm k})]$. Then, the corresponding $Q$-matrix is given by 
\beq
Q({\bm k}) = \sum^4_{j=1} \hat{m}_j({\bm k})\gamma _j .
\label{eq:h2}
\eeq
In the above representation of $\gamma$ matrices, the chiral operator $\Gamma$ is written as $\Gamma \!=\!\gamma_5
\!\equiv\!\gamma_1\gamma_2\gamma_3\gamma_4$. The four-dimensional spinor $\hat{m}_{\mu}({\bm k})$ satisfies
$\hat{\bm m}\cdot\hat{\bm m} \!=\! 1$, so it defines a three-sphere $S^3$ with unit radius. 

The parametrization of the $Q$-matrix in Eq.~\eqref{eq:h2} implies that for single-band spin-triplet superfluids/superconductors, there exists a simpler expression for $w_{\rm 3d}$. By substituting the $Q$-matrix to Eq.~(\ref{eq:windingQ}), the winding number is recast into the following form:~\cite{satoPRB09}
\begin{eqnarray}
w_{\rm 3d}=\int \frac{d{\bm k}}{12\pi^3}
\epsilon^{\mu\nu\eta}\epsilon^{ijkl}
\hat{m}_i({\bm k})
\partial_\mu\hat{m}_j({\bm k})
\partial_\nu\hat{m}_k({\bm k})
\partial_\eta\hat{m}_l({\bm k}). 
\end{eqnarray}
This expression for $w_{\rm 3d}$ indicates that $w_{\rm 3d}$ counts how many the unit vector
$\hat{m}_j({\bm k})$ wraps the three-dimensional sphere when one
sweeps the entire momentum space.

Using the topological nature of the integral and $\hat{\bm m}({\bm k})=\left({\bm d}({\bm k}), \varepsilon_0({\bm k})\right)/|E({\bm k})|$, the above $w_{\rm 3d}$ can 
be calculated as~\cite{satoPRB09}
\begin{align}
w_{\rm 3d}=&-\frac{1}{2}\sum_{{\bm k}_0} 
{\rm sgn}[\varepsilon_0({\bm k}_0)] 
{\rm sgn}[{\rm det}\{\partial_\mu d_\nu({\bm k}_0)\}]
\nonumber\\
&+\frac{1}{2}
{\rm sgn}[\varepsilon_0(\infty)] 
{\rm sgn}[{\rm det}\{\partial_id_j(\infty)\}],
\end{align}
where the summation is taken for all ${\bm k}_0$ satisfying ${\bm d}({\bm
k}_0)=0$ and ${\rm det}\{\partial_\mu d_\nu({\bm k}_0)\}$ denotes the
determinant of the $3\times 3$ matrix
$\partial_\mu d_\nu({\bm k}_0)$. 
The second term in the right-hand side
is the contribution from ${\bm k}_0\rightarrow \infty$.
From the above formula, the winding number of the B-phase is evaluated
as $w_{\rm 3d}=1$.

The BW state realized in the bulk $^3$He serves a prototype of
three-dimensional topological superfluids and superconductors, where the
nontrivial topology is protected by PHS and TRS~\cite{schnyderPRB08}. After a pioneering work
by Grinevich and Volovik~\cite{grinevichJLTP88},
the topological superfluidity of the BW state was discussed by
Schnyder {\it et al}.~\cite{schnyderPRB08}, Roy~\cite{roy08}, Qi {\it et al}.~\cite{qiPRL09}, Volovik~\cite{volovikJETP09v2}, and Sato~\cite{satoPRB09}. The nontrivial topological invariant ensures the existence of surface helical Majorana fermions with a linear dispersion, which brings about novel quantum phenomena~\cite{chungPRL09,nagatoJPSJ09,volovikJETP09,volovikJETP09v2,shindouPRB10,silaevPRB11,mizushimaPRL12,silaevJETP12v2,mizushimaJPCM15,silaevJETP14}.

\subsection{Helical Majorana fermions in $^3$He-B}
\label{sec:MF3HeB}

The direct consequence of the nontrivial topological invariant is the emergence of helical Majorana fermions on the surface. Since the BW state with a specular surface is invariant under the ${\rm SO}(2)_{{\bm L}+{\bm S}}$ rotation about the surface normal axis, in addition to the PHS, the BdG Hamiltonian holds the magnetic point group symmetry with $A_{\rm spin} \!=\! -i \tilde{\sigma}_z$ in Eq.~\eqref{eq:aspin}, where $\tilde{\sigma}_{\mu}$ is the spin Pauli matrices associated with $R^{({\rm s})}_{\mu\nu}$. It is then obvious from Sec.~\ref{sec:chiral} that a pair of topologically protected zero modes in the BW state possesses the Majorana Ising spins as a generic consequence of the chiral symmetry.

Let us show the dispersion and Majorana nature of topologically protected gapless states bound to the surface of the BW state by analytically solving the Andreev equation \eqref{eq:a}. Here, we set a specular surface to be normal to the $\hat{\bm z}$-axis and the region $z>0$ is occupied by the $^3$He-B. We also assume the spatially uniform isotropic energy gap $\Delta _{\rm B}$. It is worth mentioning that using the ${\rm SU}(2)$ spin rotation $U(\hat{\bm n},\varphi)$ and the unitary matrix $S_{\phi_{\bm k}} \!\equiv\! (\sigma _x+\sigma _z)e^{i\vartheta\sigma _z}/\sqrt{2}$ with $\vartheta = \frac{\phi _{\bm k}}{2} - \frac{\pi}{4}$, one finds that the  pair potential of the BW state can be transformed into the diagonal representation
\begin{align}
{\Delta}(\hat{\bm k},{\bm r}) 
= U(\hat{\bm n},\varphi) S_{\phi_{\bm k}}
\left( 
\begin{array}{cc}
\Delta _{\rm B}e^{i\theta _{\bm k}} & 0 \\
0 & -\Delta _{\rm B} e^{-i\theta _{\bm k}}
\end{array}
\right)
S^{\rm T}_{\phi_{\bm k}}U^{\rm T}(\hat{\bm n},\varphi),
\label{eq:unitaryB} 
\end{align}
where we set ${\bm k}_{{\rm F}} \!=\! k_{\rm F}(\cos\!\phi _{\bm k}\sin\!\theta _{\bm k},\sin\!\phi _{\bm k}\sin\!\theta _{\bm k},\cos _{\bm k}\!\theta _{\bm k})$. 
This implies that the unitary transformation effectively maps the Andreev equation \eqref{eq:a} with the BW order parameter onto spin-polarized chiral $p$-wave superconductors~\cite{mizushimaPRB12,mizushimaJPCM15,nagaiJPSJ08}.

In accordance with the consequence of the index theorem in Sec.~\ref{sec:index2}, the bound state solution with $|E({\bm k}_{\parallel})| \!\le\! \Delta _{\rm B}$ has an energy dispersion linear in the momentum ${\bm k}_{\parallel} \!=\! (k_x,k_y)$ as
\beq
E_0({\bm k}_{\parallel}) = \pm \frac{\Delta _{\rm B}}{k_{\rm F}} |{\bm k}_{\parallel}|.
\label{eq:E0a}
\eeq 
This expression is independent of the orientation of $\hat{\bm n}$ and the angle $\varphi$. The wavefunction for the positive energy branch is given by~\cite{mizushimaPRB12,mizushimaJPCM15,nagaiJPSJ08,nagatoJPSJ09,chungPRL09}
\beq
{\bm \varphi}^{(+)}_{0,{\bm k}_{\parallel}} ({\bm r}) = N_{\bm k}
e^{i{\bm k}_{\parallel}\cdot{\bm r}_{\parallel}}f(k_{\perp},z)\mathcal{U}(\hat{\bm n},\varphi)
\left(
{\bm \Phi}_+
- e^{i\phi _{\bm k}}{\bm \Phi}_-
\right), 
\label{eq:varphi1}
\eeq
where $N_{\bm k}$ is the normalization constant and $\mathcal{U} \!\equiv\! {\rm diag}(U,U^{\ast})$. The PHS in Eq.~(\ref{eq:PHS}) ensures the one-to-one correspondence between the two branches of the energy eigenstates through ${\bm \varphi}^{(-)}_{0,{\bm k}_{\parallel}}({\bm r}) \!=\! \mathcal{C}{\bm \varphi}^{(+)}_{0,-{\bm k}_{\parallel}}({\bm r})$. In Eq.~(\ref{eq:varphi1}), we also set $f(k_{\perp},z)\!=\! \sin\left( k_{\perp}z\right)e^{-z/\xi}$ with $k^2_{\perp} \!\equiv\! k^2_{\rm F}-k^2_{\parallel}$. The spinors,  
${\bm \Phi}_+ \!\equiv\! (1,0,0,-i)^{\rm T}$ and ${\bm \Phi}_- \!\equiv\! (0,i,1,0)^{\rm T}$, are the eigenvectors of the spin operator $S_{z}\equiv \frac{1}{2}{\rm diag}(\sigma _z, - \sigma^{\rm T}_z)$ in the Nambu space, 
\beq
S_z {\bm \Phi}_{\pm} = \pm \frac{1}{2} {\bm \Phi}_{\pm}. 
\label{eq:sz}
\eeq

Following the same procedure as in Sec.~\ref{sec:chiral}, we expand the quantized field ${\bm \Psi}\!=\! (\psi _{\uparrow}, \psi _{\downarrow},\psi^{\dag}_{\uparrow}, \psi^{\dag}_{\downarrow})^{\rm T}$ in terms of the positive energy states of the surface Andreev bound states. For low-temperature regimes $T\!\ll\!\Delta _{\rm B}$, the field operator can be constructed from the contributions of only the surface Andreev bound states as 
\beq
{\bm \Psi}({\bm r}) 
= \sum _{{\bm k}_{\parallel}} 
\left[ {\bm \varphi}^{(+)}_{0,{\bm k}_{\parallel}}({\bm r})\eta _{{\bm k}_{\parallel}}
+\mathcal{C}{\bm \varphi}^{(+)}_{0,{\bm k}_{\parallel}}({\bm r})\eta^{\dag}_{{\bm k}_{\parallel}}
\right] , 
\label{eq:psi1}
\eeq
where the continuum states with $E\!>\! \Delta _{\rm B}$ are omitted. By substituting Eq.~(\ref{eq:varphi1}) into the expanded form of ${\bm \Psi}$, the quantized field operator contributed from the surface Andreev bound states obeys the Majorana Ising condition \eqref{eq:miscond} 
\beq
\left(
\begin{array}{c}
\psi _{\uparrow}({\bm r}) \\
\psi _{\downarrow}({\bm r})
\end{array}
\right) = i \sigma _{\mu}R_{\mu z}(\hat{\bm n},\varphi)
\left(
\begin{array}{c}
\psi^{\dag}_{\downarrow}({\bm r}) \\
-\psi^{\dag}_{\uparrow}({\bm r}) 
\end{array}
\right).
\label{eq:Mcondition2}
\eeq

The condition in Eq.~\eqref{eq:Mcondition2} clarifies that the Majorana fields constructed from the surface Andreev bound states reproduce the Ising spin property and the surface bound states are not coupled to the local density operators, as generically derived in Eq.~(\ref{eq:MIS1}). Using Eq.~(\ref{eq:Mcondition2}), one first finds that the surface states do not contribute to the local density operator,
\beq
\rho^{({\rm surf})} ({\bm r}) = 0.
\label{eq:rho0}
\eeq 
The surface helical Majorana fermions cannot be coupled to the local density fluctuation and thus are very robust against nonmagnetic impurities. Similarly, the local spin operator is constructed from the surface Majorana fermion in Eq.~(\ref{eq:Mcondition2}) as 
\beq
S^{({\rm surf})}_{\mu} = R_{\mu z}(\hat{\bm n},\varphi) S^{\rm M}_z,
\label{eq:MIS3}
\eeq
where $S^{\rm M}_z$ is the logical spin operator in the case of $\hat{\bm n}=\hat{\bm z}$ and $\varphi = 0$. Equation (\ref{eq:MIS3}) implies that only the $S_z$-component is nonzero while the other components are identically zero when $\hat{\bm n}=\hat{\bm z}$ and $\varphi = 0$. With the Majorana Ising spin in Eq.~\eqref{eq:MIS3}, the dynamical spin susceptibility is obtained as 
\beq
\chi _{\mu\nu}({\bm r}_1,{\bm r}_2; \omega) 
= \chi^{\rm M}_{zz}({\bm r}_1,{\bm r}_2;\omega) R_{\mu z}(\hat{\bm n},\varphi) R_{\nu z} (\hat{\bm n},\varphi),
\label{eq:chiM}
\eeq
where $\chi^{({\rm M})}_{zz}({\bm r}_1,{\bm r}_2;\omega)\!\equiv\! \langle S^{\rm M}_{z}({\bm r}_1)S^{\rm M}_{z}({\bm r}_2)\rangle _{\omega}$. The property of the dynamical spin susceptibility was discussed in Refs.~\citeonline{chungPRL09}, \citeonline{mizushimaJLTP11}, \citeonline{silaevPRB11}, and \citeonline{kallinPRB15}. 

We emphasize that the whole branch of the surface Andreev bound states can approximately retain the Ising spin character. The Majorana nature of the whole branch is a consequence of the Andreev approximation within $k_{\rm F}\xi \gg 1$, and only the exact zero energy mode at ${\bm k}_{\parallel} \!=\! {\bm 0}$ is rigorously protected by the topology number. The approximated Majorana nature may enable the realization of the macroscopic Ising-like spin correlation. This will be discussed in Sec.~\ref{sec:spt}.

Using Eqs.~\eqref{eq:psi1} and \eqref{eq:Mcondition2}, one can obtain the effective Hamiltonian for the surface Majorana fermions as~\cite{qiPRL09,volovikJETP10}
\beq
\mathcal{H}_{\rm surf} = \sum _{{\bm k}_{\parallel}} \psi^{\rm T}_{-{\bm k}_{\parallel}} c \left( {\bm k}_{\parallel} \times \tilde{\bm \sigma} \right)\cdot\hat{\bm s}\psi _{{\bm k}_{\parallel}},
\label{eq:Hsurf}
\eeq
where we set $\tilde{\sigma}_{\mu} = R_{\mu \nu}(\hat{\bm n},\varphi)\sigma _{\nu}$. The momentum parallel to the surface is ${\bm k}_{\parallel}$ and the unit vector normal to the surface is denoted by $\hat{\bm s}$. This effective surface Hamiltonian yields the gapless relativistic spectrum with the velocity $c = \Delta _{\rm B} /k_{\rm F}$, which is protected by the TRS. In addition, the field operator $\psi$ in the Nambu space is self-conjugate, that is, the Majorana fermion. 

Only the perturbation that generates an effective mass in Eq.~\eqref{eq:Hsurf} turns out to be an external potential coupled to the Ising spin of surface Majorana fermions ${\bm S}^{({\rm surf})}$ in Eq.~\eqref{eq:MIS3},
\beq
\mathcal{H}_{\rm mass} = M \sum _{{\bm k}_{\parallel}}  \psi^{\rm T}_{-{\bm k}_{\parallel}} \tilde{\bm \sigma}\cdot\hat{\bm s}\psi _{{\bm k}_{\parallel}}.
\label{eq:Hsurfm}
\eeq
Since the mass $M$ changes its sign under the time-inversion $\mathcal{T}$, the effective Hamiltonian \eqref{eq:Hsurf} with the mass term \eqref{eq:Hsurfm} breaks the chiral symmetry. The resultant surface spectrum is of the massive Majorana fermion,
\beq
E_{\rm surf}({\bm k}_{\parallel}) = \sqrt{c^2 k^2_{\parallel} + M^2}. 
\eeq
The manifestation of three-dimensional topological superconductors is the coupling to the gravitational field through the gravitational instanton term~\cite{wangPRB11,ryuPRB12}. Owing to the nontrivial topological property, the gapped Majorana cone is responsible for the quantization of thermal Hall conductivity. It has also been predicted that the coupling of the Majorana fermion to the gravitational field gives rise to cross-correlated responses~\cite{nomuraPRL12}. In Sec.~\ref{sec:sptb}, we clarify that in  $^3$He-B confined in a slab geometry, the mass term $M$ of Eq.~\eqref{eq:Hsurf} is associated with the order parameter manifold $(\hat{\bm n},\varphi)$ and thus the acquisition of the effective mass is driven by the spontaneous symmetry breaking.

\subsection{$^3$He-A: A prototype of Weyl superconductors}
\label{sec:abm}

\begin{figure}[tb!]
\begin{center}
\includegraphics[width=80mm]{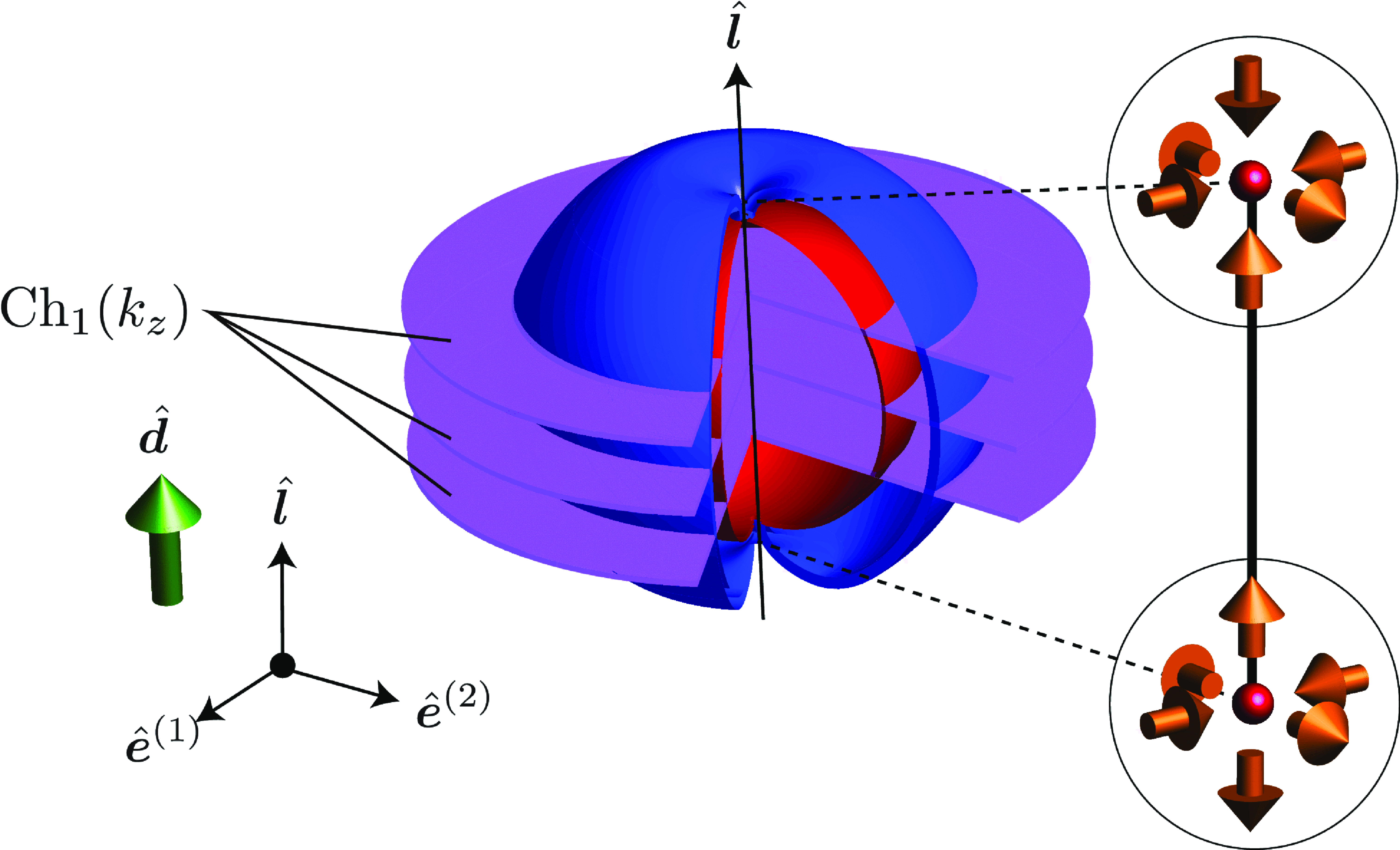} 
\end{center}
\caption{(Color online) (Left) Triad $(\hat{\bm e}^{(1)},\hat{\bm e}^{(2)},\hat{\bm l})$ that characterizes the orbital component of the ABM state. (Right) Gap function and $\hat{\bm d}$ of the ABM state on the Fermi sphere and sliced planes in which the first Chern number is nontrivial. In the zero field, $\hat{\bm d}$ is locked by the dipole interaction to the $\hat{\bm l}$-vector. We also depict the $\hat{\bm g}$-texture near the Fermi points.}
\label{fig:abm}
\end{figure}

The A-phase, which appears in the high-$P$ and high-$T$ region of Fig.~\ref{fig:phase_bulk}, corresponds to the Anderson-Brinkman-Morel (ABM) state~\cite{abm1,abm2}. The order parameter of the ABM state is given by the complex form
\beq
d_{\mu j}({\bm k}) = \Delta _{\rm A} \hat{d}_{\mu} e^{i\varphi}(\hat{e}^{(1)}_{j} + i\hat{e}^{(2)}_j) .
\label{eq:abm}
\eeq
This indicates that the ABM state is composed of the antiferromagnetic spin state ($\hat{S}\Delta ({\bm k}) \!=\! 0$) and orbital ferromagnet ($(\hat{L}_z-1)\Delta ({\bm k}) \!=\! 0$). Hence, the ABM state spontaneously breaks the TRS. The orbital part of the order parameter is characterized by a set of three unit vectors, which form the triad $(\hat{\bm e}^{(1)},\hat{\bm e}^{(2)},\hat{\bm l})$, as shown in Fig.~\ref{fig:abm}. The $\hat{\bm \ell}$-vector is defined as
\beq
\hat{\bm l} = \hat{\bm e}^{(1)} \times \hat{\bm e}^{(2)}.
\eeq
The $\hat{\bm l}$-vector physically represents the orientation of the nodal directions at which the bulk excitation is gapless. The bulk excitation energy in the ABM state is given by
\beq
E({\bm k}) = \pm \sqrt{\varepsilon^2({\bm k}) + \Delta^2_{\rm A}\sin^2\theta _{\bm k}},
\label{eq:Eabm}
\eeq
where $\theta _{\bm k}$ is the tilting angle from the $\hat{\bm l}$-direction. 

The order parameter of the ABM state in Eq.~\eqref{eq:abm} has two remarkable symmetry properties: (i) spontaneous gauge-orbital symmetry breaking and (ii) twofold discrete symmetry. For (i), the ${\rm U}(1)$ phase rotation by $\delta \varphi$ in Eq.~\eqref{eq:abm} is compensated by the continuous rotation of the orbital part about the $\hat{\bm l}$-axis by $-\delta\varphi$, 
\beq
\varphi \mapsto \varphi + \delta \varphi, \quad 
\hat{e}^{(1)}_{j} + i\hat{e}^{(2)}_j \mapsto  
e^{-i\delta\varphi}(\hat{e}^{(1)}_{j} + i\hat{e}^{(2)}_j). 
\eeq
In other words, the ${\rm U}(1)$ phase rotation involves the rotation of the orbital state and vice versa. Hence, the ABM state spontaneously breaks the relative rotation symmetry between the global ${\rm U}(1)$ phase and the orbital state, which is called gauge-orbital symmetry breaking. In addition, the ABM state in Eq.~\eqref{eq:abm} is invariant under the discrete rotation of three unit vectors, $(\hat{\bm d},\hat{\bm e}^{(1)}, \hat{\bm e}^{(2)}) \!\mapsto\! (-\hat{\bm d},-\hat{\bm e}^{(1)}, -\hat{\bm e}^{(2)})$. The symmetry topologically stabilizes the half-quantum vortex, as we will discuss in Sec.~\ref{sec:HQV}.

The ABM state remains invariant under the two-dimensional rotation symmetry in the spin space, ${\rm SO}(2)_{S_z}$, and combined gauge-orbital symmetry, ${\rm SO}(2)_{L_z-\varphi}$,
\beq
H_{\rm A} = {\rm SO}(2)_{S_z} \times {\rm SO}(2)_{L_z-\varphi} \times {\bm Z}_2 .
\eeq
The mod-2 discrete symmetry ${\bm Z}_2 $ originates from the discrete symmetry $(\hat{\bm d},\hat{\bm e}^{(1)}, \hat{\bm e}^{(2)}) \!\mapsto\! (-\hat{\bm d},-\hat{\bm e}^{(1)}, -\hat{\bm e}^{(2)})$. The manifold of the order parameter degeneracy is given by 
\beq
\mathcal{R}_{\rm A} = G/H_{\rm A} = S^2_{\bm S} \times {\rm SO}(3)_{{\bm L},\varphi}/{\bm Z}_2,
\label{eq:RA}
\eeq
where $S^2_{\rm S}$ is the two-sphere associated with the variation of the spin vector $\hat{\bm d}$. ${\rm SO}(3)_{{\bm L},\varphi}$ is interpreted as $S^2_{\bm L}\times {\rm SO}(2)_{L_z,\varphi}$, where ${\rm SO}(2)_{L_z,\varphi}$ denotes the spontaneous breaking of the relative symmetry with respect to rotations in orbital space and global ${\rm U}(1)_{\varphi}$ phase. Because of the gauge-orbital symmetry, ${\rm SO}(3)_{{\bm L},\varphi}$ can be regarded as the degeneracy space with respect to the three-dimensional rotation of $(\hat{\bm e}^{(1)},\hat{\bm e}^{(2)},\hat{\bm l})$. 

{\it Weyl superconductors.}--- Owing to the absence of the TRS and the presence of the PHS, the ABM state is always accompanied by a pair of Fermi points at which the bulk quasiparticle excitation is gapless. The Fermi points in the ABM state are characterized by the nontrivial first Chern number and thus cannot be removed by any perturbations~\cite{volovik87,volovik88,volovik}. This is known as a prototype of Weyl superconductors~\cite{volovik,balentsPRB12,sauPRB12,dasPRB13,goswami13,xuPRL14,yangPRL14}, a superconducting analogue to Weyl semimetals~\cite{murakamiNJP07,wanPRB11,burkovPRL11,zyuzinPRB12,fangPRL12,balentsPRB12-2,lin13}. Weyl superconductors possess chiral Weyl fermions residing in the vicinity of the Fermi points. 

The Weyl points are also regarded as the magnetic monopole in the momentum space~\cite{volovik} and the pairwise Weyl points are connected by the singular ``vortex'' line or Dirac string where the quasiparticles acquire the Berry phase $\gamma(C)=2\pi$ along the path $C$ enclosing the ``vortex'' line. As shown in Fig.~\ref{fig:abm}, therefore, the Weyl superconductors can be regarded as the layered structure of ``weak'' topological superconductors that have a nontrivial first Chern number in each sliced two-dimensional momentum plane,
\beq
{\rm Ch}_1 (k_z) = 2\hat{l},
\label{eq:weak}
\eeq
for $|k_z| \!<\! k_{\rm F}$; Otherwise, it is trivial. The bulk-edge correspondence indicates that each sliced plane is responsible for the existence of the zero energy states. They form a zero-energy flat-band structure along the ${\bm k}$-direction connecting two Weyl points. Hence, the Fermi points of Weyl superconductors correspond to the end point of the zero energy flat band, leading to the ``Fermi arc'' protected by pairwise Weyl points~\cite{silaev:2012,silaevJETP14}. 

In unconventional superconductors, the nodal direction is usually forced by the effect of crystalline symmetries and strong spin-orbit interaction. In contrast, such constraint on $\hat{\bm l}$ is absent in the bulk ABM state. The energy degeneracy with respect to the orientation of $\hat{\bm l}$ is an obstacle to realizing topological defects that host Majorana fermions in the following two senses: (i) In accordance with the Mermin-Ho relation associated with the gauge-orbit symmetry ${\rm SO}(3)_{{\bm L},\varphi}$~\cite{merminPRL76}, the continuous twist of $\hat{\bm l}$ induces the ${\rm U}(1)$ phase slippage without forming a vortex core, which generates the superfluid mass flow. The ``coreless'' vortices, such as the Mermin-Ho vortex, are not accompanied by a definite vortex core and thus have a spatially uniform superfluid density~\cite{merminPRL76,andersonPRL77,salomaaRMP87}. Indeed, although the Mermin-Ho vortex may have nontrivial low-lying quasiparticle structures, they cannot have exactly zero energy because of the absence of a topological invariant (see Sec.~\ref{sec:continuous})~\cite{ichiokaPRB10}. (ii) The second reason is that the ABM state has a strict boundary condition on $\hat{\bm l}$, that is, $\hat{\bm l}$ must be perpendicular to the surface. The ABM state with $\hat{\bm l}$ tilted from the surface normal direction gives rise to a strong pair breaking effect and thus wastes the condensation energy. For $\hat{\bm l}$ normal to the surface, each sliced momentum plane is parallel to the surface and the bulk-edge correspondence is absent. In Secs.~\ref{sec:spt} and \ref{sec:vortices}, we discuss this in further detail and propose definite ways of realizing Majorana fermions in the ABM state confined to an appropriate geometry that controls the $\hat{\bm l}$-texture. 

{\it Thermodynamics.}--- 
Let us share some thermodynamic properties of the ABM state within the GL theory, which offers a key to understanding the thermodynamic stability of the half-quantum vortex as discussed in Sec.~\ref{sec:vortices}. We start with the GL free energy functional $\mathcal{F}_{\rm GL}$~\cite{vollhardt} for the bulk superfluid $^3$He,
\beq
\mathcal{F}_{\rm GL} = \int d{\bm r} \left( f_{\rm bulk} + f^{(1)}_{\rm mag} + f^{(2)}_{\rm mag} 
+ f_{\rm dip}\right),
\label{eq:fgl}
\eeq
which holds the symmetry group $G$ in Eq.~(\ref{eq:full}). The bulk free energy is given by
\begin{align}
f_{\rm bulk} &= \alpha d^{\ast}_{\mu i}d_{\mu i}
+ \beta _1 d^{\ast}_{\mu i} d^{\ast}_{\mu i} d_{\nu j} d_{\nu j}
+ \beta _2 d^{\ast}_{\mu i} d_{\mu i} d^{\ast}_{\nu j} d_{\nu j} \nn \\
&+ \beta _3 d^{\ast}_{\mu i} d^{\ast}_{\nu i} d_{\mu j} d_{\nu j}
+ \beta _4 d^{\ast}_{\mu i} d_{\nu i} d^{\ast}_{\nu j} d_{\mu j}
+ \beta _5 d^{\ast}_{\mu i} d_{\nu i} d_{\nu j} d^{\ast}_{\mu j}.
\label{eq:gl4th}
\end{align}
The phenomenological parameters satisfy the following relations in the weak coupling limit, 
$-2\beta _1 \!=\! \beta _2 \!=\! \beta _3 \!=\! \beta _4 \!=\! - \beta _5 \!=\! \frac{6}{5}\beta _0
= 7 \zeta(3)N_{\rm F}/[120(\pi k _{\rm B}T_{\rm c0})^2]$,
where we have introduced $\alpha=-\frac{1}{3}\mathcal{N}_{\rm F}(1-T/T_{\rm c0})$ and $\xi_0 = \frac{\hbar v_{\rm F}}{\pi k_{\rm B}T_{\rm c0}}\sqrt{7\zeta (3)/48}$. 

By substituting the order parameters of the BW and ABM states, the GL free energy is found to be always lower for the BW state than that for the ABM state in the weak coupling limit
$f^{\rm BW}_{\rm bulk} < f^{\rm ABM}_{\rm bulk}$.
Therefore, the strong coupling correction is indispensable for the thermodynamic stability of the ABM state. To take account of the strong coupling correction into the GL function without breaking the full symmetry $G$, it has been proposed that the GL parameters of the bulk fourth-order terms $\beta _i$ are corrected to~\cite{abm2} 
$\beta _1/\beta _0 \!=\! -\frac{3}{5}(1+0.1\delta)$, 
$\beta _2/\beta _0 \!=\! \frac{6}{5}(1+0.1\delta)$,
$\beta _3/\beta _0 \!=\! \frac{3}{5}(2-0.05\delta)$, 
$\beta _4/\beta _0 \!=\! \frac{3}{5}(2-0.55\delta)$,
and $\beta _5/\beta _0 \!=\! -\frac{3}{5}(2+0.7\delta)$.
The corrections originate from the spin-fluctuation feedback effect, and the parameter $\delta \equiv \delta(P)$ increases as $P$ increases~\cite{vollhardt}. The ABM state becomes more thermodynamically stable than the BW state when
$\delta (P) > 0.465$, corresponding to the high-pressure regime. The microscopic calculation for $\beta _i$ was made by Sauls and Serene~\cite{saulsPRB81}. More recently, the coefficients $\beta _{j}$ have been determined by Choi {\it et al.} through the analysis of the precise measurements of the spin susceptibility and NMR spectra~\cite{choiPRB07,choiPRB13E}.  

The magnetic field energy relevant to $^3$He in the equilibrium is the quadratic Zeeman energy that is given by~\cite{merminPRL73}
\beq
f^{(2)}_{\rm mag} = g_{\rm m}H_{\mu}d^{\ast}_{\mu i} H_{\nu}d_{\nu i}.
\label{eq:FM}
\eeq
The factor $g_{\rm m}$ in Eq.~(\ref{eq:FM}) is given as 
$g_{\rm m} \!=\! \frac{2}{3}\beta _0 ( \frac{\mu _{\rm n}}{1+F^{\rm a}_0} )^2 
\!=\! \frac{7\zeta(3)\mathcal{N}_{\rm F}\gamma^2}{48[(1+F^{\rm a}_0)\pi k_{\rm B}T_{\rm c}]^2}$.
Another magnetic field energy $f^{(1)}_{\rm mag}$ appears in the GL functional, which is linear in the applied magnetic field,
\beq
f^{(1)}_{\rm mag} = i \eta \epsilon ^{\mu\nu\eta}H_{\mu}A_{\nu j}A^{\ast}_{\eta j}. 
\eeq
This describes the high-order correction on $T_{\rm c}/T_{\rm F}$ that originates from the splitting of the Fermi surfaces. This term slightly shifts the pair interaction and $\mathcal{N}_{\rm F}$ for opposite spins~\cite{vollhardt,ambegaokar}. Hence, $f^{(1)}_{\rm mag}$ is negligible in the weak coupling limit at which the quadratic Zeeman term is dominant. 

For the ABM state in Eq.~\eqref{eq:abm}, the quadratic Zeeman term is 
\beq
f^{(2)}_{\rm mag} = 2 g_{\rm m} \left( \hat{\bm d}\cdot{\bm H}\right)^2.
\label{eq:quad}
\eeq
Since $g_{\rm m}$ is a positive constant, the quadratic Zeeman term favors the situation that the ${\bm d}$-vector is perpendicular to the applied magnetic field, ${\bm d}\!\perp\! {\bm H}$. This indicates that the magnetic field strongly suppresses the $|\uparrow\downarrow + \downarrow\uparrow\rangle$ pair, and resultant Cooper pairs in the ABM state are composed of equal distributions into $|\uparrow\uparrow\rangle$ and $|\downarrow\downarrow\rangle$ states, which is called the equal spin pair state. In the equal spin pair state, the quadratic Zeeman field does not affect the ABM state and the linear Zeeman effect becomes important. The linear Zeeman effect suppresses the Cooper pair distributions into the $|\downarrow\downarrow\rangle$ pair state, while it enhances the $|\uparrow\uparrow\rangle$ pair. The order parameter deviated by the linear Zeeman shift is obtained by replacing the $\hat{\bm d}$-vector with the complex variable as
\beq
d_{\mu j} = \left( \hat{d}_{\mu} + i\eta \hat{d}^{\prime}_{\mu}\right) \left( \hat{e}^{(1)}_j + i\hat{e}^{(2)}_j \right),
\eeq
where $\hat{\bm d}^{\prime} \!=\! \hat{\bm H}\times \hat{\bm d}$ and $0 \le \eta \le 1$. The state with $\eta > 0$ is called the A$_2$ state, in which $\Delta _{\uparrow\uparrow}\!\neq\! \Delta _{\downarrow\downarrow}$. The limit of $\eta = 1$ corresponds to the A$_1$ state, where the $|\downarrow\downarrow\rangle$ completely vanishes. The topological properties are the same as those of the spinless chiral $p$-wave systems in Sec.~\ref{sec:spinless}. Both A$_1$ and A$_2$ states are nonunitary states.

We finally briefly comment on the last term in the GL functional \eqref{eq:fgl}, $f_{\rm dip}$. In $^3$He, the magnetic dipole-dipole interaction originates from the magnetic moment of $^3$He nuclei. The Hamiltonian for the nuclear-dipole interaction is given as 
\beq
\mathcal{H}_{\rm D} = \frac{\gamma^2}{2} \int d{\bm r}_1\int d{\bm r}_2 
Q_{\mu\nu}({\bm r}_{12}) S_{\mu}({\bm r}_1)S_{\nu}({\bm r}_2),
\label{eq:Hdip}
\eeq
where we have introduced the local spin operator $S_{\mu}({\bm r}) = \psi^{\dag}_a({\bm r})(\sigma _{\mu})_{ab}\psi _b ({\bm r})/2$. The dipole interaction potential $Q_{\mu\nu}({\bm r})$ is defined as 
$Q_{\mu\nu}({\bm r}) \!=\!  (\delta _{\mu\nu}- 3\hat{r}_{\mu}\hat{r}_{\nu})/r^3$.
Since the dipole interaction is much weaker than the pair interaction, it acts as a small perturbation on the order parameter. 
The dipole energy within the GL regime, $f_{\rm dip}$, is derived from the Hamiltonian (\ref{eq:Hdip}) as $\langle\langle \mathcal{H}_{\rm D}\rangle\rangle$, where $\langle\langle \cdots \rangle\rangle$ is the thermal average. The dipole energy density is given as~\cite{leggettRMP,vollhardt,leggett73,leggett74}
\beq
f_{\rm dip} = \frac{1}{5}\lambda _{\rm D}\mathcal{N}_{\rm F} \left( 
d^{\ast}_{\mu\mu}d_{\nu\nu} +d^{\ast}_{\mu\nu}d_{\nu\mu} 
- \frac{2}{3}d^{\ast}_{\mu\nu}d_{\mu\nu}
\right).
\label{eq:fdip}
\eeq
Here, $\lambda _{\rm D}$ is a dimensionless dipole coupling parameter and approximately independent of pressure. The value is estimated as $\lambda _{\rm D} \approx 5 \times 10^{-7}$~\cite{vollhardt}.

For the BW state, the spin-orbit interaction does not change the overall structure and remaining symmetry, while it imposes a constraint on the order parameter degenerate space, in particular, on $\varphi$. By minimizing $f_{\rm dip}$, the angle $\varphi$ is fixed to the Leggett angle as in Eq.~\eqref{eq:leggett}. For the ABM state, the dipole interaction energy $f_{\rm dip}$ is rewritten with Eq.~\eqref{eq:abm} as 
\beq
f_{\rm dip} = - 2 g_{\rm d} \left[ 
( \hat{\bm l}\cdot \hat{\bm d})^2 - \frac{1}{3}
\right],
\eeq
where $g_{\rm d}$ is a positive constant related to $\lambda _{\rm D}$. The configuration, $\hat{\bm l}\parallel \pm \hat{\bm d}$, can minimize the dipole interaction energy. This means that the dipole interaction gives rise to the locking of the orbital part $\hat{\bm l}$ to the spin part $\hat{\bm d}$ in the ABM state. 

\subsection{Planar state: Symmetry and topology}
\label{sec:planar}

In addition to the BW and ABM states, the planar state that is the two-dimensional analogue to the BW state is energetically competitive with the ABM state in $^3$He. This is because both ABM and planar states possess similar quasiparticle structures with point nodes. It is, however, recognized that the spin-fluctuation feedback effect always favors the time-reversal broken ABM state over the planar state, and the planar state cannot be thermodynamically stable even in a restricted geometry. Nevertheless, understanding the topological aspect provides very useful information for nodal superconducting materials with TRS. Indeed, the planar state can be a prototype of the $E_u$ state in Cu$_x$Bi$_2$Se$_3$~\cite{fuPRL10,sasakiPRL11,hashimotoJPSJ13,sasakiPC15}, the $E_{1u}$ state in UPt$_3$~\cite{tsutsumiJPSJ12-2,tsutsumiJPSJ13,mizushimaPRB14}, and other helical pairing states.

The simple form of the order parameter for the planar state is then given by
\beq
d_{\mu\nu} = \Delta _{\rm P}\left(
\begin{array}{ccc}
1 & 0 & 0 \\ 0 & 1 & 0 \\ 0 & 0 & 0
\end{array}
\right).
\eeq
The planar order parameter is composed of the $(\hat{k}_x-i\hat{k}_y)$ and $(\hat{k}_x+i\hat{k}_y)$ orbital pairs, which are equally distributed to the $S_z=+1$ and $-1$ spin states, respectively. The bulk planar state holds the subgroup of the symmetry group $G$,~\cite{volovik,vollhardt}
\beq
H_{\rm planar} = {\rm U}_{J_z,\phi} \times {\rm P}_z \times {\rm T}.
\eeq
The group ${\rm U}_{J_z,\phi}$ comprises (i) the group of joint rotations about the $z$-axis, ${\rm U}(1)_{S_z+L_z}$, and (ii) the combined symmetry ${\rm O}^{(x,\pi)}_{J}{\rm U}^{(\pi/2)}_{\phi}$ is a joint rotation by $\pi$ about the $x$-axis in the spin and orbital spaces. The generator of the rotation groups ${\rm U}(1)_{L_z+S_z}$ is $J_z$.  Hence, the Casimir operator of ${\rm U}_{S_z,\phi}$ is $J^2_z$, where $J^2_z = 0$, $1$, and $2$ are possible in spin-triplet states. The operator of the group ${\rm P}_z$ is $e^{i\pi}e^{i\pi J_z}$, which describes the discrete rotation consisting of the $\pi$-rotation about the $z$-axis in the spin and orbital spaces and the gauge rotation by $\pi$. 

The quasiparticle excitations become gapless at the south and north poles of a three-dimensional Fermi sphere, 
\beq
E({\bm k}) = \sqrt{\varepsilon^2({\bm k}) + \Delta^2_{\rm P}\sin\theta _{\bm k}},
\eeq
which is the same as that of the ABM state in Eq.~\eqref{eq:Eabm}. Hence, in the weak coupling limit, the thermodynamic behaviors are indistinguishable from those of the ABM state. 

{\it Topological stability of point nodes.}---
For the topological structure, however, the planar state becomes distinct from the ABM state.~\cite{volovik} For instance, as discussed in Sec.~\ref{sec:abm}, the point nodes in the time-reversal broken ABM state are regarded as the Weyl points having a definite monopole charge, which are protected by the first Chern number. Contrary to the {\it Weyl superconductivity}, the first Chern number defined in $S^2$ embracing the point nodes becomes zero in the planar state owing to the TRS. 

However, this does not always imply that the point nodes in the planar state are unstable and immediately gapped out by any disturbances. Kobayashi {\it et al.}~\cite{kobayashiPRB14} have recently presented the classification of the topological stability of nodes in odd-parity superconductors, which is a topological generalization of Blount's theorem. In accordance with the theory, point nodes in the planar state are protected by the mirror reflection symmetry $M_{xy}=i\sigma _z$ in the $xy$-plane, where the mirror operator changes $\hat{\bm k}$ and ${\bm d}$ to $(\hat{k}_x,\hat{k}_y,-\hat{k}_z)$ and $(d_x,d_y,-d_z)$, respectively. The planar order parameter is, therefore, even under $M_{xy}$ and the BdG Hamiltonian obeys the mirror reflection symmetry 
\beq
\mathcal{M}_{xy}\mathcal{H}({\bm k})\mathcal{M}^{\dag}_{xy} = \mathcal{H}(k_x,k_y,-k_z),
\eeq
where $\mathcal{M}_{xy} = {\rm diag}(M_{xy},M^{\ast}_{xy})$. Since the mirror operator is commutable with $\mathcal{T}$ and $\mathcal{C}$, the topological classification of the point node in the planar state is categorized into the ``P$+$DIII'' class with the additional discrete symmetry $M^{++}$ in the notation of Ref.~\citeonline{kobayashiPRB14}, while the ABM state belongs to class ``D'', that is, the point node is always topologically stable. Therefore, the stability of the point nodes in the planar states manifests the topology and symmetry background, which is essentially different from that of the ABM state.

{\it Bulk $\mathbb{Z}_2$ topological invariant.}---
In addition to the topological aspect of point nodes, the topology of quasiparticles in the planar state is also distinct from that in the BW and ABM states. Although the definition of a topological invariant requires the bulk excitation to be fully gapped for all momenta, topological invariant can be introduced even in nodal superconductors. 

The strategy for introducing the bulk topological invariant is as follows: For time-reversal-invariant odd-parity superconductors, point nodes are not topologically protected and are removed by introducing a small perturbation $\delta$ that makes a finite gap while preserving the fundamental discrete symmetries, $\mathcal{T}$ and $\mathcal{C}$~\cite{sasakiPRL11}. Removing the point nodes enables one to introduce the three-dimensional winding number $w_{\rm 3d}$ as in Eq.~\eqref{eq:w3d} for class DIII. The $\mathbb{Z}$ number has an ambiguity in choosing the perturbation $\delta$, and $w_{\rm 3d}$ is not invariant under a gauge transformation. As clarified in Refs.~\citeonline{satoPRB10}, \citeonline{fuPRL10}, and \citeonline{sasakiPRL11}, the parity of $w_{\rm 3d}$, 
\beq
\nu = (-1)^{w_{\rm 3d}}=-1,
\eeq
remains well-defined as a gauge invariant number. 


The $\mathbb{Z}_2$ number in odd-parity superconductors was first introduced by Sato~\cite{satoPRB09} in the case of a single-band system and subsequently extended to a general case by Sato~\cite{satoPRB10} and Fu and Berg~\cite{fuPRL10}, independently. The $\mathbb{Z}_2$ number provides a sufficient criterion for realizing time-reversal-invariant odd-parity topological superconductors. Let us now assume an odd-parity superconductor with inversion symmetry
\beq
\mathcal{P}\mathcal{H}({\bm k})\mathcal{P}^{\dag} = \mathcal{H}(-{\bm k}), 
\label{eq:invBdG}
\eeq
where $\mathcal{P}\!=\!P\tau _z$. The inversion symmetry simplifies the parity of $w_{\rm 3d}$ to~\cite{fuPRL10,satoPRB10,sasakiPRL11}
\beq
\nu = (-1)^{w_{\rm 3d}} = \prod _{i,m} {\rm sgn}\left[ 
\varepsilon _{2m}({\bm \Lambda}_i)
\right],
\label{eq:z2-1}
\eeq
where $\varepsilon _{m}({\bm \Lambda}_i)$ is the energy dispersion of the normal state at time-reversal-invariant momenta. Owing to the TRS, the energy band of the normal $^3$He has a Kramers pair as $\varepsilon _{2m}({\bm \Lambda}_i) = \varepsilon _{2m+1}({\bm \Lambda}_i)$. 
Equation~(\ref{eq:z2-1}) indicates that the mod-2 winding number is determined by counting the number of Fermi surfaces enclosing ${\bm k} \!=\! {\bm \Lambda}_i$. For the planar state in $^3$He, the topological invariant is nontrivial, $(-1)^{w_{\rm 3d}}\!=\!-1$, because the single Kramers-paired Fermi surface encloses the time-reversal-invariant momenta ${\bm \Lambda}_i={\bm 0}$. The nontrivial value of the bulk $\mathbb{Z}_2$ number gives sufficient conditions for realizing topological superconductivity and helical Majorana fermions.


\section{$^3$He in Confined Geometries: Symmetry-Protected Topological Phases}
\label{sec:spt}

In this section, we will show that the superfluid $^3$He confined to a restricted geometry can offer an ideal situation to study the intertwining of symmetry and topology and to realize Majorana fermions. After briefly overviewing the recent progress on understanding the superfluid phases of $^3$He in a restricted geometry, we will explain the rich topological aspects of the BW, ABM, and planar states, which are energetically competitive in a slab geometry. 

\begin{figure}[tb!]
\begin{center}
\includegraphics[width=75mm]{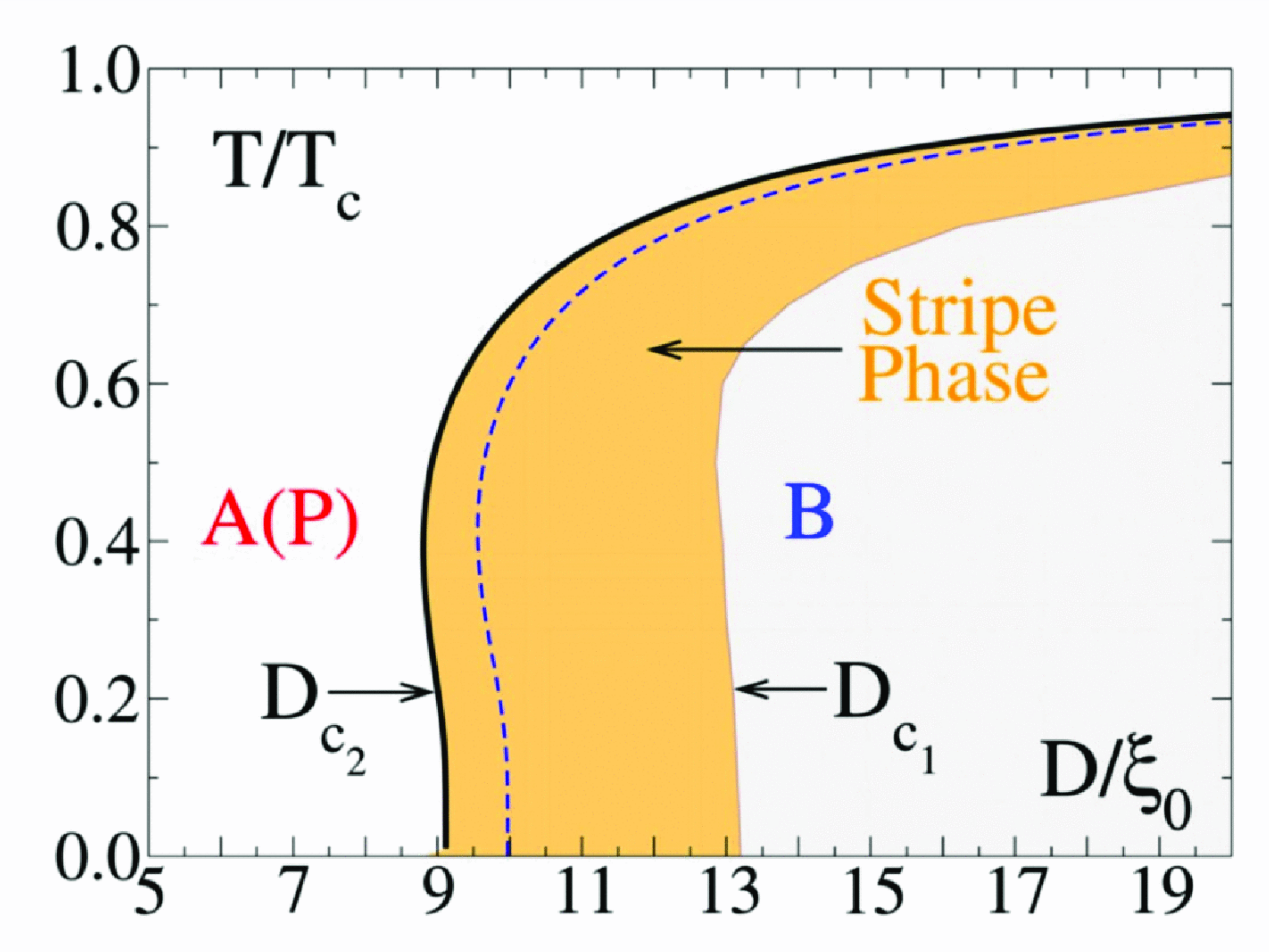} 
\end{center}
\caption{(Color online) Phase diagram of $^3$He in a slab with thickness $D$, where $\xi_0$ denotes the superfluid coherence length. Taken from Ref.~\citeonline{vorontsovPRL07}.}
\label{fig:phase_slab}
\end{figure}

\subsection{$^3$He in a slab geometry: Overview}

Let us start with the symmetric properties of $^3$He confined in a slab geometry, where $^3$He is sandwiched by two parallel surfaces. Let $\hat{\bm z}$ be normal to the two parallel surfaces. Owing to such confinement, the three-dimensional continuous rotational symmetry in the coordinate space is explicitly broken and the resultant symmetry group of the normal state is reduced to 
\beq
G_{\rm slab} = {\rm SO}(2)_{L_z} \times {\rm SO}(3)_{\bm S} \times {\rm U}(1)_{\phi} \times {\rm T}.
\label{eq:slab}
\eeq
The BW state in a slab geometry remains invariant under the two-dimensional rotation of the spin and orbital spaces about the surface normal axis, and thus the remaining symmetry is 
$
H_{\rm slab} = {\rm SO}(2)_{L_z+S_z} \times {\rm T} \times {\rm PU}_{\pi/2}$.
The BW order parameter relevant to this situation is parameterized as
\beq
d_{\mu}({\bm k},z) = R_{\mu\nu}(\hat{\bm n},\varphi)d^{(0)}_{\nu\eta}(z)\hat{k}_{\eta},
\label{eq:dvec_slab}
\eeq
where 
\beq
d^{(0)}_{\mu\nu}(z) = \Delta _{\parallel}(z)\left( 
\delta _{\mu,\nu} - \hat{z}_{\mu}\hat{z}_{\nu}
\right) + \Delta _{\perp}(z)\hat{z}_{\mu}\hat{z}_{\nu}.
\label{eq:dvec_b2_initial}
\eeq
Hence, the quasiparticle excitation gap is distorted by confinement. For $\Delta _{\parallel}=\Delta _{\perp}$, Eq.~(\ref{eq:dvec_slab}) recovers the isotropic BW order parameter, while it turns to the planar state when $\Delta _{\perp}$ vanishes. 

The spatial profile of the pair potentials, $\Delta _{\parallel}(z)$ and $\Delta _{\perp}(z)$, is determined by solving the self-consistent equations for the pair potential and underlying quasiparticles subject to a surface boundary condition. A quasiparticle incoming to the surface along the trajectory of ${\bm k}$ is specularly scattered by the wall to the quasiparticle state with 
\beq
\underline{\bm k}= {\bm k} - 2\hat{\bm z}(\hat{\bm z}\cdot{\bm k}).
\label{eq:specular}
\eeq 
The specular scattering of quasiparticles on the surface imposes a strong constraint on the wavefunction of the Cooper pairs, 
$\Delta ({\bm k},z) = \Delta (\underline{\bm k},z)$,
at the surface $z=z_{\rm surf}$. This implies that while the Cooper pairs with zero perpendicular momentum are insensitive to the surface condition, the pairs having a nonzero momentum perpendicular to the surface must vanish at the surface, namely, 
\beq
\frac{d}{dz}\Delta _{\parallel}(z)\bigg|_{z=z_{\rm surf}} = 0, \quad
\Delta _{\perp} (z_{\rm surf}) = 0.
\label{eq:bcgl}
\eeq
Hence, the surface induces an anisotropic pair-breaking mechanism that deviates $\Delta _{\parallel}$ and $\Delta _{\perp}$. 

Within the GL theory, the spatial shapes of $\Delta _{\parallel}(z)$ and $\Delta _{\perp}(z)$ are determined by solving the GL equation, that is, the nonlinear Schr\"{o}dinger-like equations. 
Under the simple assumption of $\Delta _{\parallel} = \Delta _{\rm B} $ for $D\gg \xi$, one may construct the solution that satisfies the boundary condition in Eq.~\eqref{eq:bc} as
\beq
\Delta _{\perp}(z) = \Delta _{\rm B}(T)\tanh\left(\frac{z}{\xi}\right) \tanh\left( \frac{D-z}{\xi}\right),
\label{eq:ff}
\eeq
where we set $\xi \equiv \sqrt{6}\xi _{\rm GL}$ with the coherence length in the GL regime, $\xi _{\rm GL} (T) \equiv \xi _0/\sqrt{1-T/T_{\rm c}}$. Within the GL theory, the confinement effect and the phase diagram in $^3$He confined in a slab geometry have been quantitatively discussed by numerous researchers~\cite{ambegaokar,brinkman,barton,smith,kjaldman,fujita,jacobsen,takagi,fetter,li,ullah,lin-liu,salomaa}. The full numerical calculation indicates that the parallel component $\Delta _{\parallel}$ is slightly enhanced in the surface region by the gain of the condensation energy.

Apart from the GL regime, a microscopic study on the gap structure of $^3$He-B near the surface was first initiated by Buchholtz and Zwicknagl~\cite{buchholtz:1981} on the basis of quasiclassical Eilenberger theory. The surface boundary condition~\eqref{eq:bcgl} on $\Delta$ is recast into that on the $4\times 4$ quasiclassical propagator $g(\hat{\bm k},{\bm r};\omega _n)$,
\beq
g(\hat{\bm k},{\bm r}_{\rm surf};\omega _n) = g(\underline{\hat{\bm k}},{\bm r}_{\rm surf};\omega _n),
\label{eq:bc}
\eeq
when the surface is fully specular, as in Eq.~\eqref{eq:specular}. The gap structure on the surface qualitatively reproduces that obtained from the GL theory. The microscopic calculation, however, gives extra information that the surface distortion is accompanied by the emergence of surface Andreev bound states, which are currently known as surface helical Majorana fermions.

As discussed in Sec.~\ref{sec:MF3HeB}, the BdG equation for systems with a specular surface is reduced to the one-dimensional Dirac equation with the domain wall of an effective mass~\cite{jackiw,ohashiJPSJ95}. This effective theory gives a unified description for low-energy quasiparticle states and has an important consequence that the zero-energy Andreev bound state always appears in the surface of unconventional superconductors and superfluids, when the pair potential which the quasiparticles are subjected to changes its sign as in Eq.~\eqref{eq:piphase}. Hara and Nagai~\cite{haraPTP86} analyzed a $p$-wave polar state as an exactly soluble model for studying the interplay between the surface distortion and the surface bound states. Making use of an exact self-consistent solution, they demonstrated that the spatial profile of $\Delta$ is essentially the same as the order parameter in Eq.~\eqref{eq:ff}. The spatial profile is similar to that of the continuum model of polyacetylene with a soliton lattice and Fulde-Ferrell-Larkin-Ovchinnikov superconductors~\cite{machidaPRB1984,brazovskii,mertsching,horovitz,nakaharaPRB81}. 

Studies on the confinement effect and phase diagram were followed by numerous works based on the quasiclassical theory~\cite{mizushimaPRB12,buchholtz:1981,haraPTP86,haraJLTP88,vorontsovPRB03,vorontsovPRL07,nagaiJPSJ08,tsutsumi:2010b,tsutsumi:2011b}. These works underline the role of the surface Andreev bound states in thermodynamics. Boundary conditions that describe quasiparticle scattering from an atomically rough surface were developed in several ways~\cite{zhangPRB87,thuneberg92,buchholtz79,buchholtz86,buchholtz91,buchholtz93,nagatoJLTP96,nagatoJLTP98,nagatoJLTP07}.
These include the scattering of quasiparticles from a thin layer of atomic-size impurities~\cite{zhangPRB87}, a distribution of a randomly oriented mirror on the surface~\cite{thuneberg92}, and a randomly rippled wall~\cite{buchholtz79,buchholtz86,buchholtz91,buchholtz93}. Nagato {\it et al.} implemented boundary conditions that describe a partially diffusive surface, by using a random $S$-matrix~\cite{nagatoJLTP96,nagatoJLTP98,nagatoJLTP07}. Vorontsov and Sauls~\cite{vorontsovPRB03} developed a more tractable scheme to implement the diffusive boundary condition into the quasiclassical theory. The surface roughness causes the diffusive scattering of quasiparticles and markedly changes the surface structure of order parameters and low-lying bound states~\cite{vorontsovPRB03,nagaiJPSJ08,nagatoJLTP98,nagatoJPSJ11}. It is worth noting that in $^3$He, the surface specularity is experimentally controllable by coating the container with $^4$He layers~\cite{okuda}, and the roughness may change the surface structure of the superfluid $^3$He-B. The recent progress in the understanding of surface Andreev bound states and roughness effects in $^3$He is summarized in Ref.~\citeonline{nagaiJPSJ08}.

Vorontsov and Sauls~\cite{vorontsovPRL07} revealed that a confined geometry induces a new quantum phase, which is the so-called stripe phase. The stripe phase spontaneously breaks the translational symmetry along a surface parallel direction. The stable region of the stripe phase covers the A-B transition line as shown in Fig.~\ref{fig:phase_slab}.

It is now widely accepted that such a stripe ordered phase can be stabilized in various situations, including in $^3$He confined in narrow cylinders~\cite{aoyamaPRB14}, unconventional ($d$-wave) superconducting thin films~\cite{vorontsovPRL09,hachiyaPRB13,miyawakiPRB15}, and superconducting mesoscopic thin-walled cylinders under external magnetic fields~\cite{aoyamaPRL13}. The stability of the stripe phase and the mechanism of the spontaneous breaking of the translational symmetry in a confined geometry are essentially different from those in bulk superconductors. In bulk superconductors, the emergence of the Fulde-Ferrell-Larkin-Ovchinnikov states is a consequence of the competition between the two different sources of pair breaking: the Pauli paramagnetic and orbital depairing effects. In restricted geometries, however, the orbital depairing effect is absent, and an extra key for the stability of stripe states is provided by the intertwining of the order parameter with the emergence of surface Andreev bound states. In particular, a ground state realized in $d$-wave superconducting thin films~\cite{vorontsovPRL09} can spontaneously break the TRS in addition to the translational symmetry, which develops a supercurrent flow along the film. The time-reversal broken stripe state is accompanied by a large ``backflow''. The large backflow is a manifestation of the odd-frequency pairing which is another facet of surface Andreev bound states. This is responsible for a negative superfluid density~\cite{higashitaniJPSJ97} and makes spatially uniform BCS states thermodynamically unstable. In accordance with the symmetry argument, the odd-frequency pairings emergent in spin-singlet superconductors are fragile against surface roughness, and thus the stripe state becomes unstable as surface specularity decreases~\cite{higashitaniJPSJ15}.

\subsection{Symmetry-protected topological phase in the BW state under a magnetic field}
\label{sec:sptb}

As mentioned in Secs.~\ref{sec:bw} and \ref{sec:MF3HeB}, $^3$He-B is a prototype time-reversal-invariant topological superfluid (class DIII), and its hallmark is the appearance of surface helical Majorana fermions~\cite{schnyderPRB08,qiPRL09,chungPRL09,volovikJETP09,nagatoJPSJ09}. We, however, notice that the topological property of $^3$He-B is not always protected by the TRS, and extra discrete symmetries, such as $P_2$ and $P_3$, may protect a topological invariant. Indeed, it was uncovered in Ref.~\citeonline{mizushimaPRL12} that the topological phase survives in $^3$He-B confined in a slab geometry even under a magnetic field, where the field is applied along the surface. As shown in Fig.~\ref{fig:phaseDH}, the B-phase undergoes the phase transition from the B$_{\rm I}$- to B$_{\rm II}$-phase, where the former (latter) is identified as the topological phase protected by $P_3$ (nontopological phase with broken $P_3$). Below, we overview the intertwining of spontaneous symmetry breaking with the topological phase transition after discussing the structure of the phase diagram. A more detailed review with a focus on $^3$He-B was made in Ref.~\citeonline{mizushimaJPCM15}.

\subsubsection{Topological phase diagram}

In Ref.~\citeonline{mizushimaPRL12}, it was emphasized that the nuclear-magnetic dipole interaction in Eq.~\eqref{eq:Hdip} plays a key role in determining the thermodynamic stability of the topological phase. Hence, we have to take into account the effects of the magnetic Zeeman field and the dipole interaction into the self-consistent quasiclassical framework equally (see Appendix). We here take the pairing interaction $\mathcal{V}^{cd}_{ab}(\hat{\bm k}, \hat{\bm k}^{\prime})$ as the combination of the ${\rm SO}(3)_{\bm S}\!\times\!{\rm SO}(3)_{\bm L}\!\times\!{\rm U}(1)$ symmetric $p$-wave interaction and the dipole-dipole interaction \eqref{eq:Hdip}. The dipole potential $Q_{\mu\nu}({\bm k},{\bm k}^{\prime}) \!=\! \tilde{g}_{\rm D} R \int Q_{\mu\nu}({\bm r})e^{-i({\bm k}-{\bm k}^{\prime})\cdot{\bm r}}d{\bm r}$ is composed of the series of $p$-, $f$-, and higher partial waves. Since the pairing interaction in $^3$He is dominated by the former channel, only the $p$-wave contribution is taken into account. To this end, the gap equation with the anomalous quasiclassical propagators $f _{\mu}(\hat{\bm k},{\bm r};\omega _n)$ is reduced to~\cite{mizushimaPRL12,mizushimaPRB14,mizushimaJPCM15}
\begin{align}
d_{\mu\nu}({\bm r}) 
= & 3|g|\left\langle \hat{k}_{\nu}f_{\mu} \right\rangle _{\hat{\bm k},n}
-\tilde{g}_{\rm D}
\left( 1 + 3\delta _{\mu\nu} \right)
\left\langle \hat{k}_{\nu}{f}_{\mu} \right\rangle _{\hat{\bm k},n} \nn \\
& - 3\tilde{g}_{\rm D}\left[
\left\langle \hat{k}_{\mu}{f}_{\nu} \right\rangle _{\hat{\bm k},n}
- \left\langle \hat{k}_{\nu}{f}_{\mu} \right\rangle _{\hat{\bm k},n}
\right],
\label{eq:gapv3}
\end{align}
where $g$ is the coupling constant associated with ${\rm SO}(3)_{\bm S}\!\times\!{\rm SO}(3)_{\bm L}$ symmetric pair interaction and $\tilde{g}_{\rm D}$ denotes the coupling constant of the dipole interaction renormalized with the contributions of high-energy quasiparticles. 
The gap equation \eqref{eq:gapv3} coupled to the quasiclassical equation \eqref{eq:eilen} subject to \eqref{eq:norm} now serves  as a self-consistent framework to quantitatively study the energetics and thermodynamics of $^3$He. The closed form of the self-consistent equations is accomplished by taking into account the quasiclassical self-energy that describes the ferromagnetic exchange field associated with the Fermi liquid parameter $F^{\rm a}_0=-0.7$. 

The effect of the dipole interaction in the superfluid $^3$He-B has been emphasized in different contexts, such as spin and orbital dynamics~\cite{leggett73,leggett74,tewordtPLA76,tewordtJLTP79,schopohlJLTP82,fishmanPRB87}. The self-consistent calculations with the dipole interaction were initiated by Tewordt and Schopohl~\cite{tewordtJLTP79,schopohlJLTP82} to clarify the field dependences of the spin susceptibility and collective modes in the bulk $^3$He-B. Fishman~\cite{fishmanPRB87} analyzed in detail the nonlinear field-dependences of the static and dynamic spin susceptibilities in the bulk $^3$He-B in connection with NMR experiments.

The thermodynamic stability of $^3$He-B confined in a slab in the presence of the dipole interaction was first examined in Ref.~\citeonline{mizushimaPRL12} (see Fig.~\ref{fig:phaseDH}). It turns out that under a parallel field, the B-phase is further subdivided into two phases, B$_{\rm I}$ and B$_{\rm II}$, at the critical magnetic field $H^{\ast}$, as shown in Fig.~\ref{fig:phaseDH}. To characterize the subphases, we here introduce a new quantity, $\hat{\ell}_z$, defined as
\beq
\hat{\ell}_{\mu}(\hat{\bm n},\varphi) = \hat{h}_{\nu}R_{\nu\mu}(\hat{\bm n},\varphi)
\eeq
which is associated with the order parameter of the BW state, $(\hat{\bm n},\varphi)$. Here, $\hat{h}_{\nu}$ denotes the orientation of the applied field. Using this quantity, the two subphases B$_{\rm I}$ and B$_{\rm II}$ are identified as the states with $\hat{\ell}_z=0$ and nonzero $\hat{\ell}_z$, respectively. As discussed below, $\hat{\ell}_z(\hat{\bm n},\varphi)$ is regarded as the order parameter associated with the spontaneous breaking of the $P_3$ symmetry. 

\begin{figure}[tb!]
\begin{center}
\includegraphics[width=80mm]{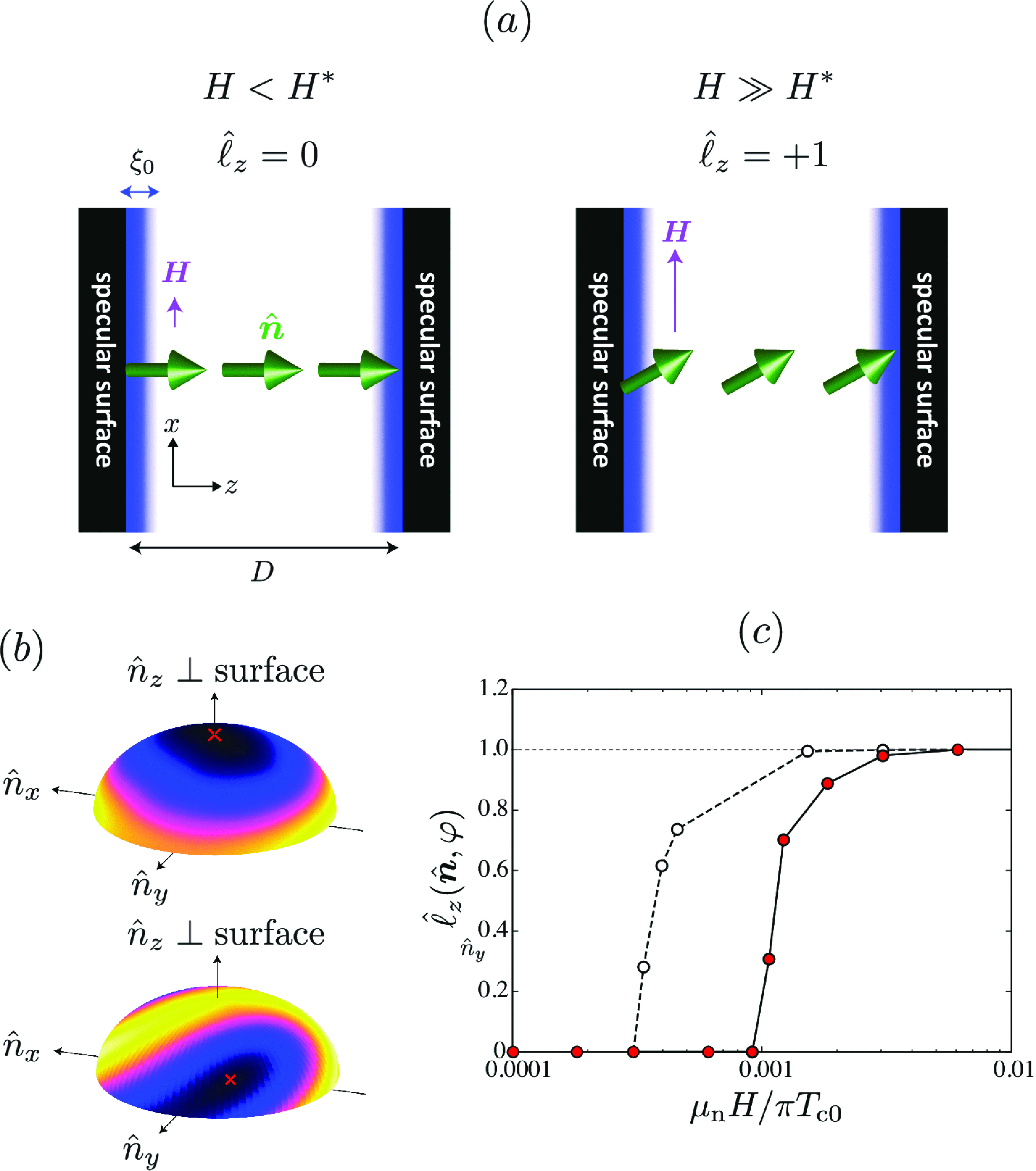}
\end{center}
\caption{(Color online) (a) Schematic picture of the orientation of the $\hat{\bm n}$-vector in the presence of a parallel magnetic field ${\bm H}$: $H< H^{\ast}$ (left) and $H\gg H^{\ast}$ (right), where $H^{\ast}$ is the critical field. In the case of $H \gg H^{\ast}$, the $\hat{\bm n}$-vector is oriented to the direction that maximizes $\hat{\ell}_z$. (b) Thermodynamic potential $\Omega$ in Eq.~\eqref{eq:omega} for $H<H^{\ast}$ (upper panel) and $H>H^{\ast}$ (lower panel) as a function of $\hat{\bm n}$, where $\varphi$ is fixed so as to minimize the dipole energy. (c) Field dependence of $\hat{\ell}_z$ at $T=0.2T_{\rm c0}$ and $D=20\xi _0$. Figures adapted from Ref.~\citeonline{mizushimaPRL12}}
\label{fig:nvec}
\end{figure}

To understand the phase structure in Fig.~\ref{fig:phaseDH}, let us consider the energetics within the GL theory. The GL analysis initiated theoretical studies for understanding the pair breaking effect and $\hat{\bm n}$-texture in restricted geometries~\cite{ambegaokar,brinkman,barton,smith,kjaldman,fujita,takagi,jacobsen,fetter,li,ullah,lin-liu,salomaa}. The order parameter of the BW state in a slab is composed of $(\hat{\bm n},\varphi)$ in addition to $\Delta _{\parallel}(z)$ and $\Delta _{\perp}(z)$, as in Eq.~\eqref{eq:dvec_slab}. For the regime wherein $H$ is much weaker than the dipolar field $H_{\rm d}$, the order parameters $(\hat{\bm n},\varphi)$ in the equilibrium are determined by minimizing the dipole energy $\int dz f_{\rm dip}(z)$ as 
\beq
\hat{\bm n} = (0,0,1),
\label{eq:n}
\eeq
and 
$\varphi = \cos^{-1}( 
- \frac{1}{4}\langle \Delta _{\parallel}(z) \Delta _{\perp}(z)\rangle/\langle \Delta^2_{\parallel}(z)\rangle)$,
where $\langle\cdots\rangle$ denotes the spatial average over the slab. This corresponds to the situation that $\hat{\ell}_z=0$, and thus the ${\rm B}_{\rm I}$-phase is favored by the dipole interaction when the magnetic field is applied along the surface [See Fig.~\ref{fig:nvec}(a)]. 

For $H\gg H_{\rm d}$, the GL free energy is dominated by the magnetic energy $f_{\rm mag}$. Substituting the order parameter defined in Eq.~(\ref{eq:dvec_slab}), one obtains the magnetic energy density in the GL regime as
\beq
f_{\rm mag} = g_{\rm m}H^2 \Delta^2_{\parallel} \left[
1 - \hat{\ell}^2_z(\hat{\bm n},\varphi) \left( 
1 - \eta^2(z)
\right)
\right].
\eeq
The function $\eta(z)\equiv\Delta _{\perp}(z)/\Delta _{\parallel}(z)$ denotes the local distortion of the isotropic BW state, where $\eta =1 $ is the isotropic BW state and $\eta = 0$ denotes the planar state. Since the pair breaking effect results in $\Delta _{\perp}(z) \le \Delta _{\parallel}(z)$ locally, one finds that $0\le 1-\eta^2 \le 1$. This indicates that the magnetic field energy, $\mathcal{F}_{\rm mag}\equiv \int f_{\rm mag}(z)dz$, is minimized when $\hat{\ell}_z(\hat{\bm n},\varphi) = \pm 1$, namely, the B$_{\rm II}$-phase.

In accordance with the GL analysis~\cite{vollhardt}, the critical magnetic field $H^{\ast}_{\rm GL}$ beyond which $\hat{\bm \ell}_z$ becomes nonzero is given by 
\beq
H^{\ast}_{\rm GL} = \alpha \lambda^{1/2}_{\rm D}\frac{\pi k_{\rm B}T_{\rm c0}}{\gamma \hbar}(1+F^{\rm a}_0),
\label{eq:HGL}
\eeq
where $\alpha =\sqrt{54/7\zeta(3)}$. The characteristic field $H^{\ast}_{\rm GL}$ is about the dipolar field $25$G. This field is temperature-independent in the GL regime but slightly depends on pressure. We notice that in the case of a perpendicular field ${\bm H}\parallel\hat{\bm z}$, both the dipole field and the magnetic Zeeman field favor $\hat{\ell}_z=1$, regardless of $H$. This implies that an infinitesimal field destroys the topological phase and the surface Majorana fermions acquire an effective mass proportional to the Zeeman energy~\cite{mizushimaPRB12}.

The GL theory does not take into account low-lying quasiparticles that may alter the $\hat{\bm n}$ configuration. The quasiclassical formalism will offer a tractable and quantitative scheme for understanding the interplay of the order parameters and quasiparticles. The field dependence of $\hat{\ell}_z$ is shown in Fig.~\ref{fig:nvec}(d) by carrying out the full self-consistent calculation of the quasiclassical equations~\cite{mizushimaPRL12}. Similarly to the GL regime, in the case of a parallel field ${\bm H}\parallel\hat{\bm x}$, there exists a critical field $H^{\ast}$ below which $\hat{\bm \ell}=0$ survives. Figures~\ref{fig:nvec}(b) and \ref{fig:nvec}(c) depict the energy landscape $\delta \Omega$ on the unit sphere of $\hat{\bm n}$. The thermodynamic functional $\delta \Omega$ is obtained from the Luttinger-Ward thermodynamic functional with the quasiclassical operators~\cite{vorontsovPRB03}
\beq
\delta \Omega [{g}]
= \frac{1}{2} \int^{1}_0 d\lambda {\rm Sp}^{\prime} \left\{ \left(\underline{\nu}+\underline{\Delta}\right)
\left( {g}_{\lambda} - \frac{1}{2}{g}  \right)
\right\} ,
\label{eq:omega}
\eeq
where we set $
{\rm Sp}^{\prime}\{ \cdots\} 
= {\mathcal N}_{\rm F} \int d{\bm r} \langle {\rm Tr}_4 \{ \cdots\} \rangle _{\hat{\bm k},\omega _n}$. 
The quasiclassical auxiliary function ${g}_{\lambda}$ is obtained from Eq.~(\ref{eq:eilen}) by replacing $\underline{\nu}\!\rightarrow\! \lambda \underline{\nu}$ and $\underline{\Delta}\!\rightarrow\! \lambda \underline{\Delta}$ ($\lambda \!\in\![0,1]$). Equation (\ref{eq:omega}) includes the effects of the condensation energy and quasiparticle excitations as well as the Fermi liquid corrections. It turns out that $\varphi$ is insensitive to $H$ in low fields. We therefore fix $\varphi$ to be the value that minimizes the thermodynamic potential at zero fields. For $H<H^{\ast}$, as shown in Fig.~\ref{fig:nvec}(b), $\delta \Omega$ for a weak field has a minimum point when $\hat{\bm n}\parallel\hat{\bm z}$, while the $\hat{\bm n}$-vector tends to tilt from the surface normal direction for $H\gg H^{\ast}$. 
As shown in Fig.~\ref{fig:nvec}(c), $\hat{\ell}_z$ is
locked to be $\hat{\ell}_z \!=\! 0$ for $H<H^{\ast}$ up to the critical
value $\mu _{\rm n} H^{\ast}/\pi T_{\rm c0} \!\approx\! 0.001$, which is
consistent with the GL analysis in Eq.~\eqref{eq:HGL}. The critical field 
is estimated as $H^{\ast}\approx 20$--$30$G depending on pressure.

\subsubsection{Symmetry-protected topological phase and Majorana fermions}
\label{eq:hidden}


{\it Discrete symmetries in $^3$He-B.}--- 
Having clarified the phase diagram of $^3$He in a slab under a parallel field, we now turn to examine the characteristics and hallmarks of both subphases B$_{\rm I}$ and B$_{\rm II}$. We start with the group symmetry subject to the confined $^3$He under a magnetic field, 
\beq
G_{\rm slab, H, D} = P_2 \times P_3 \times {\rm U}(1)_{\phi},
\eeq
which is composed of additional order-two discrete symmetries ($P_2$ and $P_3$) and ${\rm U}(1)$ gauge symmetry. The $P_3$ ($P_2$) symmetry is constructed by the combination of the TRS and joint $\pi$-rotation in spin and orbital spaces (mirror reflection), where the rotation about the surface normal ${\bm H}\rightarrow(-H_x,-H_y,H_z)$ [the mirror reflection in the $xz$-plane ${\bm H}\rightarrow(-H_x,H_y,-H_z)$] is compensated by the TRS when $H_z=0$ ($H_y=0$). The additional discrete symmetries can be maintained even if each symmetry is explicitly broken. For $^3$He-B, since the $P_2$ symmetry does not play a major role in the topological superfluidity, we focus our attention on the $P_3$ symmetry from now on.

The operator of the $P_3$ symmetry can be constructed by combining the TRS $\mathcal{T}$ in Eq.~\eqref{eq:TRS} and the $\pi$ spin rotation $\mathcal{U}(\pi)\equiv {\rm diag}(U_z(\pi),U^{\ast}_z(\pi))$ as 
\beq
\mathcal{P}_3 \equiv \mathcal{T}\mathcal{U}(\pi), \quad \mathcal{P}^2_3 = +1.
\eeq
The corresponding $\pi$-rotation matrix in the spin space is introduced as
\beq
U_z(\pi) \equiv U(\hat{\bm n},\varphi) U(\hat{\bm z},\pi) U^{\dag}(\hat{\bm n},\varphi),
\eeq
where $U(\hat{\bm z},\pi)$ denotes the ${\rm SU}(2)$ spin rotation matrix about the $\hat{\bm z}$-axis (surface normal) by the angle $\pi$. The pair potential in Eq.~\eqref{eq:dvec_slab} remains invariant under the joint $\pi$-rotation, 
$U_z(\pi)\Delta ({\bm k})U^{\rm T}_z(\pi) = \Delta (-k_x,-k_y,k_z)$, as well as the TRS, independently.


In the presence of a magnetic field, the single-particle Hamiltonian ${\varepsilon}({\bm k})$ in Eq.~(\ref{eq:Hbdg}) contains the magnetic Zeeman term, 
\beq
\varepsilon({\bm k})=\varepsilon _0 ({\bm k}) - \frac{\gamma \tilde{H}}{2}\hat{\bm h}\cdot{\bm \sigma},
\eeq
where $\tilde{H}$ is an effective magnetic field including the self-energy corrections due to the ferromagnetic exchange interaction. The $\pi$-rotation $U_z(\pi)$ changes the magnetic Zeeman term as 
$\hat{\bm h}\cdot{\bm \sigma} \rightarrow -\hat{\bm h}\cdot{\bm \sigma} +2\hat{\ell}_z(\hat{\bm n},\varphi)\tilde{\sigma}_z$,
where $\tilde{\sigma}_z \equiv \sigma _{\nu}R_{\nu z}(\hat{\bm n},\varphi)$. The additional term induced by the joint $\pi$-rotation is characterized by the quantity $\hat{\ell}_z$ defined as  
\beq
\hat{\ell}_{\mu}(\hat{\bm n},\varphi) \equiv \hat{h}_{\nu}R_{\nu\mu}(\hat{\bm n},\varphi). 
\label{eq:ell}
\eeq
Similarly to the $\pi$-rotation, $\mathcal{T}=\Theta K$ changes the sign of the magnetic Zeeman term. Therefore, by combining the $\pi$-rotation $U_z(\pi)$ with $\mathcal{T}$, the $\mathcal{P}_3$ operator transforms the BdG Hamiltonian with the order parameter of the generalized BW state as
\begin{align}
\mathcal{P}_3 \mathcal{H}({\bm k})\mathcal{P}^{-1}_3
= \mathcal{H}(k_x,k_y,-k_z)  - \gamma H \hat{\ell}_z \left(
\begin{array}{cc} 
\tilde{\sigma}_z & 0 \\ 0 & - \tilde{\sigma}^{\ast}_z \end{array} \right).
\label{eq:Z2-2}
\end{align}
The term associated with $\hat{\ell}_z (\hat{\bm n},\varphi)$ can be regarded as the breaking term of the magnetic $\pi$-rotation $P_3$ symmetry.

The ${P}_3$ symmetry in Eq.~\eqref{eq:Z2-2} indicates that there are two possible subgroups of $G_{\rm slab, H, D}$, $H_{\rm I}$ and $H_{\rm II}$, namely, two subphases B$_{\rm I}$ and B$_{\rm II}$ in a confined $^3$He-B under a magnetic field, depending on $\hat{\ell}_z$. The B$_{\rm I}$-phase with $\hat{\ell}_z=0$ maintains the $P_3$ symmetry, 
\beq
{H}_{\rm I} = P_3.
\label{eq:HI}
\eeq
Another competing phase, B$_{\rm II}$, which is characterized by a definite $\hat{\ell}_z\neq 0$, breaks the $P_3$ symmetry, 
$H_{\rm II}= \mathbb{Z}_1$, where $\mathbb{Z}_1$ denotes the trivial group.

The condition that preserves the magnetic $\pi$-rotation $P_3$ symmetry is found to be
\beq
\hat{\ell}_z(\hat{\bm n},\varphi) = 0,
\label{eq:ellz}
\eeq
which is associated with the order parameter $(\hat{\bm n},\varphi)$. 
This implies that altering $\hat{\ell}_z$ induces the phase transition from the $P_3$ symmetric B$_{\rm I}$ to the $P_3$ symmetry-breaking B$_{\rm II}$-phase. The quantity $\hat{\ell}_z$ mentioned above is transformed nontrivially by the $P_3$ operator as 
\beq
\mathcal{P}_3:~ \hat{\ell}_z \mapsto -\hat{\ell}_z.
\label{eq:ellz2}
\eeq 
Therefore, $\hat{\ell}_z$ can be interpreted as an order parameter of the magnetic $\pi$-rotation $P_3$ symmetry, and it should be zero unless the discrete symmetry is spontaneously broken~\cite{mizushimaPRL12}. 

Notice that the $P_3$ symmetry in the BW state remains even in the presence of superfluid flow~\cite{wuPRB13}, where the phase bias that generates the flow field is parallel to the surface. The flow field explicitly breaks the ${\rm SO}(2)_{L_z+S_z}$ symmetry and the TRS simultaneously. The BdG Hamiltonian, however, still remains invariant under the combined symmetry of the time-inversion and the joint $\pi$-rotation about the surface normal.

{\it Symmetry-protected topological phase.}---
Remarkably, one can introduce a topological invariant as long as the $P_3$ symmetry is not spontaneously broken, namely, $\hat{\ell}_z = 0$. Combining it with the PHS in Eq.~\eqref{eq:PHS}, one obtains the so-called chiral symmetry in the confined $^3$He-B under a magnetic field as
\beq
\{{\Gamma}_1 , \underline{\cal H}(0,0,k_z)\} = 0,
\quad \Gamma _1 \equiv \mathcal{C} \mathcal{P}_3.
\label{eq:chiral1}
\eeq
Since the chiral symmetry is maintained by the ${Q}$ matrix as well, it imposes a constrain on the target space of ${Q}$ in Eq.~\eqref{eq:h2}. The target space $\hat{\bm m}\in S^3$ is reduced to $S^1$, and thus ${Q}$ is regarded as the projector that maps the chiral symmetric momenta ${\bm k}_{\perp}=(0,0,k_z)\in S^1$ to $\mathcal{M}=S^1$. The topological invariant relevant to the fundamental group $\pi _1(S^1)=\mathbb{Z}$ is the one-dimensional winding number~\cite{satoPRB09,satoPRB11,mizushimaPRL12}
\beq
w_{\rm 1d} = -\frac{1}{4\pi i}\int_{-\infty}^{\infty}dk_z\left. 
{\rm tr}[{\Gamma}_1{\cal H}^{-1}({\bm k})\partial_{k_z}{\cal H}({\bm k})]
\right|_{{\bm k}_{\parallel}\!=\!{\bm 0}},
\label{eq:winding}
\eeq
which is evaluated as $w_{\rm 1d} \!=\! 2$ for $\gamma H \! <\! E_{\rm F}$ ($\Delta_{\perp}>0$). 
The bulk-edge correspondence shown in Sec.~\ref{sec:index} implies that in the case of $\hat{\ell}_z=0$, the surface bound state remains gapless as $E({\bm k}_{\parallel}) \!=\! 0$ at ${\bm k}_{\parallel}\!=\! {\bm 0}$ even in the presence of the magnetic field. Hence, the chiral symmetry and the bulk-edge correspondence still have the physical consequence that the gapless bound states yield the Ising anisotropic magnetic response.

\begin{figure}[t!]
\begin{center}
\includegraphics[width=80mm]{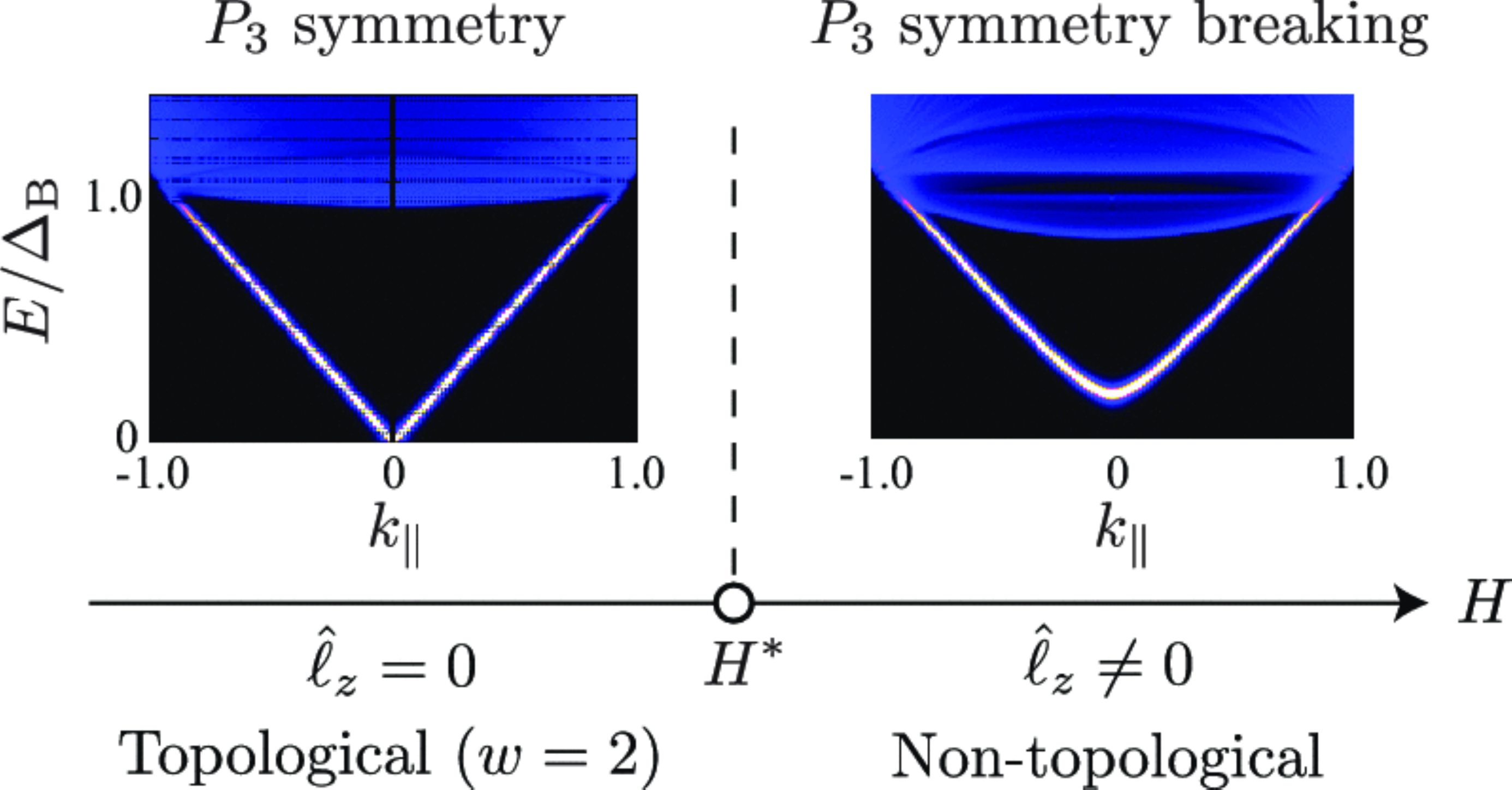}
\end{center}
\caption{(Color online) Schematic phase diagram of $^3$He-B under a parallel magnetic field, where $H$ involves the topological phase transition with spontaneous symmetry breaking. The panels show the momentum-resolved surface density of states, $\mathcal{N}(\hat{\bm k},z;E)$ for the symmetry-protected topological B$_{\rm I}$ and non-topological B$_{\rm II}$ (see also Figs.~\ref{fig:phaseDH} and \ref{fig:nvec}). }
\label{fig:mag}
\end{figure}

The B-phase under a magnetic field is classified into two subphases: The $P_3$ symmetry-protected topological phase, ${\rm B}_{\rm I}$, and the $P_3$ symmetry-broken nontopological phase, ${\rm B}_{\rm II}$. As mentioned in Eqs.~\eqref{eq:ellz} and \eqref{eq:ellz2}, these two phases are characterized by $\hat{\ell}_z$, where the former (latter) corresponds to $\hat{\ell}_z=0$ ($\hat{\ell}_z\neq 0$). The phase diagram is shown in Fig.~\ref{fig:phaseDH}, where there exists a topological quantum critical point at a weak field, beyond which the $P_3$ symmetry spontaneously breaks and it simultaneously triggers off the topological phase transition (see also Fig.~\ref{fig:phaselz} and the text in Sec.~\ref{sec:summary}). Figure \ref{fig:mag} summarizes the phase diagram and the momentum-resolved surface density of states in Eq.~\eqref{eq:dosk} at the surface ${\bm r}\rightarrow z$.
The topological phase transition concomitant with spontaneous symmetry breaking can occur without closing the bulk energy gap. The acquisition of the mass of surface Majorana fermions is generated by the spontaneous symmetry breaking.

The hallmark of the symmetry-protected topological phase ${\rm B}_{\rm I}$ is the existence of surface helical Majorana fermions. In the absence of a magnetic field, as mentioned in Sec.~\ref{sec:MF3HeB}, the surface Majorana fermions are the consequence of the TRS. Similarly, it is demonstrated that for ${\rm B}_{\rm I}$, the characteristics of surface helical Majorana fermions in Eqs.~\eqref{eq:rho0} and \eqref{eq:MIS3} are protected by the chiral symmetry \eqref{eq:chiral1} and PHS \eqref{eq:PHS}. This is because the Majorana nature in Eqs.~\eqref{eq:rho0} and \eqref{eq:MIS3} is a general consequence of the chiral symmetric BdG Hamiltonian and is thus maintained unless the $P_3$ symmetry is spontaneously broken. The chiral symmetry protection of the Majorana Ising spin was first pointed out in Ref.~\citeonline{mizushimaPRL12} for $^3$He-B, and Ref.~\citeonline{shiozakiPRB14} presented the proof generalized to an antiunitary operator relevant to the magnetic point group symmetry, as briefly mentioned in Sec.~\ref{sec:chiral}.

The gapless dispersion is protected by $w_{\rm 1d}$ or $w_{\rm 3d}$, while the interference between surface Majorana cones distorts the conelike spectrum when two specular surfaces at $z \!=\! 0$ and $D$ approach each other. The hybridization of two surface states exponentially splits the zero energy state at $|{\bm k}_{\parallel} |\!=\! 0$ with quantum oscillation on the scale of $k^{-1}_{\rm F}$ as $\delta E({\bm k}_{\parallel} \!=\! {\bm 0}) \!\sim\! e^{-D/\xi}\sin(k_{\rm F}D)$~\cite{mizushimaPRA10-2,kawakamiJPSJ11,chengPRL09,chengPRB10}. The numerical calculation based on the quasiclassical theory confirmed the appearance of a mini-gap generated by the hybridization when $^3$He is confined in two directions~\cite{tsutsumi:2011b}. For $^3$He confined in one direction, e.g., in parallel plates, however, no finite excitation gap is generated by the hybridization even for strong confinement~\cite{wuPRB13}.

{\it Effective action of surface Majorana fermions.}---
Although only the exact zero energy states at ${\bm k}_{\parallel}={\bm 0}$ are rigorously protected by $w_{\rm 1d}$ and possess the Majorana nature, the whole branch of surface bound states turns out to have the characteristic of Majorana fermions within the Andreev approximation $\Delta _{\rm B}\ll E_{\rm F}$. We here present the effective action of such helical Majorana fermions with an effective mass and briefly comment on the nontrivial topological properties. 

Within the Andreev approximation, the surface bound states maintain the characteristic of helical Majorana fermions, similar to those in Sec.~\ref{sec:MF3HeB}. The effective Hamiltonian of surface bound states in Eq.~(\ref{eq:Hsurf}) is now extended to contain the mass term $M\equiv M(\hat{\bm n},\varphi)$ as~\cite{mizushimaJPCM15}
\begin{align}
\mathcal{H}_{\rm surf} = \sum _{{\bm k}_{\parallel}} {\psi}^{\rm T}_{\rm M}(-{\bm k}_{\parallel}) \left[
c \left( {\bm k}_{\parallel} \times {\bm \sigma} \right)\cdot\hat{\bm z}
+M\sigma _z
\right] {\psi}_{\rm M}({\bm k}_{\parallel}),
\label{eq:Hsurf2}
\end{align}
where $c = \Delta _0/k_{\rm F}$. The Majorana field ${\psi}_{\rm M}$ is associated with the original quantized field for surface states, $\psi$, as  
$\psi({\bm k}) \equiv U(\hat{\bm n},\varphi){\psi}_{\rm M}({\bm k})$, 
which obeys $\{ \psi _{a}, \psi _b \} = \delta _{ab}$. It is remarkable that the effective mass of the surface Majorana fermion, $M$, is determined by the single parameter $\hat{\ell}_z$ as 
\beq
M(\hat{\bm n},\varphi) = \frac{\gamma H}{2}\hat{\ell}_z(\hat{\bm n},\varphi).
\label{eq:Mmass}
\eeq 
The equation of motion for the surface Majorana fermions ${\psi}_{\rm M}(x)$ is governed by the $2+1$-dimensional Majorana equation,
\beq
\left(-i \gamma^{\mu}\partial _{\mu} + M(\hat{\bm n},\varphi) \right){\psi}_{\rm M} (x) = 0,
\label{eq:Majoranaeq}
\eeq
where we replace $(k_x,k_y)$ with $(-i\partial _x, -i\partial _y)$ and set $x\equiv ({\bm r},t)$. Without loss of generality, we set $M/c\rightarrow M$. The $\gamma$-matrix is introduced as $(\gamma^0,\gamma^1,\gamma^2) = (\sigma_z, i\sigma _x, i\sigma _y)$, which satisfies $\{\gamma^{\mu},\gamma^{\nu}\}=2g^{\mu\nu}$ with the metric $g^{\mu\nu} = g_{\mu\nu} = {\rm diag}(+1,-1,-1)$ ($\mu, \nu = 0, 1, 2$). The effective action for the $2+1$-dimensional Majorana equation (\ref{eq:Majoranaeq}) is given as
\beq
\mathcal{S}_{\rm surf} = \int dx^3 \bar{\psi}_{\rm M}(x)\left( -i \gamma^{\mu}\partial _{\mu} + M(\hat{\bm n},\varphi) \right) {\psi}_{\rm M}(x).
\label{eq:action}
\eeq
As shown in Eq.~(\ref{eq:Mmass}), the effective mass $M$ is parameterized with $\hat{\ell}_z$, which is the order parameter associated with the $P_3$ symmetry breaking. 

At the quantum critical point $H^{\ast}$, the quantum fluctuation of the order parameter $\hat{\ell}_z$ may be enhanced. Grover {\it et al.}~\cite{grover} proposed the $2+1$-dimensional effective action that describes the Majorana fermion coupled to the Ising field. They found the emergence of supersymmetry (SUSY) at the quantum critical point $H^{\ast}$. Contrary to previous works~\cite{friedan,fendley,huijse,yu,bauer} in which the emergence of SUSY requires the fine tuning of two or more parameters, the terms that break SUSY become irrelevant at the critical point and SUSY emerges without enforcing the conditions microscopically. 

Park {\it et al.}~\cite{parkPRB15} proposed another scenario that spontaneously generates a finite mass $M$ without breaking the TRS. They found that surface Majorana fermions are not coupled to the $J=0$ Nambu-Goldstone mode (phase mode), while they can be coupled to one of the $J=1$ Nambu-Goldstone modes (spin wave modes). The coupling to spin wave modes mediates the interaction between surface Majorana fermions, which generates an effective mass. Park {\it et al.}~\cite{parkPRB15} also predicted that in realistic situations the coupling constant of Majorana fermions to the spin wave mode lies in the vicinity of the quantum critical point beyond which the Majorana fermion acquires an effective mass. It is also remarkable that the quantum mass acquisition takes place in the first-order transition, which is in contrast to the second-order transition in the case of the magnetic-field-driven topological phase transition at $H^{\ast}$ [see Fig.~\ref{fig:nvec}(d)].


For $\hat{\ell}_z \neq 0$, the Majorana cone in Eq.~(\ref{eq:Hsurf2}) has a finite energy gap. In the effective Hamiltonian of such a quasi-two-dimensional system, the topological invariant can be introduced as~\cite{volovik}
\beq
N_{\rm 2} = \frac{1}{4\pi^2}\int d{\bm k}_{\parallel} \int d\omega {\rm tr}\left[G \partial _{k_x} G^{-1} G \partial _{k_y} G^{-1}
G\partial _{\omega} G^{-1}\right].
\label{eq:topologicalN}
\eeq
Here, the Green's function of the surface Majorana fermion is defined as $G^{-1}\equiv i\omega - \mathcal{H}_{\rm surf}({\bm k}_{\parallel})$, where $\mathcal{H}_{\rm surf}({\bm k}_{\parallel})\!\equiv\! c ( {\bm k}_{\parallel} \times {\bm \sigma} )\cdot\hat{\bm z} +M\sigma _z$. It turns out that this winding number is equivalent to the topological invariant introduced in Eq.~(\ref{eq:topologicalN}), and thus the topological invariant is estimated for massive Majorana fermions as~\cite{mizushimaJPCM15}
\beq
N_{2} = \frac{{\rm sgn}(\hat{\ell}_z)}{2}.
\label{eq:m2d}
\eeq
In the case of three-dimensional topological insulators, the Dirac fermion acquires a mass $M$ when magnetic impurities are sprinkled in the surface region. Similarly to Eq.~\eqref{eq:m2d}, the nontrivial topology characterized by $N_{2}={\rm sgn}(M)/2$ is responsible for the half-quantum Hall effect~\cite{qiRMP11}.

For superconductors and superfluids, however, the ${\rm U}(1)$ gauge symmetry is spontaneously broken, and thus the Hall conductivity is not quantized. Even though the ${\rm U}(1)$ symmetry is absent, massive surface Majorana fermions carry the Hall component of the nondissipative thermal transport~\cite{read,wangPRB11,ryuPRB12,nomuraPRL12,shiozakiPRL13},
\beq
\kappa _{\rm H} = N_{2} \frac{\pi^2 k^2_{\rm B}}{6h}T,
\eeq
as a consequence of the energy conservation. The direct relation between $\kappa _{\rm H}$ and the bulk winding number of three-dimensional topological superconductors was clarified by Shiozaki and Fujimoto~\cite{shiozakiPRL13}. This quantized transport quantity is a hallmark of the nontrivial topological aspect of massive Majorana fermions emergent in $^3$He-B. It is however pointed out that the surface Majorana fermion may not be stable against the combined effects of weak disorder and weak interactions~\cite{foster1,foster2,foster3}. The quantized thermal (class DIII) and spin (class AIII) Hall effects can be predicted if the surface states are stable~\cite{foster4}. We also notice that contrary to superconductors, Nambu-Goldstone modes remain gapless in the case of superfluids. The order parameter fluctuation modes contribute to the thermal conductivity through the vertex corrections~\cite{graf}, which might deviate $\kappa _{\rm H}$ from the quantized value. We also notice that the thermal response of the $^3$He-A in a thin film is characterized by the Chern number~\cite{sumiyoshiJPSJ13}. Furthermore, if superconductors maintain the spin-rotation symmetry, which is not the case in $^3$He, the spin Hall conductivity is quantized~\cite{read,senthil}.

\subsection{Spin susceptibility and odd-frequency pairing}
\label{sec:odd}

We have illustrated that the superfluid $^3$He-B confined in a slab is accompanied by the critical field $H^{\ast}$ at which the topological phase transition is concomitant with the spontaneous $P_3$ symmetry breaking. The topological order $\hat{\ell}_z$ is directly associated with the Majorana Ising spin \eqref{eq:MIS3} as
\beq
\hat{\bm h}\cdot{\bm S}_{\rm surf} = \hat{\ell}_z(\hat{\bm n},\varphi) S^{\rm M}_z,
\label{eq:MIS4}
\eeq
where $\hat{\bm h}$ is the orientation of an applied field. This tells us that the surface spin can be coupled to the applied field only when $\hat{\ell}_z$ becomes nonzero, while the massless Majorana fermions do not contribute to the local spin susceptibility. Hence, it is naturally expected that the reorientation of $\hat{\ell}_z$ at the topological phase transition $H^{\ast}$ is accompanied by the anomalous magnetic response of surface Majorana fermions, which is a hallmark of the reorganization of the surface bound states. To fully understand the anomalous behavior, we also introduce the concept of odd-frequency even-parity pairing, which is another facet of surface Andreev bound states~\cite{tanakaJPSJ12,mizushimaPRB14}. 

\subsubsection{Spin susceptibility}

\begin{figure}[t!]
\begin{center}
\includegraphics[width=80mm]{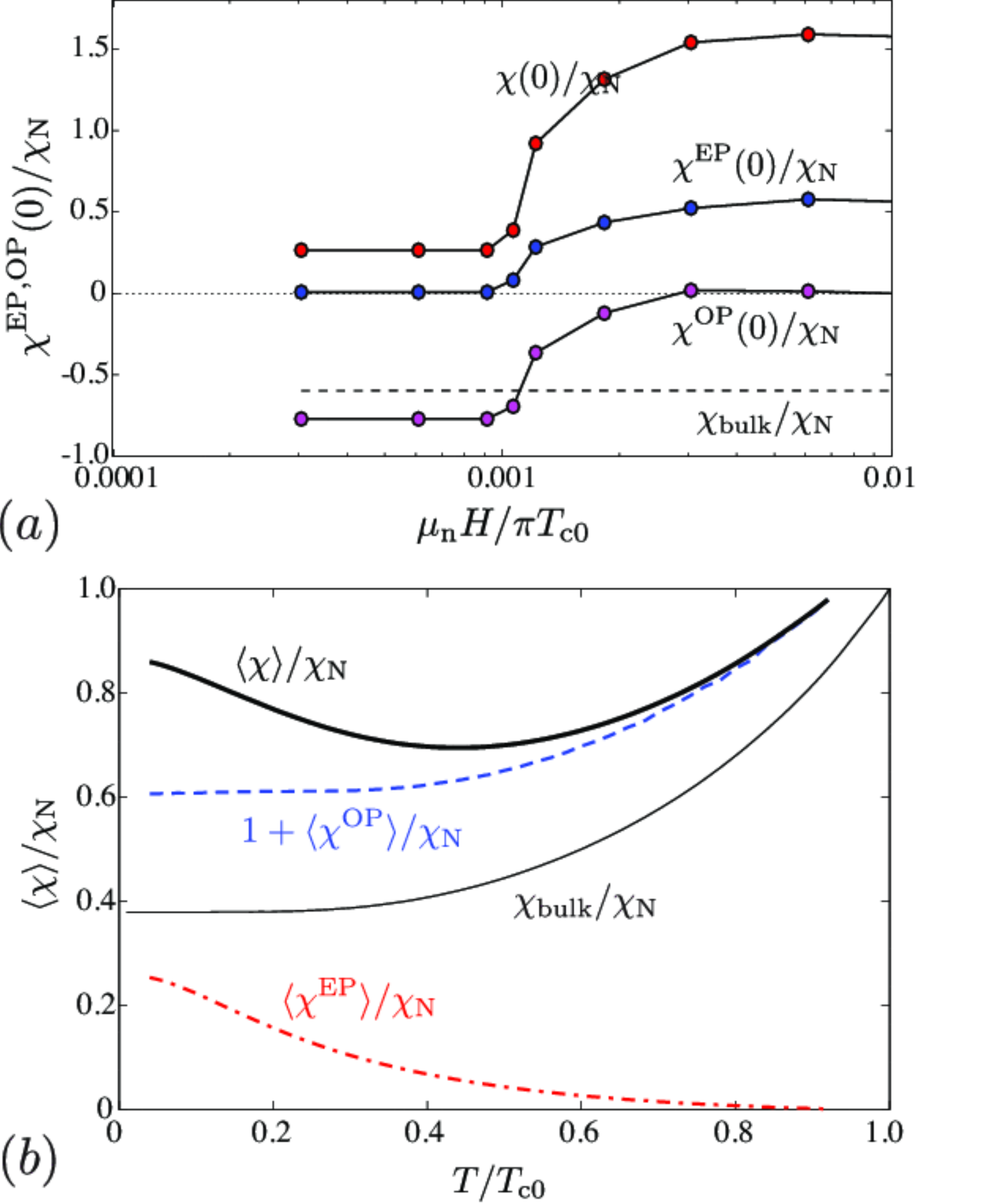}
\end{center}
\caption{(Color online) (a) Field dependence of surface spin susceptibilities $\chi (0)$, $\chi^{\rm OP}(0)$, and $\chi^{\rm EP}(0)$ at $T=0.2T_{\rm c0}$. (b) Temperature dependence of the spatially averaged spin susceptibilities $\langle\chi \rangle$, $\langle\chi^{\rm OP}\rangle$, and $\langle\chi^{\rm EP}\rangle$ in the nontopological B$_{\rm II}$-phase at $\mu _{\rm n} H = 0.009 \pi T_{\rm c0}$ and $D=20\xi _0$. Here, $\chi^{\rm OP}(0)$ and $\chi^{\rm EP}(0)$ denote the contribution of even-frequency spin-triplet odd-parity and odd-frequency spin-triplet even-parity pair amplitudes to spin susceptibilities. We also plot the spin susceptibility of the bulk $^3$He-B in Eq.~(\ref{eq:chibw}), $\chi _{\rm bulk}$. Figures adapted from Ref.~\citeonline{mizushimaPRB14}.}
\label{fig:chi}
\end{figure}

We start to display in Fig.~\ref{fig:chi} the field dependence of the local spin susceptibility on the surface, ${\chi}(z \!=\! 0)$, where $\chi (z)$ is defined as ${\chi}(z)/\chi _{\rm N} \!\equiv\! M(z)/M_{\rm N}$ with the local spin susceptibility
\beq
\chi(z) = \chi_{\rm N}\left[ 
1 + \frac{2}{\gamma H}\left\langle \hat{h}_{\mu}g_{\mu}(\hat{\bm k},z;\omega _n)
\right\rangle _{\hat{\bm k},n}
\right].
\label{eq:chi}
\eeq 
The magnetic field is applied along the surface ${\bm H}\parallel\hat{\bm x}$. The spin susceptibility of the bulk BW states is defined in Eq.~\eqref{eq:chibw}, which can be estimated as $\chi _{\rm bulk}\approx 0.4 \chi _{\rm N}$ for $F^{\rm a}_0=-0.7$ at low temperatures. For a parallel field (${\bm H} \!\parallel\! \hat{\bm x}$), it turns out that the spin susceptibility $\chi(z)$ on the surface is sensitive to the orientation of $\hat{\bm \ell}$. The strong anisotropy of the surface spin susceptibility is attributed to the Majorana Ising nature of surface bound states: Surface Majorana fermions are not coupled to a parallel field unless the Majorana fermion acquires a mass. Nagato {\it et al.}~\cite{nagatoJPSJ09} demonstrated that the surface spin susceptibility is anomalously enhanced once the surface state has a finite energy gap. By carrying out a fully self-consistent calculation that takes into account the dipole interaction and Zeeman energy, Ref.~\citeonline{mizushimaPRL12} uncovers that the anomalous behavior of the local spin susceptibility is accompanied by the topological phase transition concomitant with spontaneous $P_3$ symmetry breaking. 

The field dependence of the surface spin susceptibility in the confined $^3$He-B is plotted in Fig.~\ref{fig:chi}(a), where $D=20\xi _{\rm 0}$ and $T=0.2T_{\rm c0}$. In the B$_{\rm I}$-phase within $H<H^{\ast}$, $\hat{\bm \ell}=0$ implies that the Majorana Ising spin points to the surface normal direction ($\hat{\bm z}$), which is perpendicular to the applied field. Hence, as expected in Eq.~\eqref{eq:MIS4}, the surface Majorana fermions do not contribute to the magnetic response, and the surface spin susceptibility does not deviate from $\chi _{\rm bulk}$. In Ref.~\citeonline{mizushimaPRL12}, it is also found that in the vicinity of $H^{\ast}$ the orientation of the surface magnetization density is tilted from the direction of the applied field and ${\bm H}\parallel \hat{\bm x}$ significantly induces the local magnetization ${\bm M}=(M_x,0,M_z)$. This is attributed to the fact that in the vicinity of $H^{\ast}$, the order parameter $\hat{\ell}_z(\hat{\bm n},\varphi)$ is reoriented from the dipolar-field-favored state $\hat{\ell}_z=0$ to the Zeeman-field-favored state $\hat{\ell}_z=1$, and the competition between these two energy scales gives rise to neither the gapless surface state with $\hat{\ell}_z=0$ nor the fully gapped state with $\hat{\ell}_z=1$. The emergence of $M_z(z)$ on the surface is a hallmark of the reorientation of the BW order parameter in the vicinity of $H^{\ast}$.

In the B$_{\rm II}$-phase with $H > H^{\ast}$, however, $\chi(z)$ is enhanced around the surface, where the surface Majorana fermions are massive. The direct relation between the mass acquisition of surface bound states and the anomalous enhancement of the surface spin susceptibility was first pointed out by Nagato {\it et al.}~\cite{nagatoJPSJ09}. The surface spin susceptibility in the B$_{\rm II}$-phase amounts to $1.6\chi _{\rm N}$, and thus the total susceptibility averaged over the slab may yield the anomalous behavior. To capture this, we plot in Fig.~\ref{fig:chi}(b) the temperature dependence of the spatially averaged spin susceptibility, $\langle \chi \rangle \equiv \frac{1}{D}\int^{D}_0 \chi (z) dz$, 
at $\mu _{\rm n}H = 0.009\pi T_{\rm c0}$ corresponding to the nontopological B$_{\rm II}$-phase. It is seen that the $T$-dependence of $\langle \chi \rangle $ in a slab exhibits nonmonotonic behavior, where there exists a critical temperature below which $\langle \chi \rangle$ increases as $T$ decreases. We demonstrate below that the anomalous enhancement and the nonmonotonic behavior are fully understandable with the concept of odd-frequency pairing amplitudes, another facet of the surface bound states.

According to the sum rule, the static spin susceptibility $\langle\chi \rangle$ is obtained by integrating the absorptive part of the dynamical spin susceptibility over all frequencies~\cite{leggett73,leggett74}. Hence, the temperature and field dependences of $\langle \chi \rangle$ are detectable through NMR experiments~\cite{ahonenJLTP76}. The field and temperature dependences of $\langle \chi \rangle$ may unveil the surface state of the symmetry-protected topological superfluid $^3$He-B.

\subsubsection{Odd-frequency pairing and paramagnetic response}

To capture the direct relation between the odd-frequency Cooper pair amplitudes and the local spin susceptibility, we first recast the definition of the spin susceptibility \eqref{eq:chi} into~\cite{mizushimaPRB14}
\beq
\chi({\bm r}) = \chi _{\rm N}\left[ 1 + \frac{1}{\mu _{\rm n}H}
\left\langle \frac{f_0 \bar{f}_{\mu} + \bar{f}_0 f_{\mu}}{2g_0} \right\rangle _{\hat{\bm k},n}\right].
\label{eq:m2}
\eeq
We here utilize the fundamental symmetries of the propagator in Eqs.~\eqref{eq:sym1}, \eqref{eq:sym2}, and \eqref{eq:trs2} and the normalization condition in Eq.~\eqref{eq:norm}, $g_{\mu} \!=\! ( f_0 \bar{f}_{\mu} + \bar{f}_0 f_{\mu}
+ i\epsilon ^{\mu\nu\eta}f_{\nu}\bar{f}_{\eta})/2g_0$. The anomalous propagator, $f=i\sigma _y f_0 + i{\bm \sigma}\cdot{\bm f}\sigma _y$, contains all the information on the Cooper pair correlation of quasiparticles. Equation \eqref{eq:m2} indicates that only the mixing term of spin-singlet and triplet Cooper pair amplitudes contributes to the spin susceptibilities. The expression in Eq.~\eqref{eq:m2} is a generic form for superfluids and also applicable to the surface region of type-II superconductors. This was derived in Ref.~\citeonline{higashitaniPRL13} for the aerogel-superfluid $^3$He-B system and in Ref.~\citeonline{mizushimaPRB14} for time-reversal-invariant superfluids and superconductors.

In accordance with the Fermi-Dirac statistics, a wavefunction of Cooper pairs must change its sign after a permutation of two paired fermions. Then, as summarized in Table~\ref{table:odd}, the symmetry of Cooper pairing in a single-band centrosymmetric superconductor is naturally categorized into four types: Even-frequency spin-singlet even-parity (ESE),  spin-triplet odd-parity (ETO), odd-frequency spin-singlet odd-parity (OSO), and spin-triplet even-parity (OTE) pairings. The former two are conventional pairings in the sense that they do not change the sign of the Cooper pair wavefunction by exchanging the times of paired fermions. In general, the Cooper pair amplitudes are separated to even-frequency and odd-frequency components, 
\beq
f_{j}(\hat{\bm k},{\bm r};\omega_n)= f^{{\rm EF}}_{j}(\hat{\bm k},{\bm r};\omega_n) + f^{{\rm OF}}_{j}(\hat{\bm k},{\bm r};\omega_n),
\eeq
where we set $f^{\rm EF}_j(\omega _n) = [f_j(\omega _n)+f_j(-\omega _n)]/2$, $f^{\rm OF}_j(\omega _n) = [f_j(\omega _n)-f_j(-\omega _n)]/2$, and $j=0,x,y,z$. 

Although conclusive evidence of odd-frequency pairing in bulk materials has not yet been observed experimentally since the first prediction by Berezinski~\cite{berezinskiiJETP74}, the OSO and OTE pair amplitudes can emerge ubiquitously in spatially nonuniform systems accompanied by Andreev bound states and the anomalous proximity effect~\cite{bergeretPRL01,tanakaPRB05,tanakaPRL07,tanakaPRL07v2,eschrigJLTP07,asanoPRL07,asanoPRB07,eschrigNP,linderPRL09,higashitaniJLTP09,linderPRB10,yokoyamaPRL11,asanoPRL11,blackPRB12,blackPRB13,blackPRB13v2,parhizgar,stanev,asanoPRB14,suzukiPRB14,higashitaniPRB14}. In Table~\ref{table:odd}, we summarize the four possible classes of Cooper pair amplitudes in bulk superconductors and superfluids, and the additional Cooper pairs induced by a symmetry-breaking field~\cite{tanakaJPSJ12,dainoPRB12,tanakaPRL07,tanakaPRB07,yokoyamaPRB08,yokoyamaJPSJ10}. In the case of spin-triplet superconductors and superfluids, ETO components $f^{{\rm EF}}_{\mu}$ exist in the bulk, where $f^{{\rm EF}}_{\mu}({\bm k})=-f^{{\rm EF}}_{\mu}(-{\bm k})$. A time-reversal-breaking perturbation, such as a magnetic Zeeman field, can induce the mixing of spin-singlet Cooper pair amplitudes. Since the induced spin-singlet pairing must have an odd parity unless the translational symmetry is broken, the Cooper pairs residing in bulk ETO superconductors and superfluids have OSO pairing in addition to ETO pairing. A translational symmetry breaking field, such as a surface boundary condition and vortices, induces the OTE components $f^{\rm OF}_{\mu}$ in bulk ETO superconductors and superfluids. Note also that all four pairings can emerge when both TRS and translational symmetry are broken. 

The concept of odd-frequency pairing has succeeded in extracting the anomalous charge and spin transport, electromagnetic responses, and proximity effects via Andreev bound states. The odd-frequency pairing is found to yield anomalous paramagnetic responses and negative superfluid density~\cite{asanoPRB14,higashitaniPRB14}. In addition, it has recently been demonstrated that Majorana fermions are identical to the odd-frequency pairing~\cite{dainoPRB12,higashitaniPRB12,tsutsumi:2012c,asanoPRB13,mizushimaPRB14,ebisu}. A more comprehensive review on odd-frequency pairing was made by Tanaka {\it et al.}~\cite{tanakaJPSJ12} and Eschrig~\cite{eschrigJLTP07}. 

\begin{table}
\centering
\begin{tabular}{c|ccc|cc}
\hline\hline
& \multicolumn{3}{c}{Parity} 
& \multicolumn{2}{c}{Broken symmetry} \\
$\Delta$ & frequency & spin & parity & time & trans. \\
\hline
ESE & $+$ & $-$ & $+$ & OTE & OSO \\
ETO & $+$ & $+$ & $-$ & OSO & OTE \\
OSO & $-$ & $-$ & $-$ & ETO & ESE \\
OTE & $-$ & $+$ & $+$ & ESE & ETO \\
\hline\hline
\end{tabular}
\caption{Classification of possible Cooper pairings in bulk superconductors: ESE, ETO, OSO, and OTE pairs. The fifth and sixth columns show the Cooper pair amplitudes emergent in systems with the breaking of time-reversal symmetry and translational symmetry, respectively.  
}
\label{table:odd}
\end{table}

When the system is assumed to maintain the TRS at zero fields, the general form of $\chi$ in Eq.~\eqref{eq:m2} is further recast into a more convenient form in terms of Cooper pair amplitudes. For a weak magnetic field, $\mu _{\rm n}H/\Delta \!\ll\! 1$, one can formally expand $g_0$, $f_0$, and $f_{\mu}$ in powers of $\mu _{\rm n}H/\Delta$: $g_0 \!=\! g^{(0)}_0 + g^{(1)}_0 +\cdots$, $f_0 \!=\! f^{(1)}_0 + \cdots$, and $f_{\mu} \!=\! f^{(0)}_{\mu} + f^{(1)}_{\mu} + \cdots$. To this end, the spin susceptibility $\chi \equiv \hat{h}_{\mu}\chi _{\mu\nu}\hat{h}_{\nu}$ is composed of odd- and even-parity Cooper pair amplitudes~\cite{mizushimaPRB14,mizushimaJPCM15}, 
\beq
\chi(z) = \chi _{\rm N} 
+ \chi^{\rm OP}(z) +\chi^{\rm EP}(z).
\label{eq:chi2}
\eeq
The odd-parity contribution $\chi^{\rm OP}(z)$ is given by the mixing term of the OSO pair amplitude $f^{\rm OF}_0$ and the ETO pair ${\bm f}^{\rm EF}$,
\beq
\frac{\chi^{\rm OP}(z)}{\chi _{\rm N}} \equiv
\frac{1}{\mu _{\rm n}H}{\rm Re}\left\langle 
\frac{f^{{\rm OF}(1)}_0\hat{h}_{\mu}f^{{\rm EF}(0)\ast}_{\mu}}{g^{(0)}_0}
\right\rangle _{\hat{\bm k},n}. \label{eq:chiOP} 
\eeq
The even-parity contribution $\chi^{\rm EP}(z)$ is given by the mixing term of the ESE pair amplitude $f^{\rm EF}_0$ and the OTE pair ${\bm f}^{\rm OF}$,
\beq
\frac{\chi^{\rm EP}(z)}{\chi _{\rm N}} \equiv
-\frac{1}{\mu _{\rm n}H}{\rm Re}\left\langle 
\frac{f^{{\rm EF}(1)}_0\hat{h}_{\mu}f^{{\rm OF}(0)\ast}_{\mu}}{g^{(0)}_0}
\right\rangle _{\hat{\bm k},n}. \label{eq:chiEP}
\eeq
We here utilize the relation $g^{(0)}_0(\hat{\bm k},z; \omega _n) = - g^{(0)}_0(\hat{\bm k},z; -\omega _n)$, which can be maintained by time-reversal-invariant systems. Equations \eqref{eq:chiOP} and \eqref{eq:chiEP} indicate that only the spin-triplet pairings $f^{(0)}_{\mu}$ at zero fields can be directly coupled to the applied field. 

In the bulk of spin-triplet superfluids and superconductors, ETO pairings $f^{{\rm EF}(0)}_{\mu}$ are simply the ${\bm d}$-vector and thus the behavior of $\chi^{\rm OP}$ is then understandable with the rotation of the ${\bm d}$-vector, where $\chi^{\rm OP}=0$ for $\hat{\bm h}\perp{\bm d}$ and $\chi^{\rm OP} \le 0$ for ${\bm d}\cdot\hat{\bm h}\neq 0$. In contrast,  OTE Cooper pairs $f^{{\rm OF}(0)}_{\mu}$ are absent in the bulk and induced by the breaking of translational symmetry at surfaces, interfaces, or vortices. As a result, the total spin susceptibility at surfaces is determined by the OTE pairing $f^{{\rm OF}(0)}_{\mu}$ directly coupled to the applied field in addition to the ordinary contribution from the relative orientation of the ${\bm d}$-vectors to $\hat{\bm h}$.

The OTE pairing $f^{{\rm OF}(0)}_{\mu}$ emergent in the surface region is subject to the discrete symmetries that $^3$He-B maintains. For the pair amplitudes $f^{(0)}_{\mu}$ at zero fields, the ${\rm SO}(2)_{L_z+S_z}$ rotation symmetry and the boundary condition \eqref{eq:bc} impose a strong constraint on the OTE pairing as~\cite{mizushimaPRB14}
\beq
\hat{\bm h}\cdot {\bm f}^{{\rm OF}(0)}(\hat{\bm k},z_{\rm surf};\omega _n) = \hat{\ell}_z(\hat{\bm n},\varphi) \tilde{f}^{{\rm OF}(0)}_z(\hat{\bm k},z_{\rm surf};\omega _n) .
\label{eq:fOF}
\eeq
The OTE Cooper pair in the case of $\hat{\bm n}=\hat{\bm z}$, $\tilde{f}^{{\rm OF}(0)}_{z}$, is identical to the momentum-resolved surface density of states,~\cite{higashitaniPRB12,tsutsumi:2012c}
\beq
\mathcal{N}(\hat{\bm k},z;E)
\approx \frac{1}{\pi}
\left|{\rm Re}\tilde{f}^{{\rm OF}(0)}_z(\hat{\bm k},z;\omega _n \rightarrow -iE+0_+)\right|.
\label{eq:equality}
\eeq
Hence, the OTE Cooper pair amplitudes always appear on the surface of $^3$He-B when the surface Andreev bound state exists. Similarly, the ETO pairing at the surface is given by 
\beq
f^{{\rm EF}(0)}_{\mu} = \left( R_{\mu x}(\hat{\bm n},\varphi)\cos\phi _{\bm k} 
+ R_{\mu y}(\hat{\bm n},\varphi)\sin\phi _{\bm k}\right)
\tilde{f}^{{\rm EF}(0)}_{\parallel}.
\label{eq:fEF} 
\eeq
Hence, the orientation of the emergent OTE pairing relative to the applied field is parameterized with the order parameter $\hat{\ell}_z(\hat{\bm n},\varphi)$ associated with the $P_3$ symmetry breaking. In the B$_{\rm I}$ phase with $\hat{\ell}_z=0$, only the ETO pairing exists and thus the spin susceptibility is unchanged from that of the bulk. In the case of B$_{\rm II}$ with $\hat{\ell}_z=1$, however, the surface states turn into the OTE pairing, which gives rise to a paramagnetic response. 

By substituting the expression for pairing amplitudes into Eqs.~(\ref{eq:chiOP}) and (\ref{eq:chiEP}), the spin susceptibility in Eq.~(\ref{eq:chi2}) is recast into the following form:
\beq
\chi _{\rm surf} 
= \chi _{\rm N} + \sqrt{1-\hat{\ell}^2_{z}}\tilde{\chi}^{\rm OP}_{\rm surf}
+ \hat{\ell}_z \tilde{\chi}^{\rm EP}_{\rm surf},
\label{eq:chi_final}
\eeq
where $\tilde{\chi}^{\rm OP}$ and $\tilde{\chi}^{\rm EP}$ denote the spin susceptibility for the $\hat{\bm n}=\hat{\bm z}$ configuration. We notice that the odd-frequency even-parity contribution to the spin susceptibility is estimated within the GL regime as~\cite{mizushimaPRB14,mizushimaJPCM15}
\beq
\chi^{\rm EP}_{\rm surf} = \frac{7\zeta(3)}{12(1+F^{\rm a}_0)}\left( \frac{\Delta _{\perp}}{\pi T}\right)^2   > 0,
\eeq
where $\zeta(3)$ is the Riemann zeta function. Therefore, Eq.~\eqref{eq:chi_final} clearly shows that only the OTE pairs contribute to the surface spin susceptibility when $\hat{\ell}_z=0$, while $\chi$ for $\hat{\ell}_z=1$ is composed of only the ETO Cooper pairs,
\beq
\chi = \left\{
\begin{array}{ll}
\displaystyle{\chi _{\rm N}+ \tilde{\chi}^{\rm OP} < \chi _{\rm N}} & \mbox{for $\hat{\ell}_z = 0$} \\
\\
\displaystyle{\chi _{\rm N}+ \tilde{\chi}^{\rm EP} > \chi _{\rm N}} & \mbox{for $\hat{\ell}_z = 1$} 
\end{array}
\right. .
\eeq
In the case of the bulk superfluid $^3$He-B, since the OTE pairing is absent, the spin susceptibility is given as $\chi = \chi _{\rm N}+\chi^{\rm OP}$, where $\chi^{\rm OP}<0$ suppresses the spin susceptibility. In contrast, the spin susceptibility contributed from the OTE pairs, $\chi^{\rm EP}$, is expected to increase the spin susceptibility to $\chi > \chi _{\rm N}$~\cite{higashitaniPRB14}.

The symmetry consideration described above implies that although the OTE pair amplitudes always exist in the surface of ETO superconductors and superfluids and yield a paramagnetic response, they do not necessarily couple to the applied magnetic field. The topological order $\hat{\ell}_z$ determines the contribution of odd-parity Cooper pairs to the surface spin susceptibility. This behavior is clearly observed in Fig.~\ref{fig:chi}(a), where the even-parity contribution $\chi^{\rm EP}$ abruptly increases at $H=H^{\ast}$ and $\chi^{\rm OP}$ disappears. Figure~\ref{fig:chi}(a) shows that in B$_{\rm II}$ with $\hat{\ell}_z\neq 0$, the OTE pair amplitudes maintain the paramagnetic response even at low temperatures beyond the GL regime. Hence, the surface spin susceptibility is anomalously enhanced at $H^{\ast}$ by the change in the Cooper pair amplitudes that couple to the applied field. Owing to the paramagnetic response of the OTE pair amplitudes, the resultant spin susceptibility at the surface exceeds $\chi _{\rm N}$.

We now identify that the increase in $\langle \chi\rangle$ in the low-temperature regime reflects the coupling of the applied field to the OTE pairing that yields a paramagnetic response. In the high-temperature regime, as shown in Fig.~\ref{fig:chi}(b), the continuum states with $E>\Delta$ dominate the spin susceptibility. As $T$ decreases, however, the OTE pairing gradually grows, while the contributions from the continuum states exponentially decrease. Hence, the increase in the averaged spin susceptibility $\langle \chi \rangle$ in the low-$T$ region of Fig.~\ref{fig:chi}(b) indicates the enhancement of the local magnetization density at the surface. This nonmonotonic behavior of $\langle\chi\rangle$ may be observable only in the nontopological phase. Since the OTE pairing is not responsible for the susceptibility in the symmetry-protected topological phase within $H<H^{\ast}$, the $T$-dependence follows that of $\chi _{\rm N}+\langle \chi^{\rm OP}\rangle$ in Fig.~\ref{fig:chi}(b).

\subsection{Observations of surface bound states in $^3$He-B}
\label{sec:exp}

In $^3$He-B, the pair-breaking effect and surface Andreev bound states have been observed by several experimental groups. We here briefly summarize the experimental observations and key issues on the surface distortion of the gap and the hunting of exotic quasiparticles.

{\it Vibrating wires.}--- 
The Lancaster group~\cite{castelijnsPRL86,carneyPRL89,fisherPRL01} measured the damping rate of a vibrating wire embedded in superfluid $^3$He-B. Contrary to the naive expectation that the moving object gives rise to the pair breaking at the Landau critical velocity, $v_{\rm L}=\Delta_{\rm B}/k_{\rm F}$, the measured critical velocity where the damping rate starts to increase was $v_{\rm L}/3$. The radii of the vibrating wire, $2$ and $50 \mu{\rm m}$, are much larger than the superfluid coherence length $\xi _{\rm B}\lesssim 80 {\rm nm}$. Hence, the vibrating wire can be regarded as a diffusive surface moving with a velocity, giving rise to pair breaking and the formation of surface bound states. The solution to the puzzle was first presented by Lambert~\cite{lambert}. Lambert found the reduction of the critical velocity to $v_{\rm L}/5$ by assuming a strong gap distortion around the moving object, which, however, underestimates the measured value. A further scenario for explaining the measured value $v_{\rm L}/3$ is that as the velocity of the wire slowly increases from $v_{\rm L}/5$, the energy of quasiparticles bound to the wire matches the threshold energy of the scattering state at $v_{\rm L}/3$. Hence, the measured $v_{\rm L}/3$ is regarded as the critical velocity beyond which the moving wire loses momentum via the emission of quasiparticles. The scenario was further developed by Calogeracos and Volovik~\cite{calogeracos} in connection with the Zel'dovich mechanism of positron nucleation. 

{\it Heat capacity.}---
The manifestations of surface Andreev bound states have also been captured by several experimental groups through the deviation of the heat capacity~\cite{choiPRL06,bunkov,bunkov1}. In Ref.~\citeonline{choiPRL06}, they reported the clear deviation of $C(T)$ from that of the bulk $^3$He-B in the vicinity of $T_{\rm c}$, which is attributed to the sufficient zero-energy density of states at the diffusive surface~\cite{vorontsovPRB03,nagaiJPSJ08,nagatoJLTP98}. For a specular surface, however, the two-dimensional relativistic dispersion is responsible for the linear behavior of the low-energy density of states, $
\mathcal{N}(z=z_{\rm surf},E) \propto |E|$. The local density of states gives rise to a power-law behavior of the specific heat, $C(T) \propto T^2$, for low temperatures $T\!\ll\! T_c$~\cite{mizushimaJLTP11}. The measurement of $C(T)$ at lower $T$ down to $T=135~\mu{\rm K}$ was performed by Bunkov {\it et al.}~\cite{bunkov,bunkov1}. They observed a $10~\%$ deviation from the heat capacity of the bulk superfluid $^3$He-B at $T=135~\mu{\rm K}$. The deviation is attributed to the contribution from the surface Majorana cone.

{\it Transverse sound and impedance.}--- 
As a generic property of Majorana fermions, they cannot couple to the local density fluctuation. This implies that the surface bound states might not be detectable through the attenuation of the longitudinal sound wave which propagates the density fluctuation. In strongly correlated Fermi liquids, however, there exists another type of sound wave, the transverse sound. This is the wave propagating the transverse current fluctuation and thus its coupling to the surface state is not forbidden by the Majorana nature. Contrary to normal Fermi liquids, the transverse current in $^3$He-B propagates as a sound wave owing to the coupling to low-lying bosonic collective modes, that is, $J=2$ imaginary squashing modes~\cite{moores,sauls99}. The transverse sound was first observed by Lee {\it et al.}~\cite{lee1999} through the acoustic Faraday effect that is peculiar to the coupling with the squashing modes. The transverse sound with the Faraday effect has been established as a high-resolution spectroscopy for low-lying bosonic and fermionic excitations in $^3$He-B~\cite{davis1,davis2,davisPRL08,davis4}. Among the series of experiments, Davis {\it et al.}~\cite{davisPRL08} observed the unexpected behavior of the attenuation in the frequency range of $1.6\lesssim \omega/\Delta _0 \lesssim 2.0$. The attenuation becomes saturated to the temperature-independent value at low $T$, which is anomalously larger than that expected from theoretical calculation~\cite{moores,sauls99}. Since the fraction of thermally excited quasiparticles exponentially decreases at low temperatures, the damping mechanism due to the coupling to bulk quasiparticles can be ruled out. Hence, the experimental observation in Ref.~\citeonline{davisPRL08} suggests that the anomalous attenuation might be attributed to the coupling of transverse sound waves with surface Majorana fermions. 

The surface acoustic impedance measurement also provides another powerful tool for probing the surface structure~\cite{aokiPRL05,saitoh,wada,murakawaPRL09,murakawaJLTP10,wasaiJLTP10,murakawaJPSJ11}. The complex transverse acoustic impedance is defined as the ratio of the shear stress $\Pi _{xz}$ to the wall velocity $u_x$, $Z=Z^{\prime}+iZ^{\prime\prime}\equiv \Pi _{xz}/u_x$, where the surface is set to be in the $xy$-plane. In the case of the diffusive surface, the surface density of states has a very sharp edge at $E=\Delta^{\ast}$, which separates the bound states $|E|\le \Delta _{\rm B}-\Delta^{\ast}$ and the scattering states $|E|\ge \Delta _{\rm B}$. Recently, Nagato {\it et al.}~\cite{nagatoJPSJ11} elucidated the physical origin of the subgap $\Delta^{\ast}$ which is a consequence of the level repulsion between the bound states and scattering states. The two different energy scales, $\Delta^{\ast}$ and $\Delta _{\rm B}$, were observed as a kink and peak in the complex impedance $Z(T)$~\cite{aokiPRL05,saitoh,wada,murakawaPRL09,murakawaJLTP10,wasaiJLTP10,murakawaJPSJ11,okuda}. The surface specularity is well controllable by coating the wall of the transducer with thin $^4$He layers. With increasing specularity, they observed the change in the dispersion of the surface bound states, such as the reduction in the zero-energy density of states and the behavior of $\Delta^{\ast}-\Delta _{\rm B} \rightarrow 0$~\cite{murakawaPRL09,murakawaJPSJ11,okuda}. All the measurements are well explained by the quasiclassical Keldysh theory with the random $S$-matrix model for surface roughness~\cite{nagatoJLTP07,nagaiJPSJ08} and consistent with the surface density of states characteristic to the Majorana cone.

{\it Spin dynamics.}---
Probing the spin dynamics has been a fingerprint for determining the order parameters of $^3$He, such as $(\hat{\bm n},\varphi)$. The spin dynamics is governed by the phenomenological theory developed by Leggett and Takagi~\cite{leggett74,leggett-takagi1,leggett-takagi2}. The theory, which is composed of the coupled equations for ${\bm d}$ and the quasiparticle spin ${\bm S}$, has succeeded in explaining the NMR properties of the bulk $^3$He~\cite{leggettRMP,wheatleyRMP}. The key is the spontaneous breaking of relative spin-orbit rotation symmetry. In the normal $^3$He, since the spin rotation symmetry is preserved, the contributions of all the spins that have no correlation to their directions average to zero, and the resulting magnetic field generated by the dipole interaction vanishes in the lowest order of the nuclear dipole constant $g_{\rm D}\equiv \mu^2_{\rm n}/a^3$, where $a$ is the mean interatomic distance. As a result, the NMR frequency in the normal $^3$He is the Larmor frequency $\omega _{\rm L}=\gamma H$. In the superfluid phases, however, the spin rotation symmetry is spontaneously broken. The symmetry breaking generates the nuclear dipolar field, which is responsible for the large shift of the NMR frequency~\cite{leggett73}. Indeed, the longitudinal frequency shift in $^3$He-B is given by $\Omega _{\rm B}\equiv\omega - \omega _{\rm L}  = (3\pi\gamma^2\Delta^2_{\rm B}(T)/2g^2\chi _{\rm B})^{1/2}$, which is distinguishable from that of $^3$He-A, $\Omega _{\rm A}$, as $\Omega^2_{\rm B}/\Omega^2_{\rm A}=5/2$~\cite{leggett74}.




A remarkable observation was made by Webb {\it et al.}~\cite{webbJLTP77}, who experimentally determined the critical field above which the $\hat{\bm n}$-vector is tilted from the surface normal, namely, $H^{\ast}$. Here, liquid $^3$He is confined in a single slab cavity (a long rectangular cavity of $1.0 \times 10.0 \times 23~{\rm mm}^3$ size) with a thickness of $1.0$ mm, and an applied magnetic field is parallel to the walls. A sudden change in the applied field, $H\rightarrow H+\delta H$, gives rise to the ``ringing'' of magnetization, which is the nontrivial nonlinear phenomenon described by the Leggett equation. Webb {\it et al.}~\cite{webbJLTP77} observed the ``wall-pinned'' ringing mode in $^3$He-B, which is peculiar to the configuration of the $\hat{\bm n}$ normal to the surface. Indeed, they observed the wall-pinned mode for $H<H^{\ast}$, while such a mode was no longer observed for $H>H^{\ast}$. 

The linear and nonlinear ringing phenomena without damping were theoretically studied by Maki and Tsuneto for the A-phase~\cite{makiPTP74}, and Brinkman~\cite{brinkmanPL74} and Maki and Hu~\cite{makiJLTP75v1,makiJLTP75v2}, independently, for the B-phase. For the bulk $^3$He-B, the change in the magnetization after a sudden application of $\delta H$ generates a torque on the spin axis, while $\hat{\bm n}$ is fixed to be parallel to the field ${\bm H}$, i.e.,  $\partial _t \hat{\bm n}(t) = 0$. The angle $\varphi(t)$ oscillates with the longitudinal resonance frequency $\Omega _{\rm B}$ when $\delta H \ll \Omega _{\rm B}/\gamma$~\cite{makiJLTP75v1}. In the opposite limit, where $\delta H \gg \Omega _{\rm B}/\gamma $, the ringing frequency approaches $\delta \omega =\gamma \delta H$.  

The wall-pinned ringing mode in $^3$He-B was first predicted by Brinkman~\cite{brinkmanPL74} and Maki and Hu~\cite{makiJLTP75v1,makiJLTP75v2}, independently. The damping effects were taken into account by Leggett~\cite{leggettPRL75} and Maki and Ebisawa~\cite{makiPRB76}. To be specific, let $\hat{\bm z}$ be normal to the surface and $\hat{\bm x}$ be the direction of the applied field. In the low field regime, $H \ll H^{\ast}$, $\hat{\bm n}$ and $\varphi$ are forced by the dipole interaction energy to be $\hat{\bm n}\parallel\hat{\bm z}$ and $\varphi _{\rm L} = \cos^{-1}(-1/4)$, while the spin is parallel to the applied field ${\bm S}\parallel\hat{\bm x}$. This configuration corresponds to the symmetry-protected topological ${\rm B}_{\rm I}$ phase with $\hat{\ell}_z=0$ (see Fig.~\ref{fig:nvec}). The wall-pinned ringing mode is generated by a sudden removal of the static field. In the limit of $H,\delta H \ll \Omega _{\rm B}/\gamma$, since the gain of the kinetic energy provided by the field change, $E_{\rm kin}=\frac{1}{2}\chi (\delta H)^2$, is much smaller than the dipole interaction energy (\ref{eq:fdip}), the spin dynamics is constrained to be on the local minima of the dipole interaction energy so that $\varphi (t) = \varphi _{\rm L}$ is fixed for all periods. The Leggett equation with the constraint has a solution with $\partial _t (\hat{\bm n}\cdot{\bm S})=0$ and $\partial _t {\bm S}\propto \hat{\bm n}$~\cite{brinkmanPL74}. The wall-pinned mode corresponds to the mutual rotation of $\hat{\bm n}$ and ${\bm S}$ where the total magnetization is conserved. In the weak field limit, the magnetization harmonically oscillates with the ringing frequency $\omega _r = \sqrt{2/5} (\gamma \delta H)$.

The observation in Ref.~\citeonline{webbJLTP77} indicates that $\hat{\bm n}$ is tilted by the magnetic field energy from the surface normal ($\hat{\ell}_z = 0$), and the nontopological phase with the broken $P_3$ symmetry ($\hat{\ell}_z = +1$) is realized. The critical field observed in the experiment is around 10 G in the vicinity of $T_{\rm c}$, which is the same order as $H^{\ast}\sim 20$-$30$ G obtained from the microscopic calculation in Sec.~\ref{sec:sptb}. However, we would like to mention that the experiments were carried out in the narrow temperature range $1-T/T_{\rm c} \lesssim 0.015$ and the surface is not coated by the $^4$He layer, i.e., the surface was diffusive. In addition, the Leggett equation does not take into account the surface bound states and thus misses the mutual dynamics of surface Majorana fermions with the BW order parameter. The observation of $H^{\ast}$ in the whole temperature region and the effect of the surface condition remain as unresolved problems. 



For $H\gg H^{\ast}$, the $\hat{\bm n}$-vector texture that satisfies $\hat{\ell}_z = +1$ was observed by transverse NMR measurements in a parallel plate geometry~\cite{ahonenJLTP76,ishikawaJLTP89}. In both the experiments, the magnetic field is applied along the plates. NMR techniques have been developed to reveal the phase diagram in a restricted geometry~\cite{kawae,miyawakiPRB00,kawasakiPRL04,levitinJLTP10,bennettJLTP10,levitin13,levitinPRL13}. Most recently, Levitin {\it et al.}~\cite{levitin13,levitinPRL13} have succeeded in uncovering the $\hat{\bm n}$-textures and the confinement-induced order parameter distortion in a thin slab with well-controlled surface condition. In these experiments, a magnetic field is perpendicular to the surface, where the $P_3$ symmetry is explicitly broken and only the B$_{\rm II}$ with $\hat{\ell}_z = +1$ is thermodynamically stable. Using a sensitive SQUID NMR spectrometer, Levitin {\it et al.}~\cite{levitin13,levitinPRL13} observed positively and negatively shifted NMR signals in the B-phase. The former is attributed to the configuration of $\hat{\ell}_z = +1$, while the negative shift is explained by the configuration of $\hat{\ell}_z = -1$, where $\hat{\bm n} \perp \hat{\bm H}\parallel\hat{\bm z}$ and $\varphi = \pi$. The latter configuration does not minimize the nuclear dipole energy, but may be nearly degenerate with the ground state when $H \gg H_{\rm D}\sim 30$G.

{\it Mobility of ions embedded in the surface of $^3$He-B.}---
Ikegami {\it et al.}~\cite{ikegamiJPSJ13} have recently measured the mobility of both negative and positive ions embedded in the free surface of $^3$He-B at low temperatures down to $250\mu{\rm K}$. A potential well generated by applying an electric field confines ions at the minimum point of the potential well. The minimum point $d$ is located at $20$-$60{\rm nm}$ from the free surface, which is comparable to the superfluid coherence length $\xi _{\rm B}$. Ikegami {\it et al.}~\cite{ikegamiJPSJ13} observed that the mobility of ions markedly increases with decreasing $T$, while the measured mobility shows no depth dependence within $20\le d\le 60 {\rm nm}$. 

When the ions are embedded in a bath of thermally excited quasiparticles, the mobility $\mu$ is determined by the momentum transfer from the ions to thermal quasiparticles~\cite{baym}. Since the excitation energy of the bulk B-phase is fully gapped, the fraction of thermal quasiparticles exponentially decreases at low temperatures. In accordance with the simple analysis within the constant-differential-cross-section approximation, the mobility of ions in the bulk $^3$He-B is given by $\mu /\mu _{\rm N} = (e^{\Delta_{\rm B}/k_{\rm B}T}+1)$, where $\mu _{\rm N}$ denotes the mobility in the case of the normal $^3$He~\cite{bowley1,bowley2}. 

The central issue is the contribution of surface bound states to the momentum transfer of ions. In Ref.~\citeonline{ikegamiJPSJ13}, Ikegami {\it et al.} estimated the contribution, on the basis of the simple assumption that the surface bound states behave as {\it conventional} fermions but not as Majorana fermions, which may give the lower limit of the mobility in the surface region. Contrary to the simple estimation of $\mu$ that clearly shows the depth dependence at low temperatures, the experimentally measured $\mu$ is insensitive to the change in the depth of injected ions from the free surface. The discrepancy might be resolved by taking into account the Majorana nature of surface bound states in the $t$-matrix that describes the quasiparticle scattering at the ions.

\subsection{Weyl superfluidity and mirror Chern number in $^3$He-A}
\label{sec:abmslab}

In the thin slab limit $D\sim \xi _0$, the strong pair breaking effect at surfaces gives rise to the quantum phase transition from the ``isotropic'' BW state to the anisotropic ABM state through the stripe phase. As discussed in Sec.~\ref{sec:abm}, the ABM state with definite $\hat{\bm l}$ can be a prototype of Weyl superconductors, while the formation of the $\hat{\bm l}$-texture is an obstacle for realizing exotic quasiparticles. We here discuss an ideal situation to realize the Weyl superfluidity. Indeed, Tsutsumi {\it et al.}~\cite{tsutsumi:2010b,tsutsumi:2011b} demonstrated that a restricted geometry with thickness $D$ shorter than the dipole coherence length $\xi _{\rm d} \sim 10\mu{\rm m}$ provides forces $\hat{\bm l}$ uniformly in the surface normal direction. Indeed, such a restricted geometry with $D\lesssim\xi_{\rm d}$ has been accomplished by several experimental groups~\cite{ahonenJLTP76,freemanPRL88,ishikawaJLTP89,xuPRL90,freemanPRB90,miyawakiPRB00,kawasakiJLTP02,saitoh03,kawasakiPRL04,levitinJLTP10,levitin13,levitinJLTP14,rojasPRB15}. In this situation, the gapless edge states exist, whose dispersion at the edge $x=0$ of Fig.~\ref{fig:lvec}(a) is given as~\cite{tsutsumi:2010b}
\beq
E_0(k_y,k_z) = \frac{\Delta _{\rm A}}{k_{\rm F}}k_y.
\label{eq:Eedge}
\eeq
Below, we mention that the existence of gapless edge states is a hallmark of Weyl superconductivity and responsible for the spontaneous mass flow. Depending on the orientation of ${\bm d}$, the gapless edge states exhibit the aspect of either the Majorana fermion or Dirac fermion.~\cite{sato14}

\begin{figure}[tb!]
\begin{center}
\includegraphics[width=80mm]{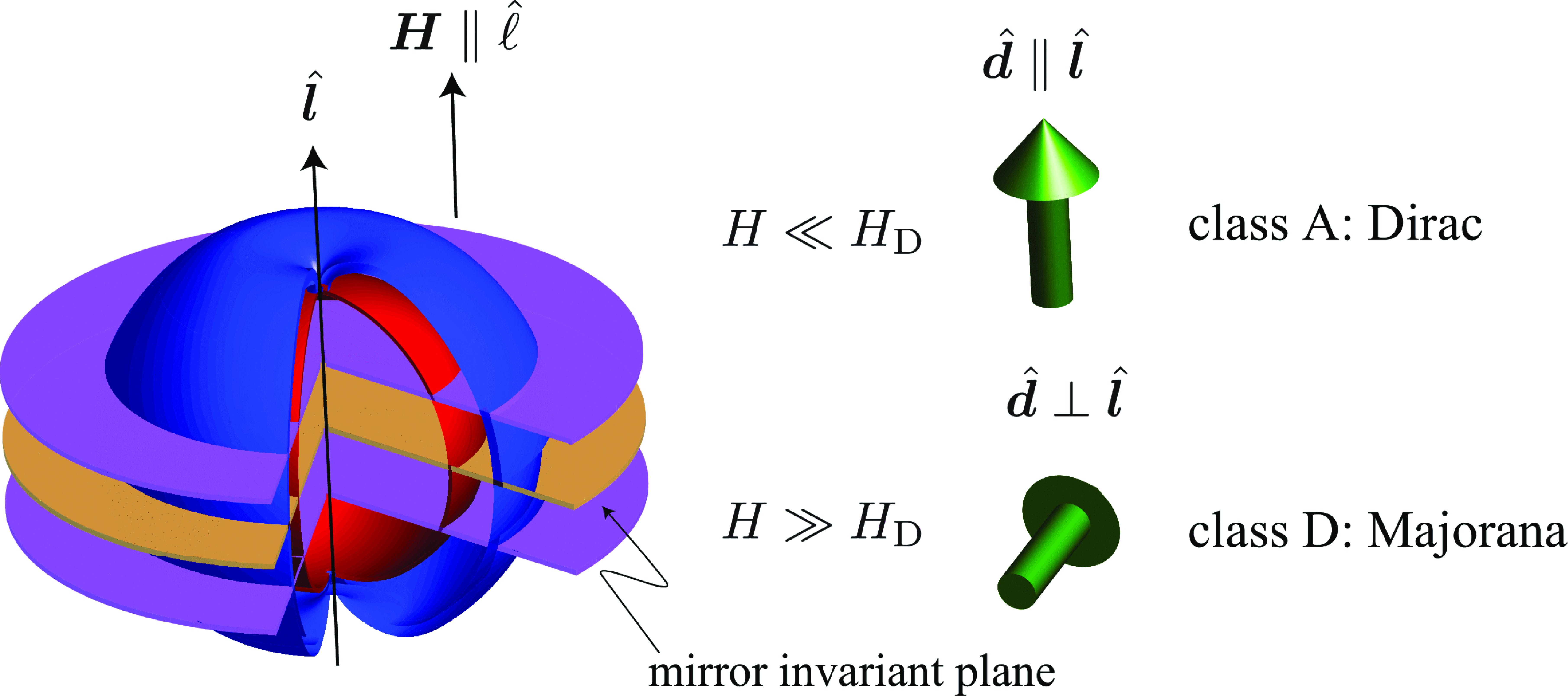} 
\end{center}
\caption{(Color online) Gap function of the ABM state on the Fermi sphere and sliced planes in which the first Chern number is nontrivial. Depending on the relative angle between ${\bm d}$ and $\hat{\bm l}$, the resultant gapless ``edge'' states behave as either Majorana or Dirac fermions.}
\label{fig:abm2}
\end{figure}

The emergence of the chiral Weyl fermions can be observed from the BdG Hamiltonian for the bulk ABM state
\beq
\mathcal{H}_{\rm ABM}({\bm k})
= \varepsilon ({\bm k})\tau _z 
+ \frac{\Delta _{\rm A}}{k_{\rm F}}k_x\sigma _x\tau _x - \frac{\Delta _{\rm A}}{k_{\rm F}}k_y\sigma _x\tau _y,
\label{eq:Habm}
\eeq
where the single-particle energy is $\varepsilon ({\bm k}) \!=\! k^2/2m - \mu$. As in Eq.~\eqref{eq:Eabm}, the AMB state has the Fermi point at ${\bm k} \!=\!{\bm k}_0 \!=\! k_{\rm F}\hat{\bm l}$. Owing to the PHS, the Fermi point must appear as a pair and the counterpart exists at ${\bm k}\!=\!-{\bm k}_0$. The effective Hamiltonian \eqref{eq:Habm} in a spin subsector near Fermi points is then recast into the $2\!\times \! 2$ Hamiltonian for chiral Weyl fermions,
\beq
\mathcal{H}_{\pm}({\bm k}) = c{\bm g}_{\pm}({\bm k})\cdot {\bm \tau},
\label{eq:Hpm}
\eeq
where the three-dimensional vector ${\bm g}$ is given by 
${\bm g}_{\pm}({\bm k}) = ( \tilde{k}^{(\pm)}_x, -\tilde{k}^{(\pm)}_y, \pm\tilde{k}^{(\pm)}_z )$.
We now introduce the new coordinate centered at the Fermi points as $\tilde{\bm k}^{(\pm)} \!=\! (k_x\mp k_{0,x},k_y\mp k_{0,y},\frac{v_{\rm F}}{c}(k_z\mp k_{0,z}))$, where $\tilde{\bm k}^{(+)}$ ($\tilde{\bm k}^{(-)}$) corresponds to the Fermi point residing on the north (south) pole of the Fermi sphere.

In the ABM state with aligned $\hat{\bm l}$, the Fermi point corresponds to the singular point of $\hat{\bm g}\!=\! {\bm g}/|{\bm g}|$~\cite{volovik}. The characteristic feature of the Weyl fermions is that they are expressed by the two-dimensional spinor with a well-defined notion of the left- or right-handed coordinate, while the Dirac fermion is a combination of left- and right-handed Weyl fermions. The left- or right-handed Weyl fermion is accompanied by the topologically different hedgehog texture of $\hat{\bm g}$ around the Fermi point, which is given by the two-dimensional winding number 
\beq
w_{\rm 2d} = \frac{1}{8\pi} \int _{S^2} \hat{\bm g}\cdot\left(
\partial _{k_{\mu}}\hat{\bm g} \times \partial _{k_{\nu}}\hat{\bm g} 
\right) dk_{\mu} \wedge dk_{\nu}.
\eeq
This characterizes the mapping of $S^2$ embracing the Fermi point to the target space $\hat{\bm g}\in S^2$. This is equivalent to the first Chern number on $S^2$ embracing the Fermi point, which counts the ``magnetic monopole'' at the Fermi point. The Weyl fermion residing on the north (south) pole is characterized by $w_{\rm 2d} \!=\! -1$ (+1), which corresponds to the left-handed (right-handed) fermion. 

Since the ABM state is the equal spin pairing state, the total winding number is $w_{\rm 2d} \!=\! \pm 2$. Hence, the Fermi point in the ABM state is topologically protected and cannot be removed by any disturbances. As mentioned in Sec.~\ref{sec:abm}, the pairwise Weyl points are connected by a ``vortex line'' that generates the change in the Berry phase $\gamma (C) = 2\pi$ along a path enclosing the vortex line. The Berry phase is responsible for the nontrivial Chern number, ${\rm Ch}_1 (k_z) = 2\hat{l}$ for $|k_z| \!<\! k_{\rm F}$, as in Eq.~\eqref{eq:weak}, which is defined in each sliced plane perpendicular to the ``vortex line'' connecting the Weyl points (see also Fig.~\ref{fig:abm}). Hence, at the side edge parallel to the aligned $\hat{\bm l}$, there exists a zero-energy flat-band for $|k_z| \!<\! k_{\rm F}$ and the end points are protected by the pairwise Weyl points~\cite{silaev:2012,silaevJETP14}.

Among the zero energy flat band, the zero energy state at $k_z=0$ turns out to be protected by the mirror Chern number introduced in Eq.~\eqref{eq:mirrorCh}. To be specific, consider that $\hat{\bm l}$ is aligned along the $\hat{\bm z}$-axis and the side ``edge'' is located at $x=0$. The BdG Hamiltonian is invariant under the mirror reflection symmetry on the $xy$-plane 
\beq
\left[ \mathcal{M}^{\eta}_{xy},\mathcal{H}(k_x,k_y,k_z=0) \right]=0,
\label{eq:mirrorabm}
\eeq
when ${\bm d}$ is either perpendicular or parallel to $\hat{\bm l}$. The mirror operator in the Nambu space is defined in Eq.~\eqref{eq:mirror} with $M_{xy}=i\sigma _z$. The gap function in the case of ${\bm d}\parallel\hat{\bm l}\parallel\hat{\bm z}$ has even mirror parity ($\eta = +$), while ${\bm d}\perp\hat{\bm l}\parallel\hat{\bm z}$ has odd mirror parity ($\eta = -$). 
Therefore, the mirror operator is given as ${\mathcal M}_{xy}^+$ in the former case, but it is ${\mathcal M}_{xy}^{-}$ in the latter. 

As mentioned in Sec.~\ref{sec:abm}, the ${\bm d}$-vector is locked to the orientation of $\hat{\bm l}$ by the dipole interaction in the zero field. In contrast, as in Eq.~\eqref{eq:quad}, the quadratic Zeeman energy forces ${\bm d}$ to be perpendicular to the orientation of the applied field. When the magnetic field is applied along the $\hat{\bm z}$-axis, as shown in Fig.~\ref{fig:abm2}, the ${\bm d}$-vector rotates at the critical field, which is on the order of the dipolar field $H_{\rm D}\sim 30{\rm G}$, and the ${\bm d}$-vector lies in the $xy$-plane for $H\gg H_{\rm D}$.

When the BdG Hamiltonian maintains the mirror reflection symmetry in Eq.~\eqref{eq:mirrorabm}, the quasiparticle states are simultaneous eigenstates of $\mathcal{M}^{\eta}_{xy}=\pm i$ and one can
introduce the mirror Chern number in Eq.~\eqref{eq:mirrorCh}, which is indeed nontrivial, 
\beq
{\rm Ch}^{(\pm i)}_1(k_z=0) = 1.
\eeq
This ensures that there exists a single zero mode in each mirror subsector. Although the mirror Chern number is nontrivial for the two possible mirror symmetries, ${\cal M}^{\pm}_{xy}$, the corresponding gapless edge states can
be Majorana only for ${\cal M}_{xy}^{-}$. As mentioned in Sec~\ref{sec:mirror}, this is because only ${\cal M}_{xy}^{-}$ (${\bm d}\perp\hat{\bm l}$) supports its own PHS and the corresponding Hamiltonian belongs to class D, while ${\cal M}_{xy}^{+}$ (${\bm d}\parallel\hat{\bm l}$) is categorized into class A. This means that the mirror-symmetry-protected Majorana fermions are possible only in the former case, and in the latter case, only Dirac fermions can be obtained. The consequence of the connection between the orientation of ${\bm d}$ and the Majorana fermions is also applicable to the $^3$He-A with a quantized vortex, as discussed in Sec.~\ref{sec:IQV}.

\begin{figure}[t!]
\begin{center}
\includegraphics[width=80mm]{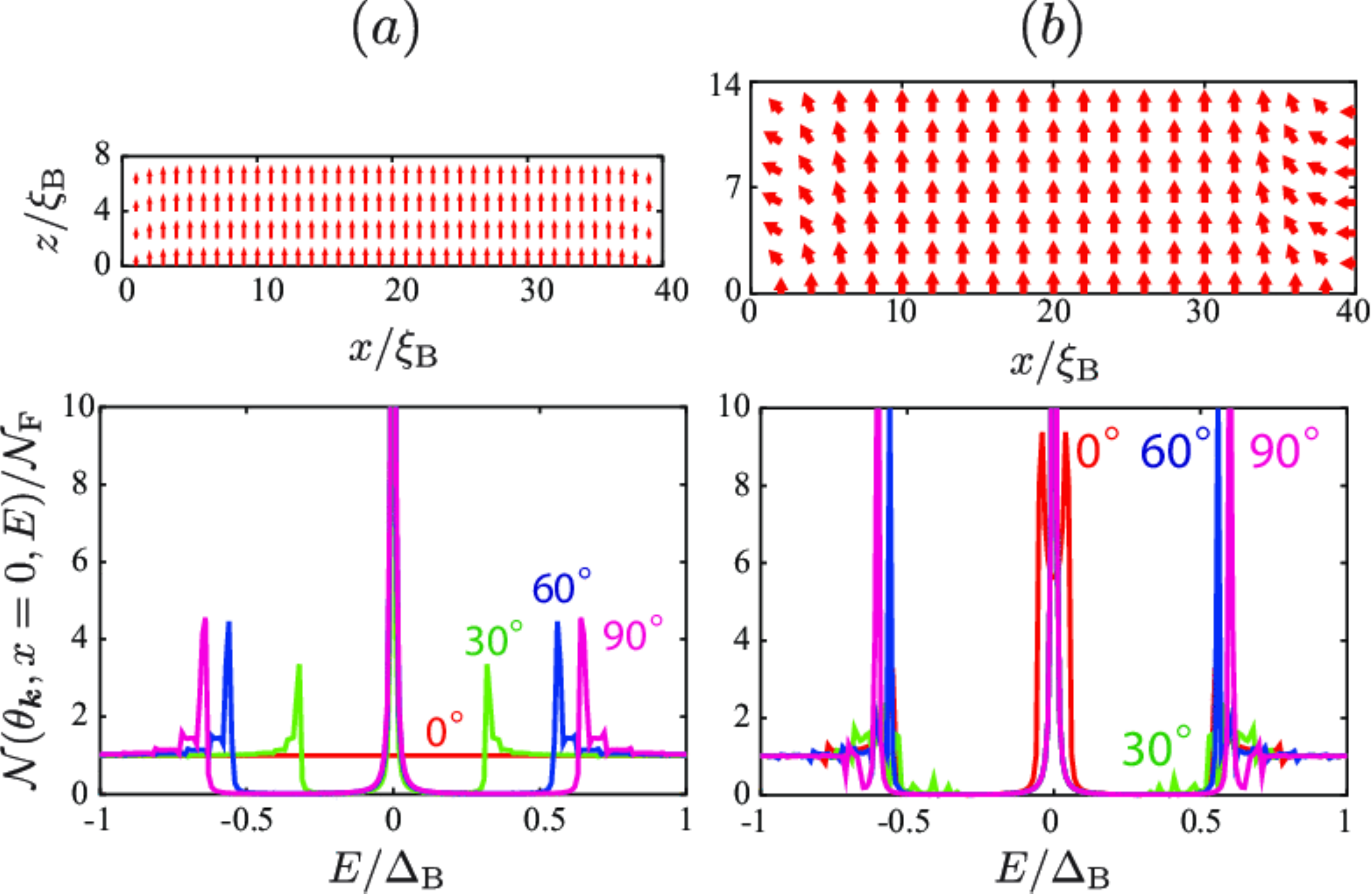}
\end{center}
\caption{(Color online) (top) Stable $\hat{\bm l}$-texture in the $xz$-plane of slabs with thickness $D=8\xi _0$ (a) and $14\xi_0$ (b). (bottom) Angle-resolved edge density of states, $\mathcal{N}(x=0,\phi _{\bm k}=0, \theta _{\bm k},E)$, for the $\hat{\bm l}$-texture.
All the data are obtained from the self-consistent calculation of the quasiclassical theory~\cite{tsutsumi:2011b} at the temperature $T/T_{\rm c}=0.2$. Figures adapted from Ref.~\citeonline{tsutsumi:2011b}.}
\label{fig:lvec}
\end{figure}

In Fig.~\ref{fig:lvec}, we display the $\hat{\bm l}$-texture in the $xz$-plane and the angle-resolved edge density of states, $\mathcal{N}(x=0,\phi _{\bm k}=0, \theta _{\bm k},E)$, for slabs with thickness $D=8\xi _0$ and $14\xi_0$, where $\theta _{\bm k}$ is the bending angle from $\hat{\bm z}$ and $\phi _{\bm k}$ denotes the azimuthal angle. These are obtained from the self-consistent calculation of the quasiclassical theory~\cite{tsutsumi:2011b}. For the thinner slab, all $\hat{\bm l}$ are forced to be parallel to $\hat{\bm z}$ by the strong pair-breaking effect at the upper and lower surfaces. The angle-resolved density of states at the edge $x=0$ has a pronounced zero energy peak which turns to be dispersionless along $\theta _{\bm k}$. This is consistent with the consequence of the Weyl superconductivity, that is, the edge Fermi arc protected by the Weyl points. 

When the thickness $D$ is comparable to the dipole coherence length $\xi _{\rm d}\sim 10 {\mu}{\rm m}$, which characterizes the length scale of $\hat{\bm l}$, owing to the gain of the condensation energy, $\hat{\bm l}$ is oriented in the $\hat{\bm x}$ direction in the vicinity of both the side edges. This results in the splitting of the zero energy density of states at the side edges, as shown in Fig.~\ref{fig:lvec}(b).

\subsection{Planar state: mirror chiral symmetry and Fermi arc}
\label{sec:mirrorchiral}

The planar state confined in a slab geometry provides a good platform for studying the topological properties of time-reversal-invariant nodal superconductors. In Sec.~\ref{sec:planar}, we introduced the $\mathbb{Z}_2$ number that is obtained from the dimensional reduction of the second Chern number on the four-dimensional momentum space. In addition to the bulk topological number, one can introduce the topological $\mathbb{Z}_2$ invariant on the layers of sliced two-dimensional momentum planes which are perpendicular to the nodal direction. Since the BdG Hamiltonian for the planar state in the absence of a time-reversal-breaking field is categorized into class DIII, the topological property is characterized by the time-reversal-symmetry-protected $\mathbb{Z}_2$ number~\cite{fuPRB06}. For the planar state, however, two additional discrete symmetries enable one to introduce two different types of $\mathbb{Z}$ topological number rather than the $\mathbb{Z}_2$ number.

Makhlin {\it et al.}~\cite{makhlinPRB14} revealed that in each sliced two-dimensional plane ($xy$-plane), the combined discrete symmetry $\tilde{\mathcal{C}}$ ensures the $\mathbb{Z}$ topological number. The discrete symmetry $\tilde{\mathcal{C}}=\sigma _z$ is constructed from the combination of spin $\pi$ rotation about the $\hat{\bm z}$-axis and global phase rotation by $\pi/2$. The BdG Hamiltonian in each plane holds the symmetry
\beq
\tilde{\mathcal{C}}\mathcal{H}(k_x,k_y,k_z)\tilde{\mathcal{C}}^{\dag} 
= \mathcal{H}(k_x,k_y,k_z).
\eeq
The combined symmetry $\tilde{\mathcal{C}}$ is responsible for the $\mathbb{Z}$ topological number in each sliced $k_xk_y$-plane. Makhlin {\it et al.}~\cite{makhlinPRB14} also found the intrinsic connection between the topological invariant in the two-dimensional planar state and $w_{\rm 3d}$ in the bulk $^3$He-B. This allows one to infer the bulk-surface correspondence for $^3$He-B from the $\mathbb{Z}$ topological number protected by $\tilde{\mathcal{C}}$ in the two-dimensional planar state. 

Another topological invariant is defined in the planar state and nodal superconductors, even if the TRS is explicitly broken. The additional discrete symmetry relevant to the three-dimensional planar state is the $P_2$ symmetry, which is a combination of the mirror reflection symmetry and TRS \eqref{eq:TRS} (see Fig.~\ref{fig:discrete}). Let us now set the surface to be normal to the $\hat{\bm z}$-axis and the planar order parameter is given by
\beq
{\bm d}({\bm k}) = \Delta _{\rm P}\left(\hat{k}_x\hat{\bm x} + \hat{k}_z\hat{\bm z}\right).
\eeq
The superfluid and superconducting states retain the mirror symmetry if the gap function $\Delta({\bm k})$ is even or odd under the mirror reflection, $M\Delta({\bm k})M^{\rm T} = \pm \Delta (-k_x,k_y,k_z)$, where without losing generality, the $yz$-plane is taken as the mirror plane. Then, the BdG Hamiltonian holds the mirror reflection symmetry 
\beq
\mathcal{M}^{\pm}\mathcal{H}({\bm k})\mathcal{M}^{{\pm}\dag} = \mathcal{H}(-k_x,k_y,k_z),
\label{eq:mirror2}
\eeq
when an external field is absent. The mirror reflection operator $\mathcal{M}^{{\pm}}$ is defined as $\mathcal{M}^{\pm} \!=\! {\rm diag}(M,\pm M^{\ast})$.

Applying a magnetic field, ${\bm H}$, may break the mirror reflection symmetry \eqref{eq:mirror2} and TRS, but the $P_2$ symmetry, which is a combination of the mirror reflection symmetry and TRS, may be maintained. The $P_2$ discrete symmetry rotates the magnetic field ${\bm H}\!\rightarrow\!(-H_x,H_y,H_z)$. Consequently, the Hamiltonian $\mathcal{H}({\bm k})$ with $H_x = 0$ holds the following ${\bm Z}_2$ symmetry,
\beq
\mathcal{P}_2\mathcal{H}({\bm k})
\mathcal{P}^{-1}_2 = \mathcal{H}(k_x,-k_y,-k_z), \quad
\mathcal{P}_2 = \mathcal{T}\mathcal{M}^{\pm}.
\label{eq:z2}
\eeq
By combining the $P_2$ symmetry with the PHS in Eq.~\eqref{eq:PHS}, 
one can introduce the chiral operator $\Gamma _1$, which is anticommutable with the Hamiltonian for the planar state, 
\beq
\left\{ \Gamma _1, \mathcal{H} (0,k_y,k_z)\right\} =0, \quad
\Gamma _1 = \mathcal{C}\mathcal{P}_2.
\eeq
Similarly to $w_{\rm 3d}$ in Eq.~(\ref{eq:winding}), the one-dimensional winding number is defined as~\cite{mizushimaJPCM15,satoPRB11,mizushimaPRL12}
\beq
w_{\rm 1d}(k_y) \!=\! - \frac{1}{4\pi i}\int d k_z[ 
\Gamma _1\mathcal{H}^{-1}({\bm k})\partial _{k_z}\mathcal{H}({\bm k})
]_{k_x=0}.
\label{eq:w1dmirror}
\eeq
The winding number is estimated as
\beq
w_{\rm 1d}(k_y) = \left\{  
\begin{array}{ll}
2 & \mbox{ for $|k_y|<k_{\rm F}$}, \\
\\
0 & \mbox{ otherwise}
\end{array}
\right. .
\eeq
From the generalized index theorem in Sec.~\ref{sec:index}~\cite{satoPRB11}, the non-zero value of $w_{\rm 1d}(k_y)$ is equal to the number of zero energy states that are bound to the surface. Hence, doubly-degenerate zero energy states appear along the chiral symmetric plane $k_x \!=\! 0$. The mirror-symmetry-protected topological invariant ensures the existence of a surface Fermi arc that connects two point nodes. Such a surface Fermi arc can also be realized in the $E_u$ state of superconducting topological insulators~\cite{sasakiPC15} and the $E_{1u}$ scenario of the heavy-fermion superconductor UPt$_3$~\cite{tsutsumiJPSJ13,mizushimaPRB14} (see also Sec.~\ref{sec:materials}).

\begin{figure}[t!]
\begin{center}
\includegraphics[width=80mm]{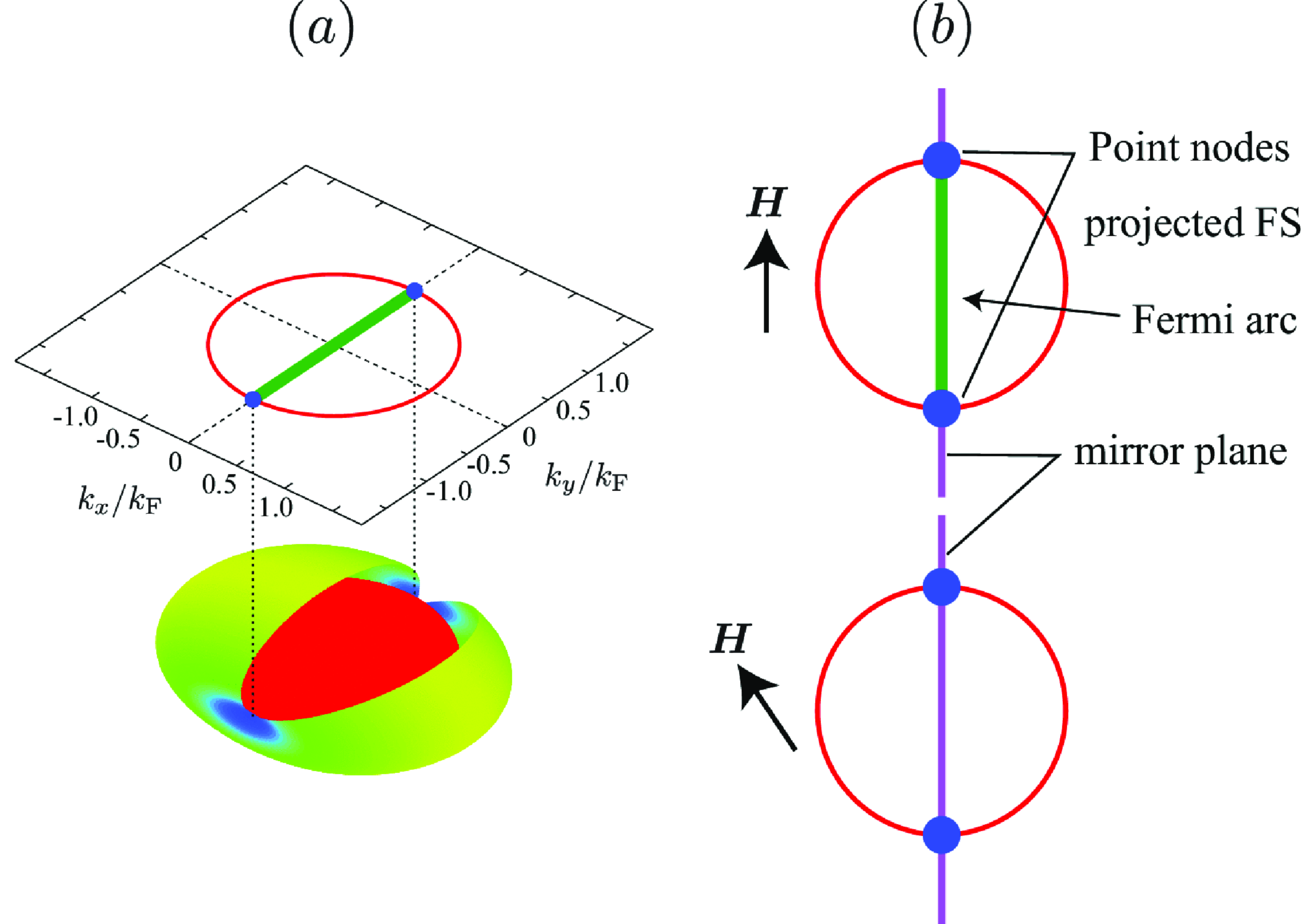}
\end{center}
\caption{(Color online) (a) Stereographic view of the energy gap of the planar state. The Fermi surface and point nodes are projected onto the surface in the $k_xk_y$-plane, where the topologically protected surface Fermi arc connects to two point nodes. (b) Schematic pictures of the $P_2$ symmetry protection of the topological Fermi arc connecting to two point nodes.}
\label{fig:planar}
\end{figure}

Now, following Refs.~\citeonline{shiozakiPRB14}, \citeonline{mizushimaPRL12}, \citeonline{mizushimaJPCM15}, and \citeonline{mizushimaNJP13}, one can show that multiple Majorana zero modes with chiral symmetry ensure the Ising character of the topologically protected zero energy states. The topologically protected surface Fermi arc does not contribute to the local density operator,
$\rho^{({\rm surf})} ({\bm r})  \!=\! 0$.
This indicates that the chiral-symmetry-protected Majorana fermions cannot be coupled to the local density fluctuations and thus are very robust against nonmagnetic impurities. Similarly, the local spin operator ${\bm S}$ is constructed from the surface Majorana fermions as 
\beq
{\bm S}^{({\rm surf})} \!=\! (S_x,0,0). 
\eeq
This implies that the surface Majorana fermions yield an anisotropic magnetic response, as shown in Fig.~\ref{fig:planar}. The surface Majorana fermions and topological Fermi arc are insensitive to a magnetic field along the mirror reflection plane, i.e., the $yz$-plane, but fragile against a magnetic field perpendicular to the plane. Hence, this is the Fermi arc protected by the discrete $P_2$ symmetry, and its physical origin is essentially different from that of the Fermi arc in $^3$He-A, as mentioned in Sec.~\ref{sec:abmslab}, where the Fermi arc is a consequence of pairwise Weyl points. 

The $P_2$-symmetry-protected topological number is useful from the following two viewpoints: First, the $P_2$ symmetry is robust even if the mirror symmetry or TRS is explicitly broken. Second, since the mirror reflection symmetry may be an element of the crystalline point group, the topological number is applicable to superconductors with an appropriate crystalline symmetry. In Sec.~\ref{sec:materials}, we discuss in detail the application of the $P_2$-symmetry-protected topological number and extract the generic properties of topological ``crystalline'' superconductors.


\section{Spontaneous Mass and Spin Currents}
\label{sec:mass}

In $^3$He-B confined in a slab geometry, as discussed in the previous section, helical Majorana fermions exist as a consequence of the intertwining of the topology and discrete symmetries. In the case of $^3$He-A, the motion of the Cooper pair is confined in the two-dimensional $xy$-plane, and the chiral Majorana fermions with a topological Fermi arc appear at the edge of the slab. This is attributed to the spontaneous breaking of the TRS in the bulk ABM order parameter. Following Refs.~\citeonline{tsutsumi:2012c} and \citeonline{tsutsumi:2012}, we here demonstrate that such chiral and helical Majorana fermions emergent in $^3$He-A and B give rise to the spontaneous mass and spin current along the edge and the surface of the container, respectively. In particular, in the case of the A-phase, we examine the edge mass current carried by the chiral edge Majorana fermions and quantify the total angular momentum that generates the mass current in connection with the paradox of the {\it intrinsic angular momentum}~\cite{leggettbook}.

\subsection{Edge mass current in $^3$He-A}

The intrinsic angular momentum paradox is a fundamental question: {\it what is the true value of the total angular momentum in $^3$He-A or more generally time-reversal-broken superconducting states?} There exist three different calculations on the total angular momentum $L_z$ of $^3$He-A composed of $N$ $^3$He atoms, which result in three different values of $L_z$. Their results are summarized as
\beq
L_z = \alpha\left( \frac{\Delta _{\rm A}}{E_{\rm F}}\right)^n\frac{N\hbar}{2}, \quad 
\mbox{$n = 0$, $1$, or $2$},
\eeq
where $\alpha$ is a constant of order unity. Ishikawa~\cite{ishikawa:1977} and McClure and Takagi~\cite{mcclure:1979} predicted $n=0$, which implies that $^3$He-A acquires the macroscopic angular momentum through spontaneous symmetry breaking. In contrast, Anderson and Morel~\cite{anderson:1961} and Cross~\cite{cross:1975} independently predicted $n=1$ and $2$, respectively, which state that the intrinsic angular momentum of $^3$He-A is vanishingly small.

Apart from the intrinsic angular momentum paradox, Stone and Roy~\cite{stone} calculated the angular momentum associated with the chiral edge Majorana fermions in two-dimensional chiral $p$-wave systems confined in a cylindrical container, which turned out to be the macroscopic value, $L_z=N\hbar /2$. Kita~\cite{kita:1998} also arrived at the same conclusion even in three-dimensional systems. The spontaneous edge mass current, therefore, seems to be related to the intrinsic angular momentum of Cooper pairs. 

Here, as the start of a brief review on the ``intrinsic angular momentum paradox'', which still remains unresolved over the last four decades, we examine the total angular momentum carried by the chiral edge Majorana fermions. Using the quasiclassical theory, we analytically derive the temperature dependence of the total angular momentum in $^3$He-A confined in a cylindrical container, which is compared with the temperature dependence of superfluid densities. It is clarified that the temperature dependence of $L_z$ associated with the chiral edge Majorana fermions does not follow that of the superfluid density of the bulk A-phase and thus the coincidence of the edge mass current with the intrinsic angular momentum might be accidental. In order to focus on the orbital angular momentum, we omit spin indices for simplicity in Sec.~6.1.1.

\subsubsection{Intrinsic angular momentum paradox}

The angular momentum of chiral $p$-wave Cooper pairs was first discussed by Anderson and Morel~\cite{anderson:1961} before the discovery of the superfluid phase in $^3$He.
They derived the angular momentum from the density-current correlation function $\langle n({\bm r}_1){\bm j}({\bm r}_2)\rangle$, where $n({\bm r})=\psi^*({\bm r})\psi({\bm r})$ is the density operator, ${\bm j}({\bm r})=-(i/2)\hbar[\psi^*({\bm r}){\bm \nabla}\psi({\bm r})-\psi({\bm r}){\bm \nabla}\psi^*({\bm r})]$ is the current density operator, and the BCS ground state is
\begin{align}
|{\rm BCS}\rangle=\prod_{\bm k}(u_{\bm k}+v_{\bm k}c_{\bm k}^{\dagger}c_{-\bm k}^{\dagger})|0\rangle.
\end{align}
If we focus on only the off-diagonal long-range correlation, the density-density correlation function gives
\begin{align}
\langle n({\bm r}_1)n({\bm r}_2)\rangle_{\rm av}=\left|\sum_{\bm k}u_{\bm k}v_{\bm k}^*e^{i{\bm k}\cdot{\bm r}}\right|^2\equiv|I({\bm r})|^2,
\end{align}
where ${\bm r}={\bm r}_1-{\bm r}_2$ is the relative coordinate and $\langle\cdots\rangle_{\rm av}$ indicates the spatial average of the center-of-mass coordinate ${\bm R}=({\bm r_1}+{\bm r}_2)/2$, namely, $\langle\cdots\rangle_{\rm av}\equiv\int d{\bm R}\langle\cdots\rangle/V$.
For the expression $I({\bm r})=|I({\bm r})|\exp[i\gamma({\bm r})]$, the density-current correlation function gives
\begin{align}
\langle n({\bm r}_1){\bm j}({\bm r}_2)\rangle_{\rm av}=\hbar\langle n({\bm r}_1)n({\bm r}_2)\rangle_{\rm av}{\bm \nabla}\gamma({\bm r}).
\end{align}
For the simple phase factor $\gamma({\bm r})=\varphi$ for chiral $p$-wave superfluids,
$
\langle n({\bm r}_1)j_{\varphi}({\bm r}_2)\rangle_{\rm av}=\frac{\hbar}{|{\bm r}|}\langle n({\bm r}_1)n({\bm r}_2)\rangle_{\rm av}$,
where $\varphi$ is the azimuthal angle of ${\bm r}$.
Since the density-density correlation function gives the density of condensed particles,
$n_{\rm c}\equiv\int d{\bm r}\langle n({\bm r}_1)n({\bm r}_2)\rangle_{\rm av}\sim n\Delta/E_{\rm F}$,
where $n$ is the total particle density, the angular momentum of Cooper pairs is given by
\begin{align}
L_{\rm AM}=\int d{\bm R}\int d{\bm r}|{\bm r}|\langle n({\bm r}_1)j_{\varphi}({\bm r}_2)\rangle\sim N\hbar\frac{\Delta}{E_{\rm F}},
\end{align}
with the number of total particles $N=nV$.

The angular momentum of the Cooper pairs in $^3$He-A was also calculated by Cross~\cite{cross:1975} from the Green's function $G_{\omega_n}({\bm k})$ in the Gor'kov equation.
The mass current density is given by the Green's function as
%
${\bm j}=k_{\rm B}T\sum_n\int\frac{d{\bm k}}{(2\pi)^3}{\bm p}G_{\omega_n}({\bm k})$.
%
The Green's function can be performed as a gradient expansion:~\cite{werthamer:1963},
$G_{\omega_n}({\bm k})=G_{\omega_n}^{(0)}({\bm k})+G_{\omega_n}^{(1)}({\bm k})+G_{\omega_n}^{(2)}({\bm k})+\cdots
$,
where $G_{\omega_n}^{(m)}({\bm k})$ is the $m$-th order of $(k_{\rm F}\xi)^{-1}$.
The mass current density is obtained from its definition in terms of $G_{\omega_n}^{(1)}({\bm k})$,
\begin{align}
{\bm j}_{\rm C}={\bm \rho}_{\rm s}{\bm v}_{\rm s}+\frac{\hbar}{4m}\rho_{\rm s\parallel}{\bm \nabla}\times\hat{\bm l}-\frac{\hbar}{2m}\rho_{\rm s\parallel}\hat{\bm l}(\hat{\bm l}\cdot{\bm \nabla}\times\hat{\bm l}),
\label{eq:jCross}
\end{align}
where ${\bm v}_{\rm s}$ is the superfluid velocity, $m$ is the mass of a condensed particle, and  $\hat{\bm l}$ is the local direction of the ${\bm l}$-vector.
The superfluid density tensor $({\bm \rho}_{\rm s})_{ij}=\rho_{\rm s\perp}\delta_{ij}-\rho_0\hat{l}_i\hat{l}_j$ is obtained from
$\rho_{\rm s\perp}=\frac{3}{2}\rho\langle\phi(\theta)\sin^2\theta\rangle _{\hat{\bm k}}$ and
$\rho_{\rm s\parallel}\equiv \rho_{\rm s\perp}-\rho_0$,
where the total mass density is given by $\rho=mn$, and we introduced the Yosida function extended to the $p$-wave states, 
$Y(\theta)=1-\phi(\theta)=\int_0^{\infty}d\epsilon\frac{1}{2k_{\rm B}T}{\rm sech}^2\frac{\sqrt{\epsilon^2+\Delta(T)^2\sin^2\theta}}{2k_{\rm B}T}$.
If Cooper pairs have intrinsic angular momentum, a term proportional to ${\bm \nabla}\times{\bm L}$ emerges in the expression of mass current density, which is absent in Eq.~\eqref{eq:jCross}.
This term is found in the mass current density derived from the higher order of the expanded Green's function, which gives the magnitude of the intrinsic angular momentum at $T=0$,
\begin{align}
L_{\rm C}\sim N\hbar\left(\frac{\Delta}{E_{\rm F}}\right)^2\frac{\mathcal{N}_1}{\mathcal{N}_0}\ln\frac{\omega_{\rm c}}{\Delta}
\sim N\hbar\left(\frac{\Delta}{E_{\rm F}}\right)^2,
\end{align}
where $\omega_{\rm c}$ is the cutoff energy of the Matsubara frequency $\omega_n$.
In this calculation, the density of states in the normal state is assumed to be $\mathcal{N}(\epsilon)=\mathcal{N}_0+\mathcal{N}_1(\epsilon/E_{\rm F})$;
therefore, $L_{\rm C}$ vanishes for particle-hole symmetric systems.

The intrinsic angular momentum was also evaluated by Ishikawa~\cite{ishikawa:1977} by taking the expectation value of the angular momentum operator for the ground state at $T=0$.
The BCS ground state projected on the $N$-particle state is described by
\begin{align}
|N\rangle=\frac{1}{A_N}\left(Q^{\dagger}\right)^{N/2}|0\rangle,
\label{eq:creation_pair}
\end{align}
where $A_N$ is a normalization constant and
$
Q^{\dagger}=\int\int d{\bm r}_1d\bm{r}_2\phi({\bm r}_2,\bm{r}_1)\psi^{\dagger}({\bm r}_1)\psi^{\dagger}({\bm r}_2)
$
is the creation operator of a Cooper pair with the Cooper pair wavefunction $\phi({\bm r}_2,\bm{r}_1)$.
On the basis of the $N$-particle state, Ishikawa defined the angular momentum 
\begin{align}
\langle N|L|N\rangle=\int\int d{\bm r}_1d{\bm r}_2\delta({\bm r}_2-{\bm r}_1)\hat{L}^{(1)}_z\rho({\bm r}_2,{\bm r}
_1)
\label{eq:lz}
\end{align}
as the expectation value with the one-particle density matrix
\begin{align}
\rho({\bm r}_1,{\bm r}_2)\equiv\langle N|\psi^{\dagger}({\bm r}_1)\psi({\bm r}_2)|N\rangle
=\int d{\bm r}_1'\phi^*({\bm r}_1',{\bm r}_1)\Phi({\bm r}_1',{\bm r}_2),
\label{eq:density_matrix}
\end{align}
where $\Phi({\bm r}_1,{\bm r}_2)\equiv-\langle N-2|\psi({\bm r}_1)\psi({\bm r}_2)|N\rangle$ is the order parameter
and $\hat{L}^{(1)}_z\equiv{\bm r}_1\times(-i\hbar{\bm \nabla}_1)$, and $\Phi$ denotes the chiral $p$-wave order parameter,
$\hat{L}_z\Phi({\bm r})=\hbar\Phi({\bm r})$.
We focus on the angular momentum from the internal motion of Cooper pairs, which is described by the relative coordinate ${\bm r}={\bm r}_1-{\bm r}_2$. We notice that the order parameter $\Phi({\bm r}_2,{\bm r}_1)$ decreases rapidly for $|{\bm r}_1-{\bm r}_2|\gg\xi$. By substituting Eq.~\eqref{eq:density_matrix} into Eq.~\eqref{eq:lz}, the intrinsic angular momentum is then recast into 
\begin{align}
L_{\rm I}=\frac{N}{2}\hbar.
\end{align}
Hence, contrary to the other two arguments, the angular momentum is on the order of the total particle number. In the same manner based on the BCS ground state, McClure and Takagi~\cite{mcclure:1979} demonstrated that the total angular momentum, including the internal motion and center-of-mass motion of Cooper pairs, is given as $L_{\rm tot}=N\hbar/2$ for axial symmetric systems.

Ishikawa~\cite{ishikawa:1977} pointed out the oversight in the calculation of $L_{\rm AM}$ that the short-range two-body correlation is not taken into account. However, the reason for the discrepancy between $L_{\rm C}$ and $L_{\rm I}$ has not been clarified. Mermin and Muzikar remarked on the difference of their expressions for the mass current density~\cite{mermin:1980}.
Cross~\cite{cross:1975} derived the mass current density as in Eq.~\eqref{eq:jCross}, while Ishikawa {\it et al.}~\cite{ishikawa:1980} calculated it with the one-particle density matrix in Eq.~\eqref{eq:density_matrix} as
\begin{align}
{\bm j}({\bm R})=\frac{i\hbar}{2}\left(\frac{\partial}{\partial{\bm R}}-\frac{\partial}{\partial{\bm R}'}\right)\rho({\bm R},{\bm R}')|_{{\bm R}'={\bm R}}.
\end{align}
This results in 
\begin{align}
{\bm j}_{\rm IMU}=\rho{\bm v}_{\rm s}+\frac{\hbar}{4m}{\bm \nabla}\times(\rho\hat{\bm l}).
\label{eq:jIMU}
\end{align}
Mermin and Muzikar pointed out that the singular behavior in the Cooper pair wavefunction $\phi({\bm r}_2,{\bm r}_1)$ in Eq.~\eqref{eq:creation_pair} induces an extra term into Eq.~\eqref{eq:jIMU}~\cite{mermin:1980}, that is,
\begin{align}
{\bm j}_{\rm MM}={\bm j}_{\rm IMU}-\frac{\hbar}{2m}c_0\hat{\bm l}(\hat{\bm l}\cdot{\bm \nabla}\times\hat{\bm l}),
\label{eq:jmm}
\end{align}
where
$c_0=m\int_{-\infty}^{\infty}\frac{dk_z}{\pi^2}n(0,0,k_z)k_z^2$
with the k-space occupation number $n({\bm k})$ defined by
$n=\int\frac{d{\bm k}}{(2\pi)^3}n({\bm k})$.

If the Cooper pair wavefunction $\phi$ is localized in a short range as the Bose-Einstein condensation of diatomic molecules, $n(0,0,k_z)$ is identically zero, and thus ${\bm j}_{\rm MM}={\bm j}_{\rm IMU}$.
In the BCS limit, however, $n(0,0,k_z)$ is identical to the Fermi distribution function in the normal state at $T=0$, since $\hat{\bm z}\parallel\hat{\bm l}$ is directed to the point nodes of the superfluid gap.
In this limit, $c_0$ reduces to $mk_{\rm F}^3/3\pi^2=\rho$, and thus the last term of ${\bm j}_{\rm MM}$ in Eq.~\eqref{eq:jmm} becomes equivalent to ${\bm j}_{\rm C}$ at $T=0$.
In addition, the term $\frac{\hbar}{4m}{\bm \nabla}\times(\rho\hat{\bm l})$ in ${\bm j}_{\rm MM}$ is recast into the second term in ${\bm j}_{\rm C}$ with a small correction associated with $L_{\rm C}$.
Therefore, Mermin and Muzikar concluded that Bose-Einstein condensed diatomic molecules in the chiral $p$-wave state have  the mass density current ${\bm j}_{\rm IMU}$ and the intrinsic angular momentum $L_{\rm I}=N\hbar/2$, while Cooper pairs in $^3$He-A have ${\bm j}_{\rm C}$ and a very small $L_{\rm C}$.

However, we should mention that taking into account the last term in Eq.~\eqref{eq:jmm} gives the extra contribution that deviates the resultant angular momentum from that of the McClure-Takagi result $L_{\rm tot}=N\hbar/2$~\cite{mcclure:1979}.
Moreover, Kita~\cite{kita:1996} demonstrated the intrinsic angular momentum $L_{\rm I}=N\hbar/2$ and the mass current density ${\bm j}_{\rm IMU}$ by the one-particle density matrix formalism, which is independent of the Cooper pair radius. Since Kita considered the kinetic-energy operator for a charged system, the internal motion and center-of-mass motion of Cooper pairs could be perfectly separated.

To sum up, the issue on the intrinsic angular momentum remains as a paradox~\cite{leggettbook}, whether the Cooper pairs' relative angular momentum is $L_{\rm I}=N\hbar/2$ or $L_{\rm C}=0$ for particle-hole symmetric systems. The discrepant results are derived from the equivalent microscopic formalisms on the basis of the one-particle density matrix for the BCS ground state and the Gor'kov formulation with the gradient expansion. In addition, the question arises: If $L_{\rm I}=N\hbar/2$ gives the correct intrinsic angular momentum at $T=0$, how does it decrease to the normal value, namely, zero, at the critical temperature?

Recently, Stone and Roy discussed the angular momentum carried by the edge mass current in two-dimensional chiral $p$-wave superfluids~\cite{stone}. They derived $L_{\rm tot}=N\hbar/2$ for an axisymmetric disk system at $T=0$ from the eigenfunction of the BdG Hamiltonian within the Andreev approximation. As shown in Sec.~\ref{sec:vortex1}, the discussion on $L_{\rm tot}$ was extended to the BCS-to-BEC topological phase transition regime, where the system possesses the constant $L_{\rm tot}=N\hbar/2$ in both the BCS and BEC phases~\cite{mizushimaPRA10}. Further studies on the edge or bulk mass current have been performed by numerous researchers~\cite{stonePRL85,volovikJETP85,combescotPRB86,goryoPLA98,mizushimaPRL08,stoneAP08,tsutsumi:2012,sauls:2011,tsutsumi:2012c,tada:arXiv,kallinPRB14,shitadePRB14,volovikJETP15,kallinPRB15-2,shitade:arxiv,morozPRB14,morozPRB15,tsuruta} in connection with the intrinsic angular momentum paradox. In the next section, we review the edge mass current in $^3$He-A, following Ref.~\citeonline{tsutsumi:2012c}.

\subsubsection{Andreev bound states and edge mass current}

{\it Edge currents in $^3$He-A confined in a slab.}---
In this section, we consider $^3$He-A in a slab with a small thickness $D$ along the $z$-direction.
In a sufficiently thin slab within $k^{-1}_{\rm F}\ll D \sim \xi \ll \xi _{\rm d}$, the dipole-locked ${\bm l}\parallel{\bm d}$ vectors are aligned toward the $z$-direction. As shown in Sec.~\ref{sec:abmslab}, this setup can be realized by using a slab geometry with sub-$\mu$m thickness~\cite{tsutsumi:2011b}.
The ${\bm d}$-vector relevant to this geometry is given in Eq.~\eqref{eq:abm} with $\hat{\bm e}^{(1)}\mapsto \hat{\bm x}$ and $\hat{\bm e}^{(2)}\mapsto \hat{\bm y}$.

Here, we discuss a side edge of the slab at $x=0$, where the region $x>0$ is occupied by $^3$He-A. Assuming that the side edge is specular, i.e., uniformity in the $yz$-plane, only the $k_x$-component of the ${\bm d}$-vector is suppressed at the edge and it recovers to the bulk $\Delta _{\rm A}$ within the coherence length $\xi$. In order to study Andreev bound states at the edge, accordingly, we solve the Eilenberger equation, Eq.~\eqref{eq:eilen}, under the pair potential
\begin{align}
d_z(\bm{k},x)=\Delta_{\rm A}\left[\hat{k}_x\tanh\left(\frac{x}{\xi_{\rm A}}\right)+i\hat{k}_y\right],
\label{OPA}
\end{align}
where the coherence length is defined by $\xi_{\rm A}\equiv\hbar v_{\rm F}/\Delta_{\rm A}$.
The quasiclassical Green's function is analytically obtained as
\begin{multline}
g_0(\bm{k},x;\omega_n)=\frac{1}{\sqrt{\omega_n^2+\Delta_{\rm A}^2\sin^2\theta _{\bm k}}}\\
\times\left[\omega_n+\frac{\Delta_{\rm A}^2\sin^2\theta _{\bm k}\cos^2\phi _{\bm k}}
{2(\omega_n+i\Delta_{\rm A}\sin\theta _{\bm k}\sin\phi _{\bm k})}\ {\rm sech}^2\left(\frac{x}{\xi_{\rm A}}\right)\right],
\label{Green}
\end{multline}
which is the scalar component of $g(\hat{\bm k},{\bm r};\omega _n)$, $g_0\equiv \frac{1}{4}{\rm tr}[g]$. The retarded propagator with the analytic continuation $i\omega _n\rightarrow E+i0_+$ has poles at the dispersion of the gapless edge states in Eq.~\eqref{eq:Eedge}. The vectorial components $g_{\mu}=\frac{1}{4}{\rm tr}[\sigma _{\mu}g]$ vanish. The off-diagonal component $f$, which describes the Cooper pair amplitudes, is given by 
\begin{multline}
f_z(\bm{k},x;\omega_n)=\frac{1}{\sqrt{\omega_n^2+\Delta_{\rm A}^2\sin^2\theta _{\bm k}}}\\
\times\left[d_z(\bm{k},x)-\frac{\Delta_{\rm A}^2\sin^2\theta _{\bm k}\cos^2\phi _{\bm k}}
{2(\omega_n+i\Delta_{\rm A}\sin\theta _{\bm k}\sin\phi _{\bm k})}\ {\rm sech}^2\left(\frac{x}{\xi_{\rm A}}\right)\right].
\label{eq:fabm}
\end{multline}
In Eqs.~\eqref{Green} and \eqref{eq:fabm}, the first term denotes the Green's function for the bulk ABM state, while the second term describes the pair breaking effect at the edge. We notice that Eqs.~\eqref{OPA} and \eqref{eq:fabm} offer a self-consistent solution in the weak-coupling limit.

\begin{figure}
\begin{center}
\includegraphics[width=85mm]{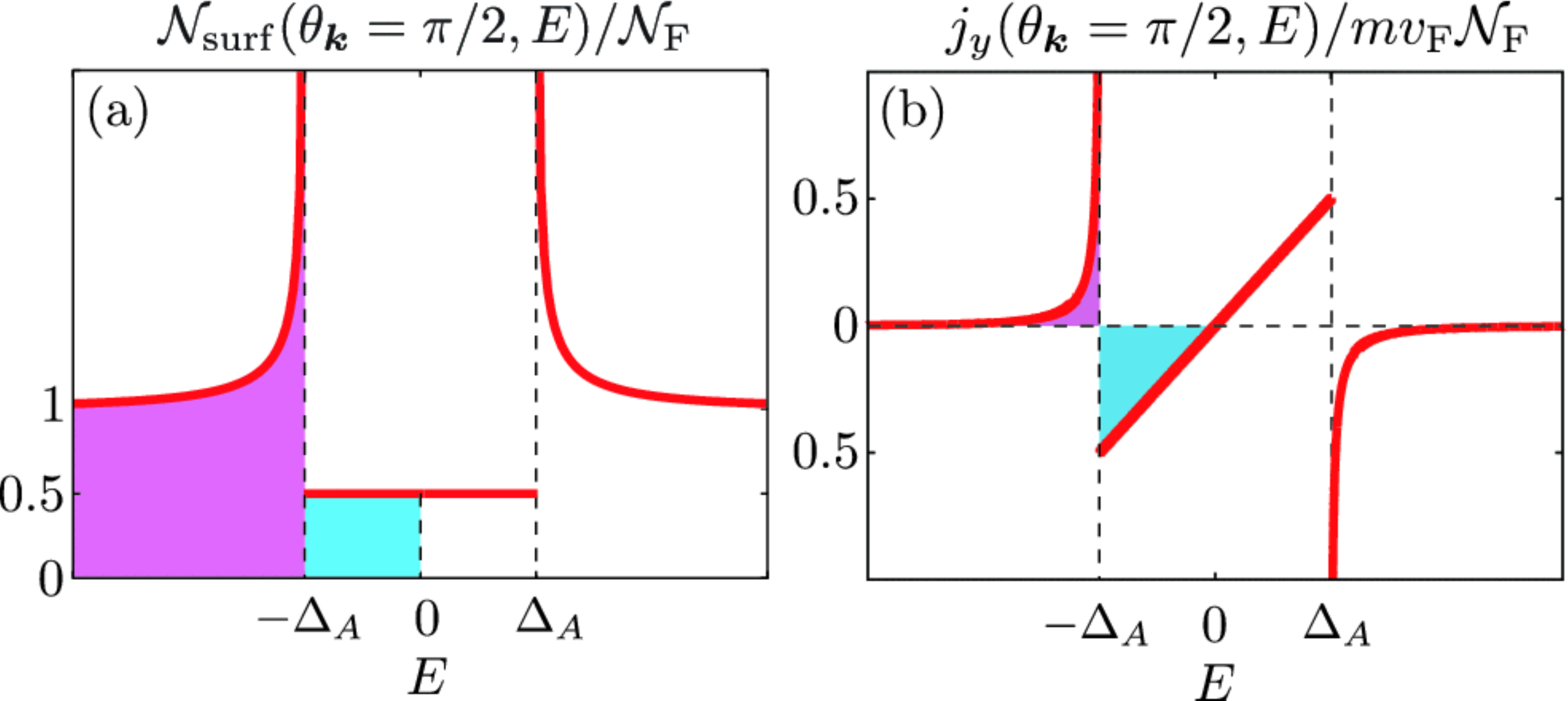}
\end{center}
\caption{\label{energyA}(Color online) 
Energy spectra of $\theta_{\bm k}$-angle-resolved local density of states (a) and mass current along the edge (b) at $x=0$ for $\theta_{\bm k}=\pi/2$ in the A-phase.
At the zero temperature, quasiparticles fill the colored (shaded) states in (a).
The mass currents from the bound and continuum states are derived 
by integrating the blue (light gray) and pink (gray) regions in (b), respectively.
Figures adapted from Ref.~\citeonline{tsutsumi:2012c}.
}
\end{figure}

From Eqs.~\eqref{Green} and \eqref{eq:dosk}, the $\theta _{\bm k}$-angle-resolved local density of states is obtained as
\begin{align}
\mathcal{N}(\theta_{\bm k},x,E)\equiv\int\frac{d\phi _{\bm k}}{2\pi }\mathcal{N}(\bm{k},x,E)
=\frac{\mathcal{N}_{\rm F}}{2}\ {\rm sech}^2\left(\frac{x}{\xi_{\rm A}}\right),
\label{LDOSMJA}
\end{align}
for the bound states $|E|<\Delta _{\rm A}\sin\theta _{\bm k}$, and
\beq
\mathcal{N}(\theta _{\bm k}, x,E)=
\left[\mathcal{N}_{\rm A}-\frac{1}{2}\left(\mathcal{N}_{\rm A}-\mathcal{N}_{\rm F}\right)
\ {\rm sech}^2\left(\frac{x}{\xi_{\rm A}}\right)\right],
\label{LDOScontA}
\eeq
for the continuum states $|E|>\Delta_{\rm A}\sin\theta _{\bm k}$. Here, we introduce the density of states of the bulk ABM state, $\mathcal{N}_{\rm A} \equiv \mathcal{N}_{\rm F}|E|/\sqrt{E^2-\Delta _{\rm A}^2\sin^2\theta _{\bm k}}$; $\theta_{\bm k}$ is the polar angle from a point node. 
Owing to the formation of gapless edge states, the continuum state is deviated from that of the bulk ABM state having no surface. The $\theta _{\bm k}$-angle-resolved local density of states at the edge $x=0$ for $\theta _{\bm k}=\pi/2$ is shown in Fig.~\ref{energyA}(a).
A finite edge density of states exists at $E=0$, which is estimated as $\mathcal{N}_{\rm F}/2$. 
It turns out that the contribution of the bound states to $\mathcal{N}(\theta_{\bm k},x,E)$ is independent of $\theta _{\bm k}$, since the edge bound states form the flat band along $\hat{k}_z$, which is a consequence of Weyl superconductors~\cite{tsutsumi:2010b,tsutsumi:2011b} (see also Secs.~\ref{sec:abm} and \ref{sec:abmslab}). 

It is also worth mentioning that the density of states in the quasiclassical theory holds the sum rule $\frac{1}{2E_{\rm c}}\int^{E_{\rm c}}_{-E_{\rm c}}\mathcal{N}(z,E)dE = \mathcal{N}_{\rm F}$ for $E_{\rm c}\gg \Delta _0$. The sum rule implies that the edge density of states for $|E|>\Delta _{\rm A}$ is also modified from $\mathcal{N}_{\rm A}$, which compensates the increase in the low-energy density of states owing to the formation of the edge bound states. We will show below that the deviation of the density of states of the continuum states at the edge affects the edge mass current and thus the macroscopic angular momentum.

The $\theta _{\bm k}$-angle-resolved mass current spectrum is obtained with Eq.~\eqref{Green} as
\begin{align}
j_y(\theta _{\bm k},x,E)\equiv\int\frac{d\phi _{\bm k}}{2\pi }j_y(\hat{\bm k},x,E)
=\frac{mv_{\rm F}\mathcal{N}_{\rm F}}{2}\frac{E}{\Delta_{\rm A}}\ {\rm sech}^2\left(\frac{x}{\xi_{\rm A}}\right),
\label{jEMJ}
\end{align}
for $|E|<\Delta _{\rm A}\sin\theta _{\bm k}$, and as
\begin{multline}
j_y(\theta _{\bm k},x, E)=-\frac{mv_{\rm F}\mathcal{N}_{\rm F}}{4}\frac{E}{|E|}\left[\frac{\sqrt{E^2-\Delta_{\rm A}^2\sin^2\theta _{\bm k}}}{\Delta_{\rm A}}\right.\\
\left.+\frac{E^2}{\Delta_{\rm A}\sqrt{E^2-\Delta_{\rm A}^2\sin^2\theta _{\bm k}}}-2\frac{|E|}{\Delta_{\rm A}}\right]\ {\rm sech}^2\left(\frac{x}{\xi_{\rm A}}\right),
\label{jE}
\end{multline}
for the continuum state $|E|>\Delta_{\rm A}\sin\theta_{\bm k}$. The mass current spectrum at the edge is shown in Fig.~\ref{energyA}(b) for $\theta_{\bm k}=\pi/2$.
The energy spectrum of the mass current carried by the Andreev bound states turns out to be linear on $E$, while the mass current generated by the continuum state has the opposite sign to the mass current carried by the bound states, which is the backaction to the Andreev bound states and absent in the bulk ABM without boundaries. It turns out that the edge mass current spectrum decays as $\sim E^{-3}$ for $|E|\gg\Delta_{\rm A}\sin\theta _{\bm k}$.
%
%

The total edge mass current carried by the bound states is estimated as
%
$J_y^{\rm bound}\equiv\int_0^{\infty }dxj_y^{\rm bound}(x)=-n\hbar/2$,
%
and that from the continuum state is
%
$J_y^{\rm cont}\equiv\int_0^{\infty }dxj_y^{\rm cont}(x)=n\hbar/4$,
%
where the density of $^3$He atoms, $n$, emerges from the normal density of states $\mathcal{N}_{\rm F}=(3/mv_F^2)n$. Since the energy states below the Fermi energy are occupied at $T=0$, the edge mass current carried by the bound state is obtained by integrating $j_y({\bm k},x,E)$ over $E \!\in\! [-\Delta _{\rm A}\sin \theta _{\rm A},0]$ and $\hat{\bm k}$ as
$j_y^{\rm bound}(x)=-\frac{mv_{\rm F}\mathcal{N}_{\rm F}\Delta_{\rm A}}{6}\ {\rm sech}^2(x/\xi_{\rm A})$, 
and that from the continuum state is obtained as
$j_y^{\rm cont}(x)=\frac{mv_{\rm F}\mathcal{N}_{\rm F}\Delta_{\rm A}}{12}\ {\rm sech}^2(x/\xi_{\rm A})$.
%
Since these currents flow in opposite directions, the total mass current induced by the edge state is
\beq
J_y=J_y^{\rm bound}+J_y^{\rm cont}=-\frac{n\hbar }{4}.
\label{eq:Jy}
\eeq

The mass current can be regarded as localized at the edge in a disk with a large radius $R\gg\xi_{\rm A}$.
Then, the angular momentum from each state is obtained as 
%
$L_z^{\rm bound}=N\hbar$ and $L_z^{\rm cont}=-N\hbar/2$.
%
To this end, the total angular momentum at zero temperature is simply related to the total number of particles as  
\begin{align}
L_z=L_z^{\rm bound}+L_z^{\rm cont}=\frac{N\hbar }{2}.
\label{Lz}
\end{align}
The total angular momentum attributable to the edge mass current coincides with that in two-dimensional chiral superfluids under a uniform pair potential~\cite{stone,sauls:2011}. Interestingly, half of the angular momentum carried by the bound states is canceled out by that from the continuum states.

\begin{figure}
\begin{center}
\includegraphics[width=85mm]{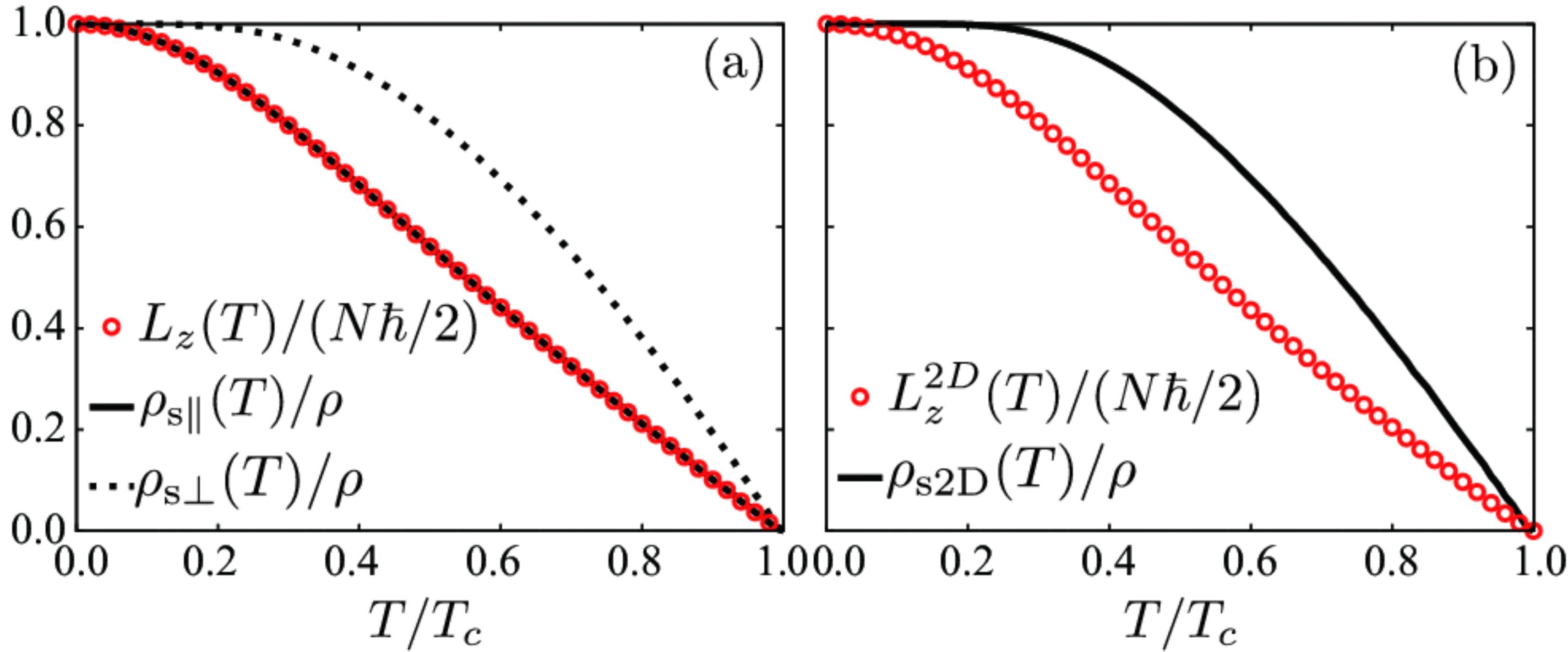}
\end{center}
\caption{\label{temperatureA}(Color online)
Temperature dependence of angular momentum (open circles) 
in the A-phase (a) and two-dimensional chiral $p$-wave state (b) with the superfluid density.
The components of the superfluid density tensor $\rho_{{\rm s}\parallel }$ (solid line) and $\rho_{{\rm s}\perp }$ (dotted line) are shown in (a) 
and the superfluid density $\rho_{\rm s2D}$ (solid line) is shown in (b). In (a), $L_z(T)$ perfectly traces $\rho _{{\rm s}\parallel}(T)$. Figures adapted from Ref.~\citeonline{tsutsumi:2012c}.
}
\end{figure}

The $T$-dependence of the angular momentum, $L_z(T)$, is shown in Fig.~\ref{temperatureA}(a) with open circles. 
%
%
The superfluid densities $\rho_{{\rm s}\parallel}$ and $\rho_{{\rm s}\perp}$, are also depicted with the solid and broken lines, where the former (latter) denotes the superfluid density parallel (perpendicular) to the ${\bm l}$-vector. It is seen that $L_z(T)$ traces the same $T$-dependence of $\rho _{{\rm s}\parallel}(T)$, which was also reported by Kita~\cite{kita:1998}. However, the depletion of $L_z(T)$ at low $T$ can be explained only by taking into account edge bound states, while $\rho _{{\rm s}\parallel}(T)$ reflects the quasiparticles in the bulk ABM state. Hence, the $T$-dependences of $L_z(T)$ and $\rho _{{\rm s}\parallel}(T)$ may be accidental coincidences in chiral $p$-wave superfluids with a three-dimensional Fermi sphere. We also demonstrate below that in the two-dimensional Fermi surface model, $L_z(T)$ has no connection with that of the superfluid density [see Fig.~\ref{temperatureA}(b)].

The low-temperature behavior of $L_z(T)$ is obtained as 
\begin{align}
L_z(T)=\frac{N\hbar }{2}\left[1-\left(\frac{\pi k_BT}{\Delta_A}\right)^2+\mathrm{O}\left(\frac{\pi k_BT}{\Delta_A}\right)^4\right].
\label{LzTlow}
\end{align}
As shown in Fig.~\ref{energyA}(b), the edge bound states for $|E|<\Delta _{\rm A}$ yield the linear $E$-dependence of $j_y(x,\theta,E)$, which leads to the $T^2$-depletion of $L_z(T)$ at low $T$. Hence, although $^3$He-A in a slab possesses gapless quasiparticles bound to point nodes in addition to edge bound states, the low-temperature behavior in Eq.~(\ref{LzTlow}) is a direct manifestation of edge bound states. The contribution of quasiparticles bound to point nodes is derived from 
the energy spectrum of the mass current in the continuum state. 
The low-energy behavior of the mass current spectrum $j_y(x=0,E)$ is estimated as
\begin{align}
j_y(x=0,E)\approx -\frac{mv_{\rm F}N_0}{12}\left(\frac{E}{\Delta_{\rm A}}\right)^3,
\end{align}
for $|E|\ll\Delta_{\rm A}$. Thus, the excitations at point nodes contribute to the angular momentum as the fourth order of temperature ($\sim T^4$) in Eq.~\eqref{LzTlow}. 

{\it Two-dimensional chiral superfluids.---}
In the two-dimensional Fermi surface model, where point nodes are absent, quasiparticle excitations in the chiral superfluid are fully gapped with $\Delta_{\rm 2D}$. The local density of states $N(x,E)$ is then obtained from Eqs.~\eqref{LDOSMJA} and \eqref{LDOScontA} by replacing $\Delta_{\rm A}\sin\theta$ with $\Delta_{\rm 2D}$.
The total mass current and angular momentum are also given by Eqs.~\eqref{eq:Jy} and \eqref{Lz}, respectively. 

In Fig.~\ref{temperatureA}(b), we plot the full temperature dependence of the total angular momentum, $L_z(T)$, and the superfluid density in the 2D chiral $p$-wave state, which is given by 
\begin{align}
\rho_{\rm s2D}(T)=\rho\left[1-\int_0^{\infty}d\epsilon\frac{1}{2k_{\rm B}T}{\rm sech}^2\frac{\sqrt{\epsilon^2+\Delta_{\rm 2D}(T)^2}}{2k_{\rm B}T}\right].
\label{eq:rhos2d}
\end{align}
It is seen from Fig.~\ref{temperatureA}(b) that $L_z^{\rm 2D}(T)$ yields a different $T$-dependence from that of $\rho_{\rm s2D}(T)$, which was also pointed out by Sauls~\cite{sauls:2011}. The deviation between $L_z^{\rm 2D}(T)$ and $\rho_{\rm s2D}(T)$ originates from the fact that the depletion of $L_z(T)$ at low $T$ is attributed to the edge bound states. The asymptotic behavior of $L_z(T)$ in the low-$T$ region is given by
\begin{align}
L_z^{\rm 2D}(T)
=\frac{N\hbar}{2}
\left[
1-\frac{2}{3}\left(\frac{\pi k_{\rm B}T}{\Delta _{\rm 2D}}\right)^2
+\mathrm{O}\left(\frac{\pi k_{\rm B}T}{\Delta _{\rm 2D}}\right)^4\right].
\label{Lz2DTlow}
\end{align}
The coefficient of the $T^2$ term in Eq.~\eqref{Lz2DTlow} is different from that in Eq.~\eqref{LzTlow}, which reflects the difference in the normal density of states due to the dimensionality of the Fermi surface. This implies that the density of states on the two-dimensional Fermi surface is $2/3$ times smaller than that in the case of the three-dimensional Fermi sphere. 

%
%

For $^3$He-A with $\hat{\bm l}\parallel\hat{\bm z}$, therefore, the total mass current $J_y=-n\hbar/4$ flows along the specular edge at $T=0$. In an axisymmetric disk with a large radius $R\gg\xi_{\rm A}$, the edge mass current generates the macroscopic angular momentum, $L_z=N\hbar/2$, which coincides with that of the McClure-Takagi result~\cite{mcclure:1979}. The angular momentum $L_z(T)$ decreases as $T$ increases and yields the same $T$-dependence as $\rho_{\rm s\parallel}(T)$, while the physical meaning of the coincidence remains unclear. Although the resultant angular momentum is comparable to the intrinsic angular momentum~\cite{ishikawa:1977,ishikawa:1980}, the edge mass current is sensitive to the condition of the edge, contrary to the argument of the intrinsic angular momentum. The depletion of the edge mass current at rough edges was indeed demonstrated in Refs.~\citeonline{sauls:2011} and \citeonline{nagato:2011}, which found the deviation of $L_z(T=0)$ from $N\hbar/2$. For nodeless two-dimensional chiral $p$-wave superfluids, the angular momentum in the bulk can be obtained as $L_{\rm bulk}^{\rm 2D}=N\hbar/2$ using the Berry connection~\cite{shitade:2014}, whereas the Berry connection is not well-defined for nodal $^3$He-A. Therefore, the intrinsic angular momentum paradox of $^3$He-A still remains as an unresolved problem.

In the case of chiral superconductors, the Meissner surface current flows in a layer within the penetration depth $\lambda(T)$ which is typically much longer than the superconducting coherence length, the length scale of chiral edge states. The contribution of the Meissner current considerably screens the spontaneous edge current and no net current remains in chiral superconductors. Tsuruta {\it et al.}~\cite{tsuruta}, however, demonstrated that in multiband chiral superconductors, the contribution of the Meissner current becomes less important, making it possible to observe the spontaneous edge current. The amount of the net current is found to be sensitive to the orbital channel of chiral Cooper pairs. The total angular momentum in chiral $\ell$-wave superconductors ($\ell \ge 2$) is deviated from $L_z=\hbar N/2$, which reflects the existence of $\ell$ branches of chiral edge Majorana fermions as shown in Eq.~\eqref{eq:kx}~\cite{tada:arXiv,kallinPRB14}.

\subsubsection{Chiral domain wall and spectral flow}

$^3$He-A spontaneously breaks TRS as well as the rotational symmetry in the orbital space, and is thus regarded as the orbital ferromagnetic state. The ${\bm l}$-vector characterizes the orientation of the rotational symmetry breaking in the orbital space, and the degeneracy space ${\rm SO}(3)$ denotes the degeneracy of the ground states with respect to a orientation of $\hat{\bm l}$ as in Eq.~\eqref{eq:RA}. Hence, the transition to the A-phase may be accompanied by multiple domains where each domain has the different orientation of $\hat{\bm l}$, i.e., the chiral domain walls. Ikegami {\it et al.}~\cite{ikegami13} observed multiple domain walls as well as the chirality in a single domain through the mobility of  electron bubbles injected on the surface of the liquid $^3$He. The moving electrons on the surface of $^3$He experience the intrinsic Magnus force, which is attributed to the skew scattering of electrons by quasiparticles.

\begin{figure}
\includegraphics[width=8.5cm]{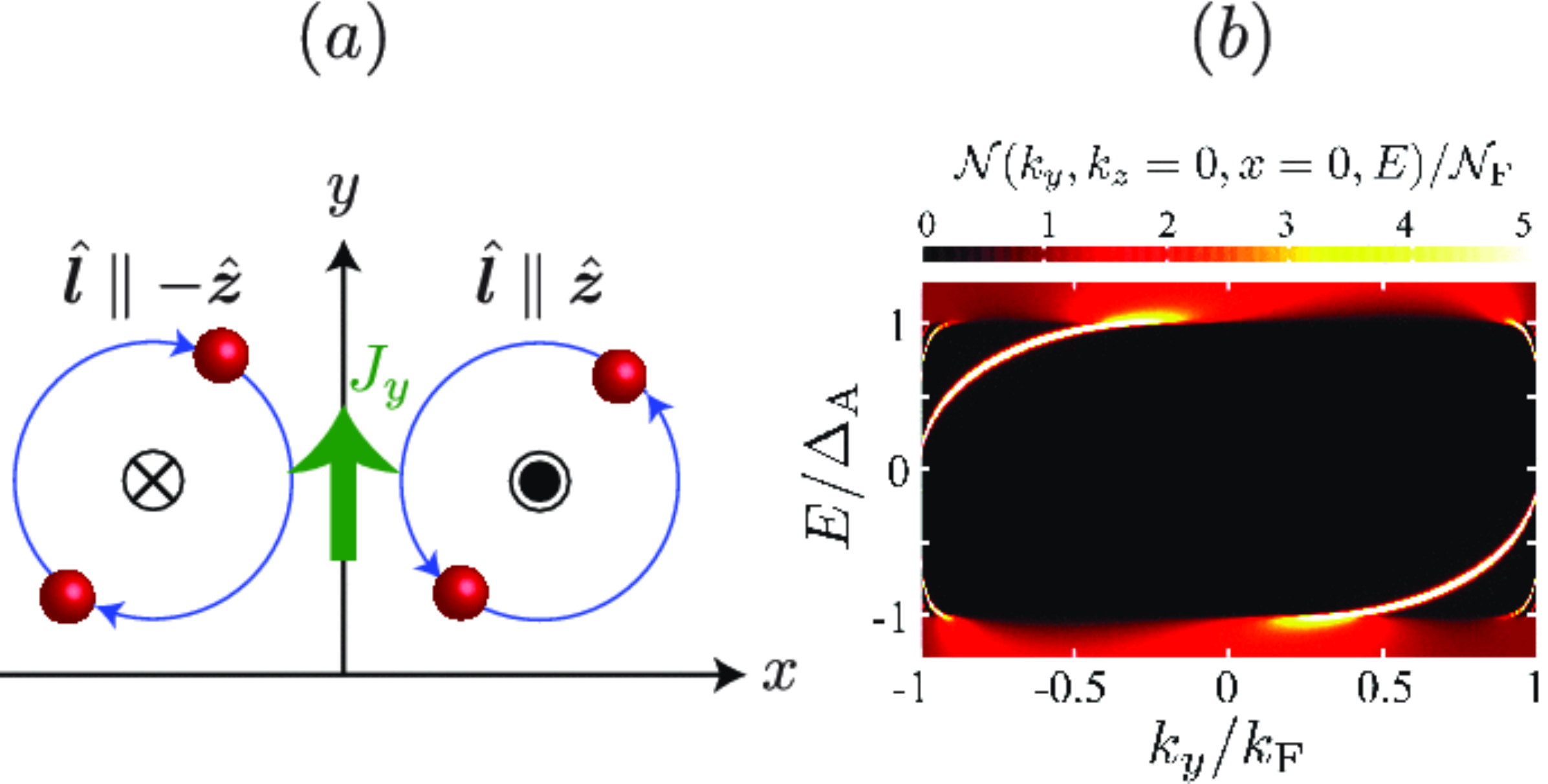}
\caption{(Color online)
(a) Schematic picture of the domain wall and (b) angle-resolved local density of states $N({\bm k},x=0,E)$ at the domain wall as a function of $k_y$ for $k_z=0$. The mass current ${\bm J}$ flows along the domain wall ($+\hat{\bm y}$-direction). 
Figures adapted from Ref.~\citeonline{tsutsumi:2014}.
}
\label{fig:domain_current}
\end{figure}

The structure of a chiral domain wall has been investigated in a $^3$He-A thin film by many researchers~\cite{ohmiPTP82,salomaa:1989,nakahara:1986,silaev:2012,tsutsumi:2014}. Nakahara~\cite{nakahara:1986} first demonstrated that a chiral domain wall composed of the A-phase with a different $\hat{\bm l}$ is accompanied by the zero energy states. Silaev and Volovik~\cite{silaev:2012} uncovered the topological aspect of the chiral domain wall. The topologically nontrivial quasiparticles bound to the domain wall form the dispersionless zero energy flat band, i.e., the topological Fermi arc, which is the manifestation of the Weyl superconductivity of $^3$He-A as discussed in Sec.~\ref{sec:abm}. Most recently, Tsutsumi~\cite{tsutsumi:2014} has investigated the mass current flowing along the domain wall. Owing to the nontrivial topological structure of the chiral domain wall, the mass current turns out to flow along the domain wall in the opposite sense from the Cooper pairs' angular momentum [see also Fig.~\ref{fig:domain_current}(a)].
 
We here show the structure of a single chiral domain wall in $^3$He-A in connection with the macroscopic angular momentum carried by the bound states~\cite{tsutsumi:2014}. The $\hat{\bm l}$-vector points to the $\hat{\bm z}$-direction in $x>0$ and is aligned to be the $-\hat{\bm z}$-direction in $x<0$ [see Fig.~\ref{fig:domain_current}(a)]. The domain wall structure and quasiparticle spectrum are determined by self-consistently solving the Eilenberger equation coupled to the gap equation with an appropriate boundary condition~\cite{tsutsumi:2014}. The resultant dispersion on the domain wall ($x=0$) with $k_z=0$ at $T=0.2T_{\rm c}$ is shown in Fig.~\ref{fig:domain_current}(b). Since the $k_y$-component of the ${\bm d}$-vector parallel to the domain wall changes its sign at $x=0$~\cite{salomaa:1989}, the branches of the bound state cross the zero energy at $k_y=\pm\sqrt{k_{\rm F}^2-k_z^2}$ and form the dispersionless flat band along $k_z$. 

The negative energy branch of the bound states with $k_y>0$ is occupied at $T=0$ and thus responsible for the generation of the nonvanishing mass current toward the $+\hat{\bm y}$-direction. The net mass current carried by the bound states and the continuum state is estimated as $J_y\approx n\hbar/2$. We notice that the quasiclassical approximation is not appropriate in the vicinity of $|k_y|\sim \sqrt{k_{\rm F}^2-k_z^2}$~\cite{nakahara:1986}; however, the deviation between the quasiclassical approximation and the full quantum mechanical BdG theory is negligible and thus the quasiclassical theory is reliable for the qualitative understanding of the mass current~\cite{silaev:2012,tsutsumi:2014}.

As depicted in Fig.~\ref{fig:domain_current}(a), the mass current flows along the domain wall in the opposite sense of Cooper pairs' angular momentum. This is interpreted with the knowledge of spectral flow~\cite{volovik:1995}. As in Eq.~\eqref{eq:Jy}, $^3$He-A touching a single hard wall has the mass current bound to the edge, $J_y=-n\hbar/4$, whose direction is in the same sense as the Cooper pairs' angular momentum. If we suppose that the domain mass current is composed of the simple summation of edge mass currents contributed from two domains, the total mass current is expected to be $J_y=-n\hbar/2$. However, when the momentum across the zero energy in a branch of the bound state moves, the number of negative energy quasiparticles carrying the mass current in the branch will change. The deviation of the actual mass current from the naive expectation is given by~\cite{volovik:1997}
\begin{align}
\Delta J_y =\frac{n_{\rm 2D}\hbar}{2}\sum _a\left(\frac{k_a}{k_{\rm F}}\right)^2{\rm sign}(c_a),
\label{eq:spectral_flow}
\end{align}
where $k_a$ is the momentum crossing $E=0$, $c_a$ is the group velocity of the bound states at $k_y=k_a$, and $\sum _a$ implies  the sum over all the zero energy states. For the bound state on the chiral domain wall, the contribution of the zero energy states for $|k_z|\le k_{\rm F}$ is estimated as $\Delta J_y=n\hbar$ in the case of a single chiral domain, while zero energy states do not contribute to the net mass current in the case of the single A-phase touching a hard specular wall. This implies that the zero energy states bound to the domain wall induce an extra contribution to the mass current, which changes the naively expected edge mass current $-n\hbar/2$ to $n\hbar/2$.


The edge current for {\it spinless} chiral $\ell$-wave pairing states described by the pair potential in Eq.~\eqref{eq:deltaell}, $\Delta=\Delta _0(\hat{k}_x+i\hat{k}_y)^{\ell}$, also has the contribution from the spectral flows for $\ell>1$. For chiral $\ell$-wave systems touching a specular hard wall, it is naively expected that each gapless branch carries the edge mass current $J_y=-\ell \times n\hbar/2$. For $\ell > 1$, however, in the same manner as the chiral domain wall, the spectral flow gives an extra contribution to the original edge current, $J_y=-\ell \times n\hbar/2$. Since all zero-energy modes with momentum in Eq.~(\ref{eq:kx}) have the same group velocity as ${\rm sign}(c_a)=+1$, the mass current deviation due to the spectral flow is given as $\Delta J_y=\ell (n\hbar/2)$ from Eq.~\eqref{eq:spectral_flow}. The extra mass flow generated by the spectral flow cancels the edge mass current $-\ell \times n\hbar/2$. The vanishing total angular momentum for $\ell>1$ was microscopically demonstrated by Tada {\it et al.}~\cite{tada:arXiv}. 
The net angular momentum for $\ell \!\ge\! 2$ is markedly suppressed to $L_z/\hbar N \!=\! \mathcal{O}(\Delta _0/E_{\rm F})\!\ll\! N$~\cite{tada:arXiv,kallinPRB14}.

The spectral flow also enables one to understand the dependence of the edge mass current on the surface roughness~\cite{nagato:2011,sauls:2011}. When the wall has microscopic roughness, quasiparticles incoming to the wall are diffusively scattered out by the wall in random directions. The random distribution of the scattering direction results in the smearing of the zero energy bound state in the momentum space. The smeared momentum gives rise to the spectral flow, which generates an extra contribution to the edge mass current as in Eq.~\eqref{eq:spectral_flow}. Since the contribution is opposite to the direction of the original edge mass current, the amplitude of the edge mass current is suppressed at the diffusive wall.


\subsection{Surface spin current in $^3$He-B}

$^3$He-B is known as a prototype of time-reversal-invariant topological superfluids and is accompanied by topologically protected surface Majorana fermions. The quasiparticles of the bulk $^3$He-B have the following two characteristics owing to the intrinsic features of the BW order parameter in Eq.~\eqref{eq:dvecB}: (i) The bulk maintains the TRS $\mathcal{T}$ as well as the inversion symmetry $\mathcal{P}$. Hence, the bulk quasiparticles are doubly degenerate, i.e., the eigenstates at an arbitrary ${\bm k}$ are $|u({\bm k})\rangle$ and $\mathcal{TP}|u({\bm k})\rangle$. (ii) In addition, the BW order parameter induces the spontaneous breaking of the spin-orbital symmetry, and thus the bulk quasiparticles are the eigenstates of the helicity operator $\hat{h}$ with the eigenvalues $h=\pm 1$. The helicity manifests the spin-momentum locking of the quasiparticles and is responsible for the spin current. Howver, since the combined symmetry $\mathcal{TP}$ changes the helicity to $\mathcal{TP}\hat{h}(\mathcal{TP})^{-1}=-\hat{h}$, the doubly degenerate quasiparticles have opposite helicity. Hence, the helicity is canceled out by two bands and the spin current is absent in the bulk. 

Confining $^3$He-B into a restricted geometry, however, markedly changes the quasiparticle structures, since the inversion symmetry is explicitly broken by the presence of the surface. This involves the degeneracy due to the symmetry $\mathcal{TP}$ disappearing in the vicinity of the surface, and the surface Majorana fermions have a definite helicity, i.e., the spin-momentum locking with a definite orientation. Hence, the spin current is a direct manifestation of the intertwining of symmetry and topology in $^3$He-B. 

%
%

In this section, we revisit the quasiparticle structures of $^3$He-B touching a specular surface at the zero field, where the surface is set at $z=0$ and the region of $z>0$ is occupied by $^3$He-B. As discussed in Sec.~\ref{sec:spt}, only the $k_z$-component of the ${\bm d}$-vector is suppressed near the surface within the coherence length and the ${\bm d}$-vector recovers to the bulk form in Eq.~\eqref{eq:dvecB} far from the surface. In this situation, the ${\bm n}$-vector is locked into the $\hat{\bm z}$-direction by the interplay between the dipole field and the pair breaking effect at the surface. The angle $\varphi $ is also locked into $\varphi _{\rm L}$ by the dipole field~\cite{vollhardt}. Without loss of generality, however, we take the $x$- and $y$-coordinates so as to satisfy $\varphi = 0$ from now on. 

Accordingly, the pair potential with a specular surface is given by 
\begin{align}
\Delta({\bm k}, z) = \Delta _{\rm B}
\begin{pmatrix}
-\sin\theta e^{-i\phi _{\bm k}} & \cos\theta _{\bm k}\tanh(z/\xi_{\rm B}) \\
\cos\theta _{\bm k}\tanh(z/\xi _{\rm B}) & \sin\theta e^{i\phi _{\bm k}}
\end{pmatrix}.
\label{OPB}
\end{align}
We here follow the same notation as in Sec.~\ref{sec:spt}. The self-consistent quasiclassical propagator is obtained by the unitary transformation, ${g}=\mathcal{S}({\phi _{\bm k}}){\tilde{g}}\mathcal{S}^{\dag}({\phi _{\bm k}})$, where $\mathcal{S}({\phi _{\bm k}})=(\sigma _x + \sigma _z)e^{i\vartheta\sigma_z}$ with $\vartheta = \frac{\phi _{\bm k}}{2}-\frac{\pi}{4}$. The reduced propagator $\underline{\tilde{g}}$ obeys the Eilenberger equation identical to the {\it spinless} chiral $p$-wave systems and is thus exactly solvable in the same manner as that in $^3$He-A~\cite{tsutsumi:2012c,mizushimaJPCM15}. The diagonal component of the quasiclassical propagator for $^3$He-B is 
\begin{gather}
g_0(\hat{\bm k},z;\omega _n) = - \frac{i\pi\omega _n}{\lambda (\omega _n)}\left[
1 + \frac{1}{2}\frac{\Delta^2_{\rm B}\cos^2\theta _{\bm k}}{\omega^2_n+E^2_0(\hat{\bm k}_{\parallel})}
{\rm sech}^2\!\left( \frac{z}{\xi _{\rm B}}\right)
\right], 
\label{Green0B}\\
\tilde{g}_{\parallel}(\hat{\bm k},z;\omega _n) = -\frac{\pi}{2\lambda (\omega _n)} \frac{\Delta^3_{\rm B}\sin\theta _{\bm k}\cos^2\theta _{\bm k}}{\omega^2_n+E^2_0(\hat{\bm k}_{\parallel})}{\rm sech}^2\!\left( \frac{z}{\xi _{\rm B}}\right), 
\label{eq:gvec}
\end{gather}
and $\tilde{g}_z(\hat{\bm k},z;\omega _n) = 0$, where we set $\lambda (\omega _n)\!\equiv\! \sqrt{\omega^2_n+\Delta^2_{\rm B}}$. The spin part of the quasiclassical propagator, ${\bm g}$, is obtained as 
$({g}_x,{g}_y,{g}_z) 
= (\tilde{g}_{\parallel}\sin\phi _{\bm k},-\tilde{g}_{\parallel}\cos\phi _{\bm k},0)$. The quasiparticle propagator derived here deviates from that of bulk $^3$He-B in the surface region within $\xi _{\rm B}$. The self-consistent solutions of the anomalous part of the quasiclassical propagator ${f}$ are also given as in Refs.~\citeonline{mizushimaJPCM15} and {tsutsumi:2012c}.

\begin{figure}
\begin{center}
\includegraphics[width=85mm]{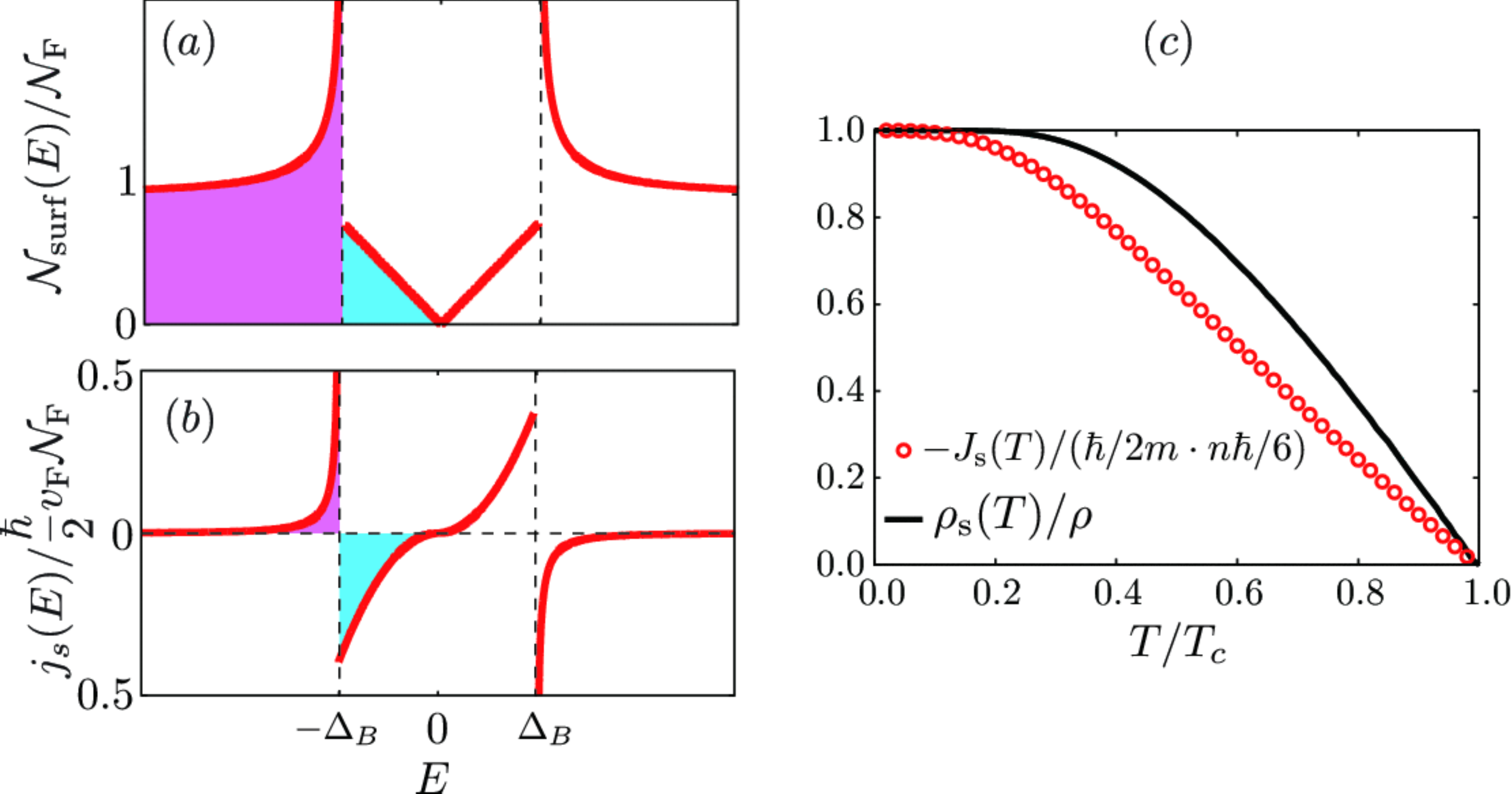}
\end{center}
\caption{\label{energyB}(Color online) 
Energy spectra of local density of states (a) and spin current (b) at $z=0$ in the B-phase.
At zero temperature, quasiparticles fill the colored (shaded) states in (a).
The spin currents from the bound and continuum states are derived 
by integrating the blue (light gray) and pink (gray) regions in (b), respectively.
(c) Total spin current $J_{\rm s}(T)$ (open circles) and superfluid density $\rho _{\rm s}(T)$ (solid line).
Figures adapted from Ref.~\citeonline{tsutsumi:2012c}.
}
\end{figure}

The retarded propagator $g^{\rm R}_0(E) \!=\! g_0(\omega _n \rightarrow -i E+ 0_+)$ has poles at the dispersion of the surface Andreev bound states, $E_{0}(\hat{\bm k}_{\parallel})$. From the exact solution Eq.~\eqref{Green0B}, the local density of states for the bound state $|E|<\Delta _{\rm B}$ is obtained as
\beq
\mathcal{N}(z,E)
=\frac{\pi}{4}\mathcal{N}_{\rm F}\frac{|E|}{\Delta _0}\ {\rm sech}^2\!\left(\frac{z}{\xi _{\rm B}}\right).
\label{LDOSMJB}
\eeq
Owing to the sum rule, the local density of states for the continuum state $|E|>\Delta _0$ is also deviated from that of the bulk B-phase by the appearance of the surface bound state. The local density of states at the surface $z=0$ is shown in Fig.~\ref{energyB}(a). This local density of states in the bound state has a linear energy dependence with slope $(\pi/4)(\mathcal{N}_{\rm F}/\Delta_{\rm B})$. The linear dependence is also obtained by numerical calculations~\cite{buchholtz:1981,nagato:1998,tsutsumi:2011b}. The linear behavior of the surface density of states is consistent with the dispersion of Majorana fermions bound to the surface region~\cite{chung:2009,nagato:2009}, which is linear on ${\bm k}_{\parallel}=(k_x,k_y)$. 

The surface mass current is prohibited by the TRS in $^3$He-B. As mentioned above, however, the emergence of spin-momentum locking is responsible for the spin current on the surface. Owing to the ${\rm SO}(2)_{J_z}$ symmetry, the nonvanishing components of the spin current tensor $j^{\rm spin}_{\mu\nu}(z,E)$ are 
%
$j^{\rm spin}_{xy}(z,E) =- j^{\rm spin}_{yx}(z,E)  \equiv  j_{\rm s}(z,E)$.
%
The spin current spectrum $j_{\rm s}(z,E)$ is
\begin{align}
j_{\rm s}(z,E)=\frac{\pi }{8}\frac{\hbar }{2}v_{\rm F}\mathcal{N}_{\rm F}
\frac{E|E|}{\Delta_{\rm B}^2}\ {\rm sech}^2\left(\frac{z}{\xi_{\rm B}}\right),
\end{align}
for the bound state $|E|<\Delta_{\rm B}$ and
\begin{multline}
j_{\rm s}(z,E)=-\frac{1}{12}\frac{\hbar }{2}
v_{\rm F}\mathcal{N}_{\rm F}\frac{E}{|E|}\left[\frac{\xi(E)}{\Delta_{\rm B}}\right.\\
\left.+2\frac{E^2}{\Delta_{\rm B}\xi(E)}
-3\frac{E^2}{\Delta_{\rm B}^2}\tan^{-1}\frac{\Delta_{\rm B}}{\xi(E)}\right]\ {\rm sech}^2\left(\frac{z}{\xi_{\rm B}}\right),
\label{jscont}
\end{multline}
for the continuum state $|E|>\Delta_{\rm B}$. This spin current spectrum at the surface is plotted in Fig.~\ref{energyB}(b), where the spin current spectrum contributed from the bound state turns out to be quadratic on $E$. The surface spin current from the continuum state reflects the deformation of the local density of states due to the sum rule, which decreases away from $\Delta _{\rm B}$.
The asymptotic behavior of the spin current spectrum in Eq.~\eqref{jscont} for $|E|\gg\Delta _{\rm B}$ indicates that the contribution of quasiparticles decreases with the same power law $\sim E^{-3}$, as in the A-phase.
%
%

In the same manner as that in $^3$He-A, the total spin currents contributed from the bound state and the continuum state
are estimated as $J_{\rm s}^{\rm bound}\equiv\int _0^{\infty }dz \int^{0}_{-\Delta _{\rm B}}dEj_{\rm s}^{\rm bound}(z,E)$ and $J_{\rm s}^{\rm cont}\equiv\int_0^{\infty }dz \int^{-\Delta _{\rm B}}_{-\infty} dE j_{\rm s}^{\rm cont}(z,E)$, respectively.
%
%
%
%
Since the spin current generated by the bound states, $J_{\rm s}^{\rm bound}$, flows in the opposite sense to $J_{\rm s}^{\rm cont}$, similarly to the A-phase, the total spin current at the surface reduces to 
\begin{align}
J_{\rm s}=J_{\rm s}^{\rm bound}+J_{\rm s}^{\rm cont}=-\frac{\hbar }{2m}\frac{n\hbar }{6}.
\label{Js}
\end{align}
The prefactor of $n\hbar$ in $J_{\rm s}$ is $2/3$ times smaller than that in the total mass current in $^3$He-A. 
This is because the Cooper pairs in the BW order parameter are equally distributed to all three spin states or three ${\bm d}$-vectors, and two of them are responsible for the spin current. 
%
%
We notice that the spin current also flows on an interface between two domains of $^3$He-B~\cite{silveriPRB14}; however, the direction of the spin flow is the opposite sense as well as the mass current on the chiral domain wall in $^3$He-A.

The temperature dependence of the total spin current, $J_{\rm s}(T)$, 
is plotted in Fig.~\ref{energyB} with open circles, compared with the superfluid density $\rho _{\rm s}(T)$, which is obtained from Eq.~\eqref{eq:rhos2d} with $\Delta _{\rm 2D}\mapsto\Delta _{\rm B}$.
The temperature dependence of $J_{\rm s}(T)$ deviates from that of $\rho _{\rm s}(T)$. To understand the deviation, we expand the exact formula on $J_{\rm s}(T)$ at low temperatures, $k_{\rm B}T\ll\Delta _{\rm B}$. The asymptotic behavior is then given by 
\begin{align}
J_s(T)=-\frac{\hbar }{2m}\frac{n\hbar }{6}\left[1-C\left(\frac{\pi k_{\rm B}T}{\Delta_{\rm B}}\right)^3+\mathrm{O}\left(\frac{\pi k_{\rm B}T}{\Delta_{\rm B}}\right)^4\right],
\end{align}
where the coefficient $C$ is a constant within $3/5\le C\le 1$ determined by using the Euler-Maclaurin formula up to the fourth Bernoulli number. The $T^3$-power behavior of the depletion manifests the contributions from the excitations of surface Majorana fermions with the gapless dispersion that is responsible for the quadratic energy dependence of the spin current spectrum $j_s(z,E)$ [see Fig.~\ref{energyB}(b)]. Hence, the low-temperature behavior of the total spin current is dominated by the quasiparticle excitations bound to the surface with no contributions from the continuum states. 

Detecting the spin current is more difficult than detecting the mass current because we have to separate spin states for the detection of the spin current.
Wu and Sauls~\cite{wuPRB13} proposed a practical way of detecting the spin current by generating the superfluid flow in a narrow channel connecting two $^3$He-B chambers. The superfluid flow explicitly breaks the TRS in the channel and induces the mass current on the surfaces of the narrow channel. As $T$ increases, the mass current decreases as $T^3$ at low temperatures, which reflects the dominant contribution of surface Majorana fermions. Hence, the surface Majorana fermions in $^3$He-B are detectable through the thermal depletion of the mass current.

\section{Vortex Structures and Symmetry-Protected Majorana Fermions in $^3$He-B}
\label{sec:vortices}

As has been emphasized in previous sections, the huge symmetry group $G$ of the liquid $^3$He ensures the competition between superfluid phases accompanied by a variety of spontaneous symmetry breaking. The superfluid phases, BW, ABM, and planar states, hold the nontrivial symmetries $H$ even in the presence of an external field, which ensures the emergence of rich topological quantum phenomena, such as the symmetry-protected topological phase and Majorana fermions. 

In addition to the topological aspect of quasiparticles, however, it is well known that superfluids and superconductors possess another type of topological property, that is, the topology of the order parameter manifold. In general, the spontaneous symmetry breaking in ordered media is characterized by the coset space $\mathcal{R}\!=\! G/H$~\cite{merminRMP,volovikJETP77}. The manifold $\mathcal{R}$ describes the degeneracy space of the order parameter and can be a source of the formation of nontrivial topological defects. After briefly reviewing the topology and symmetry classifications of vortices, we examine the topological aspect of low-lying quasiparticles bound to vortices and textures of $^3$He-B in Sec.~\ref{sec:vortexbw} and that of the A-phase in Secs.~\ref{sec:continuous} and \ref{sec:HQV}. 

\subsection{Topology and symmetry classification of vortices}
\label{sec:vortex}

To take a simple example, first consider an $s$-wave BCS superconducting state whose degeneracy space is $\mathcal{R}\!=\!{\rm U}(1)$. The order parameter residing in the manifold $\mathcal{R}\!=\!{\rm U}(1)$ gives rise to a variety of topologically distinct states. The superconducting state having a spatially uniform ${\rm U}(1)$ phase is known as the Meissner state, while the order parameter is allowed to have the spatially winding ${\rm U}(1)$ phase $e^{i\kappa\phi}$, where $\phi$ is the azimuthal angle in the spatial coordinate. The state with a definite phase winding is called the vortex state, where the single valuedness of the order parameter requires the ``vorticity'' $\kappa$ to be an integer. The state having nonzero $\kappa$ must be accompanied by a singular point at which the ${\rm U}(1)$ phase is ill-defined and the order parameter becomes zero. This is called the vortex core, and $\kappa$ implies the strength of the magnetic flux penetrating the vortex core in the superconductor. 

The topological stability of such a line defect in ordered media is represented by the first homotopy group, $\pi _1 ({\mathcal{R}})$, 
which counts the number of times that a path $S^1$ enclosing the line defect covers the degeneracy space $\mathcal{R}$. For the case of  $\mathcal{R}={\rm U}(1)$, the first homotopy group is isomorphic to the integer group, $\pi _1 (\mathcal{R})\!=\! \mathbb{Z}$, which is associated with the vorticity $\kappa$. The higher homotopy groups $\pi _2 (\mathcal{R})$ and $\pi _3(\mathcal{R})$ represent the possibility of the formation of monopole and skyrmion excitations, respectively~\cite{nakahara,vollhardt}.

{\it Linear defects in the bulk BW state.}--- 
The bulk BW state is known as the most symmetric phase and holds the joint ${\rm SO}(3)$ symmetry of the spin and orbital spaces as in Eq.~\eqref{eq:symmetrybw}. This indicates that the spin-orbit coupling emerges through the spontaneous symmetry breaking, and the order parameter possesses the ${\rm SO}(3)$ degrees of freedom in addition to the ${\rm U}(1)$ phase rotation, as in Eq.~\eqref{eq:R}. The possible linear defects in the bulk BW state are given by the homotopy group 
\beq
\pi _1 (\mathcal{R}_{\rm B}) = \pi _1 ({\rm SO}(3)_{{\bm L}-{\bm S}}) \oplus \pi _1 ({\rm U}(1)_{\varphi})
= \mathbb{Z}_2 \oplus \mathbb{Z}.
\label{eq:homotopybw}
\eeq
The former associated with the ${\rm SO}(3)$ degrees of freedom presents the topological stability of the textural structure formed by $R_{\mu i}(\hat{\bm n},\varphi)$, while the latter $\pi _1 ({\rm U}(1)_{\varphi})$ provides the possibility of a quantized vortex. We here notice that in real $^3$He, the magnetic dipole-dipole interaction originating from the nuclear magnetic moment forces $\varphi$ to the so-called Leggett angle $\varphi _{\rm L} \!=\! \cos^{-1}(-1/4)$. This reduces the order parameter manifold ${\rm SO}(3)$ to $S^2$ and the first homotopy group is trivial. Hence, an $\hat{\bm n}$-texture having a linear defect is unstable toward the spatially uniform $\hat{\bm n}$-vector, and only quantized vortices generated by the ${\rm U}(1)$ phase rotation can be topologically stable linear defects in the bulk BW state. 


{\it Linear defects in the bulk ABM state.}--- 
The linear defects realized in the bulk ABM state are essentially different from those in the BW state, because in the ABM state, the gauge symmetry is intrinsically coupled with the rotational symmetry of the orbital space. This implies that the spatially inhomogeneous configuration of $\hat{\bm l}$ generates the superfluid flow. Hence, the gauge-orbital symmetry peculiar to $^3$He-A gives rise to the topological stability of continuous vortices without a vortex core at which the phase singularity exists. 

When the dipole interaction is negligibly weak, the order parameter of the ABM state is composed of two independent vectors, $\hat{\bm l}$ and $\hat{\bm d}$. As in Eq.~\eqref{eq:RA}, the corresponding degeneracy space has an extra $\mathbb{Z}_2$ symmetry that the change from $\hat{\bm d}$ to $-\hat{\bm d}$ can be compensated by the phase rotation $\varphi \mapsto \varphi + \pi$. Owing to $\pi _1 (S^2) \!=\! 0$ and the $\mathbb{Z}_2$ symmetry of the dipole-free $\hat{\bm d}$-vector, the topologically stable linear defects in the bulk ABM state are characterized by the group of the integers modulo 4,
\beq
\pi _1 (\mathcal{R}_{\rm A}) = \pi _1 ({\rm SO}(3)/\mathbb{Z}_2) = \mathbb{Z}_4.
\eeq
This indicates that there exist four different classes of topologically protected linear defects in the dipole-free case.

\begin{figure}[tb!]
\begin{center}
\includegraphics[width=80mm]{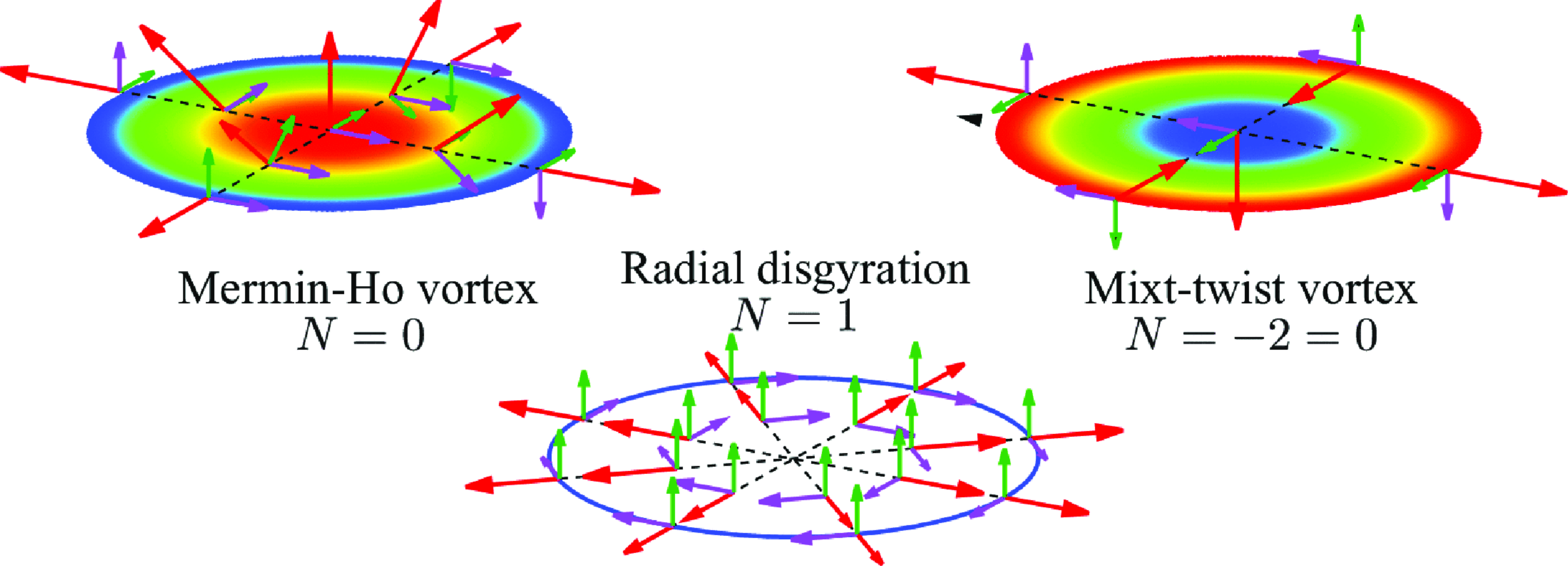} 
\end{center}
\caption{(Color online) Representatives of continuous vortices and texture formed by the $\hat{\bm l}$-texture in rotating $^3$He-A: Mermin-Ho vortex with $N=0$, mixt-twist vortex with $N=-2\equiv 0$, and radial disgyration with $N=1$. 
}
\label{fig:continuous}
\end{figure}

As for the dipole-free case, the four possible linear defects can be categorized by the fractional topological charge 
\beq
N = 0, \frac{1}{2}, 1, \frac{3}{2},
\eeq
where $N=2$ is topologically identical to $N=0$. The extra $\mathbb{Z}_2$ symmetry in the ABM state allows us sto take the half-integers of the topological charge, because the $\pi$-phase jump can be canceled out by the change in the orientation of $\hat{\bm d}$. The representatives of possible defects are the Anderson-Toulouse and Mermin-Ho vortices~\cite{merminPRL76,andersonPRL77} for $N=0$, half-quantized vortices~\cite{volovikJETP76} for $N=1/2$, and a radial $\hat{\bm l}$ disgyration~\cite{volovikJETP77} without phase winding for $N\!=\! 1$. The topological state with $N=3/2$ is identical to that with $N=-1/2$. The representatives of continuous vortices are illustrated in Fig.~\ref{fig:continuous}. 
In Sec.~\ref{sec:continuous}, we will discuss the low-lying quasiparticles bound to the Mermin-Ho vortex with $N=0$ and the ground-state texture of rotating $^3$He-A confined in a narrow cylinder. The half-quantum vortex state with $N=1/2$ is intriguing in the sense that it hosts non-Abelian Majorana fermions~\cite{ivanovPRL01}. The stability of the half-quantum vortices will be discussed in Sec.~\ref{sec:HQV}.

As already mentioned, however, the $\hat{\bm d}$-vector in the ABM order parameter is locked to $\hat{\bm l}$ by the dipole-dipole interaction originating from the nuclear magnetic moment. In the situation that the dipolar field is dominant, the $\mathbb{Z}_2$ symmetry vanishes and the degeneracy space is reduced to ${\rm SO}(3)_{{\bm L},\varphi}$. The topologically stable defect is characterized by $\pi _1 ({\rm SO}(3)) \!=\! \mathbb{Z}_2$, where only $N=0$ and $N=1$ classes are possible. The vortex structure and low-lying quasiparticles are shown in Sec.~\ref{sec:continuous}. In contrast, when the orientation of the $\hat{\bm l}$-vector is forced by a restricted geometry, the $\mathbb{Z}_2$ symmetry associated with $\hat{\bm d}$ plays an essential role in realizing a half-quantum vortex. The thermodynamical stability and topologically protected Majorana fermions are discussed in Sec.~\ref{sec:HQV}.

In the presence of a strong magnetic field, as mentioned in Eq.~\eqref{eq:quad}, the ${\bm d}$-vector tends to be confined in the plane perpendicular to the applied field. Owing to the competing effect between the magnetic field and the dipole field energy, the rotating $^3$He-A under a magnetic field possesses a rich phase diagram of exotic vortices~\cite{parts,karimakiPRB99,kitaPRL01,kitaPRB02}, including vortex sheets~\cite{partsPRL94,eltsovPRL02} and nonaxisymmetric continuous vortices with a soft core~\cite{sv,seppalaPRL84,pekolaPRL90}. Arguing the structure of various continuous vortices, however, would carry us too far away from the main scope of this paper. Comprehensive reviews on exotic vortices and textures in both $^3$He-A and B can be found in Refs.~\citeonline{volovik}, \citeonline{salomaaRMP87}, \citeonline{hakonenPB89}, and \citeonline{eltsov}. In this paper, therefore, we focus our attention on the existence of symmetry-protected Majorana fermions in $^3$He with a topological defect, and as for continuous vortices, we present in Sec.~\ref{sec:continuous} only the results on confinement-induced textural structures in a rotating narrow cylinder and low-lying quasiparticles bound to continuous vortices, which have not been discussed in previous review papers.

{\it Symmetry classification of vortices in $^3$He.}--- 
We have classified topologically stable vortex states in the superfluid $^3$He-A and B in the light of the spontaneous symmetry breaking and degeneracy space. Before closely examining each linear defect, we must draw attention to the remaining symmetry that the vortex state holds. 

As explicitly emphasized in Ref.~\citeonline{salomaaPRB85}, in general, ordered media having an isolated linear defect may maintain  continuous symmetries in addition to extra discrete symmetries. Regarding continuous symmetries in a vortex state, what first comes to mind might be translational symmetry along the linear defect. The order parameter having a perfectly straight line defect is invariant under the translational operator $\hat{p}_z \!\equiv\! -i\partial _z$ as $\hat{p}_zd_{\mu i}({\bm r}) \!=\! 0$. 

In addition to the translational symmetry along the defect line, there might be an extra continuous symmetry associated with the axial symmetry of the vortex order parameter. Although bulk superfluids maintain continuous ${\rm SO}(3)$ and $S^2$ symmetries, the phase winding emergent in a vortex state generates a superfluid flow along a path enclosing the line defect and breaks the continuous rotational symmetry in the bulk. The order parameter having a vortex acquires the extra phase $e^{i\kappa \varphi}$ under the ${\rm U}(1)$ phase rotation about the line defect by an angle $\varphi$. For $^3$He, the generator of the continuous symmetry may be expressed as $e^{i\hat{Q}\varphi}$, where the operator $\hat{Q}$ is given by a combination of two operators, 
\beq
\hat{Q} = \hat{J}_z - \kappa \hat{I}.
\eeq
The total angular momentum operator $\hat{J}_z$ is composed of the generator of rotations in the orbital space, $\hat{L}_z \!=\! \hat{L}^{\rm ext}_z + \hat{L}^{\rm int}_z$, and in the spin space, $\hat{S}_z$. The operator $\hat{I}$ denotes the generator of the ${\rm U}(1)$ phase rotation. The combined generator $e^{i\hat{Q}\varphi}$ implies that even though the order parameter is not invariant under each ${\rm U}(1)$ phase rotation and ${\rm SO}(2)$ spin and orbital rotation, their combined rotation may not change the order parameter. Hence, the generator $\hat{Q}$ offers a condition for axially symmetric vortex states. The general representation of the order parameter for axially symmetric vortex states is then given by solving the equation 
$e^{i\hat{Q}\varphi}d_{\mu i}({\bm r}) = d_{\mu i}({\bm r})$, 
\beq
\hat{Q}d_{\mu i}({\bm r}) = 0 .
\label{eq:axisymmetric}
\eeq


The discrete symmetries relevant to vortex states are given by three different types of spatial symmetry, $\{\hat{P}_1,\hat{P}_2,\hat{P}_3\}$~\cite{salomaaPRL83,salomaaPRB85}, which are composed of discrete elements of the symmetry group $G$, such as the time-reversal operator $\hat{T}$, the spatial inversion $\hat{P}$, the discrete phase rotation $\hat{U}_{(\kappa+1)\pi}\!\equiv\! e^{i\hat{I}(\kappa+1)\pi}$, and the joint $\pi$-rotation about an axis perpendicular to the linear defect $\hat{C}^{(J_x)}_{2}\!\equiv\! e^{-i\hat{J}_x\pi}$. The elements of discrete symmetries in $^3$He are given by
\beq
\hat{P}_1 \equiv \hat{P}\hat{U}_{(\kappa+1)\pi},\quad
\hat{P}_3 \equiv \hat{T}\hat{C}^{(J_x)}_{2}, \quad
\hat{P}_2 \equiv \hat{P}_1\hat{P}_3,
\eeq
where $\hat{P}_1$ is an order-two unitary operator and the others are order-two antiunitary operators. The role of the order-two antiunitary operator $\hat{P}_3$ is emphasized in Sec.~\ref{sec:spt} for $^3$He-B under a parallel magnetic field~\cite{mizushimaPRL12}. Another order-two antiunitary operator $\hat{P}_2$ is composed of the time-reversal operator and mirror reflection operator, where the mirror plane contains the linear defect. The $\hat{P}_2$ operator was originally introduced in Ref.~\citeonline{mizushimaNJP13} to understand the topological properties of quasi-one-dimensional spin-orbit coupled Fermi gases whose effective Hamiltonian is equivalent to that of superconducting nanowires. This has also been utilized to understand the topological protection of the zero-energy flat-band realized in the planar state in Sec.~\ref{sec:spt}. Schematic pictures of the action of these order-two operators are shown in Fig.~\ref{fig:discrete}.

To this end, the most symmetric vortex in the superfluid $^3$He may hold the symmetry group
\beq
H_{\rm vortex} = {\rm U}(1)_Q \times t_z \times P_2 \times P_3,
\label{eq:Hvortex}
\eeq
where ${\rm U}(1)_Q$ is the ${\rm U}(1)$ symmetry group having an element $e^{i\hat{Q}\varphi}$ and $t_z$ denotes the translational symmetry along the line defect ($\hat{z}$-axis). All the possible linear defects in $^3$He-A and $^3$He-B can be classified in terms of the spontaneous breaking of continuous discrete symmetries. Table~\ref{table:bw} summarizes the classification of possible vortices in the bulk BW state in terms of the continuous symmetry ${\rm U}(1)_{Q}$ and the discrete symmetries $\{\hat{P}_1,\hat{P}_2,\hat{P}_3\}$. The detailed structure of each vortex state will be discussed in Sec.~\ref{sec:vortexbw}. 

The most symmetric vortex that preserves $H_{\rm vortex}$ in Eq.~\eqref{eq:Hvortex} is called the $o$-vortex. There are different types of vortices, depending on the breaking of the symmetry group. Nonaxisymmetric vortices break the continuous ${\rm U}(1)_Q$ symmetry. The vortices with one of the three order-two symmetries $\{\hat{P}_1,\hat{P}_2,\hat{P}_3\}$ are called the $u$-, $v$-, and $w$-vortices, respectively. In principle, there is a possibility of realizing the $uvw$-vortex that does not have any discrete symmetries.

\begin{table*}[t!]
\begin{center}
\begin{tabular}{c|c|c|c|c|c|c}
\hline\hline
vortex & ${\rm U}(1)_{Q}$ & discrete sym. & states of vortex core & topo. $\#$ & ZES & Majorana \\
\hline
$o$ & yes & $P_1$, $P_2$, $P_3$ & normal & $w_{\rm 1d}$ & yes & yes \\
$u$ & yes & $P_1$ & normal & -- & yes & no \\
$v$ & yes & $P_2$ & ABM \& $\beta$ & -- & yes & no \\
$v$ & no & $P_2$ & ABM \& $\beta$ & -- & no & no \\
$w$ & yes & $P_3$ & ABM \& $\beta$ &  $w_{\rm 1d}\oplus{\rm Ch}_2$ & yes & yes \\
$uvw$ & yes & --  & ABM \& $\beta$ &  ${\rm Ch}_2$ & yes & no \\
\hline\hline
\end{tabular}
\end{center}
\caption{Symmetry classification of singly quantized vortices in superfluid $^3$He-B.
}
\label{table:bw}
\end{table*}


\subsection{Symmetry-protected Majorana fermions in vortices of $^3$He-B}
\label{sec:vortexbw}

{\it Vortex order parameter and symmetries.}---
We here examine whether quantized vortices in $^3$He-B can host Majorana fermions. To examine the existence of symmetry-protected Majorana fermions in rotating $^3$He-B, let us start by clarifying the boundary condition at ${\bm r}\rightarrow {\bm \infty}$ far from the vortex center. As mentioned in Sec.~\ref{sec:vortex}, quantized vortices generated by the ${\rm U}(1)$ phase winding are the only topologically stable linear defects in bulk $^3$He-B. For an isolated vortex, we here take the cylindrical coordinate ${\bm r}\!=\! (\rho, \phi,z)$ centered at the vortex core. Far from the vortex center, the asymptotic form of the vortex order parameter is obtained from Eq.~\eqref{eq:dvecB} as
\beq
d_{\mu i}(\rho \rightarrow\infty,\phi) = \Delta _{\rm B}(T) R_{\mu i}(\hat{\bm n},\varphi)e^{i\kappa \phi},
\label{eq:vortexbc}
\eeq
where $\kappa \!\in\! \mathbb{Z}$ is the vorticity and topological charge associated with the nontrivial homotopy group in Eq.~\eqref{eq:homotopybw}. The vorticity is responsible for the superfluid velocity field ${\bm v}_{\rm s}({\bm r}) \!=\! \frac{\hbar \kappa}{2m\rho}\hat{\bm \phi}$ for $\rho \rightarrow \infty$. The vortex core is centered at $\rho = 0$, and the translational symmetry along the vortex line, i.e., the $\hat{\bm z}$-axis, is assumed. The three-dimensional rotation matrix $R_{\mu i}$ denotes the relative rotation of orbital and spin spaces, originating from the spontaneous symmetry breaking ${\rm SO}(3)_{{\bm L}-{\bm S}}$ in the bulk BW state. As in Eq.~\eqref{eq:HB}, however, the rotation matrix is reduced to $\delta _{\mu i}$ by employing an appropriate ${\rm SU}(2)$ rotation in the BdG Hamiltonian. Hence, without loss of generality, we will focus our attention on the situation that the $\hat{\bm n}$-vector is polarized to the $\hat{\bm z}$-axis and the $xy$-coordinate is taken so as to satisfy $\varphi \!=\! 0$ from now on. 


To avoid the singular behavior of ${\bm v}_{\rm s}$, the BW order parameter must vanish at the core. Owing to the $3\times 3$ order parameter degrees of freedom, however, the vortex core can be filled in by other superfluid components, which must be continuously transformed from the BW state in Eq.~\eqref{eq:vortexbc}. This implies that, even if the asymptotic order parameter has the maximal symmetry represented in Eq.~\eqref{eq:Hvortex}, the emergence of the different types of order parameter in the core may lower the symmetry group, leading to the five classes of the vortex states, i.e., the $o$-, $u$-, $v$-, $w$-, and $uvw$-vortices as in Table~\ref{table:bw}. Following Ref.~\citeonline{salomaaPRB85}, we express the generic form of the vortex order parameter as 
\beq
d_{\mu i} ({\bm r}) = \sum _{m,n = 0, \pm 1} C_{mn}(\rho,\phi)\hat{e}^{(m)}_{\mu}\hat{e}^{(n)}_{i}e^{i\kappa \phi},
\label{eq:vortexgeneric}
\eeq
which is subject to the boundary condition in Eq.~\eqref{eq:vortexbc}. The bases $\hat{e}^{(m)}_{\mu}$ and $e^{(n)}_{i}$ are the eigenstates of the spin and orbital angular momentum operators as $\hat{L}_z\hat{\bm e}^{(m)} \!=\! m\hat{\bm e}^{(m)}$ and $\hat{S}_z\hat{\bm e}^{(n)} \!=\! n\hat{\bm e}^{(n)}$, with $\hat{\bm e}^{(\pm 1)}\!=\! (\hat{\bm x}\pm i\hat{\bm y})/\sqrt{2}$ and $\hat{\bm e}^{(0)}\!=\! \hat{\bm z}$. Hence, the generic form of the vortex order parameter in the BW state is given by the tensor $C_{mn}(\rho,\phi)$ in cylindrical coordinates with the boundary condition \eqref{eq:vortexbc}. 

The generators of the symmetry group $H_{\rm vortex}$ in Eq.~\eqref{eq:Hvortex} impose constraints on the tensor $C_{mn}(\rho,\phi)$. First, the condition in Eq.~\eqref{eq:axisymmetric} for an axisymmetric vortex requires that the tensor must be factorized as~\cite{salomaaPRB85,tsutsumiPRB15}
\beq
C_{mn}(\rho,\phi) = C_{mn}(\rho) e^{-i(m+n)\phi},
\label{eq:u1q}
\eeq
where $C_{mn}(\rho)$ denotes a $3\times 3$ complex-valued tensor. This implies that each order parameter component in an axisymmetric vortex must satisfy the selection rule of the phase winding, depending on the orbital ($m \!=\! 0, \pm 1$) and spin ($n \!=\! 0, \pm 1$) states.

Let ${P}_1$ be an operator acting on the spin matrix, which is associated with the combined symmetry ${P}_1$. The order-two unitary operator transforms the $3\times 3$ order parameter tensor as
\beq
{P}_1 C_{mn}(\rho, \phi) ({P}^{\ast }_1)^{-1}
= -C_{mn}(\rho, \phi+\pi).
\eeq
In accordance with Eq.~\eqref{eq:u1q}, the $P_1$ symmetry is preserved in an axisymmetric vortex when $\kappa - m -n$ is odd. The magnetic $\pi$-rotation symmetry $P_3$ changes the order parameter as
\beq
{P}_3 C_{mn}(\rho, \phi) ({P}^{\ast }_3)^{-1}
= -C^{\ast}_{mn}(\rho, \pi-\phi),
\eeq
where ${P}_3$ is the combination of the time-reversal operator $\mathcal{T}$ and the ${\rm SU}(2)$ $\pi$-rotation about an axis normal to the vortex line. The operator ${P}_2\equiv P_1P_3$ associated with the discrete symmetry $P_2$ is composed of the combination of $\mathcal{T}$ and the mirror reflection operator ${M}$, where the mirror plane contains the vortex line. The combined operator acts as 
\beq
{P}_2 C_{mn}(\rho, \phi) ({P}^{\ast}_2)^{-1}
= -C^{\ast}_{mn}(\rho, -\phi).
\label{eq:up2}
\eeq

The set of symmetry constraints indicates that for a $P_1$-preserving axisymmetric vortex ($o$- and $u$-vortices), $C_{mn}(\rho)$ must vanish for $\kappa - m -n = {\rm odd}$, while vortices with the $P_1$ symmetry breaking may have all the components of $C_{mn}$. The core of $P_1$ breaking vortices is filled in by superfluid components that are continuously connected to the boundary condition \eqref{eq:vortexbc}.

{\it Vortex structures.}---
The studies on vortex structures in rotating $^3$He-B were initiated by Ohmi {\it et al.}~\cite{ohmiPTP83} and Theodorakis and Fetter~\cite{theodorakisJLTP83} within the GL theory. They focused on the most symmetric vortex, that is, the $o$-vortex. As listed in Table~\ref{table:bw}, the $o$-vortex preserves the maximal symmetry group $H_{\rm vortex}$, whose order parameter tensor is given in Eq.~\eqref{eq:u1q} with $\kappa = 1$.
The $P_1$ symmetry imposes the constraint that some of the components must vanish $\forall \rho$, 
\beq
C_{+0}(\rho) = C_{0+}(\rho) = C_{0-}(\rho) = C_{-0}(\rho) = 0.
\eeq
The selection rule of the phase winding for an axisymmetric vortex, Eq.~\eqref{eq:u1q}, indicates that the $C_{+0}$ and $C_{0+}$ components have no phase winding and occupy the vortex core. The $P_1$ symmetry, however, requires them to be zero and the $o$-vortex core is filled in by the normal components. The $P_2$ symmetry uniquely determines the relative phase shift between the nonvanishing components of the order parameter, where $C_{mn}(\rho)$ must be real. Once some of them acquire an imaginary part, the $\hat{P}_2$ symmetry is broken and the resultant vortex is called the $u$-vortex. 

\begin{figure}[tb!]
\begin{center}
\includegraphics[width=85mm]{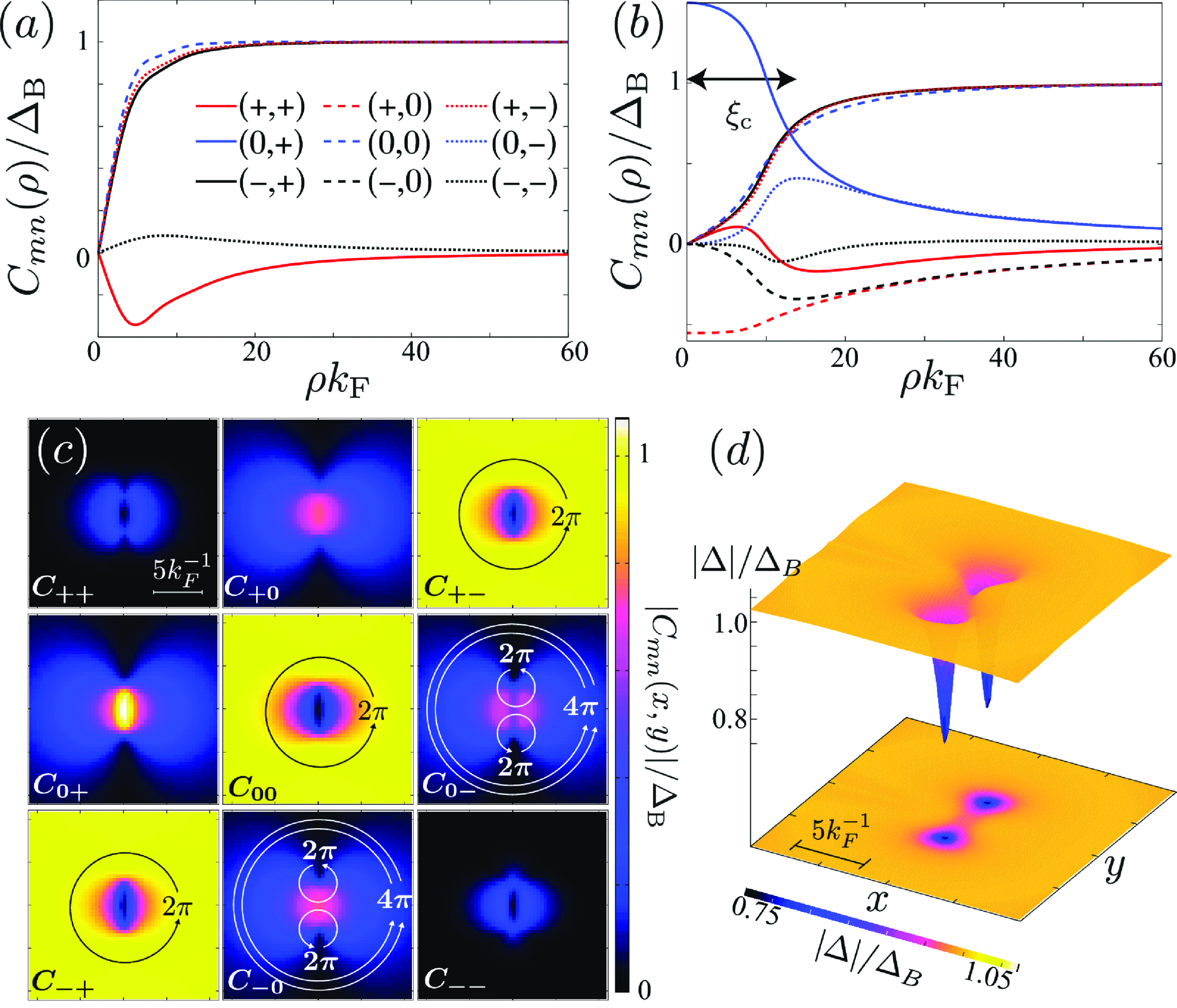} 
\end{center}
\caption{(Color online) Spatial profiles of order parameters $C_{mn}(\rho)$ for $o$-vortex (a) and axisymmetric $v$-vortex (b). Order parameters $C_{mn}(x,y)$ (c) and the root-mean-square value $|\Delta|$ for nonaxisymmetric $v$-vortex (d). All the order parameters are obtained by fully self-consistent calculations based on the quasiclassical Eilenberger theory. Figures adapted from Ref.~\citeonline{tsutsumiPRB15}.  
}
\label{fig:vortexbw}
\end{figure}

The spatial profile of the order parameters $C_{mn}(\rho)$ for the axisymmetric $o$-vortex is shown in Fig.~\ref{fig:vortexbw}(a). The order parameter is obtained by self-consistently solving the Eilenberger equation coupled to the gap equation~\cite{tsutsumiPRB15}. The order parameters for the $u$-vortex with breaking of the $P_1$ symmetry are given by rotating the relative phase of $C_{++}$ and $C_{--}$, while $C_{+-}$, $C_{00}$, and $C_{-+}$ are kept real. The set of coefficients, $\{C_{+-},C_{00},C_{-+}\}$, smoothly connects to the bulk value $C_{+-} = C_{00}=C_{-+} \!=\! \Delta _{\rm B}(T)$. The other nonvanishing components, $C_{++}$ and $C_{--}$, are induced near the vortex core and yield the asymptotic behaviors $C_{++}(\rho)\sim \rho$ and $C_{--}(\rho)\sim \rho^3$. 

In accordance with symmetry classification in Table~\ref{table:bw}, however, other classes of quantized vortices may exist in the bulk BW state, depending on the remaining discrete symmetries. The $v$-, $w$-, and $uvw$-vortices have vortex cores that are filled in by other superfluid components, such as the ABM state ($C_{0+}$) and the spontaneously ferromagnetic $\beta$ state ($C_{+0}$). Notice that the bulk $\beta$ state is given by interchanging the orbital and spin parts of the ABM state, i.e., the spin-orbital symmetry is spontaneously broken~\cite{vollhardt}. 

The axisymmetric $v$-vortex order parameter is expressed as in Eq.~\eqref{eq:vortexgeneric} with Eq.~\eqref{eq:u1q} and the boundary conditions
\begin{gather}
C_{mn}(\rho\rightarrow \infty) = \left\{
\begin{array}{ll}
\Delta _{\rm B}(T) & \mbox{ for $m+n=0$} \\
\\
0 &\mbox{ otherwise}
\end{array}
\right. ,
\end{gather}
and $\forall \rho$, the coefficients $C_{mn}(\rho)$ must be nonvanishing and real functions. The self-consistently obtained order parameter for the axisymmetric $v$-vortex state is shown in Fig.~\ref{fig:vortexbw}(b). Large amounts of the ABM state ($C_{0+}$) and $\beta$ state ($C_{+0}$) appear in the vortex core region. A similar structure is observed in the case of the $w$-vortex, where the coefficients $C_{+0}$, $C_{0+}$, $C_{0-}$, and $C_{-0}$ are replaced by pure imaginary numbers. The $w$-vortex preserves the magnetic $\pi$-rotation symmetry $P_3$, whereas it breaks the $P_2$ symmetry. Furthermore, the $uvw$-vortex has the complex coefficients $\{C_{++},C_{--},C_{+0},C_{0+},C_{0-},C_{-0}\}$, which break all the discrete symmetries.

The extra components emergent in the core of the $v$-vortex are found to be responsible for the magnetization of the vortex core. This is because the $\beta$ state is regarded as the spontaneous ferromagnetic state, and the core structures are essentially different from those in the $o$- and $u$-vortices having the ``normal core''. This novel feature of vortices was first pointed out by Salomaa and Volovik~\cite{salomaaPRL83}. Subsequently, Passvogel {\it et al.}~\cite{passvogelJLTP84} demonstrated that among five types of axisymmetric vortices, the only thermodynamically stable vortex is the $v$-vortex, and the $o$-vortex is metastable. The other axisymmetric ($u$, $w$, and $uvw$) vortices are found to be thermodynamically unstable. 

In addition to the various types of axisymmetric vortices, Thuneberg~\cite{thunebergPRL86} first revealed that there is another type of $v$-vortex that spontaneously breaks the axisymmetry ${\rm U}(1)_Q$. The order parameter coefficients, $C_{mn}(x,y)$, in the nonaxisymmetric $v$-vortex are displayed in Fig.~\ref{fig:vortexbw}(c), which are obtained from the self-consistent calculation of the quasiclassical theory. We also plot in Fig.~\ref{fig:vortexbw}(d) the root-mean-square value of the gap,  $|\Delta|\equiv\sqrt{\langle {\rm Tr}\Delta \Delta^{\dag} \rangle _{\bar{\bm k}}/2}$, which is the effective energy gap of quasiparticle excitations. It is clearly seen that the nonaxisymmetric $v$-vortex has a quadruple deformation of the core, and the core is filled in by the ABM and $\beta$ states. In contrast to the axisymmetric $v$-vortex, however, the original $4\pi$-phase singularity of $C_{0-}$ and $C_{-0}$ is split into two singularities with the $2\pi$-phase winding as shown in Fig.~\ref{fig:vortexbw}(c). This lowers ${\rm U}(1)_{\rm Q}$ to the twofold rotational symmetry about the $\hat{\bm z}$-axis. 

Using NMR techniques, Hakonen {\it et al.}~\cite{hakonenPRL83,hakonenJLTP83} reported the following two key observations on vortices in rotating $^3$He-B: (i) A spontaneous magnetization was observed in rotating $^3$He-B and (ii) the first-order phase transition takes place in the vortex cores at $P=29.3$bar and $T/T_{\rm c}\!=\! 0.6$. The former feature is attributed to the emergence of ferromagnetic vortex cores and is consistent with the feature of the axisymmetric $v$-vortex having the spontaneously ferromagnetic $\beta$ state. 

The second observation in NMR experiments was theoretically solved by Thuneberg~\cite{thunebergPRL86,thunebergPRB87}, who employed a full numerical calculation of the GL equation for rotating $^3$He-B. He identified that the high-pressure phase is the axisymmetric $v$-vortex having the ${\rm U}(1)_{Q}\times P_2$ symmetry, while the nonaxisymmetric $v$-vortex with only the $P_2$ symmetry is favored in the low-pressure regime. The change in the $v$-vortex core fully agrees with the spontaneous magnetization observed by the NMR experiments~\cite{hakonenPRL83}. We notice that a similar observation was independently reported by Salomaa and Volovik~\cite{salomaaPRL86} (see also Refs.~\citeonline{fetterPRL86} and \citeonline{thunebergPRL86-2}). The nonaxisymmetric $v$-vortex was directly observed by Kondo {\it et al.}~\cite{kondoPRL91} through the measurement of the homogeneously precessing magnetic domain mode. In the nonaxisymmetric $v$-vortex, the absorption of the NMR mode is attributed to a new soft Goldstone mode, that is, the spiral twisting mode of the anisotropic core. Most recently, Kita has uncovered a novel vortex phase diagram in $^3$He at extremely high rotation speeds~\cite{kitaPRL01,kitaPRB02}. 

Apart from the GL theory, Fogelstr\"{o}m and Kurkij\"{a}rvi~\cite{fogelstromJLTP95,fogelstromJLTP99} first demonstrated the microscopic calculation of axisymmetric and nonaxisymmetric $v$-vortices, based on the quasiclassical Eilenberger theory. The microscopic theory allows one to study the low-temperature phase diagram beyond the vicinity of $T_{\rm c}$ and low-lying quasiparticle spectra, which cannot be described in the context of the GL theory. They observed that the nonaxisymmetric $v$-vortex can be thermodynamically stabilized in the high-temperature regime, while it undergoes the phase transition to the axisymmetric $v$-vortex just above $0.5 T_{\rm c}$. The resultant phase diagram indicates the new phase boundary around $T=0.5 T_{\rm c}(P)$, while it cannot explain the phase transition near the bulk AB transition, which was observed in experiments~\cite{hakonenPRL83,hakonenJLTP83,hakonenPB89}. This is attributed to the fact that the quasiclassical theory does not take into account the strong coupling correction that is essential for the stability of the bulk ABM state relative to the BW state. Hence, the quasiclassical theory may fail to reproduce the quantitative properties in the high-pressure regime. 

Most recently, Silaev {\it et al.}~\cite{silaev15arXiv} have emphasized the important role of Fermi liquid interactions that significantly change the structure of the nonaxisymmetric $v$-vortex. The effect gives rise to the Lifshitz transition in the effective Fermi surface of core-bound quasiparticles.

{\it Symmetry-protected vortex-bound states.}---
As already mentioned, among the possible vortices, only the $o$- and axisymmetric and nonaxisymmetric $v$-vortices turn out to be local minima of the thermodynamic potential, while the others are unstable. Nevertheless, we here examine the topological properties for all the vortices listed in Table~\ref{table:bw}. This is because $^3$He-B serves as a prototype for studying topologically protected vortex bound states in the background of various types of vortices and the outcome may be applicable to topological superconductors under magnetic fields. Indeed, a recent theoretical study on the spin-triplet superconductor UPt$_3$ predicts that the phase transition from nonaxisymmetric superconducting-core to axisymmetric normal-core vortices may take place at the critical magnetic field~\cite{tsutsumiJPSJ12-2,tsutsumiJPSJ13}. Furthermore, similar vortex states may be realized in the odd-parity topological superconductor Cu$_x$Bi$_2$Se$_3$ under a magnetic field.

As mentioned in Sec.~\ref{sec:az}, the topology of ordered systems with a linear defect is determined using the semiclassical BdG Hamiltonian in the base space $S^3\times S^1$, 
\beq
\mathcal{H}({\bm k},\phi) = \left(
\begin{array}{cc}
\varepsilon ({\bm k}) & \Delta ({\bm k},\phi) \\
\Delta^{\dag} ({\bm k},\phi) & -\varepsilon^{\rm T}(-{\bm k})
\end{array}
\right),
\eeq
where $\phi$ denotes a circle enclosing the linear defect. The BdG Hamiltonian satisfies the PHS
\beq
\mathcal{C} \mathcal{H}({\bm k},\phi) \mathcal{C}^{-1} = - \mathcal{H}(-{\bm k},\phi).
\eeq
The PHS is necessary for the existence of Majorana fermions~\eqref{eq:selfcharge}.

The topological structure of a vortex is evaluated from the semiclassical BdG Hamiltonian with the asymptotic form in Eq.~\eqref{eq:vortexbc}~\cite{tsutsumiPRB15}. Although the order parameter near the vortex core is given by Eq.~\eqref{eq:vortexgeneric} subject to discrete symmetries and is different from the asymptotic form, it can be smoothly interpolated to $d_{\mu i}(\infty)$ without breaking the discrete symmetries. Hence, the topological structure of the BdG Hamiltonian with the general form of the order parameter is identical to that with $d_{\mu i}({\bm \infty})$ subject to relevant discrete symmetries.

First of all, we examine the topological number defined in the whole base space $({\bm k},\phi)$. Since nonzero $\kappa$ in the vortex order parameter breaks the TRS, the BdG Hamiltonian is categorized to class D. As in Table~\ref{table1}, the class D in the base space $S^{d}\times S^D = S^3\times S^1$ has a nontrivial $\mathbb{Z}$ topological invariant corresponding to the second Chern number ${\rm Ch}_2$ defined in Eq.~\eqref{eq:nchern}. The $\mathbb{Z}$ number is subject to discrete symmetries held by a vortex, and the $P_1$ symmetry forces ${\rm Ch}_2$ to be identically zero. Furthermore, ${\rm Ch}_2$ turns out to be zero for a vortex in $^3$He-B. 

In Sec.~\ref{sec:spt}, it is demonstrated that, even if the bulk topological number is absent, the chiral symmetry constructed from the combination of discrete symmetries may guarantee a topological invariant that is well-defined along the chiral symmetric region of the base space. To clarify this, let us assume that the single-particle energy $\varepsilon({\bm k})$ preserves the discrete symmetries $\{P_1,P_2,P_3\}$. For a vortex preserving the $P_1$ symmetry, the BdG Hamiltonian obeys 
\beq
\mathcal{P}_1 \mathcal{H}({\bm k},\phi) \mathcal{P}^{-1}_1 = \mathcal{H}(-{\bm k},\phi+\pi),
\eeq
where $\mathcal{P}_1 = {\rm diag}(P_1,P^{\ast}_1)$ denotes the inversion operator for $\kappa =1$. The $P_2$ and $P_3$ symmetries imply 
\begin{gather}
\mathcal{P}_2 \mathcal{H}({\bm k},\phi) \mathcal{P}^{-1}_2 = \mathcal{H}(-k_x,k_y,-k_z,-\phi), \\
\mathcal{P}_3 \mathcal{H}({\bm k},\phi) \mathcal{P}^{-1}_3 = \mathcal{H}(-k_x,k_y,k_z,-\phi),
\end{gather}
where $\mathcal{P}_2 = {\rm diag}(P_2,P^{\ast}_2) = i\tau_z K$ denotes the combination of $\mathcal{T}$ and the mirror reflection in the $xz$-plane, and $\mathcal{P}_3 = {\rm diag}(P_3,P^{\ast}_3) = i\sigma _z \tau _z K$ is the magnetic $\pi$-rotation about the $\hat{\bm x}$-axis. 

When some of the discrete symmetries are preserved, the topological structure of the BdG Hamiltonian is categorized into class D subject to the discrete symmetries. The general strategy for topological classification subject to discrete symmetries is provided in Ref.~\citeonline{shiozakiPRB14}, and the detailed procedure for a vortex in $^3$He-B is described in the Appendix of Ref.~\citeonline{tsutsumiPRB15}. As listed in Table~\ref{table:bw}, the topological properties of quasiparticles bound to the $o$- and $w$-vortex are characterized by $\mathbb{Z}$ and $\mathbb{Z}\oplus\mathbb{Z}$, respectively, and the others are topologically trivial. In the $o$-vortex state, the $P_1$ symmetry prohibits nonzero ${\rm Ch}_2$, while such a constraint is absent in the $w$-vortex. 

To give an intuitive understanding of such a topological property, let us consider the asymptotic BdG Hamiltonian subject to discrete symmetries. The asymptotic form with Eq.~\eqref{eq:vortexbc} is expressed in terms of the gamma matrices $(\gamma _1,\gamma _2,\gamma _3, \gamma _4) = (-\sigma _z\tau _x, -\tau _y, \sigma _x\tau _x, -\tau _z)$ as
\beq
\mathcal{H}({\bm k},\phi) = e^{i(\phi/2)\gamma _4} m_{\mu}({\bm k})\gamma _{\mu}e^{-i(\phi/2)\gamma _4},
\eeq
where the Hamiltonian is parameterized with the four-dimensional vector ${\bm m}=(ck_x,ck_y,ck_z,-\varepsilon ({\bm k}))$ with $c \!\equiv\! \Delta _{\rm B}/k_{\rm F}$.

When the magnetic $\pi$ rotation symmetry $P_3$ is preserved, the BdG Hamiltonian satisfies the chiral symmetry within the segments in the base space,
\beq
\left\{
\Gamma, \mathcal{H}(k_x,k_y=0,k_z=0,\phi=0,\pi)
\right\}=0,
\eeq
where the chiral operator is composed of $\mathcal{P}_3$ and $\mathcal{C}$ as $\Gamma = \mathcal{C}\mathcal{P}_3 = \sigma _z \tau _y$. The chiral symmetric Hamiltonian is parameterized with only the two-dimensional vector $\tilde{\bm m}(k_x)=(ck_x,-\varepsilon(k_x,0,0))$. This implies that the BdG Hamiltonian for a $P_3$-preserving vortex is expressed by the manifold $S^1$. The mapping from the chiral symmetric base space $S^1$ to the manifold $S^1$ is characterized by the fundamental group $\pi_1(S^1)=\mathbb{Z}$. The corresponding topological invariant is the one-dimensional winding number given by
\beq
w_{\rm 1d}(\phi) = - \frac{1}{4\pi i}\int dk_x {\rm tr}
\left[ 
\Gamma \mathcal{H}({\bm k},\phi)
\partial _{k_x} \mathcal{H}({\bm k},\phi)
\right]_{k_y=k_z=0}.
\eeq
Similarly to the result in Sec.~\ref{sec:spt}, the winding number is estimated as $w_{\rm 1d}(\phi = 0) = 2$ and $w_{\rm 1d}(\phi = \pi) = -2$. The difference provides the $\mathbb{Z}$ topological invariant as
\beq
w_{\rm 1d} = \frac{w_{\rm 1d}(0)-w_{\rm 1d}(\pi)}{2} = 2.
\eeq
In accordance with the index theorem in Sec.~\ref{sec:index}~\cite{satoPRB11}, the nontrivial winding number guarantees that the two zero energy states exist at $k_z=0$ as long as the vortex state holds the $P_3$ symmetry. The zero energy states can be gapped out by only a perturbation while breaking the $P_3$ symmetry, such as a magnetic field tilted from the vortex line. 

In the same manner, one can introduce different types of chiral symmetry: $\Gamma _1 \equiv \mathcal{C}\mathcal{P}_1$ and $\Gamma _2 \equiv \mathcal{C}\mathcal{P}_2$. It is straightforward to observe that the BdG Hamiltonian for only the $P_1$- ($P_2$-) preserving vortex is characterized by the manifold $S^3$ ($S^2$), and the corresponding homotopy group is trivial, $\pi _1 (S^3) = 0$ ($\pi_1(S^2)=0$). This clearly explains the topological table (table~\ref{table:bw}) in which only the $o$- and $w$-vortices have a nontrivial $\mathbb{Z}$ topological invariant associated with $w_{\rm 1d}$. 

\begin{figure}[tb!]
\begin{center}
\includegraphics[width=80mm]{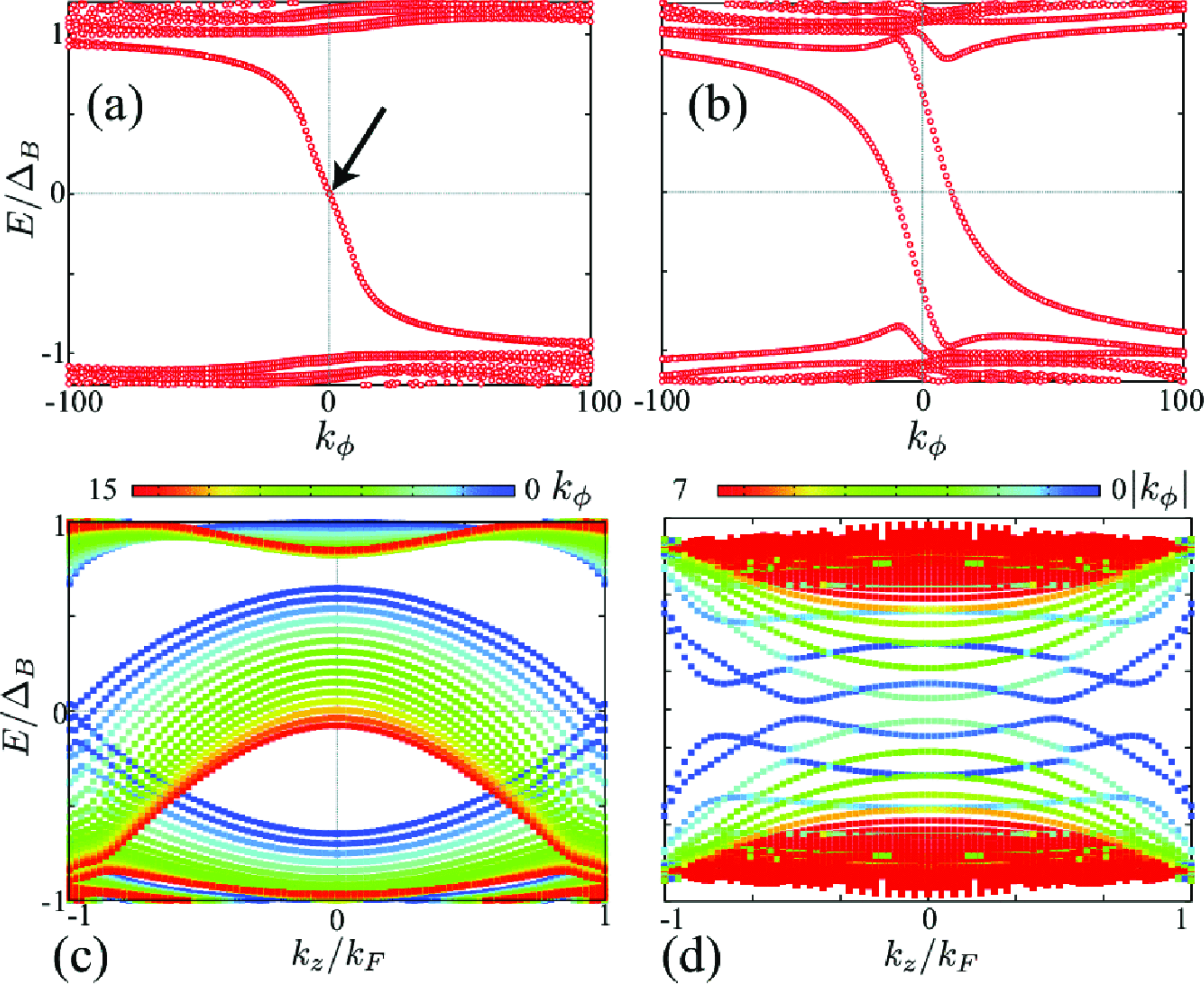} 
\end{center}
\caption{(Color online) Low-lying quasiparticle spectra for the axisymmetric $w$-vortex (a) and $v$-vortex (b) with $k_z = 0$ and $k_z$-dispersions of low-lying quasiparticles for the axisymmetric $v$-vortex (c) and nonaxisymmetric $v$-vortex (d). All the spectra are obtained from full quantum mechanical calculations based on the BdG equation. Figures adapted from Ref.~\citeonline{tsutsumiPRB15}.  
}
\label{fig:vortexspect}
\end{figure}

{\it Quasiparticle spectra and Majorana fermions.}---
To verify the topological argument described above and examine the Majorana nature of such a topologically protected zero mode, let us now turn to the numerical calculations of the BdG equation for various vortices. Full quantum mechanical quasiparticle spectra are obtained by solving the BdG equation in the real coordinate, 
\begin{align}
\left(
\begin{array}{cc}
\varepsilon (-i{\bm \partial}) & \tilde{\Delta} ({\bm r},-i{\bm \partial}) \\
-\tilde{\Delta}^{\ast} ({\bm r},-i{\bm \partial}) & -\varepsilon^{\rm T}(-i{\bm \partial})
\end{array}
\right){\bm \varphi}_{k_{\phi},k_z}({\bm r}) = E(k_{\phi},k_z){\bm \varphi}_{k_{\phi},k_z}({\bm r}),
\label{eq:bdgvortex}
\end{align}
where the off-diagonal matrix denotes the pair potential in the spin space, which is given by
\begin{gather}
\tilde{\Delta} ({\bm r},-i{\bm \partial}) =\left(
\begin{array}{cc}
\tilde{\Delta}_+ ({\bm r},-i{\bm \partial}) & \tilde{\Delta}_0 ({\bm r},-i{\bm \partial}) \\
\tilde{\Delta}_0 ({\bm r},-i{\bm \partial}) & \tilde{\Delta}_- ({\bm r},-i{\bm \partial})
\end{array}
\right), \\
\tilde{\Delta}_m ({\bm r},-i{\bm \partial}) = \frac{1}{2}\sum _{n=0,\pm 1} \left\{ C_{mn}(\rho,\phi), 
\mathcal{Y}_{1,n}(-i{\bm \partial}) \right\}.
\end{gather}
The function $\mathcal{Y}_{1,m}(-i{\bm \partial}) $ is obtained from the spherical harmonic function ${\rm Y}_{1,m}(\hat{\bm r})$ by replacing $\hat{\bm r}$ with $-i{\bm \partial}$.

For an axisymmetric vortex, the BdG Hamiltonian has the continuous symmetries, ${\rm U}(1)_Q\times t_z$ and the azimuthal and axial angular momenta $k_{\phi}$ and $k_z$ remain as well-defined quantum numbers. Then, the quasiparticle wavefunction ${\bm \varphi}_{k_{\phi},k_z}({\bm r})$ in Eq.~\eqref{eq:bdgvortex} is factorized as~\cite{volovikJPCM91,volovikPB95}
\begin{align}
{\bm \varphi}_{k_{\phi},k_z}({\bm r}) = e^{ik_{\phi}\phi}e^{ik_z z} & \left( 
u_{\uparrow}(\rho)e^{i(\kappa+1)\phi/2} , 
u_{\downarrow}(\rho)e^{i(\kappa-1)\phi/2}, \right. \nn \\
& \left. v_{\uparrow}(\rho)e^{-i(\kappa+1)\phi/2}, 
v_{\downarrow}(\rho)e^{-i(\kappa-1)\phi/2} 
\right)^{\rm T}.
\end{align}
By substituting this form into \eqref{eq:bdgvortex}, the BdG Hamiltonian can be recast into $\mathcal{H}(k_{\phi},k_z,-i\partial _{\rho})$. All eigenvalues of the BdG Hamiltonian for an axisymmetric vortex are labeled by the set of $(k_{\phi},k_z)$. In Fig.~\ref{fig:vortexspect}, we display the low-lying quasiparticle spectra for the axisymmetric $w$-vortex (a) and $v$-vortex (b) with $k_z = 0$, which are obtained by solving the BdG equation \eqref{eq:bdgvortex} with the self-consistent order parameter based on the quasiclassical theory~\cite{tsutsumiPRB15}. The single branch of the vortex-bound states in the $w$-vortex crosses the zero energy at $k_{\phi}=0$ and $k_z=0$, while two branches cross $E=0$ at finite $k_{\phi}$ in the case of the $v$-vortex. We notice that for the $o$-vortex state, the BdG equation at $k_z=0$ reduces to that for a pair of spin-polarized chiral $p$-wave superconductors. As demonstrated in Sec.~\ref{sec:spinless}, the index theorem ensures the existence of a single zero energy state when $\kappa$ is odd. 

It is noted that the PHS transforms the BdG Hamiltonian as $\mathcal{C} \mathcal{H}(k_{\phi},k_z,-i\partial _{\rho})\mathcal{C}^{-1} \!=\! - \mathcal{H}(-k_{\phi},-k_z,-i\partial _{\rho})$, which implies the symmetric relation $E(k_{\phi},k_z) \!=\! -E(-k_{\phi},-k_z)$. The $P_2$ symmetry imposes the relation $E(k_{\phi},k_z) \!=\! E(k_{\phi},-k_z)$ on the eigenvalues since $\mathcal{P}_2\mathcal{H}(k_{\phi},k_z,-i\partial _{\rho})\mathcal{P}_2^{-1} \!=\! \mathcal{H}(k_{\phi},-k_z,-i\partial _{\rho})$. The $P_3$ symmetry does not impose any constraints on the eigenvalues. In accordance with the symmetric properties, the quasiparticle spectrum in axisymmetric $o$- and $v$-vortices must satisfy the following relation:
\beq
E(k_{\phi},k_z) = E(k_{\phi},-k_z) = -E(-k_{\phi},-k_z).
\label{eq:Esymmetry}
\eeq

The topologically protected zero energy states at $k_{\phi}=k_z=0$ for the $o$-vortex are invariant under the PHS and the wavefunction satisfies the Majorana condition,
\beq 
\mathcal{C}{\bm \varphi}_{k_{\phi}=0,k_z=0}({\bm r}) ={\bm \varphi}_{k_{\phi}=0,k_z=0}({\bm r}). 
\eeq
The zero energy state emergent in the $o$-vortex behaves as chiral-symmetry-protected Majorana fermions, as discussed in Sec.~\ref{sec:chiral}. 

As shown in Fig.~\ref{fig:vortexspect}(a), the zero-energy states appear at $k_{\phi}=k_z=0$ in the case of the $w$-vortex, which are protected by the magnetic $\pi$-rotation $P_3$ symmetry. Similarly to the case of the $o$-vortex, the zero-energy states obey the Majorana condition at $k_{\phi}=k_z=0$ and thus behave as chiral-symmetry-protected Majorana fermions. Notice that the Majorana fermions are protected by the $P_3$ symmetry unless a magnetic field is tilted from the vortex line. In addition, the zero-energy states are robust against the density fluctuation and nonmagnetic impurities. These are characteristics peculiar to chiral-symmetry-protected Majorana fermions, as discussed in Sec.~\ref{sec:chiral}.

In contrast to the $o$-vortex, as shown in Fig.~\ref{fig:vortexspect}(b), the low-lying spectrum of the axisymmetric $v$ state as a function of $k_{\phi}$ is composed of two branches that cross the zero energy. As in Eq.~\eqref{eq:Esymmetry}, the spectrum is an even function of $k_z$, and Fig.~\ref{fig:vortexspect}(c) shows that the dispersion on the axial momentum $k_z$ is gapless. Since no topological invariant is defined in the $v$-vortex, however, all the zero energy states emergent in the $v$-vortex are accidental and do not satisfy the Majorana condition. Hence, the accidental zero energy states may be fragile against any perturbations. 

As shown in Fig.~\ref{fig:vortexspect}(d), indeed, the accidental gapless branches observed in the axisymmetric $v$-vortex are gapped out in the nonaxisymmetric $v$-vortex by the hybridization of $k_{\phi}$ and $k_{-\phi}$ eigenstates. The axisymmetric $v$-vortex has the well-defined quantum number $k_{\phi}$, and the symmetric relation in Eq.~\eqref{eq:Esymmetry} indicates states satisfying $E(-k_{\phi},k_z) = -E(k_{\phi},k_z)$. The nonaxisymmetric $v$-vortex, however, spontaneously breaks the ${\rm U}(1)_Q$ symmetry and $k_{\phi}$ is ill-defined, and the twofold deformation of the vortex core gives rise to the hybridization of $k_{\phi}$ eigenstates with $k_{\phi}+2m$ ($m\in\mathbb{Z}$). 

Similarly to the axisymmetric $v$-vortex, the $uvw$-vortex without any discrete symmetries possesses many gapless branches crossing the zero energy~\cite{silaevJETP09}. Since the second Chern number becomes zero and no additional topological invariant is defined, zero energy states in the $uvw$-vortex may be fragile against disturbances, such as impurities and magnetic fields.

The low-lying quasiparticle structure in the axisymmetric $v$-vortex was first clarified by Volovik~\cite{volovikJPCM91} within the semiclassical approximation. He introduced an index that characterizes the spectrum asymmetry,
\beq
N(k_{\phi},k_z) = {\rm Tr}\int \frac{d\omega}{2\pi}{G}
= - \frac{1}{2}\sum _n {\rm sgn}E_n(k_{\phi},k_z),
\label{eq:nvolovik}
\eeq
where ${\rm Tr}$ denotes the sum over all eigenstates at a given $(k_{\phi},k_z)$ and ${G}\equiv [i\omega - \mathcal{H}(k_{\phi},k_z,\rho,-i\partial _{\rho})]^{-1}$ is the Matsubara Green's function. The final expression in Eq.~\eqref{eq:nvolovik} implies that the index $N(k_{\phi},k_z)$ gives the difference between the number of positive and negative eigenstates for a given $(k_{\phi},k_z)$. The index has an abrupt jump at a particular $(k_{\phi},k_z)$ where zero energy states exist. 

The index \eqref{eq:nvolovik} can be connected to the number of zeros of the semiclassical Hamiltonian, which is obtained by replacing $-i\partial _{\rho}$ with the classical wavenumber $k_{\rho}$ in the full quantum $\mathcal{H}(k_{\phi},k_z,\rho,-i\partial _{\rho})$~\cite{volovikJPCM91}. The index \eqref{eq:nvolovik} with the semiclassical Hamiltonian provides a tractable way of determining the qualitative structure of low-lying quasiparticles bound to exotic vortices, such as the number of gapless branches. Although the number of gapless branches associated with the index remains well-defined, the zeros in the semiclassical Hamiltonian are not necessarily topologically protected and may be fragile when quantum corrections are taken into account. Hence, the one-dimensional winding number $w_{\rm 1d}$ gives a more rigorous topological invariant for the distribution of zero-energy eigenstates in $P_3$-preserving vortices.


\subsection{Continuous vortices and quasiparticles in $^3$He-A}
\label{sec:continuous}

{\it Continuous vortices.---}
As mentioned in Sec.~\ref{sec:vortex}, the vortex states realized in the bulk $^3$He-A are essentially different from those in $^3$He-B in the sense that the spontaneous breaking of the gauge-orbital symmetry gives rise to the intrinsic coupling between the ${\rm U}(1)$ phase and $\hat{\bm l}$. This feature inherent to the ABM state that the nontrivial $\hat{\bm l}$-texture can generate the superfluid velocity even without the phase winding is represented by the Mermin-Ho relation~\cite{merminPRL76}
\beq
{\bm \nabla}\times {\bm v}_{\rm s} = \frac{\hbar}{4m}\epsilon _{\mu\nu\eta}
\hat{l}_{\mu}\left( 
{\bm \nabla}\hat{l}_{\nu}\times{\bm \nabla}\hat{l}_{\eta}.
\right)
\eeq
When $^3$He-A spins up, continuous vortices become energetically competitive with singular vortices having a definite core. We here demonstrate that the continuous vortices are an obstacle to realize Majorana fermions. The most tractable way of removing such an obstacle is to confine the $^3$He-A to a restricted geometry, such as parallel plates. The $\hat{\bm l}$-vector is seriously affected by the surface depairing effect and forced to be normal to the surface. 

A representative of $N=0$ continuous vortices can be generated by continuously bending the uniform $\hat{\bm l}$-texture as  $\hat{\bm l}= \sin \eta (\rho)\hat{\bm z}+\cos\eta(\rho)\hat{\bm \rho}$, where $\eta(\rho)$ denotes the rotation angle of $\hat{\bm l}$ as a function of the distance from the vortex center $\rho$. The corresponding order parameter is given by 
\beq
d_{\mu i}({\bm r})=\Delta _{\rm A}(\rho) \hat{d}_{\mu}e^{i\phi}
\left( \cos\eta(\rho) \hat{\bm \rho} -\sin\eta (\rho) \hat{\bm z} + i\hat{\bm \phi}\right)_i .
\eeq
The order parameter indicates that the $\hat{\bm l}$-vector yields the radially flare-out texture when
\beq
\hat{\bm l}= \cos \eta (\rho)\hat{\bm z}+\sin\eta(\rho)\hat{\bm \rho}, 
\eeq
where $(\hat{\bm \rho},\hat{\bm \phi},\hat{\bm z})$ denotes the triad in the cylindrical coordinates. The textural structure generates the continuous distribution of the superfluid velocity field
${\bm v}_s(\rho) = \frac{\hbar}{M\rho}[1-\cos\eta(\rho)]\hat{\bm \phi}$,
and eliminates the singular behavior at the vortex center if $\eta(\rho)$ is a monotonic function of $\rho$. The continuous vortex with the boundary conditions $\eta(0)=0$ and $\eta(\infty)=\pi$ is called the Anderson-Toulouse or Anderson-Toulouse-Chechetkin vortex~\cite{andersonPRL77,chechetkin}. Another important configuration of $N=0$ continuous vortices is obtained as $\eta(0)=0$ and $\eta(\infty)=\pi/2$, which is called the Mermin-Ho vortex~\cite{merminPRL76}.

{\it Rotating $^3$He in a narrow cylinder.---}
As we already mentioned, owing to the boundary condition at the walls of the container, the $\hat{\bm l}$ field is forced to be perpendicular to the wall. This implies that, if $^3$He-A is confined to a restricted geometry, the surface boundary condition may induce a nontrivial textural structure. 

The ISSP group has recently succeeded in observing the textural transformation, and the formation and annihilation of a single vortex in rotating $^3$He-A confined to a narrow cylinder~\cite{ishiguroPRL04,ishiguro09}. They utilized a bundle or a single narrow cylinder whose radius is about 10 times larger than the dipole coherence length $\xi _{\rm D}\sim 10 \mu {\rm m}$. The key observation is that the spinning up $^3$He-A in a narrow cylinder undergoes a textural transition at a critical rotation speed.

The competing textural states in a narrow cylinder are the radial disgyration where the $\hat{\bm l}$ field lies in the $xy$-plane as $\hat{\bm l}=\hat{\bm \rho}$~\cite{volovikJETP77}. The order parameter is given by 
\beq
d_{\mu i}({\bm r})=\Delta _{\rm A}(\rho) \hat{d}_{\mu}(\hat{\bm z}-i\hat{\bm \phi})_ie^{i\kappa\phi},
\eeq 
where $\kappa$ gives the amplitude of the circulation, ${\bm v}_s=\kappa\frac{\hbar}{M\rho}\hat{\bm \phi}$, and a state with odd (even) $\kappa$ is categorized into $N=0$ ($N=1$). In contrast to continuous vortices, the superfluid velocity field has a singularity at $\rho = 0$ for nonzero vorticity at which the ABM order parameter must vanish.

\begin{figure}[tb!]
\begin{center}
\includegraphics[width=80mm]{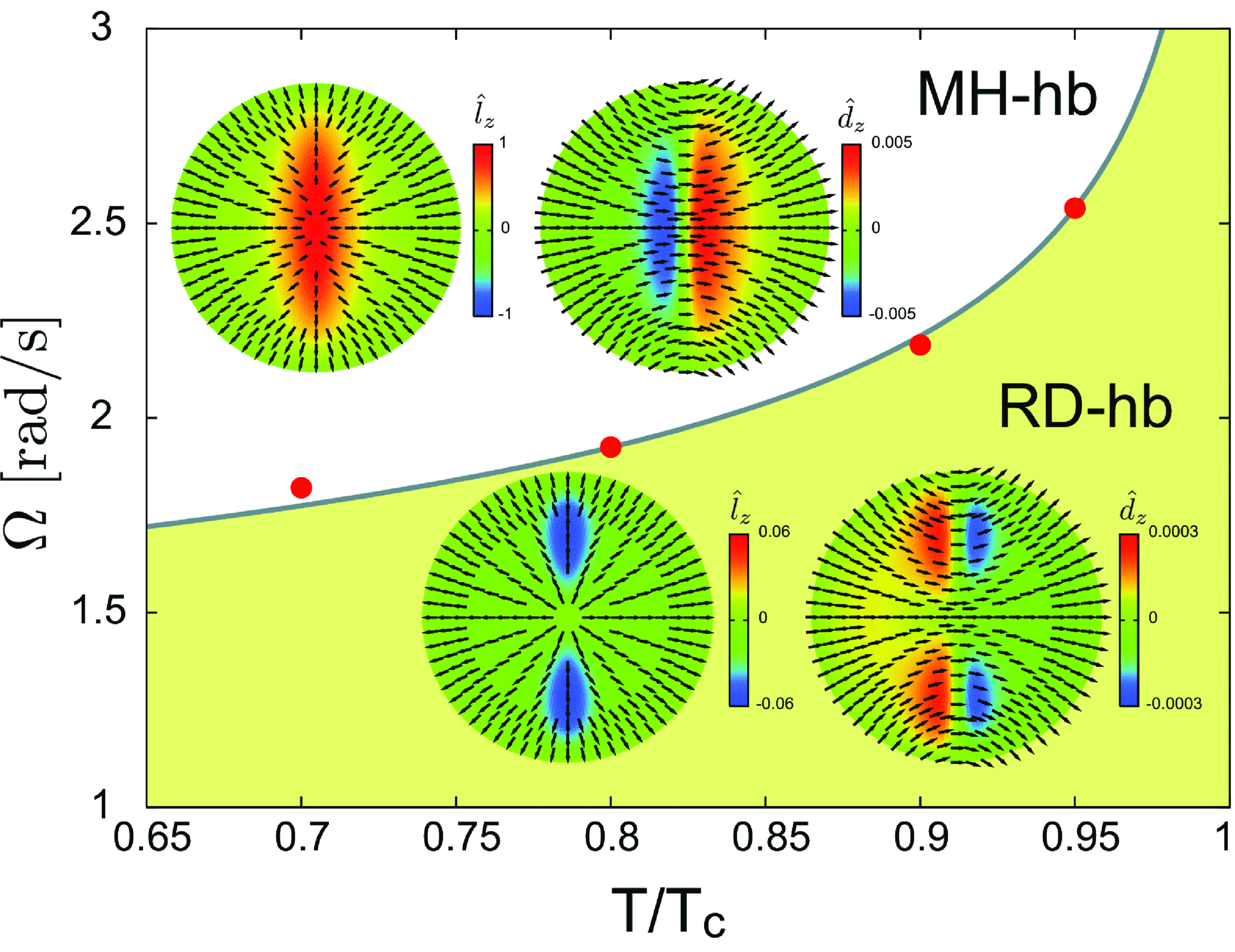} 
\end{center}
\caption{(Color online) Phase diagram of rotating $^3$He-A confined in a narrow cylinder with a radius $R=50$ $\mu{\rm m}$, where we set a pressure of 3.2 MPa and a magnetic field of 21.6 mT parallel to the cylinder. ``RD'' and ``MH'' denote the radial disgyration and Mermin-Ho texture of $\hat{\bm l}$, respectively, and ``hb'' is the hypobalic texture of $\hat{\bm d}$. The textures of the $\hat{\bm l}$-vector (left) and $\hat{\bm d}$-vector (right) are shown in the insets, whose arrows indicate the $x$- and $y$-components and the color map denotes the $z$-component. Figures adapted from Ref.~\citeonline{tsutsumiJPSJ09}.
}
\label{fig:RDMH}
\end{figure}

The temperature-rotation phase diagram of the rotating $^3$He-A confined in a narrow cylinder with a radius $R=50$ $\mu$m is displayed in Fig.~\ref{fig:RDMH}. This is obtained from the GL theory under a pressure of 3.2 MPa and a magnetic field of 21.6 mT parallel to the cylinder. At rest or under low rotation speeds, the radial disgyration with $\kappa=0$ is stable while the Mermin-Ho vortex generating the superfluid velocity field is formed under high rotations.
The critical rotation speed is enhanced toward $T_{\rm c}$ because the coherence length $\xi $ diverges and the bending energy of $\hat{\bm l}$, which is on the order of $\ln(R/\xi)$, becomes low at $T\rightarrow T_{\rm c}$. The phase diagram is helpful for understanding the recent experiment, where the radial disgyration and the Mermin-Ho texture are clearly identified by NMR spectra~\cite{kunimatsu:2014}.
By cooling from the normal state to the A-phase under rotation, the radial disgyration is observed irrespective of the rotation speed up to $\Omega=8$ rad/s, which is the maximum rotation speed in the experiment~\cite{kunimatsu:2014}.
Note that the radial disgyration does not change to the Mermin-Ho texture by the rotation at low temperatures, which reflects the difficulty of the first-order transition accompanied by the annihilation of a singularity.
The stable Mermin-Ho texture can be observed by warming from the B-phase to the A-phase under high rotation speeds.

{\it Low-lying quasiparticles bound to continuous vortices.}---
The Mermin-Ho vortex can be continuously transformed from the spatially uniform $\hat{\bm l}$-vector, since both states are categorized into the same topological class $N=0$. The Mermin-Ho vortex does not have any topological defects and thus the low-lying quasiparticle spectrum is naively expected to be the same as that of the bulk ABM state with the locally oriented $\hat{\bm l}(\rho)$. Contrary to the speculation, it was demonstrated that the continuous Mermin-Ho vortex can host low-lying quasiparticle excitations~\cite{ichiokaPRB10} as a consequence of the interplay between the vortex winding and the orbital angular momentum of Cooper pairs in the ABM state. The low-lying fermionic excitations associated with this remarkable Mermin-Ho texture are intriguing even though they are not topologically protected, because they faithfully and directly reflect the underlying topological structure of the order parameter.

In addition to a narrow cylinder, the Mermin-Ho vortex is a building block to forming a periodic array of continuous vortices in the bulk $^3$He-A under rapid rotation. The periodically arrayed Mermin-Ho vortices were first proposed by Fujita {\it et al.}~\cite{fujitaPTP78} within the GL theory, where the lattice maintains the background A-phase throughout the whole system without forming any singular points. The thermodynamic stability within the GL theory was quantitatively studied by Karim\"aki and Thuneberg~\cite{fujitaPTP78,karimakiPRB99} and Kita~\cite{kitaPRL01,kitaPRB02}.

\begin{figure}[tb!]
\begin{center}
\includegraphics[width=80mm]{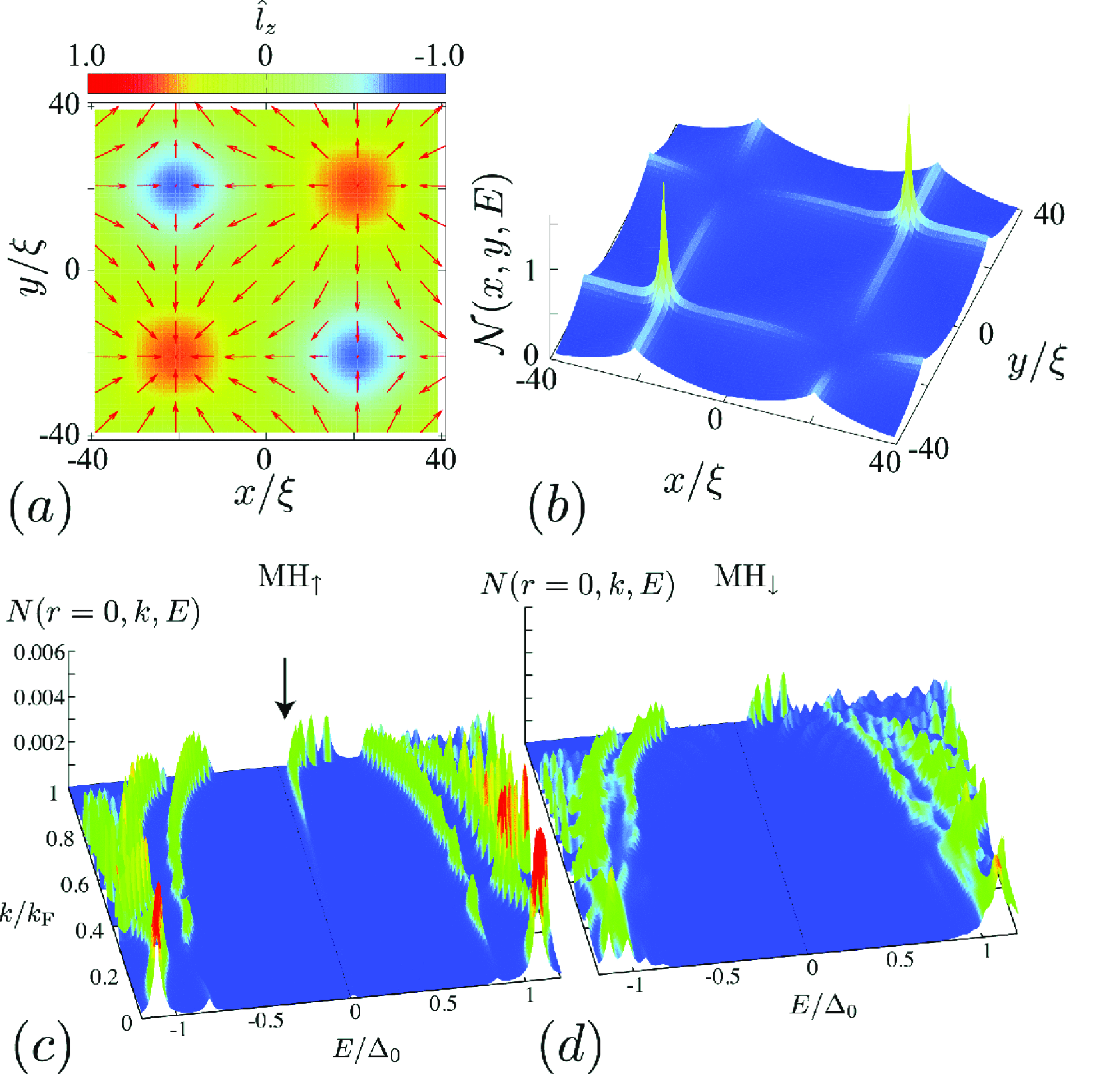} 
\end{center}
\caption{(Color online) Continuous vortex lattice formed by Mermin-Ho (MH) vortices; Spatial structure of $\hat{\bm l}$-vector (a) and zero-energy local density of states $\mathcal{N}(x,y,E=0)$ (b) within a unit cell obtained by self-consistent Eilenberger theory, where $\Omega \!=\! 0.004T_{\rm c}$ and $T\!=\! 0.9T_{\rm c}$. The unit cell is composed of two-different types of MH vortices; In MH$\uparrow$, $(\hat{l}_x, \hat{l}_y)$ has flare-in or flare-out disgyration and $\hat{l}_z$ always points to $+\hat{\bm z}$, while MH$\downarrow$ has mixt-twist disgyration and $\hat{l}_z$ points to $-\hat{\bm z}$. Full quantum-mechanical calculation of the axial-momentum-resolved local density of states at the vortex center of MH$\uparrow$ (c) and MH$\downarrow$ (d). Figures adapted from Ref.~\citeonline{ichiokaPRB10}.
}
\label{fig:MHvortex}
\end{figure}

The spatial structure of the continuous vortex lattice is shown in Fig.~\ref{fig:MHvortex}(a), which is obtained by self-consistently solving the quasiclassical Eilenberger equation coupled to the gap equation~\cite{ichiokaPRB10}. The unit cell of the periodic lattice is composed of two different types of the Mermin-Ho vortices centered at $(x/\xi,y/\xi)=(\pm 20,\pm 20)$ and two ``mixt-twist'' vortices at $(x/\xi,y/\xi)=(\pm 20,\mp 20)$. In the Mermin-Ho vortices, the $\hat{\bm l}$-texture projected onto the $xy$-plane yields the flare-out and flare-in configurations and the core has $\hat{\bm l}\parallel\hat{\bm z}$. The mixt-twist vortex is characterized by the quadrupole $\hat{\bm l}$-texture in the $xy$-plane with $\hat{\bm l}$ pointing to the $-\hat{\bm z}$-direction at the vortex center. In the whole region of the unit cell, the $\hat{\bm l}$-texture is smoothly distributed without forming any singularities.

Figure~\ref{fig:MHvortex}(b) shows the spatial profile of the zero-energy local density of states, $\mathcal{N}({\bf r},E=0)$.
It turns out that the Mermin-Ho and mixt-twist vortices possess distinctive low-lying excitation spectra owing to the different orientations of $l_z$. The Mermin-Ho vortices with $\hat{l}_z > 0$ have a sharp peak of $\mathcal{N}({\bf r},E=0)$ similar to that observed in singular vortices~\cite{ichiokaPRB97,ichiokaPRB02}, while no peak structure is observed in the mixt-twist vortices. In Figs.~\ref{fig:MHvortex}(c) and \ref{fig:MHvortex}(d), we plot the $k_z$-resolved local density of states at the vortex core, $\mathcal{N}(k_z,{\bf r},E)$, where $k_z$ denotes the axial momentum along the $\hat{\bm z}$-axis. To understand the characteristic low-lying structure, let us consider the effective pair potential $\Delta({\bm k},{\bm r})$ with  $\phi_k \rightarrow \phi+\pi/2$ that the quasiparticle is subjected to. The effective potential for the Mermin-Ho vortex is then given by 
\beq
|\Delta({\bm k},\rho)|^2 = 6|\Delta _0| (\sin^2\theta _k +\sin^2 \eta(\rho) \cos^2 \theta _k).
\eeq
This is an increasing function of ${\rho}$, and quasiparticles are subjected to the ``confinement'' potential around the core of the Mermin-Ho vortex when $|k_z| \propto |\cos\theta_k|$ is larger. On the other hand, the potential for the mixt-twist vortex is recast into 
\begin{align}
|\Delta({\bm k},\rho,\phi)|^2
\propto & \sin^2\theta_k 
-\frac{1}{2}\sin 2 \phi \sin 2 \eta(\rho) \sin 2\theta _k  \nn \\
& +\sin^2\beta(\cos^2\theta_k-\sin^2 2 \phi \sin^2 \theta _k ).
\end{align}
This effective potential breaks the continuous rotational symmetry about the $\hat{\bm z}$-axis and does not have a minimum point at vortex center. This implies that no bound states are formed inside the gap in the case of the mixt-twist vortex, as shown in Fig.~\ref{fig:MHvortex}(d). This semiclassical argument clearly explains that the low-energy peak only for the Mermin-Ho vortex with positive $\hat{l}_z$ grows as $|k_z|$ increases. Notice that the energy level bound to the Mermin-Ho vortex is roughly given as $(n+\frac{1}{2})\Delta^2/E_{\rm F}$ ($n\in\mathbb{Z}$) and cannot be exactly zero.

The essence of the origin of low-lying quasiparticles is understandable from the interplay between the orbital angular momentum of Cooper pairs and the vortex winding of each orbital component. The order parameter $C_{mn}({\bm r})$ based on the eigenstates of $\hat{L}_z$ and $\hat{S}_z$ has a definite angular momentum $\lambda^{(m)} = 0, \pm 1$, depending on the orbital state. In addition, the order parameter is factorized as $C_{mn}({\bm r})=C_{mn}(\rho)e^{i\kappa _n\phi}$ with $(\kappa _{+1}, \kappa _{0}, \kappa _{-1})=(0,1,2)$ for the Mermin-Ho vortex and $(2,1,0)$ for the mixt-twist vortex. This allows one to introduce the following symbolic algebra for the phase factors: $(\kappa _{+1}, \kappa _{0}, \kappa _{-1})+(\lambda^{(+1)},\lambda^{(0)},\lambda^{(-1)}) =(0,1,2)+(1,0,-1)\rightarrow (1,1,1)+(0,0,0)$ for the Mermin-Ho vortex and $(2,1,0)+(1,0,-1)\rightarrow (1,1,1)+(2,0,-2)$ for the mixt-twist vortex. The former $(0,0,0)$ gives rise to a vortex bound state similar to the singular hard core vortex as in the Caroli-de Gennes-Matricon state,~\cite{CdGM} whereas the latter $(2,0,-2)$ yields the fourfold symmetric angle dependence without any confinement, $e^{i2\phi}+e^{-i2\phi}\propto \cos 4\phi$.

\subsection{Half-quantum vs integer-quantum vortex in $^3$He-A thin film}
\label{sec:HQV}

\begin{figure}[tb]
\begin{center}
\includegraphics[width=80mm]{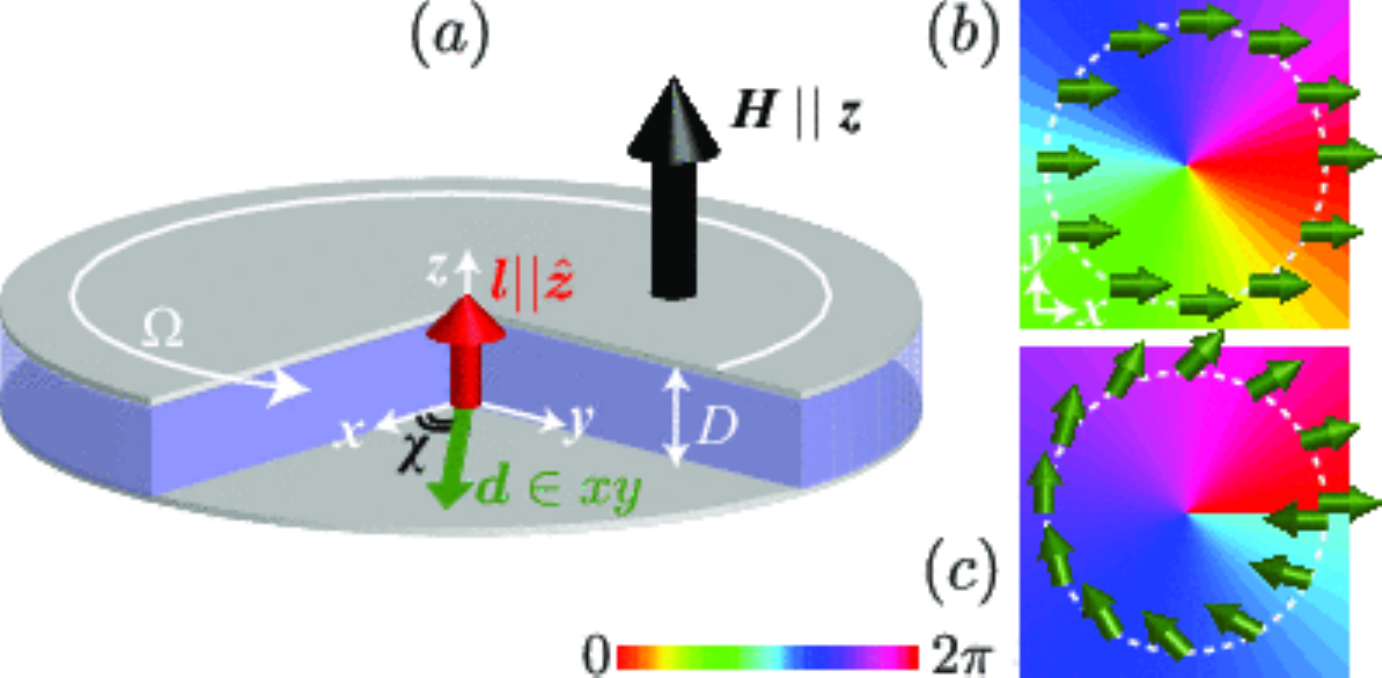}
\end{center}
\caption{(Color online) 
(a) Schematic picture of a parallel plate geometry relevant to realizing Majorana fermions in $^3$He-A. Integer quantum vortex (b) and half-quantum vortex (c), where the color map and arrows denote the ${\rm U}(1)$ phase and $\hat{\bm d}$ profiles, respectively.} 
\label{fig:geometry_pp}
\end{figure}

As discussed in Sec.~\ref{sec:continuous}, the formation of continuous vortices is an obstacle to realizing non-Abelian Majorana fermions. The $^3$He-A phase with a uniform ${\bm l}$-texture, however, can be realized in a parallel plate geometry with thickness $D\ll\xi_d$, as shown in Fig.~\ref{fig:geometry_pp}. In this geometry, the motion of Cooper pairs is confined in the $xy$-plane and the $\hat{\bm l}$-vector is locked perpendicular to the plates. Applying a magnetic field perpendicular to the plates ($\bm{H}\parallel\hat{\bm z}$ and $|\bm{H}|>H_d$) further imposes a strong constraint on $\hat{\bm d}$, where the magnetic field energy is minimized by $\hat{\bm d}\perp\hat{\bm z}$ as in Eq.~\eqref{eq:quad}. 

The parallel plate sample with $D=12.5$ $\mu$m and $R=1.5$ mm was provided by an experimental group in ISSP, Univ. of Tokyo~\cite{yamashitaPRL08}. They observed that the condition $\hat{\bm l}\perp\hat{\bm d}$ is satisfied in their sample by comparing the characteristic NMR frequency shift of the parallel plate sample with that of bulk samples~\cite{yamashitaPRL08}. Searching for a half-quantum vortex (HQV), they rotated the parallel plates up to the angular velocity $\Omega$ of $12$ rad/s in a rotating cryostat at ISSP, but no conclusive evidence for HQVs was observed.

We here mention that in such a geometry, the gauge-orbital symmetry breaking in $^3$He-A gives rise to the formation of a HQV, which turns out to be energetically competitive with an integer quantum vortex (IQV). We will show the stable condition of HQVs against IQVs in a realistic situation, by taking into account the vortex core structures within the length scale $\xi\sim 0.1$ $\mu$m. We start from the GL theory without Fermi liquid corrections~\cite{salomaaPRL85,keePRB00,chungPRL07,vakaryukPRL09} and then discuss their roles in the stability of HQVs.

\subsubsection{Half-quantum vortex}

The generic order parameter for the most symmetric vortex is given by Eq.~\eqref{eq:vortexgeneric}. The vortex state is subject to the boundary condition at $\rho \rightarrow \infty$ where the chiral pairing with $\hat{\bm l}\parallel\hat{\bm z}$ and the definite vorticity $\kappa$ is assumed as
\beq
d_{\mu i}(\infty) = \Delta _{\rm A} \hat{d}_{\mu}e^{i\kappa \phi }(\hat{\bm x}+i\hat{\bm y})_i.
\label{eq:op_pp}
\eeq
The motion of the Cooper pairs is confined to the $xy$-plane, namely, $C_{m0}=C_{0n}=0$. Hence, the order parameters relevant to this situation are given by the set of $(C_{++},C_{+-},C_{-+},C_{--})$. 

As shown in Sec.~\ref{sec:abm}, the remarkable consequence of the chiral $p$-wave pairing is the spontaneous breaking of the gauge-orbital symmetry. This reflects the fact that the order parameter in Eq.~(\ref{eq:op_pp}) is invariant under the combined discrete rotation of $\hat{\bm d}$ and $\phi$ as $\hat{\bm d}\mapsto-\hat{\bm d}$ and $\kappa \phi \mapsto \kappa \phi +\pi$. Then, as shown in Fig.~\ref{fig:geometry_pp}, there are two classes for the vorticity $\kappa$: IQVs with $\kappa = \mathbb{Z}$ [Fig.~\ref{fig:geometry_pp}(b)] and HQVs with a fractional vorticity $\kappa = \mathbb{Z}/2$ [Fig.~\ref{fig:geometry_pp}(c)]. In the HQV, both the ${\rm U}(1)$ phase and $\hat{\bm d}$ rotate by $\pi$ about the vortex center. The abrupt $\pi$ phase jump in the ${\rm U}(1)$ phase is compensated by the flip of $\hat{\bm d}$ and thus the resultant order parameter maintains its single-valuedness. The IQV is accompanied by only the mass flow around the vortex core, whereas the half-integer value of $\kappa$ and the texture of $\hat{\bm d}$ are responsible for the mass and spin currents, which are half of the mass flow of IQVs.


The asymptotic form of the order parameter of the HQV is given by Eq.~\eqref{eq:op_pp} with $\kappa = \pm 1/2$, while the IQV has $\kappa = \pm 1$. The $\hat{\bm d}$-vector of the HQV is recast into 
\beq
\hat{\bm d}({\bm r}) = \cos (\kappa^{\rm sp}\phi) \hat{\bm x} + \sin (\kappa^{\rm sp}\phi)\hat{\bm y},
\label{eq:dvecHQV}
\eeq 
where $\kappa^{\rm sp}$ denotes the winding of $\hat{\bm d}$. The HQV is characterized by $\kappa^{\rm sp}=1/2$, while the IQV has $\kappa^{\rm sp}=0$. It is remarkable to notice that since the ABM state is the equal spin pairing state, the order parameter for the HQV in the spin basis is given by
\begin{eqnarray}
{\Delta}(\infty) = \Delta _{\rm A}\left(\hat{k}_x+i\hat{k}_y\right)
\left[
\left|\uparrow\uparrow\right\rangle
+ e^{i\phi}\left|\downarrow\downarrow\right\rangle
\right]. 
\end{eqnarray}
This indicates that the $\uparrow\uparrow$-pair $C_{++}$ possesses the spatially uniform phase, while the $\downarrow\downarrow$-pair $C_{-+}$ has phase winding of $2\pi$ around the vortex as in conventional singly quantized vortices. Thus, the vortex-free state in the $\uparrow$ spin sector exhibits fully gapped quasiparticle excitations, while the low-lying structures in HQVs are effectively describable with a singly quantized vortex in the $\downarrow$ spin sector, namely, the spin-polarized chiral $p$-wave system. As discussed in Sec.~\ref{sec:spinless}, an odd-vorticity vortex in the spin-polarized chiral system hosts Majorana fermions that obey non-Abelian statistics. The existence of non-abelian anyonic zero modes in HQVs was first realized by Ivanov~\cite{ivanovPRL01}, who developed the non-Abelian braiding statistics of vortices with Majorana zero modes.

\subsubsection{Energetics within the Ginzburg-Landau theory}

To discuss the energetics in rotating $^3$He confined to parallel plates, one has to take into account the gradient energy in the GL energy functional \eqref{eq:fgl}. Since the normal $^3$He holds the ${\rm SO}(3)$ symmetry in the coordinate and spin spaces, the gradient energy terms relevant to $^3$He rotating with $\Omega$ are obtained by taking a contraction of spin and orbital indices as 
\begin{eqnarray}
f_{\mathrm{grad}} = K_1\partial _{i}^*d_{\mu j}^*\partial _{i}d_{\mu j} + K_2\partial _{i}^*d_{\mu j}^*\partial _{j}d_{\mu i} + K_3\partial _{i}^*d_{\mu i}^*\partial _{j}d_{\mu j}, 
\label{eq:fgrad}
\end{eqnarray}
where $\mu,i,j=x,y,z$, $\partial_i \equiv \nabla _i-(2im_3/\hbar)(\bm\Omega \times \bm r)_i$. In the weak coupling limit, the coefficients obey $K_1 = K_2 = K_3 = K =\frac{7\zeta (3)N(0)(\hbar v_F)^2}{240(\pi k_B T_c)^2}$.
We determine the thermodynamic stability of IQVs and HQVs by minimizing the total GL energy functional
$\mathcal{F}_{\rm GL} = \int d{\bm r} ( f_{\rm bulk} +f_{\rm grad} + f^{(1)}_{\rm mag} + f^{(2)}_{\rm mag}.
+ f_{\rm dip})$.

When the order parameter is spatially inhomogeneous, Eq.~\eqref{eq:fgrad} indicates that the so-called gradient coupling induces the minor order parameter components $C_{\pm, -}$ around the core and edge. 
For an axisymmetric vortex, the ${\rm U}(1)_Q$ symmetry imposes the following selection rule for the winding numbers:
\begin{eqnarray}
\kappa _{m, n-1} = \kappa _{m,n} + 1,
\end{eqnarray}
where $\kappa _{mn}$ denotes the vortex winding numbers of each order parameter component $C_{mn}$. Several possible textures exist with different combinations of $\kappa _{mn}$. When the phase winding of the bulk dominant component $C_{\pm +}$ has the counterclockwise sense $\kappa _{\pm+}\ge0$, the possible textures are the A-phase texture (AT) with $(\kappa _{\pm +}, \kappa _{\pm -})=(0,2)$, IQVs with $(\kappa_{\pm+}, \kappa_{\pm-})=(1,3)$, and HQVs with $(\kappa _{++},\kappa _{-+},\kappa _{+-},\kappa _{--})=(0,1,2,3)$.

The relative stability between HQVs and IQVs has been discussed by a number of authors~\cite{crossJLTP77,salomaaPRL85,salomaaRMP87,volovik}. Among them, Salomaa and Volovik first revealed the important role of the Fermi liquid parameter $F^{\rm a}_1$, which denotes the renormalization of the molecular field associated with the spin current. They found that the Fermi liquid correction stabilizes the pairwise HQVs in a rotation regime. The hydrodynamical calculation that takes into account the Fermi liquid correction shows that the HQV is energetically stable against the IQV~\cite{crossJLTP77,salomaaPRL85,chungPRL07,vakaryukPRL09}. However, we emphasize that all the hydrodynamical theories do not take into account the strong coupling corrections. Contrary to the hydrodynamical results, Kawakami {\it et al.}~\cite{kawakamiPRB09,kawakamiJPSJ10, kawakamiJPSJ11} demonstrated the critical effect of strong coupling corrections in $\beta _i$, which are indispensable for the relative stability of the bulk ABM state to the BW state at high pressures. The results indicate that the strong coupling corrections favor IQVs and are obstacles to the stability of HQVs. Following Ref.~\citeonline{kawakamiPRB09}, we  give here an overview of the quantitative calculations on the stability of HQVs in rotating $^3$He-A in the presence of strong coupling corrections but no Fermi liquid corrections. The effect of Fermi liquid corrections will also be commented on.  

\begin{figure}[tb]
\begin{center}
\includegraphics[width=80mm]{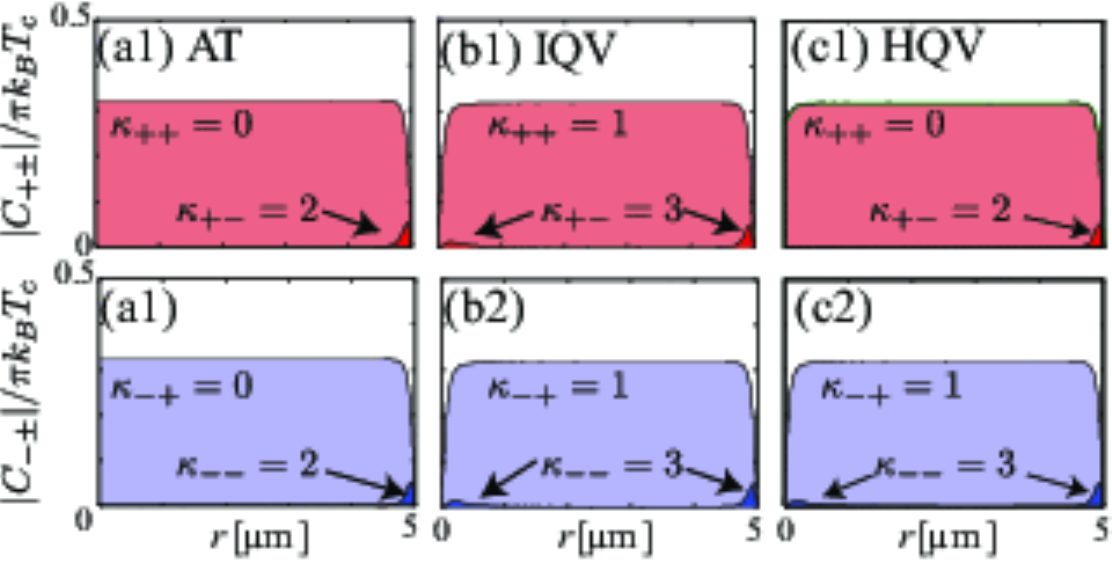}
\end{center}
\caption{(Color online) 
Spatial profiles of the order parameters for AT (a), IQV (b), and HQV (c). Upper and lower panels show the $C_{+\pm}$ and $C_{-\pm}$ pair components, respectively, where we set the system size $R=5$ $\mu$m and the temperature $T=0.95T_{\rm c}$. $\kappa _{\pm\pm}$ denotes the phase winding of $C_{\pm\pm}$.
}
\label{fig:hqvtex}
\end{figure}

\begin{figure}[tb]
\begin{center}
\includegraphics[width=80mm]{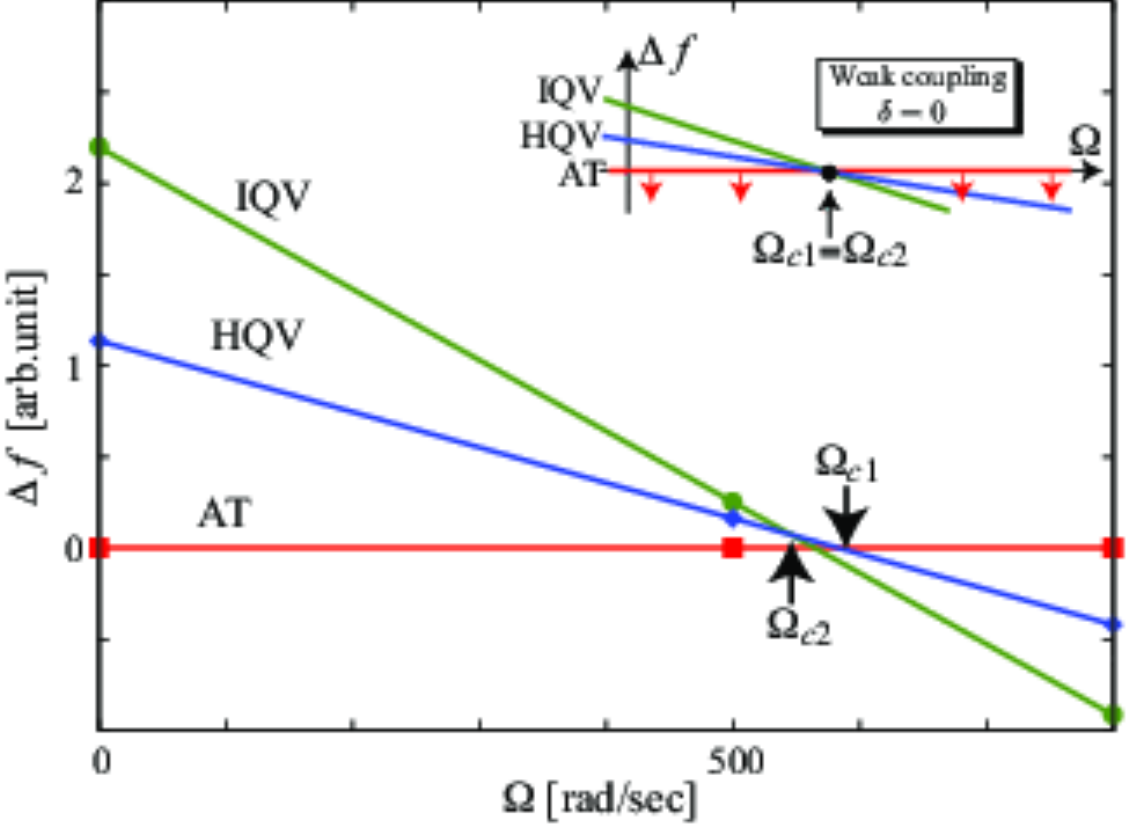}
\end{center}
\caption{(Color online) Free energies $\Delta f$ for the HQV and IQV states relative to that of the AT state as a function of $\Omega$ for $R=10$ $\mu{\rm m}$ and $T/T_{\rm c}=0.97$. In the weak coupling limit, the free energies of AT, HQV, and IQV intersect at one point $\Omega _{\rm c1}=\Omega _{\rm c2}$ (inset). Strong coupling effects lower the energy of only the AT state as denoted by red arrows in the inset. Figures adapted from Ref.~\citeonline{kawakamiPRB09}.}
\label{fig:hqv-df-omg}
\end{figure}

{\it Weak magnetic field regime}---
Let us start from a weak magnetic field regime $ H_z \ll \alpha/\eta $, where $C_{+\pm}$ and $C_{-\pm}$ have the same amplitude in the bulk. Figure~\ref{fig:hqvtex} shows the numerically obtained spatial profiles of the order parameters for AT (the spatially uniform ABM state), IQV, and HQV. In the region far away from the vortex center，$C_{+\pm}$ in the HQV [Fig.~\ref{fig:hqvtex}(c1)] has the same profile as that of AT in Fig.~\ref{fig:hqvtex}(a1), while $C_{-\pm}$ in Fig.~\ref{fig:hqvtex}(c2) has the IQV-like one as in Fig.~\ref{fig:hqvtex}(b2). The core of the HQV is filled in by the A$_1$ phase where only the $\uparrow\uparrow$ pair exists. In Fig.~\ref{fig:hqv-df-omg}，we compare the free energies of these three phases as a function of the angular velocity $\Omega$. However, there is no stable region of the HQV in the weak magnetic field regime $H_z \ll \alpha/\eta$. 

For the weak coupling limit, the GL free energy is completely decomposed into $C_{+n}$ and $C_{-n}$ sectors. To understand the relative stability, we consider the London limit disregarding the vortex core effect, which sets the amplitudes of $C_{mn}$ to be spatially constant. The amplitude of the $\hat{\bm d}$-vector is then given as 
$\left|\hat{\bm d}\right|=1$ for $r\ge\xi_0$ and $\left|\hat{\bm d}\right|=0$ for $r<\xi_0$. In the weak coupling limit and London limit, the energy differences of AT, HQV, and IQV originate from the gradient energy,
\begin{eqnarray}\label{eq:grad-london}
f_\mathrm{grad} = 2 K |\Delta_\mathrm{A}|^2|\hat{\bm d}|^2\left[\left(\bm{\nabla}\varphi + \bm{r}\times\bm{\Omega}\right)^2 + \left(\bm{\nabla}\alpha\right)^2\right].
\end{eqnarray}
The first and second terms can be understood as kinetic energies carrying mass and spin currents. By integrating Eq.~(\ref{eq:grad-london}) with Eq.~\eqref{eq:dvecHQV}, one obtains 
\beq
\int d {\bm r} f_{{\rm grad}} =
2 \pi K \left|\Delta_A\right|^2 \left\{ \left[ \kappa^2 
+ \left(\kappa^{\mathrm{sp}}\right)^2\right] \ln \frac{R}{\xi_0} - \frac{\kappa\Omega R^2}{2}\right\},
\label{eq:fgradint}
\eeq
where $R$ denotes the system size or vortex distance. The AT, HQV, and IQV are characterized by the set of winding numbers as $(\kappa,\kappa^{\rm sp})=(0,0)$, $(1/2,1/2)$, and $(1,0)$, respectively. Equation \eqref{eq:fgradint} shows that the kinetic energies of mass and spin currents are proportional to $\kappa^2$ and $(\kappa^{\rm{sp}})^2$, respectively. The mass current $\kappa$  couples to the external rotation $\Omega$ and thus decreases the energy as $\Omega$ increases in the weak coupling limit. Therefore, in the weak coupling limit, the free energies of AT, SV, and HQV intersect at one point and the HQV is never stabilized, as shown in the inset of Fig.~\ref{fig:hqv-df-omg}. 

Let us now clarify the strong coupling effect, which was first introduced by Anderson and Brinkman to explain the relative stability of the ABM state to the BW state in the bulk $^3$He. In the context of the GL theory, the correction deviates the coefficients $\beta _{i}(P)$ in Eq.~\eqref{eq:gl4th}. By substituting $\beta _i(P)$ described in Sec.~\ref{sec:abm} and employing the ansatz $C_{\pm,\pm }=\hat{d}_{\pm}C_{\pm}$, the bulk fourth-order term in Eq.~\eqref{eq:fgl} is recast into
\begin{eqnarray}\label{eq:bulk4th}
f_{\mathrm{bulk}} ^{(4)} = B_{\mathrm{d}}(|d_{+}|^2 + |d_{-}|^2) 
+ B_{\mathrm{c}}(|d_{+}||d_{-}|),
\end{eqnarray}
where
$B_{\mathrm{c}} = -\beta_0\delta[3.5(|C_{+}|^4+|C_{-}|^4)+9|C_+|^2|C_-|^2]$ and 
$B_{\mathrm{d}} = \beta_0[(4-0.35\delta)(|C_{+}|^4+|C_{-}|^4) + (16-0.55\delta)|C_+|^2|C_-|^2]$. It turns out that only the AT state, where both the OP components $|C_{+ \pm}|$ and $|C_{- \pm}|$ are finite at $r=0$，reduces the free energy in Eq.~(\ref{eq:bulk4th}) which is proportional to the spin fluctuation parameter $\delta$. As shown in Fig.~\ref{fig:hqv-df-omg}, therefore, the strong coupling effect in the weak coupling limit merely shifts the threshold rotation $\Omega_{\rm c1}$ beyond which the IQV is stabilized.

\begin{figure}[tb]
\includegraphics[width=80mm]{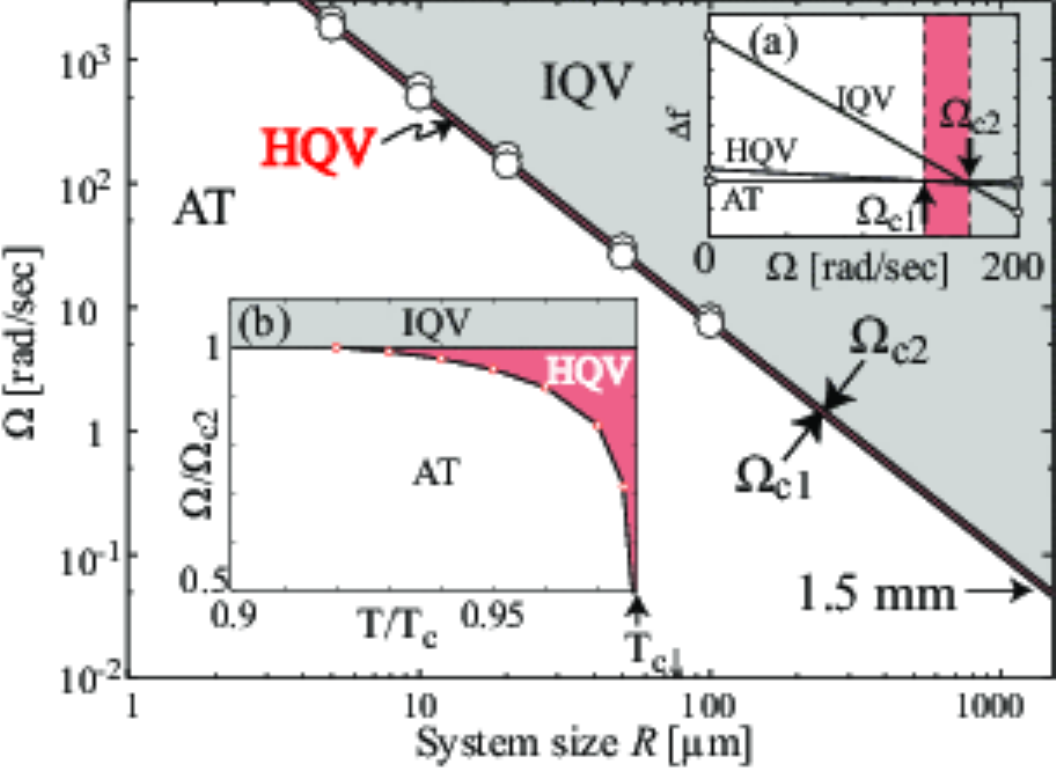}
\caption{(Color online) Phase diagram of rotating $^3$He-A in a parallel plate geometry. (Upper inset) Free energies for $R=20$ $\mu{\rm m}$. This shows the successive transitions from AT to HQV at $\Omega_{\rm c1}$ and from HQV to SV at $\Omega_{\rm c2}$. (Lower inset) Stable region of HQV in $\Omega$ versus $T/T_{\rm c}$ ($R=10$ $\mu {\rm m}$ and $\eta H_z/\alpha_0\!=\!0.05$). $T_{{\rm c}\downarrow}$ is the numerically estimated transition temperature for the $\downarrow\downarrow$-pair, where $A_{-\pm}\!=\!0$. Figures adapted from Ref.~\citeonline{kawakamiJPSJ10}.}
\label{fig:omg-t}
\end{figure}

{\it Strong magnetic field regime}---
In the strong magnetic field regime $H_z\sim\alpha/\eta$, the first-order Zeeman effect $f^{(1)}_{\rm mag}$ gives rise to the splitting of $T_{\rm c}$ for $\uparrow\uparrow$ and $\downarrow\downarrow$ pairs as $\Delta T_c \equiv T_{{\rm c},\uparrow}-T_{{\rm c},\downarrow}= \eta H_z/\alpha _0$. As shown in the upper inset of Fig.~\ref{fig:omg-t}, at $T/T_{\rm c}=0.97$ and $\Delta T_{\rm c}=0.05$, where the corresponding external magnetic field is a few kG, the HQV becomes stable in the range of $\Omega_{\rm c1}< \Omega <\Omega_{\rm c2}$ and we find that $\Omega_{\rm c1}/\Omega_{\rm c2}\simeq 0.85$. Notice that the strong coupling effect helps to stabilize the HQV under strong magnetic fields. The main panel of Fig.~\ref{fig:omg-t} shows the vortex phase diagram of the rotating $^3He$-A in the plane of the system size $R$ and angular velocity $\Omega$ at $T/T_{\rm c} = 0.97$. We observe that $\Omega_{\rm c1}/\Omega_{\rm c2}\simeq 0.85$ is insensitive to the change in $R$. By using this scaling, the estimated stable region of the single HQV state in the experiment by Yamashita {\it et al}.~\cite{yamashitaPRL08} is between $\Omega_{\rm c1}\sim 0.04$ rad/s and $\Omega_{c2}\sim 0.05$ rad/s. The rotation speed of the rotating cryostat at ISSP, Univ. Tokyo is sufficiently accurate to perform the experiment in this narrow stable region.

The window in which the HQV state is stabilized can be tuned by changing $R$ and $T$. When $R=300$ $\mu$m, for instance, one finds the critical speed $\Omega_{\rm c1}\sim 1$ rad/s which is accessible in the actual experimental setup for a rotating cryostat in ISSP. Figure~\ref{fig:omg-t}(b) shows the phase diagram spanned by  $\Omega$ and  $T$ for  $R=10$ $\mu {\rm m}$. In the high-$T$ regime, where $C_{-+}$ is suppressed by the first-order Zeeman effect, the stable region $\Omega_{\rm c2}-\Omega_{\rm c1}$ becomes wider as $T$ approaches $T_{{\rm c},\downarrow}$.

{{\it Fermi liquid correction}---}
Another important factor for the stability of HQVs is the Fermi liquid correction.~\cite{salomaaPRL85, chungPRL07, vakaryukPRL09} Using the hydrodynamic theory, one can treat the Fermi liquid correction as the effective mass of the spin current. As shown in Eq.~(\ref{eq:grad-london}), the London approximation divides the kinetic energy terms into the contributions associated with the mass and spin currents. The Fermi liquid correction generates the molecular field associated with the spin current density, which makes the effective mass of the spin current ($\rho _{\rm sp}$) less than that of the mass current ($\rho _{\rm s}$),
\beq
\rho _{\rm s} > \rho _{\rm sp}.
\eeq 
To take this discrepancy into account, Cross and Brinkman~\cite{crossJLTP77} introduced the phenomenological correction of the gradient energy as 
\begin{eqnarray}\label{eq:grad-FLcorr}
f_\mathrm{grad} = 2 K |\Delta_\mathrm{A}|^2\left\{\frac{\rho_\mathrm{s}}{\rho_\mathrm{s}^0}\left[\bm{\nabla}(\kappa\phi) + \bm{r}\times\bm{\Omega}\right]^2 + \frac{\rho_\mathrm{sp}}{\rho_\mathrm{s}^0}\left[\bm{\nabla}(\kappa^{\mathrm{sp}}\phi)\right]^2\right\}.
\end{eqnarray}
Now, we compare the energy of a pair of HQVs with that of a single IQV. By integrating Eq.~(\ref{eq:grad-FLcorr}), these gradient energies are recast into
\begin{eqnarray}\label{grad-london-integ}
\int d \bm{r} \frac{f_{\mathrm{grad}}}{2\pi K|\Delta_A|^2\rho_\mathrm{s}/\rho_\mathrm{s}^0}
=\left\{
\begin{array}{l} 
\frac{1}{2}\left(1 +\frac{\rho_\mathrm{sp}}{\rho_\mathrm{s}}\right)  
\ln\left(\frac{R}{\xi_{\sigma\sigma }}\right)
\ \mathrm{(HQV)}\\ \\
\ln\left(\frac{R}{\xi_{\sigma\sigma }}\right)  \ \mathrm{(IQV)}
\end{array}\right. 
\end{eqnarray}
when the system is at rest $\Omega=0$. It is remarkable that the gradient energy of the HQV state is contributed to by the spin current as well as the mass current, while the IQV state is not accompanied by the spin current. For $\rho _{\rm sp}<\rho _{\rm s}$, therefore, the gradient energy of the HQV state becomes lower than that of the IQV state.

Although the Fermi liquid correction is a key to stabilizing the HQV state, in order to discuss the correction, we apply the London approximation that $|C_{mn}({\bm r})|$ is spatially uniform and assume the pairing phase to be the A-phase even in the vortex core. As discussed above, however, the vortex core of the HQV is the A$_1$-phase rather than the pure A-phase. Since the effect of the Fermi liquid correction is in contrast to that of the strong coupling correction that favors the IQV state, the theory that takes into account both effects on an equal footing is indispensable for clarifying the quantitative phase diagram of rotating $^3$He-A. However, there has been no systematic way of taking into account both effects on an equal footing so far. The quantitative analysis including the Fermi liquid and strong coupling corrections on an equal footing remains as a future problem.

\subsubsection{Half-quantum vortices in superconductors}

The real-space topology of HQV was originally proposed by Volovik and Mineev~\cite{volovikJETP76}. Salomaa and Volovik~\cite{salomaaPRL85} and Kee {\it et al.}~\cite{keePRB00} found that the $\hat{\bm d}$-soliton structure in the pairwise HQVs possesses characteristic spin wave excitations, which may provide a way of identifying the HQV in experiments. In addition to their topologically nontrivial aspects, HQVs have recently attracted much more attention since they can preferentially accommodate non-Abelian Majorana fermions~\cite{ivanovPRL01}. Theoretical investigations are devoted to finding an energetically stable condition of HQVs in both superfluids and superconductors~\cite{salomaaPRL85,keePRB00,chungPRL07,kawakamiPRB09,kawakamiJPSJ10, kawakamiJPSJ11,vakaryukPRL09,kondoJPSJ12,nakaharaPRB13}.

In superconductors, however, it is difficult to stabilize the HQV state in a system of macroscopic size. Chung {\it et al.}~\cite{chungPRL07} pointed out that owing to the screening effect, the mass (charge) current exponentially decays in the length scale of the penetration depth $\lambda$. This implies that the gradient energy associated with the mass current is recast into $\ln{\lambda/\xi}$, and thus the logarithmic divergence of the gradient energy disappears in the IQV. In contrast, such a screening effect is absent in the spin current. The screening effect lowers the energy of the IQV state relative to that of the HQV state even in the absence of both the Fermi liquid and strong coupling corrections. Hence, it is naively expected that HQV states can be stabilized only in a mesoscopic system. The half-quantized flux has been experimentally observed in the superconductor Sr$_2$RuO$_4$ with a mesoscopic annular geometry~\cite{jiangSci11}, which is a candidate chiral $p$-wave superconductor. However, no firm experimental evidence for HQVs has been reported in any superconducting state.  Recently, Chung and Kivelson~\cite{chungPRB10} have proposed another scenario for stabilizing a lattice of HQVs. At finite temperatures, even if a lattice of IQVs can be favored at $T=0$, a gain in configurational entropy drives the fractionalization of IQVs into pairwise HQVs.

The half-quantum vortex is peculiar to spin-triplet chiral superfluids and superconductors. Among the possible chiral $p$-wave superfluids and superconductors, the superfluid $^3$He-A offers the best platform for studying the stability of HQVs, since it is the most established spin-triplet chiral superfluid with well-controllable parameters and geometry. The charge neutrality in $^3$He-A is also another advantage for realizing the HQV state because of the absence of the screening effect of the mass (charge) current in superconductors, as pointed out by Chung {\it et al.}~\cite{chungPRL07}.

\subsection{Braiding statistics of integer quantum vortex}
\label{sec:IQV}

In Sec.~\ref{sec:HQV}, we have demonstrated that although the HQV can be stabilized in rotating $^3$He-A by the strong magnetic field effect and the Fermi liquid corrections, the stability region turns out to be very narrow. The region is almost covered by the IQV. In contrast to the HQV, the IQV is accompanied by spinful Majorana fermions. Since a pair of Majorana fermions can be combined into a single Dirac fermion, the physics of the topological phase can be described explicitly without using Majorana fermions.  Recently, however, it has been demonstrated in Ref.~\citeonline{sato14} that IQVs can host topologically stable Majorana zero modes because of the mirror reflection symmetry. The IQVs with symmetry-protected Majorana fermions exhibit non-Abelian anyons.

We here consider the same situation as that described in Sec.~\ref{sec:HQV}, that is, the $^3$He-A confined in parallel plates. The magnetic field is applied along the $\hat{\bm h}$-axis, which can be tilted from the $\hat{\bm z}$-axis. Therefore, the $\hat{\bm d}$-vector does not necessarily lie in the $xy$-plane. Notice that the thermodynamically stable configuration of $\hat{\bm d}$ in $^3$He-A was discussed in Ref.~\citeonline{kawakamiJPSJ11}. In the limit of $H \!\gg\! H_d$, $\hat{\bm d}$ is always locked into the plane perpendicular to the applied field, while $\hat{\bm d}$ is polarized to the $\hat{\bm z}$-axis by the dipolar field for $H \!\ll\! H_d$. 


{\it Mirror-symmetry-protected topological phase.}---
The basic idea to circumvent the doubling of Majorana fermions in IQVs is the same as that in Secs.~\ref{sec:mirror} and \ref{sec:abmslab}. The topological property of an isolated IQV is determined by the semiclassical Hamiltonian $\mathcal{H}({\bm k},\varphi)$, where $\varphi$ is the azimuthal angle around the vortex line. The BdG Hamiltonian with a single IQV along the $\hat{\bm z}$-axis is invariant under the mirror reflection in the $xy$-plane, as in Eq.~\eqref{eq:Hmirror}, 
\beq
\left[
\mathcal{M}^{\eta}_{xy}, \mathcal{H}(k_x,k_y,k_z=0,\varphi)
\right] = 0,
\label{eq:Hmirrorv}
\eeq
if $\Delta({\bm k},\varphi)$ has a definite parity $\eta = \pm $
under the mirror reflection, as in Eq.~\eqref{eq:dmirror}. Similarly to the argument in Sec.~\ref{sec:abmslab}, for $^3$He-A in a thin film, the mirror symmetry is
preserved either when $\hat{\bm d}$ is parallel to the ${\bm z}$-direction
or when $\hat{\bm d}$ is normal to the ${\bm z}$-direction. 
The gap function in the former case has even mirror parity ($\eta=+$), while the
gap function in the latter case has odd mirror parity ($\eta=-$). 

When the BdG Hamiltonian maintains the mirror reflection symmetry in Eq.~\eqref{eq:Hmirrorv}, one can
introduce a novel topological number in each mirror subsector. Since each mirror subsector is regarded as
effectively ``{\it spinless}'' chiral $p$-wave systems with an isolated vortex, the relevant topological 
number is $\mathbb{Z}_2$, as in Sec.~\ref{sec:spinless}. It is clear that each mirror subsector 
has a single zero energy mode when the vorticity is odd. The IQV state always satisfies this condition.
In accordance with the generic argument in Sec.~\ref{sec:mirror}, only the mirror subsector for 
$\tilde{\cal M}_{xy}^{-}$ (${\bm d}\perp {\bm l}$) supports its own PHS, and its topological class 
is categorized to class D. In contrast, for $\tilde{\cal M}_{xy}^+$ (${\bm d}\parallel {\bm l}$), 
the mirror subsector does not maintain the PHS while the whole system does, corresponding into class A in each mirror subsector. 
Hence, only when ${\bm d}\perp {\bm l}$, does the mirror symmetry protect a single Majorana fermion in each mirror subsector. The Majorana nature may disappear when the $\hat{\bm d}$-vector is oriented from the $xy$-plane.



{\it Non-Abelian Braiding}.--- 
An immediate consequence of the Majorana zero modes is the non-Abelian
statistics.
For IQVs in a $^3$He-A thin film, each mirror subsector effectively
realizes a spinless system that supports a single Majorana zero mode in a vortex, 
and thus the zero modes obey the non-Abelian anyon statistics in the same manner as in Sec.~\ref{sec:nonabelian}.
Actually, the braiding statistics of IQVs is obtained by replacing the Majorana operator $\gamma _j$ bound to the $j$-th vortex with $\gamma ^{\lambda}_j$, which is the Majorana zero mode bound to the $j$-th vortex, in each mirror subsector labeled by $\lambda=\pm i$. We notice that no interference between the mirror subsectors
occurs during a vortex exchange process since the braiding operation does not break the mirror symmetry.
Therefore, even if we put the mirror subsectors together and consider
the entire system, the integer quantum vortices continue to obey the
non-Abelian anyon statistics~\cite{sato14}.

In contrast to a half-quantum vortex, IQVs in a rotating $^3$He-A thin film enable 
one to introduce a Dirac operator localized on a vortex core.
Indeed, since the IQV state supports pairwise Majorana zero modes at the single vortex core,
a Dirac operator $\psi_i$ localized in the $i$-th vortex can be defined as
\begin{eqnarray}
\psi_i=\frac{1}{2}(\gamma_i^{\lambda=i}+i\gamma_i^{\lambda=-i}),
\end{eqnarray}
which satisfies $\{\psi^{\dagger}_i,\psi_j\}=\delta_{ij}$.
As discussed by Yasui {\it et al.}~\cite{yasuiNPB12}, 
the Dirac operators give another expression for the non-Abelian exchange operator $\tau_i$ as
\begin{align}
\tau_i
=1+\psi_{i+1}\psi_i^\dagger
+\psi_{i+1}^{\dagger}\psi_i-\psi_i^\dagger\psi_i 
-\psi_{i+1}^\dagger\psi_{i+1}
+2\psi_{i+1}^{\dagger}\psi_{i+1}\psi_i^\dagger\psi_i. 
\end{align}
This is different from the expression for $\tau_i$ based on the Majorana fermions in Sec.~\ref{sec:nonabelian}. 

The above expression implies that the vortex exchange
process preserves the fermion number $N_{\rm
f}=\sum_i\psi_i^\dagger\psi_i$. 
We find that the conservation of the fermion number gives an alternative and simple
interpretation of the non-Abelian anyon statistics for integer quantum vortices:
For the Fock vacuum $|0\rangle$ of the Dirac operators, 
a vortex $i$ with the Dirac zero mode,
$\psi_i^{\dagger}|0\rangle\equiv|1\rangle$, has a nonzero fermion number,
while a vortex $i$ without the Dirac zero mode, $|0\rangle$, does not.
This means that we can distinguish these two vortex states, $|1\rangle$
and $|0\rangle$, by the fermion number.
Considering them as different particles, 
we obtain the non-Abelian anyon statistics naturally.
For example, let us consider the four-vortex state $|1100\rangle$ where
the first and second vortices are accompanied by the Dirac zero
modes, while the third and fourth are not.
Up to a phase factor, this state changes under $\tau_1$ and $\tau_2$
as
\begin{eqnarray}
|1100\rangle\stackrel{\tau_1}{\rightarrow} |1100\rangle
\stackrel{\tau_2}{\rightarrow}|1010\rangle,
\end{eqnarray}
while it changes under $\tau_2$ and $\tau_1$ as
\begin{eqnarray}
|1100\rangle\stackrel{\tau_2}{\rightarrow} |1010\rangle
\stackrel{\tau_1}{\rightarrow}|0110\rangle.
\end{eqnarray} 
Since $|1\rangle$ and $|0\rangle$ can be considered as different
particles, these final states are different from each other.
Therefore, we naturally obtain $\tau_2\tau_1\neq \tau_1\tau_2$.

In real systems, the mirror symmetry is easily broken locally by disorders
or ripples.
However, recent studies have suggested
that the symmetry protection is rather robust if the symmetry is
preserved macroscopically \cite{ringelPRB12, mongPRL12, fuPRL12, fulgaPRB14}.  
Indeed, we can argue that the non-Abelian anyon statistics persists if
the local breaking is weak and the mirror symmetry is preserved on average: 
Although the local breaking effects may locally lift the
degeneracy between two possible vortex states $|0\rangle$ and $|1\rangle$,
the degeneracy is recovered on average.
More importantly, because the fermion parity is preserved in
a superconductor/superfluid, no transition between $|0\rangle$ and $|1\rangle$
occurs unless a bulk quasiparticle is excited or cores of vortices are
overlapped.
Therefore, the above argument for the non-Abelian anyon braiding
works as long as the mirror symmetry is preserved on average.

\section{Topological Crystalline Superconductors}
\label{sec:materials}

The superconductivity of the heavy-fermion compound UPt$_3$ was discovered in 1984 by Stewart {\it et al.}~\cite{stewart} as one of the early trio of superconducting heavy-fermion compounds, UPt$_3$, UBe$_{13}$, and CeCu$_2$Si$_2$. As a superconducting counterpart to the superfluid $^3$He, the heavy-fermion superconductor UPt$_3$ has fascinated many physicists in the field of condensed matter. Recently, UPt$_3$ has come under the spotlight as a possible candidate topological {\it crystalline} superconductor~\cite{tsutsumiJPSJ13}. 

After the discovery of UPt$_3$, bulk unconventional superconductivity associated with spin-triplet or odd-parity pairing was reported for several materials. This includes Sr$_2$RuO$_4$~\cite{maenoRMP03}, the noncentrosymmetric superconductors CePt$_3$Si~\cite{bauer} and Li$_2$Pt$_3$B~\cite{nishiyama}, and superconducting doped topological insulators~\cite{sasakiPC15,ando15,andoJPSJ13}. The chiral $p$-wave pairing is the most plausible candidate of the gap function in Sr$_2$RuO$_4$, which is similar to that of the $^3$He-A thin film. Ueno {\it et al.}~\cite{ueno13} discussed the existence of mirror symmetry-protected Majorana fermions at the edges and dislocations. Experimental observations peculiar to the chiral pairing have been reported for Sr$_2$RuO$_4$, which include tunneling conductance characteristic to gapless edge states~\cite{kashiwayaPRL11} and half-quantum flux in a mesoscopic ring~\cite{budakian}. However, the full understanding of the gap function remains as a puzzling issue~\cite{machidaPRB,ishiharaPRB13,amanoPRB14,nakai,yanaseJPSJ}.

Apart from spin-triplet or odd parity pairing, spin-singlet superconductivity with the broken TRS has recently been an important topic in connection with Weyl superconductivity. The heavy-fermion compound URu$_2$Si$_2$ exhibits unconventional superconductivity at $T_{\rm c}\approx 1.5{\rm K}$ in the background of the so-called hidden order phase~\cite{okazakiPRL08}. Among possible pairings in a tetragonal symmetry, the field-angle-dependent specific heat measurements suggest the spin-singlet chiral $d$-wave state, $k_z(k_x\pm ik_y)$~\cite{yanoPRL08}. The broken TRS was observed by Schemm {\it et al.}~\cite{schemmPRB15} through the polar Kerr effect. Since the gap function in the $ab$-plane is similar to that of the ABM state, the topological aspect associated with Weyl superconductivity was discussed by Goswami and Balicas~\cite{goswami13}. Recently, Yamashita {\it et al.}~\cite{yamashitaNP15} have reported the observation of a colossal Nernst effect, where the transverse thermomagnetic response is enhanced by a factor of $\sim 10^{6}$. Sumiyoshi and Fujimoto~\cite{sumiyoshiPRB14} unveiled that the anomalously large thermomagnetic response is attributed to the asymmetric scattering due to chiral superconducting fluctuations. Similar asymmetric scattering due to a chiral order parameter was observed by Ikegami {\it et al.} as the intrinsic Magnus force acting on injected electrons in the surface of $^3$He-A~\cite{ikegami13}. 

Ishikawa {\it et al.}~\cite{ishikawaJPSJ13} proposed an {\it unconventional} spin-singlet pairing with broken TRS, which is called the cyclic $d$-wave pairing. The gap function is obtained from Eq.~\eqref{eq:generalD} as
$\psi ({\bm k}) \propto ( k^2_c + \omega _{\pm} k^2_a +\omega^2_{\pm}k^2_b)$, 
where $\omega _{\pm}\equiv e^{\pm i2\pi/3}$. This pairing was proposed as the most possible pairing symmetry of the superconducting phase of a spin-orbit-coupled material with the cubic $O_h$ lattice symmetry, e.g., PrOs$_4$Sb$_{12}$. The gap function is regarded as a combination of $d_{x^2-y^2}$ and $d_{3z^2-1}$ components and has $T_h$ symmetry with eight point nodes. Ishikawa {\it et al.}~\cite{ishikawaJPSJ13} observed that depending on the surface orientation, the gap function hosts the surface flat bands terminated to the point nodes. 

In this section, we briefly comment on the topological aspects of superconducting materials with a focus on UPt$_3$ and Cu$_x$Bi$_2$Se$_3$. Here, we focus our attention on the topological superconductivity of these materials and their physical consequences in connection with $^3$He. Comprehensive review papers on candidate materials of topological superconductors have been presented in Refs.~\citeonline{joynt}, \citeonline{sauls94}, and \citeonline{izawa} for UPt$_3$, Refs.~\citeonline{maenoRMP03} and \citeonline{maenoJPSJ12} for Sr$_2$RuO$_4$, and Refs.~\citeonline{sasakiPC15}, \citeonline{ando15}, and \citeonline{andoJPSJ13} for superconducting topological insulators. In addition, the topology of nodal superconductors was reviewed by Schnyder and Brydon~\cite{schnyderJPCM15} in connection with Weyl superconductors. 


\subsection{Heavy-fermion superconductor UPt$_3$}

The normal state of UPt$_3$ is qualitatively describable using the Fermi liquid theory with large effective masses and possesses a ``pseudo-spin'' degrees of freedom as a Kramers doublet. The normal state undergoes the double superconducting phase transitions at $T^{+}_{\rm c} \approx 550{\rm mK}$ into the A-phase and at $T^{-}_{\rm c}\approx 500{\rm mK}$ into the B-phase at zero field. In addition, the C-phase appears in the low-temperature and high-field region. The multiple phase transitions are a manifestation of unconventional superconductivity with multiple order parameters.  

The emergence of the double phase transitions is understandable with the splitting of nearly degenerate multiple order parameters due to a symmetry breaking field. The source of the symmetry breaking field is attributed to the antiferromagnetic order at $T_{\rm N}=5{\rm K}$, which lowers the crystalline symmetry $D_{6h}$. Two possible routes to understand the multiple phase transitions were examined in the direction of either the orbital scenario~\cite{sauls94,hess,choiPRL91,saulsJLTP94,choiPRB93,joynt88,tokuyasu} or spin scenario~\cite{machidaJPSJ89,machidaJPSJ89-2,machidaPRL91,machidaJPSJ92,ohmiPRL93,machidaJPSJ93,ohmiJPSJ96,machidaJPSJ99,tsutsumiJPSJ12-2}. 
Although there is little doubt on the unconventional superconductivity of UPt$_3$ nowadays, the pairing mechanism and the gap function have not yet been fully elucidated~\cite{joynt,sauls94,tsutsumiJPSJ12-2,izawa}. We will now provide an overview of the topological aspect of the two scenarios. 

\begin{figure}[t!]
\begin{center}
\includegraphics[width=80mm]{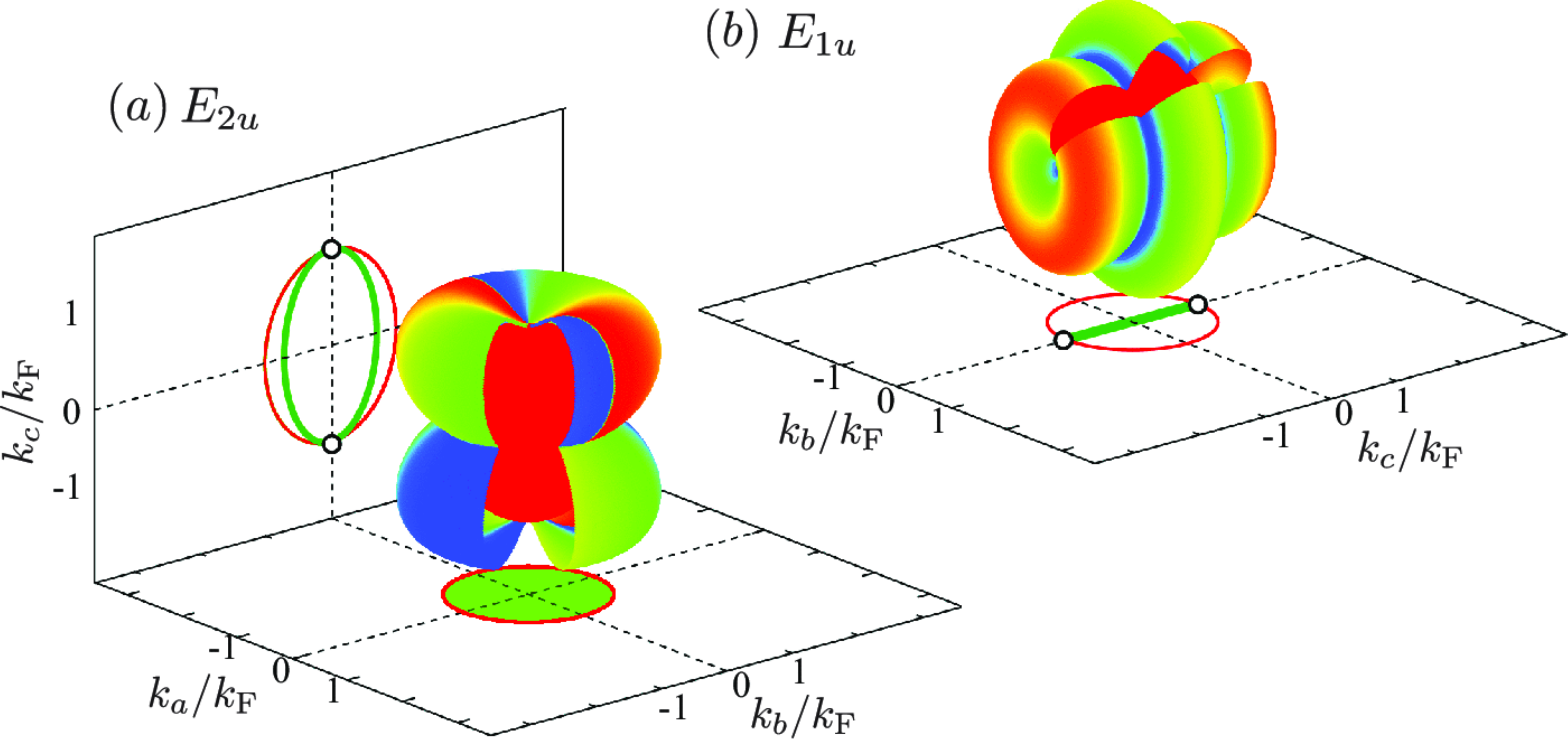}
\end{center}
\caption{(Color online) Gap functions of $E_{2u}$ (a) and $E_{1u}$ (b) and the gapless states bound to the surface. The thin solid curves denote the Fermi surface projected onto the surface and $k_a$-$k_b$ and $k_b$-$k_c$ are the momenta parallel to the surface. The shaded area (segment) in the $k_ak_b$ ($k_bk_c$)-plane depicts the momentum region where the zero energy states exist.
}
\label{fig:upt3}
\end{figure}

{\it Topology of the $E_{2u}$ state.}--- 
Among the candidates in the orbital scenario, the odd-parity pairing belonging to $E_{2u}$ was first proposed by Choi and Sauls~\cite{choiPRL91} phenomenologically to explain the anisotropic behavior of the upper critical field~\cite{shivaramPRL86}. The orbital scenario inevitably requires strong spin orbit coupling that forces the orientation of the ${\bm d}$-vector to the $c$-axis. The gap function is represented by the combination of two orbital parts~\cite{sauls94},
\beq
{\bm d}({\bm k}) = \hat{\bm c}\left[
\Delta _1(\hat{k}^2_a-\hat{k}^2_b) + \Delta _2 i\hat{k}_a\hat{k}_b\right]\hat{k}_c,
\label{eq:e2u}
\eeq
where the unit vectors $\hat{\bm a}$, $\hat{\bm b}$, and $\hat{\bm c}$ in $D_{6}$ denote the components of the ${\bm d}$-vector. The gap function is in good agreement with the thermodynamic and transport data in UPt$_3$~\cite{grafPRB00,taillefer,lussier,suderow,ellman}. In the most symmetric B-phase, the gap structure is accompanied by the line node on the equator of the three-dimensional Fermi surface and pairwise point nodes at the south and north poles [see Fig.~\ref{fig:upt3}(a)]. In the A- and C-phases, one of the orbital components vanishes and four line nodes connecting the north and south poles appear. The line nodes lower the continuous rotational symmetry in the $ab$-plane to fourfold rotational symmetry. 

The topological superconductivity of the $E_{2u}$ state with Eq.~\eqref{eq:e2u} can be inferred from the fact that the gap function is the hybridization of the chiral $d$-wave pairing $(\hat{k}_a+i\hat{k}_b)^2$ in the $ab$-plane and the polar pairing $\hat{k}_c$ along the $c$-axis. The point nodes on the north and south poles are protected by the nontrivial Chern number as Weyl points, analogous to the $^3$He-A. For the surface perpendicular to the $a$-axis, as shown in Fig.~\ref{fig:upt3}(a), the index theorem in Sec.~\ref{sec:index2} indicates that the zero energy states appear at the momentum $k_{b} = \sqrt{k^2_{\rm F}-k^2_{\rm c}}/\sqrt{2}$ in the case of a spherical Fermi surface. Hence, the Fermi arcs are terminated to two Weyl points. The polarlike gap function $\hat{k}_c$ is also responsible for the formation of the zero energy flatband in the surface perpendicular to the $c$-axis, as a consequence of the $\pi$-phase shift of the gap function. Hence, the $E_{2u}$ state might be accompanied by a pronounced zero bias conductance peak in point contact measurements. 

The topological invariants relevant to the $E_{2u}$ state for the surface states in the $k_bk_c$- and $k_ak_b$-planes are the first Chern number ${\rm Ch}_1(k_b)$ and $\mathbb{Z}_2$ number along the momentum path $k_c$ introduced in Eq.~\eqref{eq:z21d}, respectively. We notice that the normal electron state of UPt$_3$ is characterized by five Fermi surfaces, where three of them enclose the $\Gamma$ point in the Brillouin zone and the others are centered at the A point ${\bm k}=(0,0,\pm \pi/c)$~\cite{mcmullan}. Sato {\it et al.} revealed intimate relation between the topological number and the parity of Fermi surfaces~\cite{satoPRB09,satoPRB11}. This indicates that the $\mathbb{Z}_2$ number $\nu$ is nontrivial in the $E_{2u}$ state, which protects the zero energy {\it flatland} state in the surface. 

For time-reversal-breaking odd parity superconductors, however, line nodes are protected only by an additional discrete symmetry. In accordance with Blount's theorem generalized by Kobayashi {\it et al.}~\cite{kobayashiPRB14} in terms of the $K$-theory, the equatorial line node in the bulk $E_{2u}$ state is protected by spin rotation and mirror reflection symmetry in the $ab$-plane, which are elements of the crystalline symmetry $D_{6h}$. This indicates that although the line node is indispensable for the formation of the zero energy surface flatband, it is gapped out by any disturbances that break the symmetries. The discrete symmetries are indeed broken by the presence of a surface perpendicular to the $c$-axis and by an extra ``spin''-orbit coupling, which may give rise to the splitting of the zero energy density of states at the surface region. Hence, for the surface perpendicular to the $c$-axis, the zero bias conductance peak attributed to the equatorial line node is not topologically protected and might be fragile against disturbances. 



{\it Topology of the $E_{1u}$ state.}---
In the spin scenario for the multiple phase transitions observed in UPt$_3$, the gap function is composed of multiple ${\bm d}$-vectors. This requires the weak spin orbit coupling limit where the spins of Cooper pairs remain active. The key observations were provided by experiments involving Knight shift measurement~\cite{touPRL98} and angle-resolved thermal conductivity~\cite{machidaPRL12}. The former experiment showed that for ${\bm H}\parallel\hat{\bm c}$, the ${\bm d}$-vectors rotate from ${\bm d}_{\rm I}$ to ${\bm d}_{\rm II}$ at the critical field $H_{\rm rot}\sim 2$kG. The latter experiment clearly showed the twofold rotational symmetry in the $ab$-plane in the C-phase. Both observations cannot be explained by the $E_{2u}$ state and suggest a spin-triplet $f$-wavefunction in the $E_{1u}$ representation~\cite{machidaPRL12}. The $E_{1u}$ state was proposed in Ref.~\citeonline{tsutsumiJPSJ12} as 
\begin{gather}
{\bm d}_{\rm I}({\bm k}) =  (\Delta _1\hat{k}_a \hat{\bm b} + \Delta _2\hat{k}_b \hat{\bm c})(5\hat{k}^2_c-1), 
\label{eq:upt31} \\
{\bm d}_{\rm II}({\bm k}) = (\Delta _1\hat{k}_a \hat{\bm b} + \Delta _3\hat{k}_b \hat{\bm a})(5\hat{k}^2_c-1),
\label{eq:upt32}
\end{gather}
which is consistent with the observation described in Ref.~\citeonline{touPRL98}. For the B-phase where all the order parameter components are nonzero, the gap structure is accompanied by two horizontal line nodes on the northern and southern hemispheres and pairwise point nodes at the north and south poles [Fig.~\ref{fig:upt3}(b)]. For the A- and C-phases, one of the order parameter components vanishes, which lowers the rotational symmetry to the twofold rotational symmetry in the $ab$-plane. 

The $E_{1u}$ state preserving the TRS is composed of the planarlike pairing in the $ab$-plane and even-parity pairing along the $c$-axis. Hence, one can infer that the zero energy flatband connecting two point nodes appears on the surface perpendicular to the $ab$-plane. As discussed in Sec.~\ref{sec:mirrorchiral}, the zero energy flatband in the planar state is protected by the mirror chiral symmetry. Equations~(\ref{eq:upt31}) and (\ref{eq:upt32}) are invariant
under the mirror reflection ${M}_{ca}=i\sigma _b$ in the $ca$-plane which is an element of $D_{6h}$. Therefore, by combining ${M}_{ca}$ with $\mathcal{T}$ and $\mathcal{C}$, the $E_{1u}$ state holds the mirror chiral symmetry
\beq
\{\Gamma_1, 
\mathcal{H}(k_a,k_b=0,k_c)\}=0,
\label{eq:mirrorchiral}
\eeq 
with $\Gamma _1=\mathcal{T}\mathcal{C} {\mathcal{M}}^+_{ca}$ at $k_b=0$ or $\pi$~\cite{tsutsumiJPSJ13}. The mirror chiral symmetry enables us to define the one-dimensional winding number in Eq.~\eqref{eq:w1dmirror} for $k_b=0$ and $\pi$, which is evaluated as~\cite{tsutsumiJPSJ13}
\beq
|w(k_c)| =\left\{
\begin{array}{ll}
2 & \mbox{for $k_b=0$ and $|k_c| < k_{\rm F}$} \\
\\
0 & \mbox{for other $k_b$ and $k_c$}
\end{array}
\right. .
\label{eq:arc}
\eeq
Thus, the system is topologically nontrivial and the bulk-surface correspondence ensures the existence of the chiral-symmetry-protected Majorana fermions as discussed in Sec.~\ref{sec:chiral}. The remarkable consequence is the Majorana Ising anisotropy in the sense that the surface bound states are gapped only by a magnetic field along the $b$-axis. A magnetic field in the $ca$-plane or the ${\bm d}$-vector rotation in the high-field phase in the B-phase does not obscure the topological protection since the combination of the mirror reflection ${\mathcal {M}}_{ca}$ and the time reversal is not broken while each of them is not broken.

It was also pointed out in Refs.~\citeonline{tsutsumiJPSJ12-2} annd \citeonline{tsutsumiJPSJ13} that the $E_{1u}$ state possesses a nontrivial vortex phase diagram in the B-phase. This is attributed to the fact that in the $E_{1u}$ representation, three components of the ${\bm d}$-vectors, $\hat{k}_a\hat{\bm b}$, $\hat{k}_b\hat{\bm c}$, and $\hat{k}_b\hat{\bm a}$, are nearly degenerate in low magnetic fields. According to an NMR measurement~\cite{touPRL98}, the order parameter for a vortex state in low fields is composed of two ${\bm d}$-vectors with vorticity $\kappa =1$ as 
\beq
{\bm d}_{\rm I}({\bm k},{\bm r}) = \Delta({\bm r}) \left( 
\hat{k}_a\hat{\bm b}+ \hat{k}_b\hat{\bm c}
\right)(5\hat{k}^2_c-1)e^{i\kappa \phi},
\eeq
where the amplitude must vanish at the vortex center, $\Delta(|{\bm r}|\rightarrow 0)=0$. This is categorized into the $o$-vortex, which is the most symmetric vortex state preserving ${\rm U}(1)_{Q}\times P_2\times P_3$ as in Eq.~\eqref{eq:Hvortex}. In the weak spin-orbit coupling and weak fields, however, another component $\hat{k}_b\hat{\bm a}$ remains nearly degenerate and the core of the $o$-vortex can be occupied by
\beq
{\bm d}_{\rm core}({\bm k},{\bm r}) = \Delta _{\rm core}({\bm r})
\hat{k}_b\hat{\bm a}(5\hat{k}^2_c-1),
\eeq
where the boundary condition is imposed as $\Delta _{\rm core}(|{\bm r}|\rightarrow \infty)=0$. The vortex state, ${\bm d}_{\rm I}({\bm k},{\bm r})+{\bm d}_{\rm core}({\bm k},{\bm r})$, spontaneously breaks the ${\rm U}(1)_Q$ axisymmetric symmetry and $P_3$ symmetry, which is categorized to the nonaxisymmetric $v$-vortex in Sec.~\ref{sec:vortices}. The order parameter profile is also similar to that in $^3$He-B [see Fig.~\ref{fig:vortexbw}(d)]. Tsutsumi {\it et al.}~\cite{tsutsumiJPSJ12-2} demonstrated that with increasing magnetic field, the B-phase undergoes the vortex phase transition from the nonaxisymmetric $v$-vortex to the axisymmetric $o$-vortex, where the former (latter) holds $P_2$ (${\rm U}(1)_Q\times P_2\times P_3$) symmetry.

The low-lying quasiparticles bound to the vortices possess a similar topological aspect to those in the nonaxisymmetric $v$-vortex and the $o$-vortex in $^3$He-B. The nonaxisymmetric $v$-vortex is topologically trivial, while the quasiparticle bound to the $o$-vortex is characterized by the one-dimensional winding number as a consequence of the $P_3$ symmetry. In addition, in the same manner as the integer vortex in the $^3$He-A thin film, the vortex bound state changes the characteristic feature, depending on the orientation of ${\bm d}$. As long as the ${\bm d}$-vector is locked into the $ab$-plane, the mirror reflection symmetry ensures the PHS $\mathcal{C}^2=+1$ in each mirror subsector (see Sec.~\ref{sec:IQV}). Hence, with increasing applied field along the $c$-axis, the quasiparticle bound to the $o$-vortex changes from the Dirac fermion in the case of ${\bm d}_{\rm I}$ to the mirror-symmetry-protected Majorana fermion in ${\bm d}_{\rm II}$~\cite{tsutsumiJPSJ13}. These consequences are robust against the crystal field and spin-orbit coupling. 

{\it Discussion.}---
The puzzling issue concerning the gap function of UPt$_3$ has not yet been resolved. Most bulk thermodynamic and transport quantities are understandable with the $E_{1u}$ scenario as well as the $E_{2u}$ representation~\cite{grafPRB00,taillefer,lussier,suderow,ellman,tsutsumiJPSJ12-2}, while only the $E_{1u}$ state explains the angle-resolved thermal conductivity~\cite{machidaPRL12}. The Knight shift measurement~\cite{touPRL98} also supports the $E_{1u}$ state. On the other hand, a phase-sensitive experiment based on Josephson interferometry~\cite{strand,strand2} gave results consistent with the $E_{2u}$ state. Furthermore, recent experimental observation through the polar Kerr effect~\cite{schemm} indicates the breaking of the TRS. The $E_{2u}$ state spontaneously breaks the TRS, whereas the $E_{1u}$ state described in Eqs.~\eqref{eq:upt31} and \eqref{eq:upt32} maintains it. However, notice that since $E_{1u}$ is a two-dimensional representation, there is another possibility of the $E_{1u}$ pairing that spontaneously breaks the TRS.  

The characteristic structure of the surface states is a hallmark of the bulk topological property, which can be a probe for the gap functions. For instance, on the surface perpendicular to the $a$-axis, the $E_{2u}$ state has Fermi arcs protected by the first Chern number, while the mirror chiral symmetry in the $E_{1u}$ state ensures the formation of a topological Fermi arc as in Eq.~\eqref{eq:arc}. As a consequence of the chiral symmetry, the topological Fermi arc in the $E_{1u}$ state exhibits the anisotropic magnetic response: The zero energy density of states is unaffected by a magnetic field applied along the $a$- or $c$-axis, while a magnetic field along the $b$-axis breaks the mirror chiral symmetry and thus generates a finite energy gap. The flatband dispersion and Ising anisotropy peculiar to the $E_{1u}$ state give rise to a characteristic tunneling conductance in the low transparent limit. This is in contrast to that of the $E_{2u}$ state, where a magnetic field along the $b$-axis does not alter the surface states.

\subsection{Superconducting topological insulators}

The first observation of bulk superconductivity in carrier-doped topological insulators was reported by Hor {\it et al.}~\cite{horPRL10} for the carrier-doped bismuth compound Cu$_x$Bi$_2$Se$_3$. The parent material is a layered material consisting of stacked Se-Bi-Se-Bi-Se quintuple layers along the $c$-axis, which is well known as a prototype of topological insulators~\cite{andoJPSJ13,hasan}. The layers are weakly bounded by van der Waals forces. The narrow band gap and the relatively strong spin-orbit coupling give rise to the band inversion at the time-reversal-invariant momentum, the $\Gamma$ point, which is responsible for the nontrivial topology~\cite{fuPRL07}. The low-energy Hamiltonian is describable with two $p_z$ orbitals in the quintuple layer as
\beq
\mathcal{H}_{\rm TI}({\bm k}) = c({\bm k}) + m({\bm k})\sigma _x + v_z k_z \sigma _y
+ v ({\bm k}\times{\bm s})_z \sigma _z,
\label{eq:hfinal}
\eeq
where $c ({\bm k}) = c_0 + c_1 k^2_z+c_2(k^2_x+k^2_y)$ and $m ({\bm k}) = m_0 + m_1 k^2_z + m_2(k^2_x+k^2_y)$. Here, $s_{\mu}$ and $\sigma _{\mu}$ denote the Pauli matrices in the spin and $p_z$ orbital spaces, respectively. The Hamiltonian (\ref{eq:hfinal}) captures the low-energy band structure of various topological materials around the $\Gamma$ point, including Bi$_2$Se$_3$, Bi$_2$Te$_3$, and Sb$_2$Te$_3$~\cite{zhangNP09,liuPRB10}, and also around the $L$ point of the topological crystalline insulator SnTe~\cite{mitchellPR66,hsiehNC12}. The lowest-order correction to Eq.~(\ref{eq:hfinal}) is obtained as $(k^3_++k^3_-)\sigma _zs_z$ with $k_{\pm}\!\equiv\!k_x\pm ik_y$, whose role in topological insulators and Cu$_x$Bi$_2$Se$_3$ was emphasized by Fu~\cite{warping,fu14}.

Hor {\it et al.}~\cite{horPRL10} succeeded in the doping of the electron carrier density ($\sim\!10^{20}{\rm cm}^{-3}$) by intercalating Cu atoms into the van der Waals gap between quintuple layers. The intercalation does not break the original crystalline symmetry, i.e., the rhombohedral symmetry $D_{ 3d}$. The carrier-doped Bi$_2$Se$_3$ undergoes a superconducting phase transition at $T=3.8 {\rm K}$ for the optimal doping rate of $0.12<x<0.15$. Kriener {\it et al.}~\cite{krienrPRB11} developed the electrochemical intercalation technique that improves the shielding fraction up to $50\%$. The temperature dependence of the specific measurement clearly indicates the full gap behavior~\cite{krienerPRL11}, while the magnetic field dependence of the magnetization measured by Das {\it et al.}~\cite{das} cannot be explained by the scenario of conventional type-II superconductors and suggests the spin-triplet scenario of Cu$_x$Bi$_2$Se$_3$. 

{\it Gap functions.}---
A sufficient condition for realizing topological superconductivity in doped topological materials is odd parity pairing that satisfies
\beq
P\Delta ({\bm k})P^{\dag} = - \Delta (-{\bm k}),
\eeq
where the inversion operator is given as $P=\sigma _x$ for the two-orbital model. Contrary to conventional wisdom, odd-parity pairing can be realized even in an $s$-wave channel of Cooper pairs, as a spin-triplet $s$-wave orbital-singlet state, which can be favored by phonon-mediated pairing mechanism~\cite{brydonPRB14}. The possible gap functions in Cu$_x$Bi$_2$Se$_3$ were proposed by Fu and Berg~\cite{fuPRL10}, which are listed in Table~\ref{table:sti}. As clarified by Sato~\cite{satoPRB10} and Fu and Berg~\cite{fuPRL10}, all the odd-parity pairings satisfy a sufficient criterion for realizing topological superconductivity when the number of Fermi surfaces enclosing time-reversal-invariant momenta is odd. The fully gapped $A_{1u}$ state is time-reversal-invariant odd parity pairing categorized into class DIII, whose relevant topological invariant is the $\mathbb{Z}$ number, $w_{\rm 3d}$, analogous to the BW state in $^3$He. The nodal $A_{2u}$ and $E_u$ states possess similar topological superconductivity to that of the planar state and UPt$_3$-B. As clarified in Sec.~\ref{sec:planar}, although the $\mathbb{Z}$ topological number is ill-defined in nodal superconductors, the parity of $w_{\rm 3d}$ well describes their topological properties without any ambiguity. In addition to the bulk $\mathbb{Z}_2$ number, the nodal state holds the chiral mirror symmetry, as in Eq.~\eqref{eq:mirrorchiral}. This is responsible for the formation of a topological Fermi arc connecting the pairwise point nodes~\cite{sasakiPC15}.

\begin{table}
\begin{tabular}{c|c|c|c|c|c}
\hline\hline
$\Delta$ & gap & parity  & mirror & topo. & $\Gamma$ \\
\hline
$\Delta_{1a}$, $\Delta _{1b}\sigma _x$ & full & even & $+$ & -- & $A_{1g}$ \\
$\Delta_{2} \sigma_y s_z$ & full & odd & $-$ & $\mathbb{Z}$ & $A_{1u}$ \\
$\Delta_{3}\sigma_z$  & point & odd & $+$ & $\mathbb{Z}_2$ & $A_{2u}$ \\
$\Delta_{4x}\sigma_y s_{x}$ & point &  odd & $+$ & $\mathbb{Z}_2$ & $E_u$\\
$\Delta_{4y}\sigma_y s_{y}$ & full &  odd & $-$ & $\mathbb{Z}$ & $E_u$\\
\hline\hline
\end{tabular}
\caption{Pairing potentials in superconducting topological insulators, gap structures, and their parity under inversion ${P}$ and mirror reflection in the $yz$-plane. $\Gamma$ denotes the representation of the $D_{3d}$ symmetry. 
}
\label{table:sti}
\end{table}

Apart from ${\bm k}$-independent pairing, momentum-dependent pairing was examined in Refs.~\citeonline{haoPRB14} and \citeonline{chenJPCM13}. Recently, first-principles linear-response calculations predicted that a $p$-wave-like state can be favored by a conventional phonon-mediated mechanism~\cite{wan14}.

We notice that the $E_u$ state, which is the two-dimensional representation of $D_{3d},$ may have either nodal or full gap excitation. In accordance with Blount's theorem~\cite{blount,kobayashiPRB14}, the point nodes in the case of time-reversal-invariant odd parity superconductors (DIII) are protected only by mirror reflection symmetry. Although the effective Hamiltonian in Eq.~\eqref{eq:hfinal} preserves the accidentally enlarged symmetry ${\rm SO}(2)_{J_z}$, the hexagonal warping correction $(k^3_++k^3_-)\sigma _zs_z$ recovers the original crystalline symmetry, where only three mirror planes remain~\cite{fu14}. The nodal direction in the $ab$-plane can be tilted from the mirror-invariant plane by an appropriate combination of two representations $\sigma _ys_x$ and $\sigma _ys_y$, which results in a fully gapped state.


The specific heat measurement can be explained by both even- and odd-parity pairings, except for the $A_{2u}$ state~\cite{hashimotoJPSJ13,hashimoto14}. The more direct way of identifying the gap function is offered by surface-sensitive experiments, such as point contact measurements. Sasaki {\it et al.}~\cite{sasakiPRL11} first reported the pronounced zero-bias conductance peak in the soft point contact measurement on a naturally cleaved $(111)$ surface of Cu$_x$Bi$_2$Se$_3$ ($x\simeq 0.3$). A similar zero-bias conductance peak was also observed independently for $x\simeq 0.2$~\cite{kirzhnerPRB12}. The observations described in Refs.~\citeonline{sasakiPRL11} and \citeonline{kirzhnerPRB12} can be attributed to nontrivial topological odd-parity superconductivity. Most recently, however, conflicting experimental observations by the conductance and tunneling spectroscopy have been reported by Levy {\it et al.}~\cite{levyPRL13} and Peng~{\it et al.}~\cite{pengPRB13}. As shown in Ref.~\citeonline{levyPRL13} they observed a simple U-shaped scanning tunneling microscope spectrum on the (111) surface. This naively leads to a contrary statement that this material has a conventional $s$-wave (even-parity) pairing symmetry, i.e., a nontopological $A_{1g}$ state.

\begin{figure}[t!]
\begin{center}
\includegraphics[width=80mm]{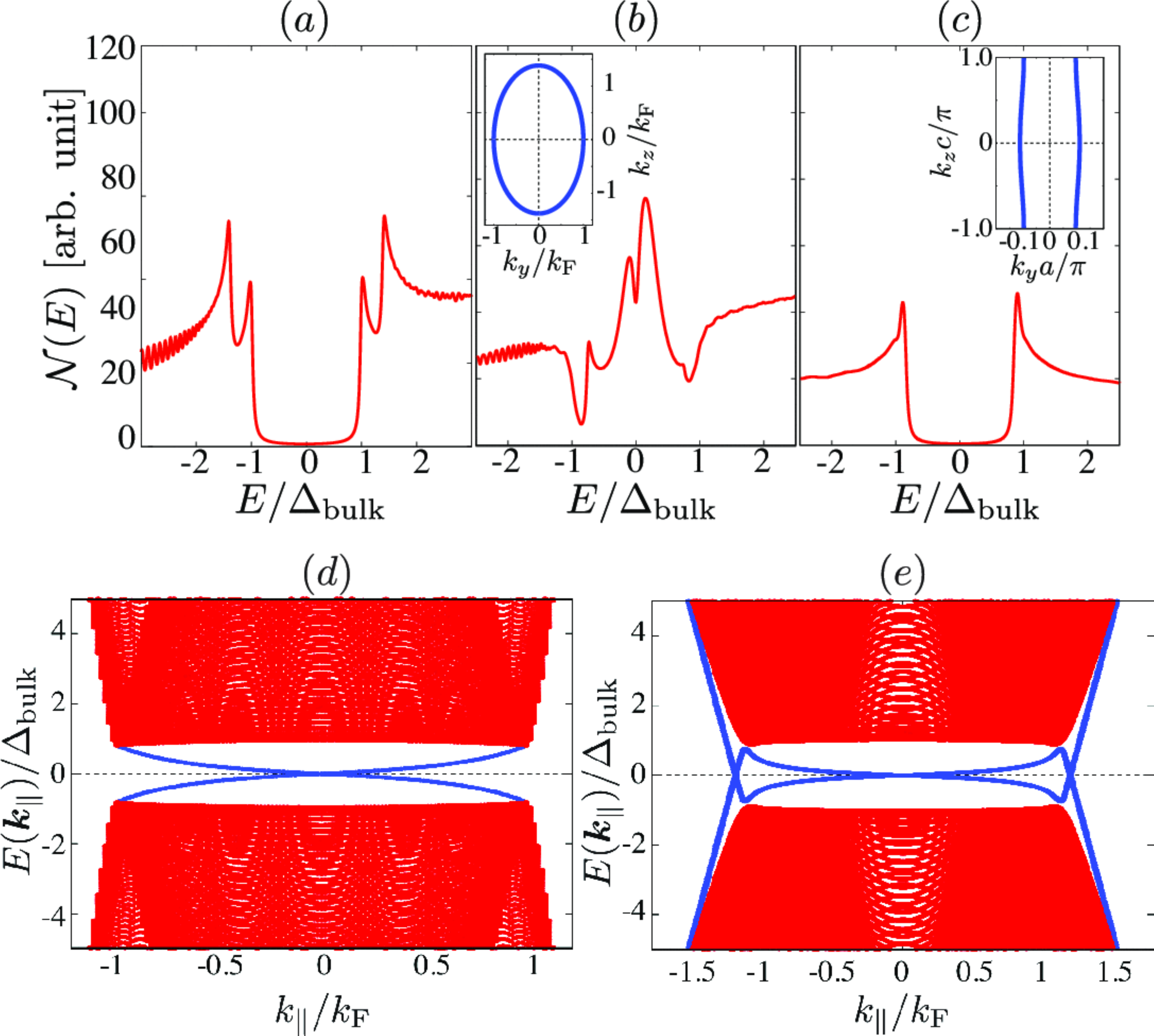}
\end{center}
\caption{(Color online) (a) Surface density of states for the bulk $s$-wave pairing ($A_{1g}$) at $(\tilde{m}_1,\tilde{m}_2)=(-0.17,-0.20)$.  Surface density of states for the $A_{1u}$ state with a spheroidal Fermi surface (b) and a cylindrical shape (c). The insets in (b) and (c) show the Fermi surface, where $a$ and $c$ are the lattice constants. Energy spectra of the $A_{1u}$ state:  ``Cone'' shape of the surface bound states (blue curve) for $(m_1m_0/v^2_z, m_2m_0/v^2_z) = (-0.59, -0.20)$ (d) and ``caldera'' shape for $(-0.17, -0.20)$ (e).}
\label{fig:fig1}
\end{figure}

{\it Surface states in even- and odd-parity pairings.}---
In Ref.~\citeonline{mizushimaPRB14-2}, Mizushima {\it et al.} demonstrated that, if the bulk of Cu$_x$Bi$_2$Se$_3$ shows a conventional $s$-wave pairing ($A_{1g}$), the interplay between the bulk superconductivity and the surface Dirac fermions, which are remnants of the parent material, gives rise to unconventional surface structures. When the surface Dirac cone is well separated from the bulk conduction band at the Fermi level, the anomalous parity mixing of odd-parity $A_{2u}$ pairing is mediated by the orbitally polarized Dirac fermions in the surface region, opening an additional surface gap larger than the bulk one. As a result, the resulting surface density of states hosts an extra coherent peak at the induced gap besides a conventional peak at the bulk gap [Fig.~\ref{fig:fig1}(a)]. Such anomalous parity mixing and a double-coherence peak structure are not observed in bulk odd-parity ($A_{1u}$ or $E_u$) states.


Another plausible scenario is offered by fully gapped odd-parity $A_{1u}$ or $E_u$ states. As mentioned above, the fully gapped $A_{1u}$ and $E_u$ states are characterized by the nontrivial $\mathbb{Z}$ topological number when the $\Gamma$ point is enclosed by an odd number of Fermi surfaces. The bulk-surface correspondence indicates the existence of gapless helical Majorana fermions. As mentioned in Sec.~\ref{sec:MF3HeB}, fully gapped topological states, such as the BW-like state, are expected to be accompanied by gapless surface bound states having a simple linear dispersion, $E_{\rm surf}({\bm k}_{\parallel})\propto |{\bm k}_{\parallel}|$, which is called the Majorana ``cone''. Such a Majorana cone in the BW-like state is known to exhibit a double-peak structure in the tunneling conductance rather than a zero-bias conductance peak~\cite{asano03}.

Contrary to conventional wisdom on the BW-like state, however, Hsieh and Fu~\cite{hsiehPRL12} and Yamakage~{\it et al.}~\cite{yamakagePRB12} have recently pointed out that the dispersion of gapless surface states twists at finite momenta $k_{\parallel} \!\equiv\! \sqrt{k^2_x+k^2_y}$. The velocity of the Majorana cone near ${\bm k}_{\parallel} \!=\! {\bm 0}$, $\tilde{v}_{\rm surf}$, is given with the parameters in $\mathcal{H}_{\rm TI}({\bm k})$ as~\cite{yamakagePRB12,hsiehPRL12}
\beq
\tilde{v}_{\rm surf} = \frac{1-\sqrt{1+4\tilde{m}_1+4\tilde{m}^2_1\tilde{\mu}^2}}{2\tilde{m}_1\tilde{\mu}^2} \frac{v\Delta}{m_0},
\label{eq:vel}
\eeq
where $\tilde{m}_1\!\equiv\!m_0 m_1 /v^2_z$. At the critical velocity that satisfies $\tilde{v}_{\rm surf} \!=\! 0$, the topological surface state undergoes a structural transition from a ``cone'' [Fig.~\ref{fig:fig1}(a)] to a ``caldera'' shape [Fig.~\ref{fig:fig1}(b)]~\cite{yamakagePRB12}. The twisting of the surface states turns out to be a consequence of the interplay between the helical Majorana fermions and the well-defined Dirac fermions at the Fermi level. The caldera shape of the surface states has a large zero-energy density of states, and thus the surface density of states has a clear peak structure, as shown in Fig.~\ref{fig:fig1}(b). 

The topological superconductivity and the existence of zero energy density of states on the surface are sensitive to the shape of the Fermi surface. For a cylindrical Fermi surface that does not enclose the $\Gamma$ point, the surface state is no longer topologically protected and thus the resultant surface density of states on the (111) surface becomes a simple U-shaped form [Fig.~\ref{fig:fig1}(c)]. This feature is commonly applicable to the $A_{1u}$ and $E_{u}$ states. 

{\it Discussion.}---
As mentioned above, there are two possible scenarios for the gap function of Cu$_x$Bi$_2$Se$_3$: Bulk conventional $s$-wave pairing ($A_{1g}$) and bulk odd parity pairing ($E_u$ or $A_{1u}$). The former is accompanied by anomalous surface structure as a consequence of the interplay between bulk superconductivity and surface Dirac fermions, giving rise to the double coherence peak structure in the surface density of states. In contrast, when the Dirac cone is well defined in the normal state, the bulk odd-parity superconductivity hosts topologically protected Majorana fermions with a twisted dispersion, which is responsible for the pronounced zero energy peak in the surface density of states. Notice that the recent ARPES measurement performed by Wray {\it et al.}~\cite{wray} and Tanaka {\it et al.}~\cite{tanakaPRB12} showed that the surface Dirac fermions are well separated at the Fermi level in Cu$_x$Bi$_2$Se$_3$. This strongly indicates that the even-parity ($A_{1g}$) scenario can be reasonably ruled out. In addition, the analysis based on the Shubnikov-de Haas measurement reported the Fermi surface evolution from a spheroidal to a cylindrical shape around the $10^{20}{\rm cm}^{-3}$ carrier density~\cite{Lahoud}. On the basis of the Fermi surface evolution, therefore, all the surface measurements~\cite{sasakiPRL11,kirzhnerPRB12,levyPRL13,pengPRB13} can be explained by the bulk odd-parity pairing ($E_u$ or $A_{1u}$)~\cite{mizushimaPRB14,sasakiPC15}.

Kriener {\it et al.}~\cite{krienerPRL11} reported that Cu$_x$Bi$_2$Se$_3$ has a short mean free path, which is comparable to the superconducting coherence length, $l\sim \xi$. Similarly to a conventional $s$-wave pairing, the $A_{1g}$ state is highly insensitive to disorder~\cite{Anderson1959}, while {\it unconventional} superconducting states have been considered to be fragile against disorders. This may suggest that unconventional superconductivity can be realized only in clean materials. In contrast to conventional wisdom, Michaeli and Fu~\cite{michaeliPRL12} have recently predicted that the odd-parity $A_{1u}$ state remains robust even in a dirty material. For a low carrier density, which is relevant to Cu$_x$Bi$_2$Se$_3$, an approximate chiral symmetry prevents the superconducting compound from pair decoherence induced by impurity scattering. The robustness of both fully gapped odd-parity $A_{1u}$ and nodal $E_u$ states was demonstrated by employing a self-consistent $T$-matrix~\cite{nagaiPRB14,nagaiPRB15}. Hence, the bulk odd-parity pairing may be robust against nonmagnetic disorders. In addition, Bay {\it et al.}~\cite{bayPRL12} have recently reported that the bulk superconductivity of Cu$_x$Bi$_2$Se$_3$ under high pressures survives up to $6.3{\rm GPa}$. A high-pressure single-crystal study in Ref.~\citeonline{bayPRL12} indicates the relatively long mean-free path $l > \xi$.

Most recently, Liu {\it et al.}~\cite{liu15} have succeeded in the intercalation of Sr atoms into the topological insulator Bi$_2$Se$_3$. The new member of superconducting doped topological insulators Sr$_x$Bi$_2$Se$_3$ possesses an upper critical field similar to that of Cu$_x$Bi$_2$Se$_3$~\cite{sbs}, while it has a high shielding fraction $\sim 90 \%$. The mean free path is estimated to be $l\sim 2\xi$. The carrier density is on the order of $10^{19}{\rm cm}^{-3}$, which forms a spheroidal Fermi surface enclosing the $\Gamma$ point. On the basis of these observations, we expect that a pronounced zero-bias conductance peak can be observed in the (111) surface for the bulk odd parity $A_{1u}$ or $E_u$ state, while the double coherence peaks appear in the surface density of states if the bulk is in the conventional $A_{1g}$ state. Hence, Sr$_x$Bi$_2$Se$_3$ may provide a new playground for studying bulk topological superconductivity based on topological insulators.

The pronounced zero-bias conductance peak was also observed in Sn$_{1-x}$In$_x$Te, which is a superconducting material based on the topological crystalline insulator SnTe~\cite{sasakiPRL12}. Recent measurements, namely, specific heat measurements~\cite{erickson,balakrishnan,novakPRB13} and muon spin spectroscopy~\cite{saghirPRB14}, show that the superconducting state of Sn$_{1-x}$In$_x$Te is likely to be fully gapped. The zero-bias conductance peak and bulk superconductivity indicates the realization of a topological superconducting state in the cubic phase.

\section{Summary and Prospects}
\label{sec:summary}

We have reviewed the current status of our knowledge on symmetry-protected topological superfluids and topological crystalline superconductors with the main focus on $^3$He. The B-phase of $^3$He is a symmetry-protected topological superfluid, while the A-phase is a superfluid counterpart to Weyl superconductors. Furthermore, the planar and polar phases, which can be competitive in a restricted geometry, possess symmetry-protected point and line nodes, respectively, which are responsible for the formation of a zero-energy flatband on the surface. We have emphasized that all these phases in $^3$He can be prototypes of topological crystalline superconductors and Weyl superconductors, such as UPt$_3$, Sr$_2$RuO$_4$, URu$_2$Si$_2$, and superconducting doped topological insulators. The knowledge established for $^3$He could also be useful for determining the topological aspect of the $^3P_2$ superfluid core of neutron stars~\cite{tamagaki,hoffberg}.

\begin{figure}[t!]
\includegraphics[width=80mm]{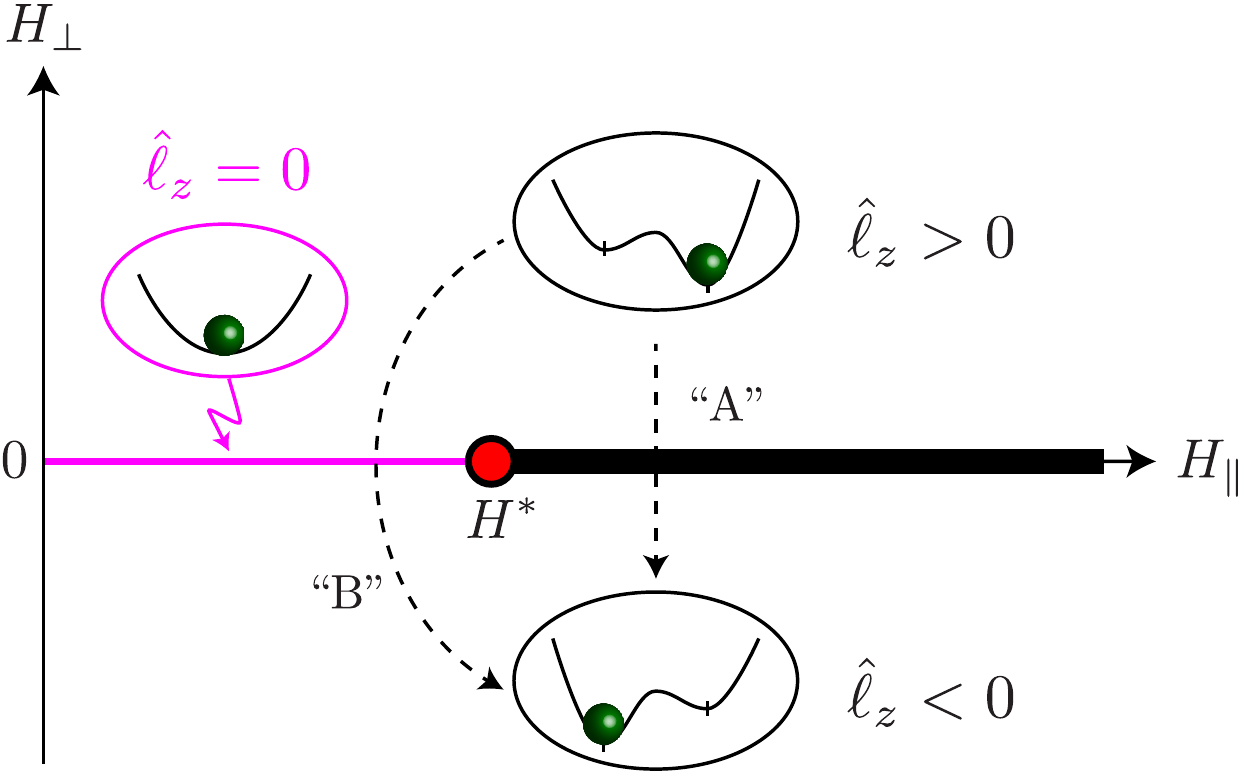}
\caption{(Color online) Phase diagram of $^3$He-B in a slab geometry under a $P_3$ symmetric magnetic field ($H_{\parallel}$) and $P_3$ symmetry breaking field ($H_{\perp}$), where $H_{\parallel}$ ($H_{\perp}$) denotes the magnetic field parallel (perpendicular) to the surface. The thin (blue) line corresponds to the symmetry-protected topological B$_{\rm I}$-phase, while the thick line is the nontopological B$_{\rm II}$-phase with breaking the $P_3$ symmetry. }
\label{fig:phaselz}
\end{figure}

Here, we would like to emphasize a new perspective on the superfluid $^3$He-B under a magnetic field. This is a unique system in the sense that the topological phase transition concomitant with spontaneous symmetry breaking takes place at the critical field $H^{\ast}$. For $H_{\perp}=0$, the B-phase is further subdivided into the symmetry-protected topological $B_{\rm I}$-phase for $H_{\parallel}<H^{\ast}$ and the nontopological $B_{\rm II}$-phase while breaking the symmetry for $H_{\parallel}>H^{\ast}$, where $H_{\parallel}$ and $H_{\perp}$ denote a magnetic field parallel and perpendicular to the surface, respectively. The phase transition is characterized by the order parameter $\hat{\ell}_z$ associated with the spontaneous breaking of the $P_3$ symmetry, and the phase diagram is illustrated in Fig.~\ref{fig:phaselz} in the plane of $H_{\parallel}$ and $H_{\perp}$. This has similarity to that of an Ising model~\cite{chaikin}: Along the path ``A'' in Fig.~\ref{fig:phaselz}, the ordered phase $\hat{\ell}_z>0$ undergoes the first-order transition at $H_{\perp}=0$, and there exists the endpoint at $H_{\parallel}=H^{\ast}$ and $H_{\perp} = 0$. For $H_{\parallel}>H^{\ast}$ and $H_{\perp}=0$, i.e., the B$_{\rm II}$-phase, the $\hat{\ell}_z>0$ ordered state is energetically degenerate with $\hat{\ell}_z<0$ and they are connected by the $P_3$ operation. In contrast to the Ising model, however, the two ordered phases are not smoothly connected to each other along the path ``B'', where the segment within $0<H_{\parallel}<H^{\ast}$ corresponds to the topological $B_{\rm I}$-phase, while the others correspond to the nontopological phase. 

Hence, $^3$He-B allows us to explore a new type of quantum critical phenomenon at $H^{\ast}$, which is associated with the intertwining of the topological phase transition and spontaneous symmetry breaking. Grover {\it et al.}~\cite{grover} predicted that supersymmetry emerges at the quantum critical point $H^{\ast}$ without enforcing the conditions microscopically. However, there exist various unresolved problems on quantum critical behaviors and their hallmarks. Uncovering new physics associated with topological quantum critical phenomena remains as an important challenge in the future. 

Detecting Majorana fermions is a pressing issue shared by the extensive field of condensed matter physics. Superfluid $^3$He serves as an ideal system for tackling this issue, since knowledge on its bulk superfluidity has been well established and a rich variety of restricted geometries can be fabricated where different topological classes are energetically competing. Recent transverse acoustic impedance measurements with an well-controlled surface condition succeeded in the spectroscopy of surface bound states, which captured the surface density of states peculiar to the Majorana cone~\cite{okuda}. The characteristics of the Majorana cone were also observed in the heat capacity measurement by Bunkov {\it et al.}~\cite{bunkov,bunkov1}. In addition, SQUID-NMR experiments reported the observation of the $\hat{\ell}_z$ domain wall~\cite{levitinPRL13}. These experimental observations are the first step to further capturing a hallmark of chiral and helical Majorana fermions. In addition, there exists another type of low-lying excitations in superfluid $^3$He, massless and massive bosonic modes. The bosonic modes associated with spontaneous symmetry breaking are well known to give rise to anomalous transport properties~\cite{vollhardt}. However, understanding their intertwining effects with surface and vortex Majorana fermions remains a a central unresolved problem, which might provide keys to extracting a hallmark of such exotic quasiparticles in topological superfluids and superconductors. 

Apart from the topological aspect, the A-phase restricted to a slab hosts the spontaneous mass current flowing along the edge, which is a direct signature of spontaneous time-reversal symmetry breaking. Although the total angular momentum that generates the edge mass current is on the order of the macroscopic value $\sim \hbar N/2$, direct observations have not yet been accomplished. Understanding the spontaneous edge current might be a milestone towards the resolution of a longstanding issue on the intrinsic angular momentum. 


The superclean quantum fluid, superfluid $^3$He, shares fundamental and essential concepts with extensive fields of physics and serves as a rich repository of novel quantum phenomena. We hope that the current review paper provides a gateway to the deep world woven by the intertwining of symmetries and topologies. 



\begin{flushleft}
{\bf Acknowledgments}\\
\end{flushleft}

We gratefully thank S. Higashitani, R. Nomura, H. Ikegami, O. Ishikawa, J. A. Sauls, J. Saunders, K. Shiozaki, Y. Tanaka, A. B. Vorontsov, A. Yamakage, K. Kono, K. Nagai, Y. Okuda, S. Sasaki, S. Fujimoto, W. P. Halperin, M. Nitta, Y. Nagato, A. Furusaki, T. Morimoto, and Y. Tada for various fruitful discussions. This work was supported by JSPS (Nos.~25800199, 25287085, 25103716, 26400360, 15K17715, and 15J05698) and ``Topological Quantum Phenomena'' (No.~22103005) and ``Topological Materials Science'' (No.~15H05855) KAKENHI on innovation areas from MEXT. T.K. is also grateful for support from the WPI Initiative on Material nanoarchitectonics, MEXT, Japan.



\appendix

\section{Basic Theories for Superconductors and Superfluids}
\label{sec:gorkov}

We here summarize the basic theories for superfluids and superconductors. We start by considering the system composed of $N$-component fermions that interact through the potential $\mathcal{V}^{c,d}_{a,b}$, where the single-particle Hamiltonian density is given by $\varepsilon _{ab} (-i{\bm \nabla})$. The quantum field operators in the Nambu space are denoted by ${\bm \Psi} \!=\! (\psi _{1}, \cdots, \psi _{N}, \psi^{\dag}_{1}, \cdots, \psi^{\dag}_{N})^{\rm T}$, where the field operators satisfy $\{ \psi _{a}({\bm r}_1), \psi^{\dag}_{b}({\bm r}_2)\} \!=\! \delta _{a,b}\delta({\bm r}_1-{\bm r}_2)$ 
and 
$\{ \psi _{a}({\bm r}_1), \psi _{b}({\bm r}_2)\} = \{ \psi^{\dag}_{a}({\bm r}_1), \psi^{\dag}_{b}({\bm r}_2) \} \!=\! 0$.  

\subsection{Gor'kov theory: Bogoliubov-de Gennes equation}

The equilibrium properties of superfluid phases at a finite temperature $T$ are governed by the Gor'kov equation as~\cite{serene,baym1,baym2}
\beq
\int dx_3
\left[
{G}^{-1}_0(x_1,x_3)
-{\Sigma}(x_1,x_3)
\right] {G}(x_3,x_2) = \delta (x_1-x_2).
\label{eq:NG}
\eeq
The Matsubara Green's functions are defined as ${G}(x_1,x_2) \!\equiv\! - \langle\langle T_{\tau} [{\bm \Psi}(x_1)\bar{\bm \Psi}(x_2)] \rangle\rangle$, which can be parameterized as
\begin{align}
{G}(x_1,x_2) =
\left(
\begin{array}{cc}
\mathcal{G}(x_1,x_2) & \mathcal{F}(x_1,x_2) \\ \bar{\mathcal{F}}(x_1,x_2) & \bar{\mathcal{G}}(x_1,x_2)
\end{array}
\right).
\label{eq:G}
\end{align}
We have introduced $x_j \!\equiv\! (\tau_j, {\bm r}_j)$, $\bar{\bm \Psi} ({\bm r},\tau) = {\bm \Psi}^{\dag}({\bm r},-\tau)$, and $\langle\langle\cdots\rangle\rangle \!\equiv\! {\rm Tr}[e^{(\Omega-\mathcal{H})/T}\cdots]$ with the thermodynamic potential $\Omega$. 
The bare Green's function ${G}^{-1}_0$ in Eq.~(\ref{eq:NG}) is given as
${G}^{-1}_0(x_1,x_2) \equiv \delta (x_{12})[ - \partial _{\tau_2} -
{\varepsilon}(-i{\bm \nabla}_2)]$, where ${\varepsilon} \equiv {\rm diag}(\varepsilon,-\varepsilon^{\ast})$.

The self-energy ${\Sigma}$ is determined by the $\Phi$-functional as ${\Sigma} [{G}] = 2 \delta \Phi[{G}]/\delta {G}^{\rm T}$, which generates the perturbation expansion for the skeleton self-energy diagrams. The thermodynamics is obtained from the Luttinger-Ward thermodynamic potential $\Omega[{G}]$~\cite{luttinger},
\beq
\Omega [{G}] = -\frac{1}{2}
{\rm Sp}\left[ \ln(-{G}^{-1}_0+{\Sigma})+ {\Sigma} {G} \right] + \Phi[{G}],
\label{eq:LW}
\eeq
where ${\rm Sp} \cdots \equiv \int dx_1\int dx_2 {\rm Tr} \cdots$. The pair potential is obtained from the gap equation
\beq
\Delta _{ab} ({\bm r}_1,{\bm r}_2)=-T\sum _n \mathcal{V}^{cd}_{ab}({\bm r}_{12})
\mathcal{F}_{dc}({\bm r}_2,{\bm r}_1;\omega _n).
\label{eq:gapf}
\eeq
Below, we expand ${G}$ using the Matsubara frequency $\omega _n = (2n+1)\pi T$ as ${G}(x_1,x_2) = T\sum _n {G}({\bm r}_1,{\bm r}_2;\omega _n)e^{-i\omega _n \tau _{12}}$.

Now, we show that the Gor'kov equation is reduced to the
BdG equation,
\begin{eqnarray}
\int d{\bm r}_2 {\mathcal{H}}({\bm r}_1,{\bm r}_2)
{\bm \varphi}_{i}({\bm r}_2)
= E_{i} {\bm \varphi}_{i}({\bm r}_1),
\label{eq:bdg3}
\end{eqnarray}
where ${\mathcal{H}}$ is a $2N\!\times\!2N$ matrix in Nambu space, ${\mathcal{H}}({\bm r}_1,{\bm r}_2) = \delta ({\bm r}_{12}){\varepsilon}(-i{\bm \nabla}_2) + {\Sigma} ({\bm r}_1,{\bm r}_2)$. 
The $N$-component eigenvector ${\bm \varphi}_{i}({\bm r})$ fulfills the orthonormal condition
$\int {\bm \varphi}^{\dag}_{i}({\bm r}){\bm \varphi}_{j}({\bm r}) d{\bm
r} = \delta _{i,j}$.  
The PHS ensures that the
positive energy solution ${\bm \varphi}_{E}({\bm r})$ is
associated with the negative energy solution,
\beq
{\bm \varphi}_{-E}({\bm r}) = \mathcal{C}{\bm \varphi}_{E}({\bm r}).
\eeq  
Therefore, the following $2N\times 2N$ unitary matrix
$\underline{u}_{i} \!\equiv\! [{\bm \varphi}^{(1)}_{i}, \cdots
{\bm \varphi}^{(N)}_{i}, 
\mathcal{C}{\bm\varphi}^{(1)}_{i}, \cdots 
\mathcal{C}{\bm \varphi}^{(N)}_{i}] 
$ diagonalizes the BdG Hamiltonian as
$ \int d{\bm r}_1 \int d{\bm r}_2\underline{u}^{\dag}_{i}({\bm r}_1) 
{\mathcal{H}}({\bm r}_1,{\bm r}_2)
\underline{u}_{i}({\bm r}_2) \!=\! \underline{E}_{i}
$, 
with
$\underline{E}_{i} \!\equiv\! {\rm diag}( 
E^{(1)}_{i}, \cdots, E^{(N)}_{i}, -E^{(1)}_{i},\cdots, -E^{(N)}_{i})$. 
The unitary matrix $\underline{u}_{i}({\bm r})$ satisfies the orthonormal
and completeness conditions 
$\int \underline{u}^{\dag}_{i}({\bm r}) \underline{u}_{j}({\bm r})
d{\bm r} \!=\! \delta _{i,j}$ and  
$\sum _{i} \underline{u}_{i}({\bm r}_1) \underline{u}^{\dag}_{i}({\bm r}_2) \!=\! \delta ({\bm r}_{12})$. 
By using the unitary matrix $\underline{u}_{i}$, the solution of the Gor'kov equation
(\ref{eq:NG}) is obtained as
\beq
\underline{G}({\bm r}_1,{\bm r}_2; \omega _n) = \sum _{i} \underline{u}_{i}({\bm r}_1) 
\left( i\omega _n  - \underline{E}_{i} 
\right)^{-1} \underline{u}^{\dag}_{i}({\bm r}_2).
\label{eq:G2}
\eeq
The BdG equation Eq.~(\ref{eq:bdg3}) coupled to the relevant gap equation offers a self-consistent framework for describing superfluid phases in equilibrium. 

\subsection{Quasiclassical theory}

The quasiclassical theory is a natural extension of Landau's Fermi liquid theory to superfluid and superconducting phases. This theory covers a vast scope of systems within the weak coupling regime $k_{\rm F}\xi\gg 1$. This theory also offers a tractable way of studying the microscopic structure of spatially inhomogeneous superconductors and superfluids.

The central object of the theory is the propagator that contains both quasiparticles and superfluidity on an equal footing. The $2N\times 2N$ matrix form of the quasiclassical propagator, ${g} \!\equiv\! {g}(\hat{\bm k},{\bm r};\omega _n)$, is obtained from ${G}$ by integrating ${G}$ over the shell $v_{\rm F}|k-k_{\rm F}| < E_{\rm c} \ll E_{\rm F}$~\cite{serene},
\beq
{g}(\hat{\bm k},{\bm r};\omega _n) = \frac{1}{a} \int^{+E_{\rm c}}_{-E_{\rm c}} d\xi _{\bm k}
{\tau}_z{G}({\bm k},{\bm r};\omega _n),
\eeq
where ${G}({\bm k},{\bm r};\omega _n) 
= \int d{\bm r}_{12} e^{-i{\bm k}\cdot{\bm r}_{12}} {G}({\bm r}_1,{\bm r}_2;\omega _n) $. The normalization constant $a$ corresponds to the weight of the quasiparticle pole in the spectral function. 

The quasiclassical propagator ${g}\equiv{g}(\hat{\bm k},{\bm r};\omega _n)$ is governed by the transport-like equation~\cite{serene,eilenberger,larkin1,eliashberg,larkin2,larkin3}. Following the procedure in Ref.~\citeonline{serene}, one obtains the quasiclassical transport equation from the Nambu-Gor'kov equation (\ref{eq:NG}) as
\beq
\left[i\omega _n {\tau}_z - \underline{v}(\hat{\bm k},{\bm r})
- \underline{\Delta}(\hat{\bm k},{\bm r}), 
{g}
\right] =- i {\bm v}_{\rm F} \!\cdot{\bm \nabla}
{g}.
\label{eq:eilen}
\eeq
The Fermi velocity is defined as ${\bm v}_{\rm F}(\hat{\bm k}) \!=\! \partial \varepsilon _0({\bm k})/\partial {\bm k}|_{{\bm k}=k_{\rm F}\hat{\bm k}}$. The self-energy term in Eq.~(\ref{eq:NG}) is replaced with ${\tau}_z{\Sigma} ({\bm k},{\bm r}) \approx{\tau}_z{\Sigma} (k_{\rm F}\hat{\bm k},{\bm r}) = [\underline{v} (\hat{\bm k},{\bm r}) + \underline{\Delta}(\hat{\bm k},{\bm r})]/a$. The term $ \underline{v}$ in Eq.~(\ref{eq:eilen}) consists of the external potential $\underline{v}_{\rm ext}$ and the quasiclassical self-energy $\underline{\nu}$ associated with Fermi liquid corrections, as $\underline{v}(\hat{\bm k},{\bm r}) = \underline{v}_{\rm ext}({\bm r}) + \underline{\nu}(\hat{\bm k},{\bm r})$, where 
$
\underline{\nu} = 
{\rm diag}(
\nu _{0} + {\sigma}_{\mu} \nu _{\mu}, \bar{\nu}_{0} + {\sigma}^{\rm T}_{\mu} \bar{\nu}_{\mu}
)
$.
The off-diagonal component of the quasiclassical self-energies is obtained from Eq.~\eqref{eq:gapf} as
\beq
\underline{\Delta}(\hat{\bm k},{\bm r}) 
= \left(
\begin{array}{cc}
0 & \Delta(\hat{\bm k},{\bm r}) \\
\Delta^{\dag}(-\hat{\bm k},{\bm r}) & 0
\end{array}
\right). 
\label{eq:underdelta}
\eeq

The quasiclassical transport equation (\ref{eq:eilen}) is a first-order ordinary differential equation along a trajectory in the direction of ${\bm v}_{\rm F}(\hat{\bm k})$. The solution of the quasiclassical transport equation (\ref{eq:eilen}) is not uniquely determined per se because $a+bg$ satisfies the same equation as $g$ ($a$ and $b$ are arbitrary constants). To obtain a unique solution for $g$, Eq.~(\ref{eq:eilen}) must be supplemented by the normalization condition on the quasiclassical propagator as~\cite{eilenberger,larkin3,shelankov2}
\beq
\left[ {g} (\hat{\bm k},{\bm r};\omega _n)\right]^2 = -\pi^2.
\label{eq:norm}
\eeq 
It is obvious that since $g^2$ is the solution of the quasiclassical transport equation (\ref{eq:eilen}), it can be parameterized as $g^2=a+bg$. In accordance with the direct calculation of Eq.~(\ref{eq:eilen}) for spatially uniform systems, the arbitrary constants $a$ and $b$ are found to be $a = -\pi^2$ and $b=0$. The general solutions for nonuniform systems should be determined without any contradiction to uniform solutions. The normalization condition (\ref{eq:norm}) was proven by Shelankov~\cite{shelankov2} in a more direct manner. 

The quasiclassical propagator ${g}$, which is a $4\times 4$ matrix in particle-hole and spin spaces, is parameterized with spin Pauli matrices $\sigma _{\mu}$ as
\beq
{g} = \left(
\begin{array}{cc}
g_{0} + {\sigma}_{\mu} g_{\mu} & i\sigma _y f_0 + i {\sigma}_{\mu} {\sigma}_y f_{\mu}  \\ 
i\sigma _y \bar{f}_0 +i \sigma _y {\sigma}_{\mu}\bar{f}_{\mu}  & \bar{g}_{0} + {\sigma}^{\rm T}_{\mu} \bar{g}_{\mu}
\end{array}
\right).
\label{eq:g}
\eeq
Here, $\sigma^{\rm T}_{\mu}$ denotes the transpose of the Pauli matrices $\sigma _{\mu}$. The off-diagonal propagators are composed of spin-singlet and -triplet Cooper pair amplitudes, $f_0$ and $f_{\mu}$. 
The quasiclassical propagators must satisfy the following relations arising from the Fermi statistics in Eq.~(\ref{eq:G}):
\beq
\left[ {g}(\hat{\bm k},{\bm r};\omega_n)\right]^{\dag} 
= {\tau}_z{g}(\hat{\bm k},{\bm r};-\omega_n) {\tau}_z
\label{eq:sym1} , \\
\left[ {g}(\hat{\bm k},{\bm r};\omega_n)\right]^{\rm T} 
= {\tau}_y{g}(-\hat{\bm k},{\bm r};-\omega_n) {\tau}_y.
\label{eq:sym2}
\eeq
From the normalization condition in Eq.~(\ref{eq:norm}), one obtains $gf = - f\bar{g}$ and $\bar{g}\bar{f} = - \bar{f}g$, leading to the relation between $\bar{g}_0$ and $g_0$ of
\beq
\bar{g}_0(\hat{\bm k},{\bm r};\omega _n) = -g_0(\hat{\bm k},{\bm r};\omega _n).
\label{eq:trs2}
\eeq 
The diagonal propagator with the analytic continuation $i\omega _n \rightarrow E+i0_+$ defines the ${\bm k}$-resolved local density of states, 
\beq
\mathcal{N}(\hat{\bm k},{\bm r};E) = -\frac{\mathcal{N}_{\rm F}}{\pi}{\rm Im}g_0(\hat{\bm k},{\bm r};\omega _n \rightarrow -iE+0_+),
\label{eq:dosk}
\eeq
where $\mathcal{N}_{\rm F}$ denotes the density of states in the normal state at the Fermi level.

The complete set of the self-consistent quasiclassical theory is composed of the transport equation (\ref{eq:eilen}) with the normalization condition in Eq.~(\ref{eq:norm}) in addition to the Fermi liquid corrections and the gap equation. The diagonal self-energies that generate the Fermi liquid corrections are determined as
\beq
\nu _0 (\hat{\bm k},{\bm r}) = \sum _{\ell} A^{({\rm s})}_{\ell}\left\langle
P_{\ell}(\hat{\bm k}\cdot\hat{\bm k}^{\prime})g_0(\hat{\bm k}^{\prime},{\bm r};\omega _n)
\right\rangle _{\hat{\bm k}^{\prime},n}, \label{eq:nu0} \\
{\bm \nu} (\hat{\bm k},{\bm r}) = \sum _{\ell} A^{({\rm a})}_{\ell}\left\langle
P_{\ell}(\hat{\bm k}\cdot\hat{\bm k}^{\prime}){\bm g}(\hat{\bm k}^{\prime},{\bm r};\omega _n)
\right\rangle _{\hat{\bm k}^{\prime},n}, \label{eq:nu}
\eeq
where $P_{\ell}(x)$ are the Legendre polynomials with $\ell = 0, 1, 2, \cdots$. The gap function is determined by anomalous propagators as
\beq
\Delta _{ab} (\hat{\bm k},{\bm r}) = \left\langle V^{cd}_{ab}(\hat{\bm k},\hat{\bm k}^{\prime})
\left[i\sigma _{\mu}\sigma _y f_{\mu}(\hat{\bm k}^{\prime},{\bm r};\omega _n)\right]_{cd} \right\rangle _{\hat{\bm k}^{\prime},n}.
\label{eq:gap_qct}
\eeq
The coefficients $A^{({\rm s})}_{\ell}$ and $A^{({\rm a})}_{\ell}$ are symmetric and antisymmetric quasiparticle scattering amplitudes, respectively, which are parameterized with Landau's Fermi liquid parameters $F^{({\rm s},{\rm a})}_{\ell}$ as 
$A^{({\rm s, a})}_{\ell} = F^{({\rm s,a})}_{\ell}/[1+F^{({\rm s,a})}_{\ell}/(2\ell+1)]$. $F^{\rm s}_{\ell =1}$ and $F^{\rm a}_{\ell = 0}$ give Fermi liquid corrections to the effective mass and spin susceptibility, respectively. In this paper, we use the following abbreviation for the Matsubara sum and the average over the Fermi surface: 
$\langle\cdots\rangle _{\hat{\bm k},n} = \frac{T}{\mathcal{N}_{\rm F}}\sum _{n}
\int \frac{d\hat{\bm k}}{(2\pi)^3 |{\bm v}_{\rm F}(\hat{\bm k})|}\cdots$, 
where $\mathcal{N}_{\rm F} \!=\! \int \frac{d\hat{\bm k}}{(2\pi)^3|{\bm v}_{\rm F}(\hat{\bm k})|}$ is the total density of states at the Fermi surface in the normal state.

Even for spatially uniform $\underline{\Delta}$ and $\underline{\nu}$, the quasiclassical transport equation (\ref{eq:eilen}) is generally accompanied by the solutions that exponentially grow and decay along trajectories in the direction of $\hat{\bm k}$. These exploding solutions are not normalizable in spatially uniform superconductors and superfluids, which make Eq.~(\ref{eq:eilen}) numerically unstable. Thuneberg {\it et al}.~\cite{thuneberg} proposed a numerically accessible method using the explosion trick, where the physical solution is constructed from the commutation relation of two exploding solutions. An alternative scheme for solving the transport equation (\ref{eq:eilen}) is based on the projection operator found by Shelankov~\cite{shelankov1,shelankov2}. This scheme was established by Eschrig {\it et al}.~\cite{eschrigPRB99,eschrigPRB00}. The projection operator maps the transport equation (\ref{eq:eilen}) having exploding solutions onto a Riccati-type differential equation that does not have non-normalizable solutions and is numerically stable. We also notice that the Riccati-type differential equation can be derived directly from the BdG equation (\ref{eq:bdg3}) within the Andreev approximation~\cite{nagatoJLTP93,nagaiJPSJ08}.

\bibliographystyle{jpsj}
\bibliography{topology}

\end{document}